\documentclass[aps,prx,onecolumn,notitlepage,superscriptaddress,showpacs,floatfix,10pt]{revtex4-2}
\usepackage{physics}
\usepackage{amsmath,amsthm,amssymb,amsfonts, color, comment}
\usepackage[caption=false]{subfig}
\usepackage{graphicx}
\usepackage{mathrsfs}
\usepackage{xcolor}
\usepackage{float}
\usepackage{dcolumn}
\usepackage{bm}
\usepackage{graphicx} 
\usepackage{hyperref}
\hypersetup{colorlinks=true, citecolor=blue, urlcolor=blue, linkcolor=blue}
\usepackage{pgffor}
\usepackage{pdfpages}
\makeatletter
\AtBeginDocument{\let\LS@rot\@undefined}
\makeatother
\UseRawInputEncoding

\usepackage{qcircuit}
\usepackage{adjustbox}
\usepackage{cancel}
\usepackage{upgreek}

\usepackage{tikz}

\usepackage[normalem]{ulem}
\usepackage{complexity}
\newcommand{\tc}[2]{\textcolor{#1}{#2}}

\newcommand{\added}[1]{\tc{blue}{#1}}

\def\be{\begin{eqnarray}}
\def\ee{\end{eqnarray}}

\newcommand{\bea}{\begin{eqnarray}}
\newcommand{\eea}{\end{eqnarray}}

\newcommand{\bp}{\mathbf{p}}

\newcommand{\btjstrw}{\mathrel{{\rotatebox[origin=c]{90}
{$\bowtie$}}\kern-0.18em\raisebox{-.95ex}{$\bullet$}
\kern-0.5em\raisebox{.97ex}{$\bullet$}
\kern-1.12em\raisebox{.97ex}{$\bullet$}
\kern-0.52em\raisebox{-.95ex}{$\bullet$}}}

\newcommand{\btjnbrR}{{\mathrel{\rotatebox[origin=c]{90}
{$\bowtie$}}\kern-0.22em\raisebox{.9ex}{$\bullet$}
\kern-1.em\raisebox{-.8ex}{$\bullet$}}}
\newcommand{\btjnbrL}{{\mathrel{\rotatebox[origin=c]{90}
{$\bowtie$}}\kern-0.22em\raisebox{-.8ex}{$\bullet$}
\kern-1.em\raisebox{+.9ex}{$\bullet$}}}

\def\x{\xi}

\newtheorem{theorem}{Theorem}
\newtheorem{corollary}[theorem]{Corollary}
\newtheorem{lemma}[theorem]{Lemma}

\newtheorem{claim}[theorem]{Claim}
\theoremstyle{remark}
\newcommand{\diag}{\text{diag}}
\newcommand{\id}{\mathbb{I}}
\newcommand{\Z}{\text{Z}}

\newcommand{\prep}{\text{PREP}}
\newcommand{\sel}{\text{SELECT}}
\newcommand{\nconst}{\mathcal{A}}
\newcommand{\vect}[1]{\mathbf{#1}}

\newcommand{\llangle}[1][]{\savebox{\@brx}{\(\m@th{#1\langle}\)}%
  \mathopen{\copy\@brx\kern-0.5\wd\@brx\usebox{\@brx}}}
\newcommand{\rrangle}[1][]{\savebox{\@brx}{\(\m@th{#1\rangle}\)}%
  \mathclose{\copy\@brx\kern-0.5\wd\@brx\usebox{\@brx}}}

\newcommand{\bx}{\mathbf{x}}

\makeatletter
\@ifundefined{commentsOn}{%
\newcommand{\egr}[1]{\textcolor{brown}{}}
\newcommand{\remarkRed}[1]{}
}{

\newcommand{\egr}[1]{\textcolor{brown}{[From Eleanor: #1]}}
}
\makeatother

\newcommand{\cmp}{\texttt{CMP}}

\newtheorem{conjecture}[theorem]{Conjecture}

\newtheorem{proposition}[theorem]{Proposition}
\begin{document}

\preprint{APS/123-QED}

\title{Scattering Processes from Quantum Simulation Algorithms for Scalar Field Theories}
 \author{Andrew Hardy}
 \email{andrew.hardy@mail.utoronto.ca}
 \affiliation{Department of Physics, University of Toronto, Toronto, ON M5S 1A7 , Canada}
\author{Priyanka Mukhopadhyay}
\thanks{These authors made equivalent contributions to this work}
\affiliation{Department of Computer Science, University of Toronto, Toronto, ON, M5S 2E4, Canada}
 \author{M. Sohaib Alam}
 \thanks{These authors made equivalent contributions to this work}
\affiliation{Quantum Artificial Intelligence Laboratory (QuAIL), NASA Ames Research Center, Moffett Field, CA, 94035, USA}
\affiliation{USRA Research Institute for Advanced Computer Science (RIACS), Mountain View, CA, 94043, USA}
\author{Robert Konik}
\affiliation{Condensed Matter and Materials Science Division, Brookhaven National Laboratory, Upton, NY 11973, USA}
\author{Layla Hormozi}
\affiliation{Computational Science Initiative, Brookhaven National Laboratory, Upton, NY 11973, USA}

\author{Eleanor Rieffel}
\affiliation{Quantum Artificial Intelligence Laboratory (QuAIL), NASA Ames Research Center, Moffett Field, CA, 94035, USA}
 \author{Stuart Hadfield} 
\affiliation{Quantum Artificial Intelligence Laboratory (QuAIL), NASA Ames Research Center, Moffett Field, CA, 94035, USA}
\affiliation{USRA Research Institute for Advanced Computer Science (RIACS), Mountain View, CA, 94043, USA}
\author{Jo\~{a}o Barata}
\affiliation{Physics Department, Brookhaven National Laboratory, Upton, NY 11973, USA}
\author{Raju Venugopalan}
\affiliation{Physics Department, Brookhaven National Laboratory, Upton, NY 11973, USA}

\author{Dmitri E. Kharzeev}
\affiliation{Center for Nuclear Theory, Department of Physics and Astronomy, Stony Brook University, Stony Brook, NY 11794-3800, USA}
\affiliation{Physics Department, Brookhaven National Laboratory, Upton, NY 11973, USA}
\author{Nathan Wiebe}
\email{nawiebe@cs.toronto.edu}
\affiliation{Department of Computer Science, University of Toronto, Toronto, ON, M5S 2E4, Canada}
\affiliation{Pacific Northwest National Laboratory, Richland, WA, 99354, USA}
\affiliation{Canadian Institute for Advanced Studies, Toronto, ON, M5G 1M1, Canada}

\begin{abstract}

We provide practical simulation methods for scalar field theories on a quantum computer that yield improved asymptotics as well as concrete gate estimates for the simulation and physical qubit estimates using the surface code. 
We achieve these improvements through two optimizations. 
First, we consider a finite volume approach for estimating the elements of  the S-matrix. 
This approach is appropriate in general for 1+1D and for certain low-energy elastic collisions in higher dimensions. 
Second, we implement our approach using a series of different fault-tolerant simulation algorithms for Hamiltonians formulated both in the field occupation basis and field amplitude basis. 
Our algorithms are based on either second-order Trotterization or qubitization. 
The cost of Trotterization in occupation basis scales as $O(\lambda N^7 |\Omega|^3/(M^{5/2} \epsilon^{3/2}))$ where $\lambda$ is the coupling strength, $N$ is the occupation cutoff, $|\Omega|$ is the volume of the spatial lattice, $M$ is the mass of the particles and $\epsilon$ is the uncertainty in the energy calculation used for the $S$-matrix determination. 
Qubitization in the field basis scales as $O(|\Omega|^2 (k^2 \Lambda +kM^2)/\epsilon)$ where $k$ is the cutoff in the field and $\Lambda$ is a scaled coupling constant.  
We find in both cases that the bounds suggest physically meaningful simulations can be performed using on the order of $4\times 10^6$ physical qubits and $10^{12}$ $T$-gates which corresponds to roughly one day on a superconducting quantum computer with surface code and a cycle time of 100 ns. This places the simulation of scalar field theory within striking distance of the gate counts for the best available chemistry simulation results.
\end{abstract}

\maketitle

\tableofcontents
\section{Introduction}

\label{sec:intro}
Quantum simulation has been one of the great success stories of quantum computing~\cite{lloyd1996universal,berry2007efficient,childs2012hamiltonian,childs2017quantum,Low2019hamiltonian}. It has led to the realization that non-relativistic quantum mechanics can, under most physically reasonable assumptions, be simulated in polynomial time on a quantum computer. 
Recent work in quantum chemistry has shown that physically meaningful problems can be simulated using fewer than a million physical qubits and a few hours worth of compute time \cite{2021_SBWetal}.  
In contrast, there have been numerous classical methods developed to simulate quantum field theories (QFT), with the state of the art approaches summarized in \cite{Bauer2022}.
However, even sign-problem free approaches relying on Hamiltonian representations \cite{Kogut1975, Drell1979}, require resources that can be  exponential in number  \cite{jordan2018}.
This opens up the possibility that quantum computing may provide the only means possible for us to numerically simulate some of the most challenging problems in quantum field theory \cite{klcoDigitizationScalarFields2019,alexeev2021quantum,bauer2023quantum,nguyen2022digital,shaw2020quantum}. \par 

The seminal work of Jordan, Lee and Preskill (JLP)~\cite{Jordan2012,jordan2014} provided a major step towards addressing the question of whether quantum computers can efficiently  simulate scattering events for scalar $\phi^4$ theory. In addition to describing aspects of the dynamics and couplings of the Higgs boson in the standard model, $\phi^4$ theory is an excellent model field theory that captures nontrivial features of the dynamics of the other fundamental quantum fields in the standard model. This is true even in 1+1D where classical $\phi^4$ theory has nontrivial bound states described by kinks and anti-kinks. The theory also contains stationary topological solitons that are periodic in time, known as ``breather"-modes \cite{Kevrekidis_2019}, 
providing insight into the nonlinear dynamics of domains walls in fields ranging from condensed matter to cosmology \cite{Kevrekidis_2019}. Further, at the very high energies now accessible at colliders,  as exemplified in the parton model of Feynman and Bjorken~\cite{Feynman_1972 ,Bjorken:1969ja}, scattering processes involving fermions or bosons become alike. Thus, the study of the $\phi^4$ scalar field theory, at high energies, provides some insight into the dynamics of more phenomenologically rich theories, which are harder to directly simulate.

Subsequent work has provided a number of optimized methods for simulating various aspects of $\phi^4$ theory 
\cite{klcoDigitizationScalarFields2019,barata2021single,Li:2022ped,kreshchuk2023simulatingscatteringcompositeparticles,Bagherimehrab:2021xlp,farrelly2020discretizingquantumfieldtheories,Klco:2019xro,Yeter-Aydeniz:2018mix}. We note that there has also been considerable progress for various other field theories \cite{Davoudi_2023,rhodes2024,hariprakash2024,Bauer:2021gup}. Despite these advances, we do not yet know the degree to which physical parameters of quantum field theories can be efficiently extracted, nor are there at present detailed estimates of the memory and  number of  quantum operations required to perform quantum simulations of field theories. 

The computation of in-vacuum to the out-vacuum transitions embedded in the $S$ matrix is, for $1+1$D, promise-\BQP\, complete \cite{Jordan2012}.  This means that if an arbitrary quantum computation can be thought of as such a scattering experiment, and  if a classical computer could efficiently compute these matrix elements, then quantum computers would be no more powerful than probabilistic classical computers. In the language of complexity theory, \BQP = \BPP. As the computation of arbitrary elements of the $S$-matrix of a QFT is exponentially difficult, we will frame the problem in the 
manner detailed below. 

\subsection{Our Contributions} 
We provide a cost-analysis of fault tolerant calculation of scattering matrix elements using L\"uscher-finite volume methods -- see \cite{Luscher1991,annurev,rusetsky2019particleslattice,Mai2021,multihadroninteractionslatticeqcd,Gabai2022,Bajnok2016,James_2018,PhysRevD.83.114508,PhysRevD.83.071504,PhysRevD.98.014507,Garofalo_2021} for a representative set of references describing these methods for both $1+1$D and higher dimensional field theories. 
With such methods, information about eigenenergies and matrix elements computed in finite volume can be converted into information about infinite volume S-matrix elements.
Classically, the needed eigenenergies and matrix elements can be accessed  both via truncated spectrum Hamiltonian methods \cite{yurov1990truncated,yurov1991truncated,PhysRevD.91.025005,James_2018,10.21468/SciPostPhys.13.2.011,chen2022,PhysRevD.102.065001,Liu2020}, a Hamiltonian based method applicable to both $1+1$D quantum field theories \cite{yurov1990truncated,yurov1991truncated,James_2018} as well as quantum field theories in higher dimensions \cite{PhysRevD.91.025005,10.21468/SciPostPhys.13.2.011,chen2022,PhysRevD.102.065001,Liu2020}, and via lattice quantum Monte Carlo \cite{annurev,rusetsky2019particleslattice,Mai2021,multihadroninteractionslatticeqcd,PhysRevD.83.114508,PhysRevD.83.071504,Garofalo_2021}.  \par
While finite volume methods have been used classically to compute scattering cross sections, the asymptotic classical complexity is not generally well established. The necessary resolution of excited states may scale exponentially either in the truncation energy of the Hilbert space or in the size of the lattice. 
In addition, there is the difficulty due to ill-posed analytic continuation to arrive at excited state information \cite{Jarrell1996, Sandvik1998, beach2004}. Precision measurements then of such low energy features as well as measurements of quantities associated with inelastic scattering information become more prohibitive with classical schemes.
Here, we extend these methods to the quantum domain, and carry out a resource analysis for computing scattering matrix elements using finite volume methods on a digital quantum computer.
We show that the asymptotic scaling is polynomial in the relevant parameters of the system. 
Specifically, we show that there is quantum advantage to be found in computing the needed eigenenergies for finite volume methods on a quantum computer, where, as we detail in Table 1, the computational costs scale polynomially.
This approach thus offers an alternative to the approach of JLP~\cite{Jordan2012,jordan2014} where scattering wave packets are initialized,  time evolved, and then measured post-scattering. 
While we focus on finite volume methods in this work, we have in mind either finite volume Hamiltonian methods or finite volume Euclidean space methods.  There are more recent works devoted to extracting scattering information from finite volume Minkoswki space computations \cite{PhysRevD.104.054507,carrillo2024inclusivereactionsfiniteminkowski,Briceno:2021jqm}.

We implement our approach using a series of different fault tolerant simulation algorithms, which form the second major contribution of this paper.
We consider primarily Hamiltonians formulated both in the field occupation basis and field amplitude basis.
The occupation basis Hamiltonian is simulated with a Trotterization algorithm, while for the amplitude basis one we describe four simulation procedures - one with Trotterization, and the three others with modern qubitization methods. 
These qubitization approaches are optimized through our introduction of unitary block encodings (implemented through LCU or linear combinations of unitaries approaches~\cite{childs2012hamiltonian}) of $\phi$, $\phi^2$ and $\phi^4$ operators and in order to reduce the resource cost.
Further, the implemented circuits have been optimized with recent optimization techniques \cite{2022_MWZ, 2023_MSW}. The Trotterization algorithm in the occupation basis performs optimally in the noninteracting limit, but is outperformed by qubitization methods (in the amplitude basis) with increasing interaction strength. However, the main appeal of the occupation basis algorithm is that it
provides a natural extension for state preparation and measurement of direct scattering calculations in higher dimensions.  

In Table~\ref{tab:compareCost} we have compared the cost of implementing the various algorithms in terms of number of qubits and T-gates used. 
We have implemented our algorithms with the Clifford+T-gate set, which is the most popular, fault-tolerant, universal gate set considered for quantum simulation algorithms~\cite{reiher2017elucidating,babbush2018low,childs2018toward,shaw2020quantum} and affords a wide variety of exact and approximate unitary synthesis methods~\cite{giles2013exact, 2015_KMM, 2016_RS, 2021_MM, gheorghiu2022quasi, 2022_GMM}. At times, it can exactly synthesize useful gates that could at best be approximately synthesized in other gate libraries~\cite{mukhopadhyay2024synthesis, 2024_Mcs, 2024_Mtof}. The resource overhead in fault-tolerant implementation of the non-Clifford T-gate is the highest in most error-correction schemes, including the surface code. A bound on the T-count can be can be used as a marker to reflect the complexity of fault-tolerant implementation of a quantum algorithm \cite{PhysRevA.86.032324}.

We then consider implementing the Trotter and qubitization-based simulation algorithms on top of the surface code and consider the overheads of magic state distillation in the simulation algorithm. The space-time volume needed to implement a $T$ gate is expected to dwarf the costs of all other gates by factors of $100$ or greater.  Our calculations for surface code costs are based on canonical approaches \cite{PhysRevA.86.032324}. This facilitates the translation for others towards any preferred modern methods,  (including those that may  exist in future) in the ever-expanding forefront of quantum error correction \cite{fowler2019, Litinski2019}.

The organization of paper is as follows. 
In Section~\ref{sec:hamiltonian} we paramaterize scalar $\phi^4$ QFT in a Hamiltonian formulation in both field amplitude and occupation bases. 
We then discuss how phase estimation can be used to estimate the scattering amplitude in Section~\ref{sec:scatteringMatrix} using extensions of L\"uscher-finite volume methods.  We  use these techniques in Section~\ref{sec:algo} to provide quantum algorithms for estimating the elements of the $S$-matrix (under the assumption of elastic collisions) in both the amplitude and the occupation basis.  This section gives not only the asymptotic scalings required but also the constant factor analysis needed to estimate the energy within fixed error.  Section~\ref{sec:FT} contains our analysis of the implementation of the quantum algorithms using the surface code along with the space-time needs for the simulation.  We finally conclude in Section~\ref{sec:Conclusion} and discuss future applications.

\begin{table}[t]
\centering
\begin{adjustbox}{width=\textwidth}
\begin{tabular}{|c|c|c|c|}
\hline
\hline
    Algorithm & Qubit $\#$ & T-gate $\mathbf{\#}$  & Reference   \\
    \hline
    Occ. Trotter & $O\left( N|\Omega| + \log{\left( \frac{\lambda N}{M \epsilon_E}\right)}\right)$ &$O\left(\frac{\lambda N^{7}|\Omega|^{3}}{M^{5/2}\epsilon_E^{3/2}}\log( \lambda \frac{N |\Omega| }{M \epsilon_E})\right)$ & Theorem \ref{thm:totalTocc}\\
    \hline
    Amp. Trotter & $O\left( |\Omega|\log(k) + \log_{2} \left( \frac{\vert \Omega \vert \Lambda k \left( \Lambda k + M^2 \right)}{\epsilon_E}\right) \right) $& $O\left(\frac{ |\Omega|^{3/2} \sqrt{\Lambda^{2} k^{5} + \Lambda M^{2} k^{4}} \log^4(k) }{\epsilon_E^{3/2}} \right)$ &  Theorem \ref{thm:ampTrotter}, Lemma \ref{lemma:anc-qubits-alg2}
 \\
    \hline
    Amp. Qubitized I  & $O \left( \vert \Omega \vert \log{k} + \log^{2}{k} + \log{\left( \frac{\vert \Omega \vert \left[ k^{2}\Lambda + k M^{2} \right]}{\epsilon_E}\right)}\right)$ & $O \left( \frac{\vert\Omega\vert^{2}}{\epsilon_E} \left[ k^{2}\Lambda + k M^{2} \right] [\log^{2}{k} ]\right)$ & Theorem \ref{thm:cost-qpe-qubitization} \\
    \hline
    Amp. Qubitized IIIa  & $O\left( \vert \Omega \vert \log{k} + \log{\left[ \frac{\vert \Omega \vert \left( k^{2}\Lambda + kM^{2}\right)}{\epsilon_E} \right]} \right)$ & $O \left( \frac{\vert\Omega\vert^{2}}{\epsilon_E} \left[ k^{2}\Lambda + k M^{2} \right] \left[ \log^{4}{k}\right] \right)$ & Theorem \ref{thm:cost-qpe-qubitization} \\
    \hline\hline
    Amp. Qubitized IIIb  & $O\left( \vert \Omega \vert \log{k} + \log{\left[ \frac{\vert \Omega \vert \left( k^{2}\Lambda + kM^{2}\right)}{\epsilon_E} \right]} \right)$ & $O \left( \frac{\vert\Omega\vert^{2}}{\epsilon_E} \left[ k^{2}\Lambda + k M^{2} \right] \left[ \log^2{k}\right] \right)^{*}$ & Proposition \ref{prop:qpe-costs-alg3b} \\
    \hline
\end{tabular}
\end{adjustbox}
\caption{Qubit counts and T-gate cost scalings for computing $\phi^{4}$ scattering matrix elements as a function of the lattice size $\vert \Omega \vert$, the (rescaled) coupling constant $\Lambda$ and mass $M$, which hide dependence on the lattice size, the amplitude cutoff $k$ in the field amplitude basis in units of the bin size, the momentum cutoff $N$ in the occupation basis, and the target precision in the energy estimate $\epsilon_E$, which dictates the precision of the S-matrix through Eq.~\eqref{eqn:error-energy-scattering-phase}. The $O$ notation hides further multiplicative factors at most logarithmic in all variables. (*) For Algorithm IIIb the estimate on the T-count has been obtained assuming Conjecture \ref{conj:phi24}.  }
\label{tab:compareCost}
\end{table}
\section{$\phi^4$ Hamiltonian}
\label{sec:hamiltonian}
\subsection{Field Amplitude Basis}

We begin by considering our Hilbert space for the theory to be of the form
\begin{equation}\label{eq:field_H_decomp}
    \mathcal{H}  =\bigotimes_{{\bf x} \in \Omega} \mathcal{H}_\phi,
\end{equation}
where $\mathcal{H}_\phi$ is the Hilbert space that describes a field at each of the lattice sites. 
The discretized lattice form of the Hamiltonian is 
\begin{equation}
H=\sum_{\bx \in \Omega} a^{d}\left[\frac{1}{2} \pi(\bx)^{2}+\frac{1}{2}\left(\nabla_{a} \phi\right)^{2}(\bx)+\frac{1}{2} m^{2} \phi(\bx)^{2}+\frac{\lambda}{4 !} \phi(\bx)^{4}\right]
\label{eqn:Hamp}
\end{equation}
where $\Omega$ is the set of all (spatial) lattice sites, $\bx = a \mathbf{n}$, where $a$ is the lattice spacing and $\mathbf{n} \in \mathbb{Z}^{\times d}$, and $d$ is the number of spatial dimensions.

We then consider the conjugate momentum operator 
\begin{equation}
    \pi(\bx) = \mathcal{F}^{\dagger} \phi(\bx) \mathcal{F},
\end{equation}
where $F$ is the discrete quantum Fourier transform acting on our truncated space.
Starting with Eq.~\eqref{eqn:Hamp}, 
we identify
\begin{eqnarray}
\left( \nabla \phi(\bx)\right)_i \rightarrow \Delta_{i} \phi(\bx) &\equiv& \frac{\phi(\bx + a \hat{x}_i) - \phi(\bx)}{a} \nonumber \\
\Rightarrow  \sum_{i=1}^{d} \left( \Delta_{i} \phi(\bx) \right)^2 &=& \sum_{i=1}^{d} \frac{\phi^{2}(\bx + a\hat{x}_i) + \phi^{2}(\bx) - 2 \phi(\bx) \phi(\bx + a\hat{x}_i)}{a^2}
\end{eqnarray}
If we assume periodic boundary conditions, then at each lattice site $\bx$, there is one contribution of $\phi(\bx)^{2}$ from each of the $d$ lattice sites at $\bx^{(i)} = \bx - a \hat{x}_i$ (without periodic boundary conditions, we would need to be careful about handling the sites at the edges), and a single additional contribution coming from the $\phi(\bx)^{2}$ term. Therefore, the Hamiltonian becomes 
\begin{equation}
H = \sum_{\bx \in \Omega} \left[ \frac{1}{2}a^{d} \pi^{2}(\bx) + \frac{1}{2}(d+1+a^2 m^2) a^{d-2} \phi^{2}(\bx) + \frac{\lambda}{4!} a^{d} \phi^{4} (\bx) -  a^{d-2} \sum_{i=1}^{d} \phi(\bx) \phi(\bx + a \hat{x}_i)\right]
\end{equation}
Let us now define
\begin{eqnarray}
\Phi := a^{\frac{d}{2} - 1} \phi, && \Pi := a^{\frac{d}{2}} \pi, \nonumber \\
M := am, \quad && \Lambda := a^{4-d} \lambda
\end{eqnarray}

Then, in terms of the scaled variables, our Hamiltonian simplifies to
\begin{equation}
H_{amp} = \sum_{\bx \in \Omega} \left[ \frac{1}{2} \Pi^{2}(\bx) + \frac{1}{2}(M^2 + d + 1) \Phi^2 (\bx) + \frac{\Lambda}{4!} \Phi^{4}(\bx) -  \sum_{i=1}^{d} \Phi(\bx) \Phi(\bx + a \hat{x}_i)\right]
\label{eqn:phi4-hamiltonian}
\end{equation}
Thus, we see that the derivative term is effectively replaced by a nearest neighbor interaction. We shall make use of this expression, which we denote by $H_{amp}$ in both the Qubitization-based and Trotterization based implementation in the field amplitude basis (Section \ref{subsec:ampBasisMain}).
The field amplitude basis is defined to be that in which the field operator is diagonal
\begin{equation}
\Phi \ket{\phi} = \phi \ket{\phi}.
\end{equation}

The strength of the field amplitude basis is that the Hamiltonian is particularly simple in the strong-coupling regime as the field operators are diagonal.  In contrast, we will see that in the weak to moderate coupling regime occupation basis provides an alternative concise representation of the Hamiltonian.

\subsection{Occupation Basis}
\label{subsec:occBasis}

The occupation basis provides an alternative way of encoding a simulation.  Rather than choosing the basis to diagonalize the field operator, the basis instead is chosen to diagonalize the Hamiltonian in the case where $\lambda=0$.  In particular, we choose a set of momenta modes $\{\mathbf{p}_i\}$ and define $a_{\mathbf{p}}^\dagger$ to be the creation operator acting on momentum mode $\mathbf{p}$ such that for field occupation state $x$, $a^{\dagger}_{\mathbf{p}} \ket{x}_{\mathbf{p}} = \sqrt{{x+1}}\ket{x+1}_{\mathbf{p}}$ and $a^\dagger_{\mathbf{p}} a_{\mathbf{p}}$ is the number operator acting on the mode. 

We show in Section \ref{subsec:occBasis} that the Hamiltonian for a simulation within volume $|\Omega|$ in real space can be written as the diagonal operator
\begin{equation}
    H_{occ} =\quad:H_0: + :H_\lambda:   \label{eqn:H_occ}
\end{equation}
where we define 
\begin{equation}
  :H_0: =\quad   \frac{1}{|\Omega| } \sum_\bp \omega_\bp a^\dagger_\bp a_\bp    \label{eqn:H0_intro}
\end{equation}
and the $H_\lambda$ term represents the $\phi^4$ and the non-normal ordered form of the term is given below.
\begin{equation}
\begin{gathered}
H_\lambda =  \frac{\lambda}{4!|\Omega|^3} \sum_{\lbrace\bp_i\rbrace} \frac{1}{4\sqrt{\omega_{\bp_1} \omega_{\bp_2} \omega_{\bp_3} \omega_{-(\bp_1 + \bp_2 + \bp_3)}}} 
 \left\lbrace \left( a_{\bp_1} + a_{-\bp_1}^{\dagger} \right) \left( a_{\bp_2} + a_{-\bp_2}^{\dagger} \right)  \left( a_{-\bp_3} + a_{\bp_3}^{\dagger} \right) \left( a_{(\bp_1 + \bp_2 + \bp_3)} + a_{-(\bp_1 + \bp_2 + \bp_3)}^{\dagger} \right)\right\rbrace 
 \label{eqn:H_int}
\end{gathered}
\end{equation}
There are two primary advantages of this representation. One is that the Hamiltonian is diagonal in the case of weak coupling. 
This means that in cases where high accuracy but weak to moderate couplings are needed, the Hamiltonian can be concisely described in this basis as a diagonally dominant Hamiltonian. In the high energy limit, the single-particle basis may also prove to be more efficient \cite{barata2021single}.  Further, \cite{barata2021single} also has appealing additional encoding improvements, such as a binary encoding of the occupation basis,  if they can be converted to a fault-tolerant regime.  
Another potential benefit is that the occupation basis can have in this limit small commutators between the terms involved, which can lead to substantial computational advantages as was noted in the simulation of the Schwinger model~\cite{shaw2020quantum}.  For this reason, we pay special attention to the Trotter method for simulations in this basis.

\section{Luscher Method and $S$-Matrix Computation}
\label{sec:scatteringMatrix}
The cost of any end-to-end quantum simulation depends on the observables that we wish to measure as well as the processing required to prepare the initial state and simulate the dynamics of the quantum system.  Unlike previous work, which directly examines scattering in field theories~\cite{Jordan2012,jordan2014}, energy estimation in finite volume via L\"uscher-like methods \cite{Luscher1991,annurev,rusetsky2019particleslattice,Mai2021,multihadroninteractionslatticeqcd,Gabai2022,Bajnok2016,James_2018,PhysRevD.83.114508,PhysRevD.83.071504,PhysRevD.98.014507,Garofalo_2021}  is computationally simple and opens up the possibility of practical simulations of quantum dynamics.

Computing the entire $S$-matrix is exponentially difficult since the number of output momenta and input momenta scales exponentially with the number of particles permitted.
We are therefore interested in possible feasible measurements of the S-matrix where 
 we can readily extract at least the elastic parts of the S-matrix through the measurement of multi-particle energies in finite volume.
We will comment on the ability to obtain inelastic information further on in this section.  
Here we will argue that the measurement of energies in finite volume leads directly to information on the S-matrix elements.   In what follows we focus on $1+1$D and introduce the notation for the 2 particle to n particle S-matrix:
\begin{equation}
    \langle p'_1\cdots p'_n|\mathcal{S}|p_1 p_2\rangle_0 = (2\pi)^2\delta^{(2)}(p_1+p_2-\sum^n_{i=1}p'_i)S(s)
\end{equation}
where $S(s)$, a function of the Mandelstam variable $s=(p_1+p_2)^2$, denotes the scattering amplitude for processes involving two initial particles.

We now provide some detail about the approach of using finite volume equilibrium data for the computation of values of the scattering matrix or $S$-matrix to give the reader a basis for comparison to the method presented in Refs. \onlinecite{Jordan2012, jordan2014}.
We consider first the simplest case where the energy determines the scattering phase.  We go into a center of mass frame where, without loss of generality, both particles have momenta $p$ and $-p$ such that $p_1=-p_2$ and by conservation of momentum and energy the output momenta also must be in $\{-p,p\}$ In finite volume, and in the absence of interactions, the momenta, $p_i,i=1,2$ of the particles are 
\begin{equation}
    p_i = \frac{2\pi n_i}{L}, ~~ n_i\in\mathbb{Z},
\end{equation}
where $n_i$ are integers indexing the two particles and $L$ is the length of the system.  The energy $E$ of the two particle state is
\begin{equation}
    E = \sum_{i=1,2} \left(m^2 + \frac{4\pi^2 n_i^2}{L^2}\right)^{1/2}.
\label{eqn:2_particle_energy}
\end{equation}
Now what happens when interactions are turned on?  Provided the energy of the two-particle state is below threshold for particle production, the quantization condition for the two momenta is altered to
\begin{eqnarray}
e^{ip_1L}S(p_1,p_2) &=& 1=e^{2\pi n_1/L} ;\cr\cr
e^{ip_2L}S(p_2,p_1) &=& 1=e^{2\pi n_2/L} ,
\end{eqnarray}
where we can write $S(p_1,p_2)=e^{i\delta(p_1,p_2)}$ and $\delta(p_1,p_2)$ is a two-body elastic scattering phase.

By taking logarithms, the quantization conditions can be written in the form
\begin{eqnarray}
\frac{2\pi n_1}{L} &= p_1 -\frac{i}{L}\log S(p_1,p_2);\cr\cr
\frac{2\pi n_2}{L} &= p_2 -\frac{i}{L}\log S(p_2,p_1).
\label{eqn:log-S-matrix}
\end{eqnarray}
If one solves the quantization condition for $p_1$ and $p_2$ one immediately knows the energy of the two-particle state via
\begin{equation}
    E = \sum_{i=1,2} (m^2 + p_i^2)^{1/2},
\end{equation}
where $p_i$ necessarily deviate from their free values.
Now if instead we start with knowledge of $E$ via measurement, we can reverse the process and infer $S(p_1,p_2)$.  And by measuring $E$ for different $L$, we can determine $S(p_1,p_2)$ over a range of $p_1$ and $p_2$ because while $p_i$ are not free, they still will behave as $1/L$ over a wide range of volumes.

{In any determination of the energy $E$ of a two particle state en route to the determination of an S-matrix element, there will be some uncertainty  $\updelta E$ associated with its determination.  If we work in the center-of-mass frame and the particles have equal and opposite momentum, i.e., $p_1=-p_2$, the associated uncertainty in the scattering phase $\delta(p,-p)$ is given by
\begin{equation}
    \updelta\delta(p,-p) = \big|\frac{LE}{8p}\updelta E\big|
\label{eqn:error-energy-scattering-phase}
\end{equation}
We can see that uncertainty in the determination of the energy affects most dramatically the determination of the scattering phase at small momentum.
}

As an illustration of this in the $\varphi^4$ theory, we will consider an example from the ordered phase of the model (i.e., $m^2<0$).  The spectrum of the theory in this regime consists of kinks and bound states of kinks (meson-like states).  As $\lambda$ is increased from 0, the bound states disappear at a $\lambda_0$ from the spectrum and only kink-like states exist.  At a large enough $\lambda=\lambda_c$, the model becomes disordered.  In our example, we will consider the regime $\lambda_0<\lambda<\lambda_c$.

\begin{figure}[htb]
    \centering
    \subfloat{
    \centering
    \includegraphics[width=.5\linewidth]{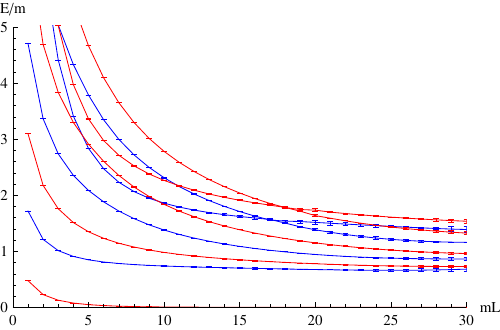}
    }
    \subfloat{
    \centering
    \includegraphics[width=.5\linewidth]{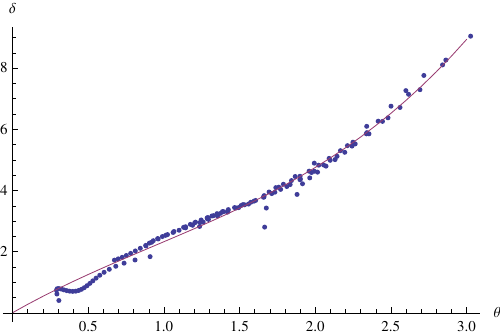}
    }
        \caption{a) The low lying spectrum of the $\varphi^4$ in its broken phase ($m^2=-0.25$, $\lambda=0.15$).  The ground state energy has been subtracted.  From Fig. 10 of \cite{Bajnok2016}.
        b) The scattering phase inferred from the energies of the two kink states presented in Fig.~\ref{EnergyLevels}.  From Fig. 11 of \cite{Bajnok2016}.} 
    \label{fig:gate_counts_cutoff}
\label{EnergyLevels}
    \end{figure}

In Fig. \ref{EnergyLevels} a), we plot the low-lying energy levels with zero total momenta as a function of system size, $L$, as computed in Ref.\cite{Bajnok2016} using Hamiltonian truncation methods.  The ground state energy has been subtracted.  The lowest lying level is the state whose energy is nearly degenerate to the ground state (we expect such a near-degeneracy in finite volume in the broken phase).  Beyond this state are energies corresponding to two-kink and four-kink states.  We can use the two-kink energies to back out the scattering phase as described above.  This phase is plotted in Fig. \ref{EnergyLevels} b) as a function of energy, $E$.  Here the energy of the two-particle state is parameterized by $\theta$ via :
$
E = 2M_{kink}\cosh(\theta)
$
$\pm M_{kink}\sinh(\theta)$.

What we have discussed so far concerns the elastic part of the $2-2$ scattering matrix.  We can also access inelastic information from the measurement of energies.  We can parameterize the S-matrix $S(p_1,p_2)$ solely in terms of $\theta$ ($\theta$ is related to the Mandelstam variable, $s$, via $s=4m^2\cosh^2(\theta/2)$).  The unitarity condition on the S-matrix then reads
\begin{equation}
    S(\theta+i\epsilon)S(-\theta+i\epsilon) = f(\theta),
\end{equation}
where $f(\theta)$ is real positive on the real line.  If $\theta_0$ marks the threshold beyond which inelastic processes are possible, $f(|\theta|<\theta_0)=1$, while $0<f(|\theta|>\theta_0)<1$.  Using the analytic structure of the S-matrix in the complex-$\theta$ plane, we can write $S(\theta)$ in the following form \cite{Gabai2022}:
\begin{equation}
S(\theta) = \pm \prod_j \frac{\sinh(\theta) + i\sin(\beta_j)}{\sinh(\theta) - i\sin(\beta_j)}\exp\bigg[-\int^\infty_{-\infty}\frac{d\theta'}{2\pi i}\frac{\log f(\theta')}{\sinh(\theta-\theta'+i\epsilon)}\bigg].\label{eq:Stheta}
\end{equation}
The first part of the parameterization involving the angles $\beta_j$ are so-called Castillejo-Dalitz-Dyson (CDD) factors.  The second part of the parameterization contains information about the inelastic part of the 2-2 scattering inasmuch as it depends on the region in $\theta$ where $f(\theta)$ differs from 1.  Notice that because of analyticity, the inelastic part of the S-matrix influences the elastic region $|\theta|<\theta_0$.
With sufficiently accurate measurements of the energies, one can use this parameterization to determine the number and values of the CDD angles $\beta_j$, as well as the function $f(\theta)$.  In practice, extracting the function $f(\theta)$ is difficult as it involves the inversion of an ill-posed integral equation.  However the integrated quantity,
\begin{equation}
\int^\infty_{-\infty}\frac{d\theta'}{2\pi i}\frac{\log f(\theta')}{\sinh(\theta-\theta'+i\epsilon)},
\label{eq:inversion}
\end{equation}
has been successfully extracted from numerical data, for example, in the case of the non-integrable Ising field theory \cite{Gabai2022}.  With sufficiently high quality data for the energies, the function $f(\theta)$ should be directly available, particularly in parameterized form.

We now provide more detail on the difficulty that classical techniques have in providing high precision data.  We first consider truncated spectrum method (TSM) computations.  For TSM the error in the energies due to the presence of a UV cutoff, $\Lambda$, in the $\phi^4$ field theory.  This error, $\delta_E$, scales with the cutoff as $\delta_E\sim \Lambda^{-3}$ \cite{PhysRevD.96.065024}. 
However, truncated spectrum methods depend on the exact diagonalization of the truncated Hilbert space in some computational basis. 
This truncated Hilbert space might be formed by considering the set of free massive states $\prod_k a^\dagger_{i_k}|0\rangle$ (where $a^\dagger_{i_k}$ creates a state with energy $(m^2+( 2\pi  i_k/L)^2)^{1/2}$ with $i_k$ some integer) that satisfy $\sum_k |i_k| \leq M$. 
Here M is an effective dimensionless cutoff. 
If we work, as is typical, in the zero momentum sector, we also require $\sum_k i_k = 0$. 
The size, $N_{\cal H}$, of this truncated Hilbert space scales as 
\begin{equation}
    N_{\cal H} = \sum_{j=1}^M p(j)^2 = \frac{\sqrt{3/2}M^{-3/2}}{48\pi}e^{2\pi\sqrt{2M/3}}(1+{\cal O}(1/M^{1/2})),
\end{equation}
where $p(j)$ is the number of partitions of the integer $j$, \added{squared for states partitioned into left and right moving sectors}.
Because exact diagonalization methods scale as $N_{\cal H}^3$, the cost of obtaining a given energy level with an error $\delta E \sim M^{-3}$ goes as 
$$
{\rm TSM~cost} = N_{\cal H}^3 \sim \delta_E^{3/2}e^{2\pi\sqrt{6}\delta_E^{-1/6}}.
$$  
Initially this scaling makes the classical truncated spectrum method computation possible. 
But for high precision data this scaling becomes prohibitive.
To obtain data with $10^{-2}$ precision is approximately 1000 times more costly than $10^{-1}$ precision.  
Asking for $10^{-6}$ precision is $10^{49}$ more costly than $10^{-1}$ precision.  \par 
Another classical computational method that could be used to compute the energies is quantum Monte Carlo (QMC).  The issue with QMC is of a different kind in comparison with TSM.  
The exponential scaling instead comes from distinguishing closely spaced levels.
In order to apply Luscher's method to determining the two-particle scattering phases via a QMC computation, one has to determine a set of of closely spaced set of two particle energy levels. 
At large volume the spacing of these levels goes as $1/L^2$.
One approach \cite{collaboration2009} to separating these levels from one another is to measure a matrix of correlation functions that couple to the 2-particle energy levels of concern (i.e. correlation functions involving operators which are parity even) and then solve a generalized eigenvalue problem  (GEVP).  
Specifically one defines a set of two particle operators $\{ O_i\}_{i=1}^M$ where $M$ is the number of targeted energy levels. 
By measuring the time evolution of the matrix, $\hat C$, of correlation functions, with entries $\hat C_{ij}(t)=\langle O_i(t)O_j(0)\rangle$, one is able to solve the GEVP:
\begin{equation}
\hat C(t)v_n(t,t_0)=\lambda_n(t,t_0)\hat C(t_0)v_n(t,t_0), ~~ v_n^\dagger C^(t_0)v_m = \delta_{nm}.
\end{equation}
The solutions of this GEVP give linear combinations of the operators ${\cal O}_i$, ${\cal O}_{v_n}=\sum^M_{i=1}v_{ni}{\cal O}_i$ that couple maximally to a single eigenstate and thus correspondingly,
the associated eigenvalues $\lambda_n$ of this GEVP depend on a single $E_n$ which can be accessed via 
$$
E_n = -\partial_t \log(\lambda_n(t,t_0)).
$$
Now this procedure to separate the $M$ energy levels from one another is `contaminated', so to speak, by potential mixing of the $M$ levels with the $M+1$-th energy level.  This mixing is controlled by the energy separation of these two levels.  If one wants to maintain an error of no more than ${\cal O}(\delta_E)$ due to this mixing, one needs to run the QMC simulation to a time, $t_s$, determined by the conditions
\begin{equation}
    \delta_{E} > e^{-t_s(E_{M+1}-E_n)} := e^{-t_{s}\Delta_n}, ~~n=1,\cdots,M.
\end{equation}
This error bound is valid provided one has chosen $t_0=t/2$ (choices of $t_0<t/2$ lead to increased errors while choices $t_0>t/2$ lead to an ill-conditioned problem).
On the other hand, the uncertainty, $\delta_E$, in the determination of energy levels is related to the uncertainty, $\delta_\lambda$, of the eigenvalues of the GEVP that arises from sampling over finite independent configurations, $N_{conf}$, in the QMC computation.  We have
\begin{equation}
\delta_E \sim \frac{\delta_\lambda}{\lambda} \sim \frac{e^{E_{min} t/2}}{\sqrt{N_{conf}}}.
\end{equation}
Here $E_{min}=2m$ is the minimum energy of the eigenstates that contribute to the correlation function $\hat C_{ij}$ (in the sense of a Lehmann expansion). 
The number of configurations needed in a QMC simulation for a desired accuracy, $\delta_E$, goes as
\begin{equation}
    N_{conf}\sim \delta_E^{-2-E_{min}/\Delta_n}.
\end{equation}
Assuming a simulation run to time $t_s$, the cost of a computing a single QMC configuration goes as
\begin{equation}
C_{conf} = c_{update}Lt_s + c_{measure}M^2t_s,
\end{equation}
where $c_{update}$ and $c_{measure}$ are constants related to the particular implementation of the QMC algorithm. 
The overall cost is then $N_{conf}C_{conf}$.
The key quantity that determines the cost of the QMC computation for a desired precision, $\delta_E$ is then $\Delta_n$. 
If M is kept fixed as L is made large, $\Delta_n$ behaves as $\Delta_n \sim 1/L^2$ for all $n$ and one sees that there is an exponential cost to the QMC simulation, going as $\delta_E^{- L^2}$. 
If one instead scales $M$ with $L$, paying the increased (but polynomial) cost of computing a single configuration, $\Delta_n$ for $n$ small will be independent of $L$. 
For these low lying energy levels you do not see exponential scaling.  However for the levels $E_n$ with $n\sim M$ we have $\Delta_n\sim 1/L$.  
Here the cost scaling returns to being exponential in volume, i.e., $\delta_E^{-L}$.  
In obtaining high precision data that will allow us to invert Eq. \ref{eq:Stheta} in order to find $f$, we require scattering data at a wide range of energies. 
We thus would require all $M$ energies to be determined with $\delta_E$ precision, not merely the low lying ones.

While we have focused our discussion here on scattering in $1+1D$, our approach can be extended to scattering in higher dimensions, i.e., the elastic scattering phases of 2-2 particle scattering, $\delta_l$, in different angular momentum channels can be connected to measurements of the two-particle energy in finite volume \cite{Luscher1991}. 
This procedure breaks down when different angular momentum channels mix, although simplifications can be made at low energies.
While originally formulated in \cite{Luscher1991} for two spinless identical particles, extensions have been made to identical particles with spin, asymmetric volumes, and amplitudes containing external current - see \cite{2018reviewLQCDScattering} for a review. 
This formalism can also be extended to energies above the inelastic scattering threshold, provided that $f(\theta)$ can parameterize the S-matrix elements
In certain cases, inelastic information involving 3 particle scattering is also available in higher dimensions \cite{annurev,rusetsky2019particleslattice,Mai2021,multihadroninteractionslatticeqcd}.

\begin{claim}
    Let $\mathcal{S}$ be the $S$-matrix for a Hamiltonian operator $H(t)$.  The value of the excited state energies can be used to infer certain elements of $\mathcal{S}$ if one of the following occurs:
    \begin{enumerate}

        \item We are interested in $2 \mapsto 2$ elastic scattering of particles in one spatial and one temporal dimension ($1+1\text{D}$) where the two-particle final state is below the threshold for particle production. 
        \item We are interested in inelastic processes in $1+1\text{D}$ as parameterized by the function $f(\theta)$ in Eq.\eqref{eq:Stheta}.

        \item We are interested in computing the scattering phases of elastic $2 \mapsto 2$ scattering in higher dimensions \cite{Luscher1991}.
        \item We are interested in certain relatively simple scattering processes that include particle decay in dimensions higher than $1+1\text{D}$. This is focused primarily on scattering involving 3-particles, an active topic of current research \cite{annurev,rusetsky2019particleslattice,Mai2021,multihadroninteractionslatticeqcd}.

    \end{enumerate}
    \label{claim:Smatrix}
\end{claim}

    Barring the cases mentioned in the above claim, there are other reasons why the energies of both the ground state and low lying excited states are independently interesting for the theory.  It can allow us to understand phase transitions in the strongly interacting theory~\cite{mattuck1968quantum,espinosa1992phase}.  For all the above reasons, we focus our attention on the computation of the energies of the low lying excited states, a well studied problem in the context of quantum algorithms~\cite{reiher2017elucidating,childs2018toward}.

 How does this approach differ from that taken by Jordan Lee and Preskill~\cite{Jordan2012,jordan2014}? In \cite{Jordan2012,jordan2014} the approach involved preparing a Gaussian wave packet in the interacting eigenbasis rather than an eigenvector of the interacting Hamiltonian via phase estimation.  Their approach has the advantage that it can be applied in circumstances where the gap is large and no prior knowledge of the eigenvectors is given.  Directly preparing the eigenstates as we discuss above can be  more computationally efficient than the approach given by JLP, but requires efficient approximations to the eigenstates which may not be available in all settings. The lack of detailed cost analysis in JLP prevents quantitative comparison however.
 
\section{Simulation Algorithms}
\label{sec:algo}
Modern simulation algorithms have converged in recent years to the point that there is no optimal single quantum simulation algorithm.  Rather, different algorithms tend to have advantages and disadvantages in different regimes~\cite{shaw2020quantum,hagan2023composite,rajput2022hybridized}.  In this spirit, we provide four simulation algorithms for 
scalar field theory in the occupation as well as field amplitude basis. 
These algorithms employ Trotter decompositions or qubitization and use optimized circuits to construct the operator exponentials or oracle calls that both methods require.  
We further observe that the methods, as expected, have advantages and disadvantages with respect to each other. 
Most obviously, the Trotter algorithm for simulations in the occupation basis is by far the most efficient for weak coupling ($\lambda\ll 1$) but the qubitized field amplitude basis is likely to scale better in the strong-coupling regime.  

The material in this section is laid out as follows.
In Section \ref{subsec:ampBasisMain} we describe an algorithm for the Hamiltonian formulated in the field occupation basis, while in Section \ref{subsec:OccBasisMain} we describe several algorithms for the Hamiltonian in the field amplitude basis. The techniques and concepts in these two sections can be followed independently of the other.   

\subsection{Amplitude Basis}  
\label{subsec:ampBasisMain}

In this section we briefly discuss a number of algorithms that have been designed to simulate the Hamiltonian $H_{amp}$ (Eq. \ref{eqn:phi4-hamiltonian}) in the amplitude basis. This provides an appealing basis for simulating dynamics in the $\lambda\gg 1$ regime. 
We use two types of algorithms - qubitization \cite{2017_LC, 2019_LC, 2019_GSLW} and Trotterization \cite{1991_S}.
We have designed three qubitization-based algorithms, which mainly differ in the LCU (linear combination of unitaries) decomposition of the operators.
Similar approaches have been considered in \cite{hariprakash2024}. 
In Algorithm I (Appendix \ref{subsecn:equal-weight-lcu}) we discuss a decomposition of $\Phi$ using an equal weight LCU decomposition of the field operator (Algorithm I).

Algorithm 1 constructs the qubitized walk operator in the field amplitude basis by exploiting the shared coefficients of the four different families of terms in the scalar field Hamiltonian. It uses an $O(k^4)$ LCU decomposition of the single-site operators appearing in the Hamiltonian. However, each of the terms in this LCU decomposition share the same coefficient, which allows us to perform a simultaneous SELECT operation by making use of a comparator circuit. This results in an $O( |\Omega| \log^{2} k)$ cost for the block encoding of the Hamiltonian using this approach, instead of $O( |\Omega| \log^{4} k)$ which would come from a naive sequential SELECT application of each of the terms in the LCU decomposition.

Next, in Appendix \ref{subsec:trotterHamp} we describe another LCU decomposition of $\Phi$ as sum of mainly Z-operators, which not only helps us in developing a qubitization-based algorithm (Algorithm IIIa) in  \ref{subsec:blockHampZ}, but it also makes the expression amenable to Trotterization (Algorithm II), as discussed in Appendix \ref{subsec:trotterHamp}. In Appendix \ref{subsec:algoIIIb} we describe another more compact LCU decomposition of all operators using binary representation of integers. With this we describe another qubitization algorithm (Algorithm IIIb).   

We have described the algorithms, the relevant circuit constructions and resource requirements in detail in Appendix \ref{subsec:algoAmp}. Here we briefly summarize the main results.

\begin{theorem}
The total cost of performing phase estimation to estimate an eigenvalue of the Hamiltonian to within error $\epsilon_E$ is given by
\begin{eqnarray}
Cost(QPE)^{(I)} &\in& O \left( \frac{\vert\Omega\vert^{2}}{\epsilon_E} \left[ k^{2}\Lambda + k M^{2} \right] \log^{2}{k} \right) \nonumber \\
Cost(QPE)^{(IIIa)} &\in& O \left( \frac{\vert\Omega\vert^{2}}{\epsilon_E} \left[ k^{2}\Lambda + k M^{2} \right] \log^{4}{k} \right) 
\end{eqnarray}
while the total number of logical qubits required, including those employed for phase estimation, are
\begin{eqnarray}
Count(Qubit)^{(I)} &\in& O \left( \vert \Omega \vert \log{k} + \log^{2}{k} + \log{\left( \frac{\vert \Omega \vert \left[ k^{2}\Lambda + k M^{2} \right]}{\epsilon_E}\right)}\right) \nonumber \\
Count(Qubit)^{(IIIa)} &\in& O\left( \vert \Omega \vert \log{k} + \log{\left[ \frac{\vert \Omega \vert \left( k^{2}\Lambda + kM^{2}\right)}{\epsilon_E} \right]} \right)
\end{eqnarray}
where the superscript denotes the algorithm employed.
\label{thm:cost-qpe-qubitization}
\end{theorem}

In Algorithm IIIa we obtain LCU decomposition of the operators $\Phi^2$ and $\Phi^4$ from a decomposition of $\Phi$, using binary representation of integers. We can also obtain LCU decomposition of $\Phi^2$ and $\Phi^4$ using binary representation of integers. This has been done in Algorithm IIIb, where we retain similar qubit count as in Algorithm IIIa. We make certain assumptions about the T-count of some special type of unitaries that are obtained from the binary representation of integers. The assumptions have been made on the basis of some proven results and observations (Appendix \ref{subsec:algoIIIb}). Thus, with Conjecture \ref{conj:phi24} we obtain a reduction of T-count, as compared to the previous two algorithms.

\begin{proposition}  
Assuming Conjecture \ref{conj:phi24}, the T-gate and qubit cost of the qubitization-based algorithm IIIb in the amplitude basis is given by
\begin{eqnarray}
Cost(QPE)^{(IIIb)} &\in& O\left( \frac{\vert \Omega \vert^{2}}{\epsilon_E} \left[ k^{2}\Lambda + k M^{2} \right] \log^{2}{k} \right) \nonumber \\
Count(Qubit)^{(IIIb)} &\in& O\left( \vert \Omega \vert \log{k} + \log{\left[ \frac{\vert \Omega \vert \left( k^{2}\Lambda + kM^{2}\right)}{\epsilon_E} \right]} \right)
\end{eqnarray}

\label{prop:qpe-costs-alg3b}
\end{proposition}

\begin{theorem}
Given an eigenstate $\ket{\psi}$ of  $H$ such that $H_{amp}\ket{\psi} = E\ket{\psi}$, 
where $H_{amp}$ is the amplitude basis Hamiltonian, as stated in Eq.~\eqref{eqn:HampIII},  then there exists a quantum algorithm that outputs with probability greater than $2/3$ a value $\hat{E}$ such that $|\hat{E} - E| \le \epsilon_E$, using a number of $T$ gates that scales as

$$O\left(\frac{ |\Omega|^{3/2} \sqrt{\Lambda^{2} k^{5} + \Lambda M^{2} k^{4}} \log^4(k) }{\epsilon_E^{3/2}} \log{\left( \frac{\vert \Omega \vert k \log{k}}{\epsilon_E} \left( \Lambda^{2}k + \Lambda M^{2} \right)\right)} \right)$$ 
T-gates and $\mathcal{O}\left( |\Omega|\log_2(2k)\right)$ qubits, plus an additional number of ancillary qubits required for phase estimation as detailed in Lemma \ref{lemma:trotter-amp-aqft-errors}.

Here the log argument is derived from $\epsilon_r = \frac{\sqrt{2}\epsilon_{E}}{8 N_{r}}\sqrt{\frac{\epsilon_E}{2^{3/2} \tilde{\alpha}_{comm}}}$, $N_r\in \mathcal{O}\left(|\Omega|\log_2^4(2k) \right)$ and 

$\tilde{\alpha}_{comm} \in \mathcal{O}\left[|\Omega| \left( \Lambda^2 k^{10}\Delta^{10} + \Lambda M^{2} k^8 \Delta^8 \right) \right] $, and $\Delta = \sqrt{\pi/k}$.
\label{thm:ampTrotter}
\end{theorem}

\subsection{Occupation Basis Algorithm}
\label{subsec:OccBasisMain}
The interaction Hamiltonian can be decomposed into four cases based on matching momentum indices. We consider the normal ordered Hamiltonian so that the unitary phases corresponding to eigenvalues that are used to construct the $S-$matrix are computed with reference to the vacuum energy. First we state some essential results that are required to map the Hamiltonian from the bosonic to qubit space. Then we synthesize quantum circuits implementing the exponentiated Hamiltonian. \par

We discretize the Hilbert space of occupation into  $N$ distinct momentum states. We have a register of $(N+1)V$ qubits to store information about the occupation states and we use the following one-hot unary encoding to map an occupation state to a qubit state. We index the qubits by a pair of integers, such as $(p,n)$, where $p$ corresponds to a momentum mode and $n$ to a momentum state. For each such pair $(p,n)$, we have a quantum state on $(N+1)V$ qubits, in which each qubit is $\ket{0}$, except the $(p, n)^{\text{th}}$ one, which is $\ket{1}$. We denote this state by $\ket{p,n}$, which is 
\begin{eqnarray}
    \ket{p,n}&=&\ket{0_{1,0},\ldots,0_{p-1,N};0_{p,0},\ldots,0_{p,n-1},1_{p,n},0_{p,n+1},\ldots,0_{p,N};0_{p+1,0},\ldots 0_{V,N}}   \nonumber \\
    &=&\left(\otimes_{j=1}^{p-1}\ket{0_{j,0}\ldots,0_{j,N}}\right)\bigotimes\ket{0_{p,0},\ldots 1_{p,n}\ldots,0_{p,N}}\bigotimes\left(\otimes_{j=p+1}^{|\Omega|}\ket{0_{j,0}\ldots,0_{j,N}}\right).
    \label{eqn:unary}
\end{eqnarray}
We emphasis that each Hilbert subspace $\mathcal{H}\bp$ is spanned only by vectors of Hamming weight 1 in order to ensure the unary encoding. We now consider a construction of the Hamiltonian that preserves this Hamming weight without restricting to  total number conservation or conservation for each $p-$mode. \par
For convenience, we denote an operator $A_{pn}$ acting on $pn^{th}$ qubit by $A_{p,n}$ or $\left(A_n\right)\bp$. The qubit mapping for the creation and annihilation operators is as follows.
\begin{eqnarray}
        a_\bp^\dagger &=& \sum_n \sqrt{n+1}\left(\sigma_{n}^{-} \sigma_{n+1}^{+}\right)_\bp  \nonumber \\
        a_\bp &=& \sum_n \sqrt{n+1}\left(\sigma_{n}^{+} \sigma_{n+1}^{-}\right)_\bp , 
        \label{eqn:ladder}
\end{eqnarray}
where 
\begin{equation}
    \sigma^+ = \frac{1}{2}(X - i Y), \quad \sigma_- = \frac{1}{2}(X + i Y).
    \label{eqn:sigma+-}
\end{equation}
and therefore
\begin{eqnarray}
    a_\bp^{\dagger}\ket{p,n}=\sqrt{n+1}\ket{p,n+1} \qquad\text{and}\qquad a_\bp\ket{p,n}=\sqrt{n}\ket{p,n-1}
    \label{eqn:ladKet}
\end{eqnarray}
Now, considering Hermitian pairing of operators, we have
\begin{eqnarray}
    a_\bp+a_\bp^{\dagger}&=&\frac{1}{2}\sum_n\sqrt{n+1}\left(X_nX_{n+1}+Y_nY_{n+1}\right)_\bp.
    \label{eqn:ferm2pauli}
\end{eqnarray}
In theory the number of momentum states range till infinity, but for our simulation, we truncate the Hilbert space and have $N$ momentum states, thus $n$ varies from $0$ to $N$ in the above summations. This truncation in the bosonic occupation is proportional to the maximum energy expected to be simulated in semi-elastic collisions $N \propto \frac{E}{\omega_\bp} $.
It is expected that the error in this truncation is exponentially small with respect to the cutoff \cite{Somma2016, Macridin_A2018, Macridin_B2018, Bauer2022}.
However, careful numerical analysis of these (and other finite-system size effects) will be required in implementation \cite{Rinaldi2022, Hanada2023}. While this unary encoding may appear inefficient as compared to binary econdings \cite{barata2021single}, it has numerous benefits in the fault-tolerant regime. The Hamiltonian implementation can be done in terms of Clifford operations alone. As our analysis shows, this means that the time-evolution operator will require a reduced number of $R_z$ and therefore $T$ gates. 

\par

We detail the explicit qubit  mappings of each term in the Hamiltonian in Appendix \ref{subsec:algoOcc}. Here we also detail explicit circuit constructions and costs for time evolution of the entire occupation basis Hamiltonian using Trotterization algorithm. We follow the techniques described in \cite{2022_MWZ} in order to minimize the number of controlled operations.
We divide the resulting sum of Paulis into groups of commuting Paulis, diagonalize each such group, derive the distinct eigenvalues of each group.
These distinct eigenvalues equal the number of controlled rotations.
With some logical reasoning and optimizations we design the diagonalizing circuit, included in the Appendix \ref{subsec:algoOcc}. The low-lying energies of the system are then extracted using quantum phase estimation. 
The resource requirements can be summarized in the following theorem.

\begin{theorem}
Given an eigenstate $\ket{\psi}$ of  $H_{occ}$ such that $H_{occ}\ket{\psi} = E\ket{\psi}$, the occupation basis Hamiltonian stated in Eq.~\eqref{eqn:H_occ}-\eqref{eqn:H_int}, then it there exists a quantum algorithm that outputs with probability greater than $2/3$ a value $\hat{E}$ such that $|\hat{E} - E| \le \epsilon_E$, using a number of $T$ gates that scales as 
$$\mathcal{O}\left(\frac{\lambda N^{7}|\Omega|^{3}}{M^{5/2}\epsilon_E^{3/2}}\log(\frac{N |\Omega| }{M \epsilon_E})\right)$$ and $\mathcal{O}\left( N|\Omega|\right)$ qubits, plus an additional $O\left( \log{\left( \frac{\lambda N}{M \epsilon_E}\right)}\right)$ ancillary qubits required for phase estimation.
Here $M$ is the particle mass for the field, the log argumenent comes from $1/\epsilon_r$ $\epsilon_r = \frac{\sqrt{2}\epsilon_{E}}{8 N_{r}}\sqrt{\frac{\epsilon_E}{2^{3/2} \tilde{\alpha}_{comm}}}$, $N_r\in \mathcal{O}\left(N^4|\Omega|^3\right)$ and $\tilde{\alpha}_{comm} \in \mathcal{O}\left(\frac{\lambda^2  N^6 }{M^5}\right) $.
\label{thm:totalTocc}
\end{theorem}

\par

\subsection{Resource Estimates for Simulation Algorithms}
\label{subsec:comparePlot}

We will now compare the $T$-count and logical qubit estimates for implementing the various algorithms described so far.  A summary of the asymptotic cost analysis has been provided in Table \ref{tab:compareCost} (Section I). In the following section and relevant appendices we provide exact costs with exact coefficients, detailed in Fig.\ \ref{fig:gate_counts_cutoff}, \ref{fig:gate_counts_momentum}, \ref{fig:qubit_counts_cutoff}, \ref{fig:qubit_counts_momentum}.
\begin{figure}[htb]
    \centering
    \subfloat{
    \centering
    \includegraphics[width=.5\linewidth]{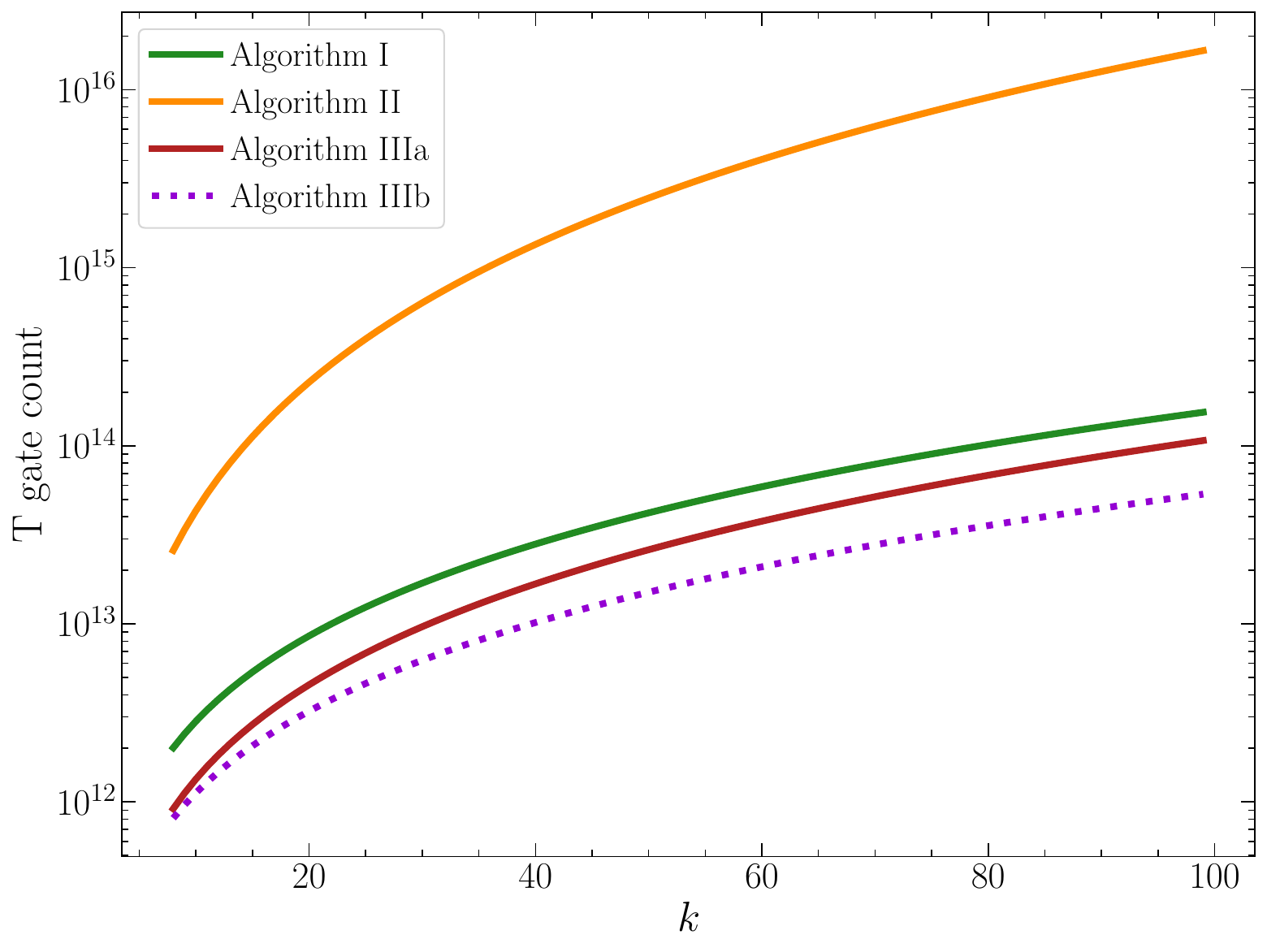}
    }
    \subfloat{
    \centering
    \includegraphics[width=.5\linewidth]{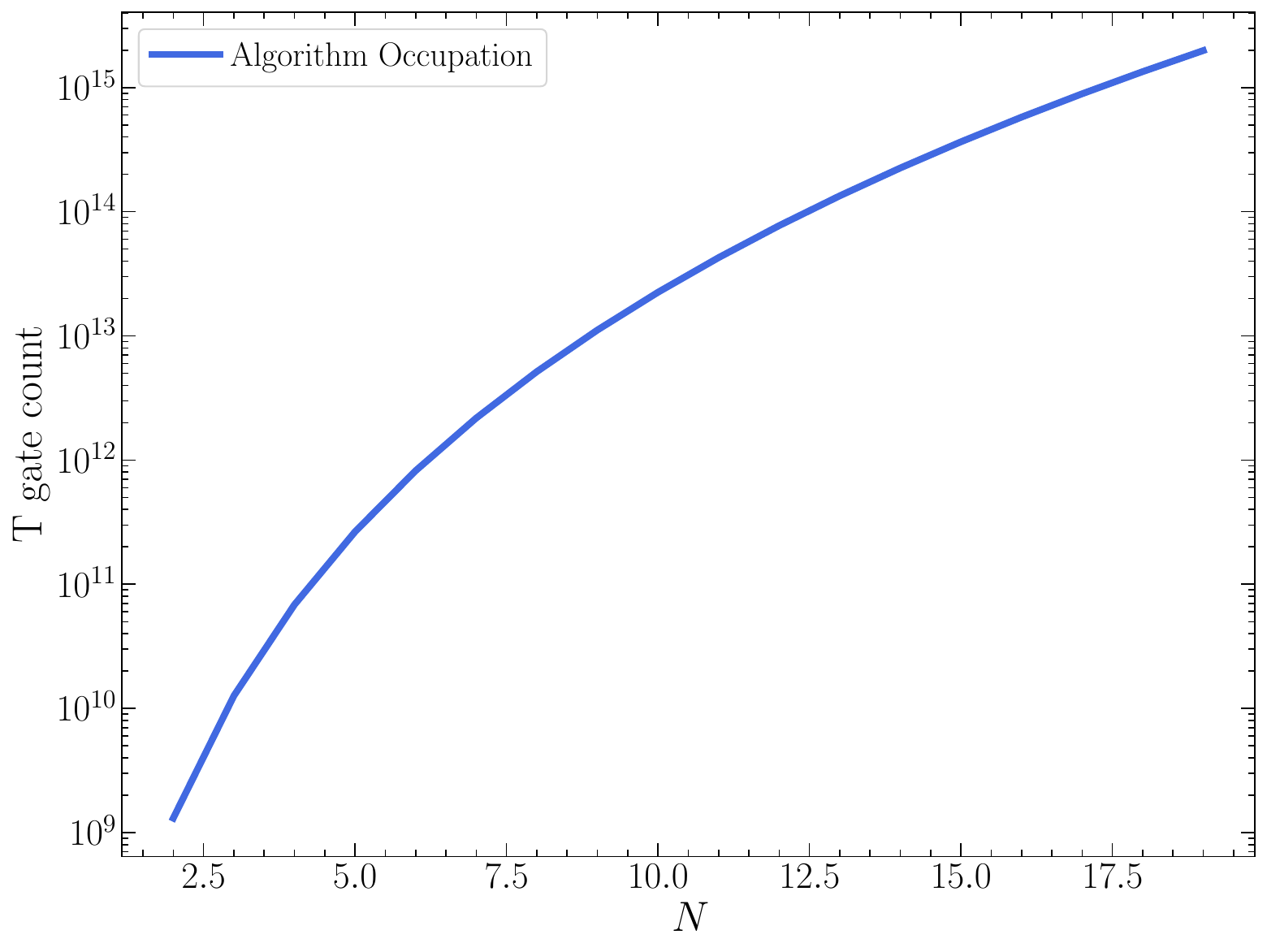}
    }
        \caption{a) The $T$ gate count on a $\log$ axis as a function of the field amplitude cutoff $k$ . Algorithm IIIb is plotted with a dotted line as the precise scaling rests upon conjectured behavior of the field operator binary decomposition (Conjecture \ref{conj:phi24})). Unknown constant prefactors for the  conjecture only for Algorithm IIIb have been set to 1 here.
        b) The $T$ gate count on a $\log$ axis. as a function of the field occupation cutoff $N$. For both, we consider a strong-coupling regime $\lambda =M =  1$, with $|\Omega| = 10^2$ and $\epsilon_E = 10^{-2}$. . } 
    \label{fig:gate_counts_cutoff}

    \end{figure}
    \begin{figure}

    \centering
    \includegraphics[width=.55\linewidth]{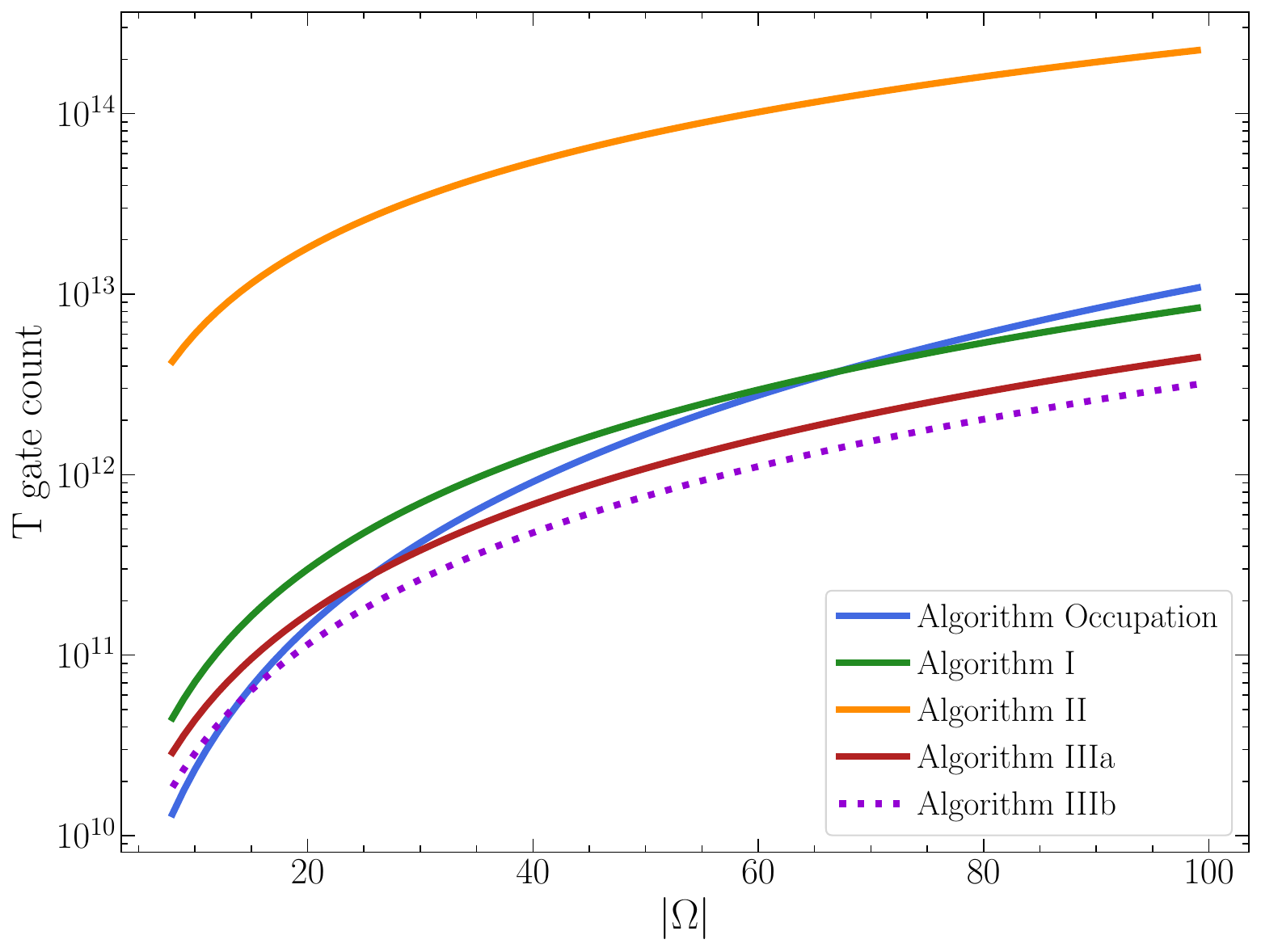}
    \caption{The $T$ gate count on a $\log$ axis as a function of the momentum volume cutoff $|\Omega|$. Here  we consider a strong-coupling regime $\lambda =M =  1$ and $\epsilon_E = 10^{-2}$. Here we have $k = 20$. 
    By dimensional analysis we expect an approximate scaling relation of $N \sim \sqrt{k}$. In order to illustrate the proper scaling relations (and optimum nature of the amplitude basis) with respect to $|\Omega|$, we select a larger (and therefore more accurate to the physics) value for $N = 9$. Algorithm IIIb is again plotted with a dotted line as the precise scaling rests upon conjectured behavior (Conjecture \ref{conj:phi24}).  }
    \label{fig:gate_counts_momentum}

\end{figure}

\begin{figure}[htb]
    \centering
    \subfloat{
    \centering
    \includegraphics[width=.5\linewidth]{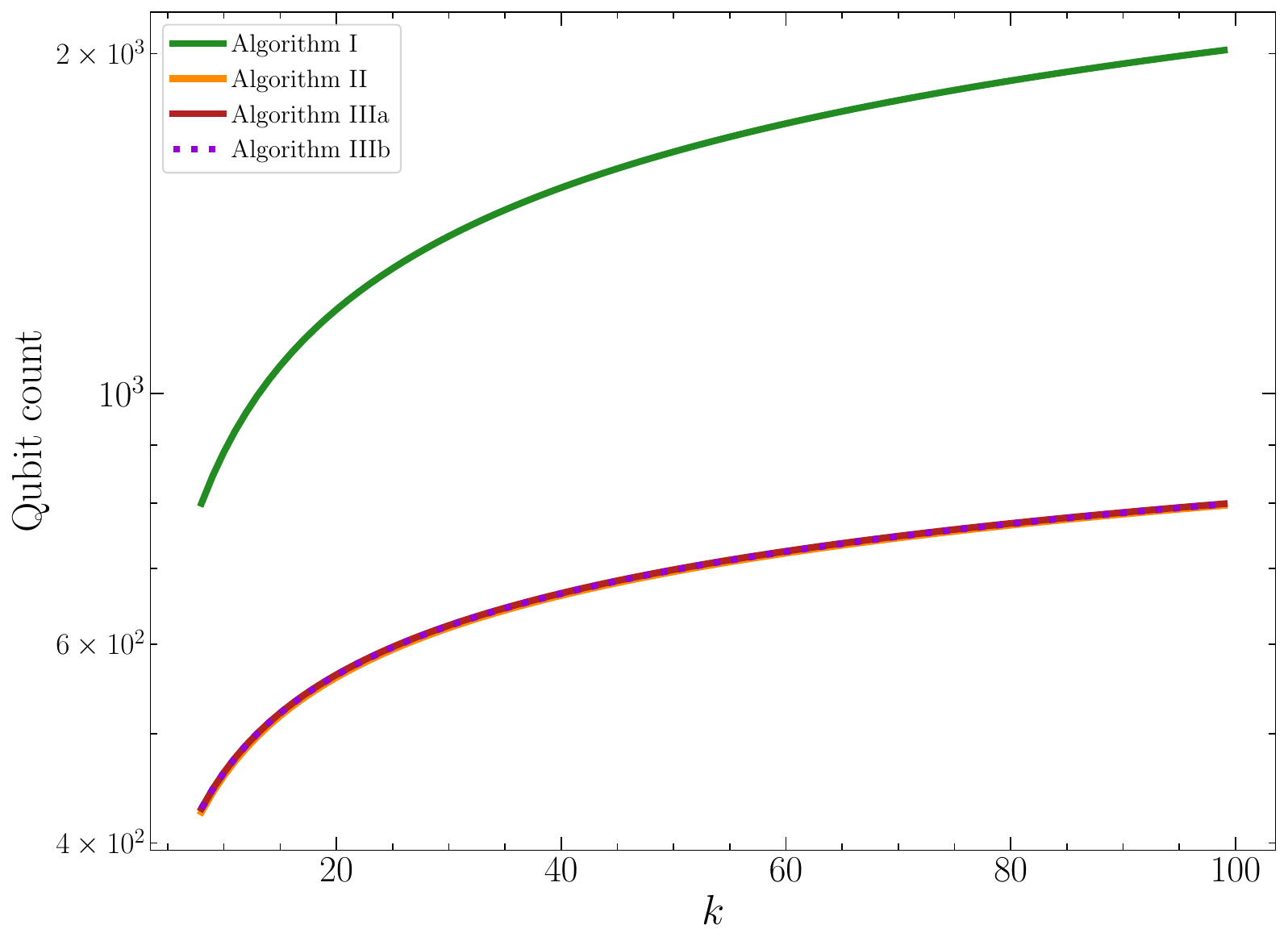}
    }
    \subfloat{
    \centering
    \includegraphics[width=.5\linewidth]{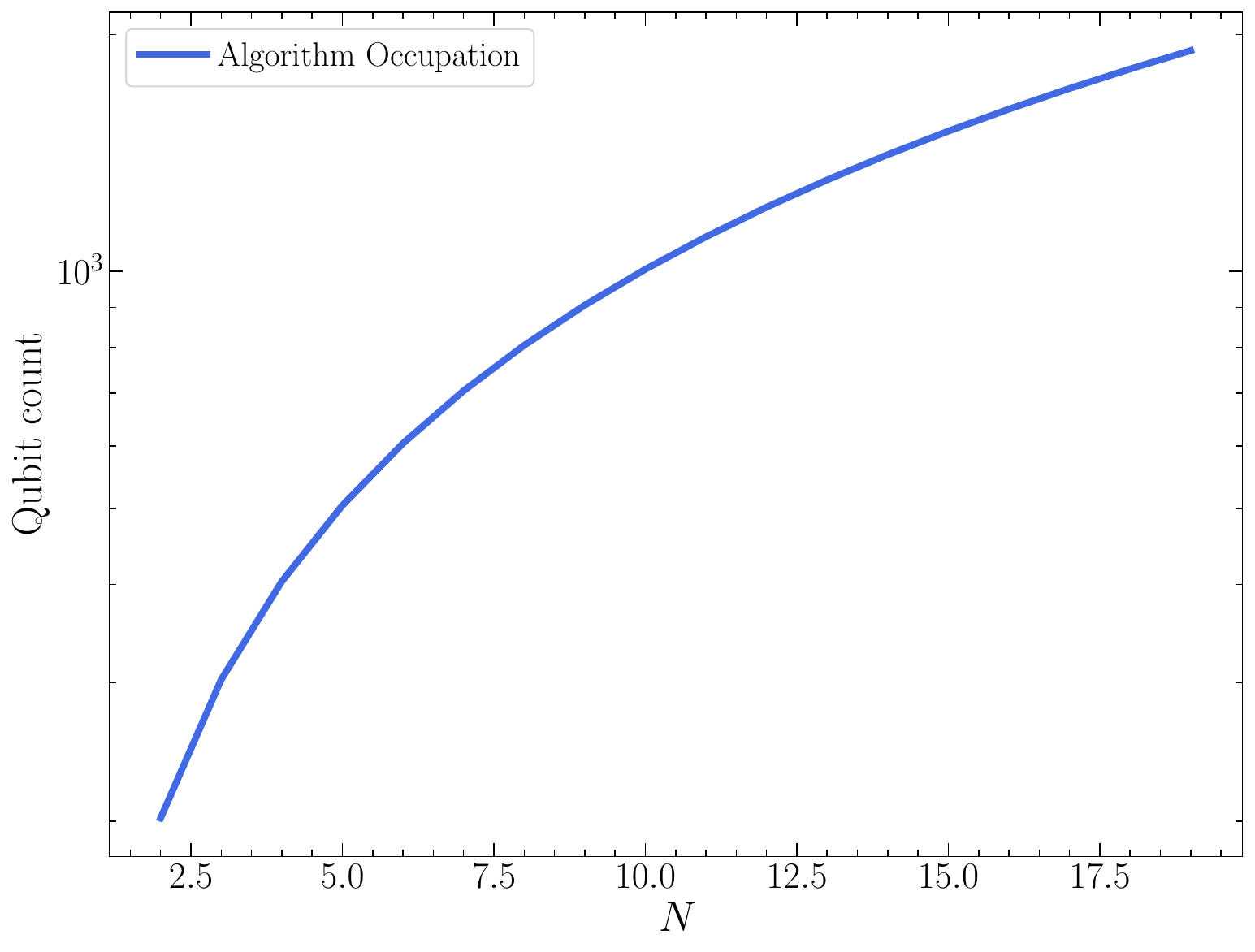}
    }
    \label{fig:qubit_counts_cutoff}
        \caption{a) The logical qubit on a $\log$ axis as a function of the field amplitude cutoff $k$ . Algorithm IIIb is plotted with a dotted line as the precise scaling rests upon conjectured behavior of the field operator binary decomposition (Conjecture \ref{conj:phi24})). Unknown constant prefactors for the conjecture have been set to 1 here.
        b) The exact logical qubiT-gate count on a $\log$ axis. as a function of the field occupation cutoff $N$. For both, we consider a strong-coupling regime $\lambda =M =  1$, with $|\Omega| = 10^2$ and $\epsilon_E = 10^{-2}$. } 
    \end{figure}

\begin{figure}
    \centering
    \includegraphics[width=.55\linewidth]{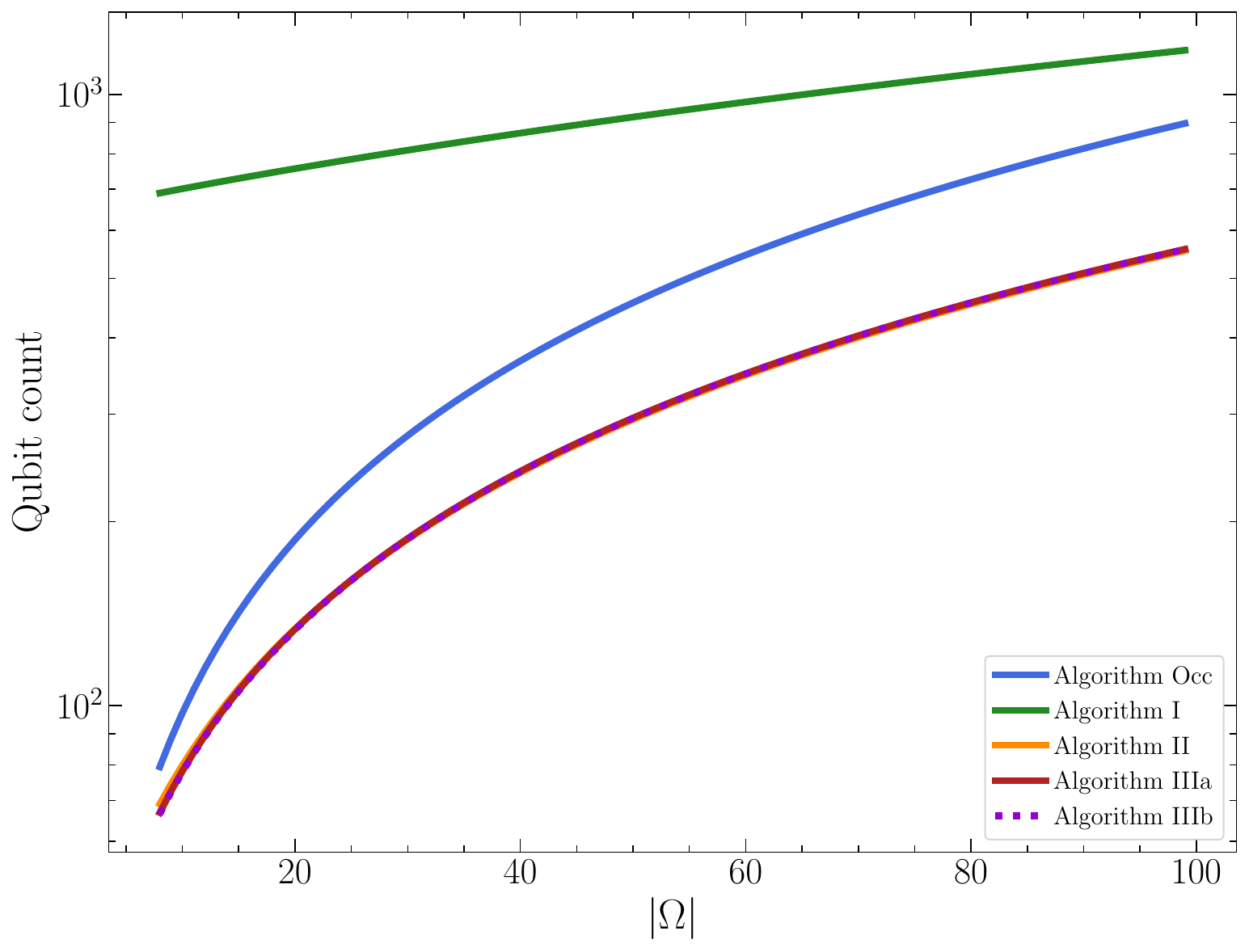}
    \caption{The logical qubit count on a $\log$ axis as a function of the momentum volume cutoff $|\Omega|$. Here  we consider a strong-coupling regime $\lambda =M =  1$ and $\epsilon_E = 10^{-2}$. Here we have $k = 20$. 
    By dimensional analysis we expect an approximate scaling relation of $N \sim \sqrt{k}$. In order to illustrate the proper scaling relations (and optimum nature of the amplitude basis) with respect to $|\Omega|$, we select a larger (and therefore more accurate to the physics) value for $N = 9$.  }
    \label{fig:qubit_counts_momentum}

\end{figure}

In Fig.~\ref{fig:gate_counts_cutoff}(a) we show the variation of T-gate count of the amplitude basis algorithms with respect to the field amplitude cutoff $k$ and in Fig.~\ref{fig:gate_counts_momentum}(b) we show the  variation of the T-gate count of the occupation basis algorithm with respect to the field occupation cutoff $N$. We consider the strong-coupling regime $\lambda = M = 1$ with $|\Omega| = 10^2$. We observe that the Trotter-based algorithms, both for the occupation basis and the amplitude basis (Algorithm II), have higher T-count than the qubitization-based algorithms (Algorithms I, IIIa and IIIb). One reason for this is the use of more number of rotation gates in Trotter-based algorithms. For a complete fault-tolerant implementation we decompose each of them further into Clifford+T. One solution to circumvent this problem can be the use of partial fault-tolerant implementation where the rotations are implemented non-fault-tolerantly but the Clifford gates have a fault-tolerant implementation \cite{2024_AMOetal}. Among the qubitization-based algorithms, Algorithm IIIb has the minimum T-gate count, but its estimates depend on Conjecture \ref{conj:phi24}. Algorithm IIIa has the minimum rigorously proved T-gate-count estimate.

In Fig.~\ref{fig:gate_counts_momentum} we plot the T-gate-count as a function of the momentum volume cutoff $|\Omega|$ in the strong-coupling regime $\lambda = M = 1$ and $\epsilon_E = 10^{-2}$. We consider the field amplitude cutoff $k = 20$ and an expected relation $N\sim \sqrt{k}$ with the field occupation cutoff. Here, we observe that the occupation basis Trotter algorithm performs better than all the amplitude basis algorithms in a small range. It performs better than Algorithms I and II for a larger range. For nearly the complete range of values considered, amplitude basis Algorithms IIIa and IIIb (with Conjecture \ref{conj:phi24}) has the minimum T-gate-count. 

In Fig.~\ref{fig:qubit_counts_cutoff}(a) we plot the logical qubit count of the amplitude basis algorithms as a function of the field amplitude cutoff $k$, while in Fig.~\ref{fig:qubit_counts_cutoff}(b) we show the logical qubit count of the occupation basis algorithm with respect to the occupation basis cutoff $N$. Again we consider the strong-coupling regime $\lambda = M = 1$ and $|\Omega| = 10^{2}$. We observe that the qubit count of Algorithms II, IIIa and IIIb is much better than the others. This is due to the LCU decomposition of $\Phi$, $\Phi^2$ and $\Phi^4$ operators based on binary representation of integers. We remember that the qubit cost of Algorithm IIIb does not depend on any conjecture.    

In Fig.~\ref{fig:qubit_counts_momentum} we show the variation of the logical qubit count of all the algorithms with respect to the momentum volume cutoff $|\Omega|$. Again, as explained before we consider the strong-coupling regime $\lambda = M = 1$ and $\epsilon_E = 10^{-2}$, $k = 20$, $N\sim\sqrt{k}$. Here we observe that the qubit count of the occupation basis Trotter algorithm is much less than the amplitude basis LCU-based algorithm I. However, again the qubit counts of the other amplitude basis algorithms - both Trotter-based Algorithm II and LCU-based Algorithms IIIa and IIIb, are the minimum among all the algorithms considered. 

In summary, the amplitude basis qubitization-based Algorithm IIIa and its improved version Algorithm IIIb have better cost estimates i.e. the minimum T-gate-count and logical qubit count compared to other algorithms. One reason for this is the particular binary representation based LCU decomposition of the operators. The Trotterization algorithm using this decomposition, though enjoys the benefit of lower qubit count, but has higher T-gate count because of more rotation gates, as mentioned earlier. The other amplitude basis qubitization-based Algorithm I has much lower T-gate-count in most cases compared to the Trotter algorithms - both in the occupation basis as well as Algorithm II. But it has much higher qubit count compared to all the algorithms, in nearly all the cases considered. The qubit cost of the occupation  basis Trotter algorithm is somewhat intermediate and its T-gate-count, though low in a small regime, soon becomes higher than the others. 

\section{Fault Tolerant Implementation}
\label{sec:FT}

To create fault-tolerant non-Clifford $T$ gates within the surface code, one needs to prepare highly accurate magic states $|A_L\rangle$ which are  then fed into the logical computation circuit to affect logical $T$ gates, $|A_L\rangle = TH|0\rangle$~\cite{Campbell2017}. Note that these states are prepared in an  area of the surface code that is separate from the area where the logical computation is taking place. The process of magic state distillation is resource intensive and is often responsible for the highest footprint of surface code-based fault-tolerant computation. 

In this section we provide an estimate for the resources required to generate sufficiently accurate $T$ gates, following Fowler's treatment in~\cite{PhysRevA.86.032324}. Assuming $N_T = 10^{12}$ logical $T$ gates and a physical gate measurement time of $t_m = 10^{-7}$ seconds, the shortest possible time to compute the $N_T$ consecutive gates is $t_c = t_m N_T = 10^5$ seconds, or around 28 hours. 

To calculate the physical qubit footprint, note that each logical $T$ gate needs one highly distilled ancilla (magic) state $|A_L\rangle$ injected into the logical circuit. The tolerable error rate per $|A_L\rangle$ state must then satisfy 
\be
P_{A_L} < 1/N_T = 10^{-12}. 
\ee
Assuming an injection error rate of $P_I = 10^{-3}$, and an error-free distillation circuit, the error rate associated with the first layer of distillation will be,
\be
P_1 = 35 P_I^3 \sim 3.5 \times 10^{-8}, 
\ee
which is greater than the required $P_{A_L}$, thus a second layer of distillation will be required. The error probability in this case will be,
\be
P_2 = 35 P_1^3 \sim 1.5 \times 10^{-21}, 
\ee
which is less than $P_{A_L}$, therefore two layers of distillation will be sufficient to purify an ancilla state with the desired accuracy. 

Using the 15-qubit Reed-Muller encoding for the distillation of $|A_L\rangle$, the first stage requires 15 sets of distillation circuits, each acting on 16 logical qubits, for a total of 240 logical qubits, operating in $8 \times 5d_1/4 = 10 \; d_1$ surface code cycles where $d_1$ is the distance associated with the first layer of distillation. To estimate $d_1$, note that the error rate for $|A_L\rangle$ after the first distillation is
\be
P_{A_{L_1}}  = 1800 \;  d_1 P_{L_1}.
\ee
Here the coefficient 1800 results from 240 logical qubits, two types of logical qubits, three types of error chains, and $5d_1/4$ surface code cycles. The logical error rate for the first layer $P_{L_1}$ is related to the distance with the following relation,
\be
P_{L_1} \simeq 3 \times 10^{-2} (\frac{P}{P_{th}})^\frac{d_1}{2},
\ee
where $P_{th} = 10^{-2}$ is the threshold error and we assume the a physical error rate of $P \simeq 10^{-3}$. 

We need to find $d_1$ such that 
\be
35 (P_{A_{L_1}})^3 < P_{A_L} < 10^{-12}. 
\ee
We find that the shortest distance for which this requirement is satisfied is for $d_1 = 15$.
Assuming the rate of $r_1 = 5/2 \times 5d_1^2/4$
physical qubits per logical qubit, we need $n_1 = 1.7 \times 10^5$ physical qubits and $t_1 = 10 \; d_1 = 150$ surface code cycles. 

The second stage of distillation requires another encoding with  16 logical qubits and $t_2 = 10 \; d_2$ surface code cycles, where again $d_2$ is the corresponding code distance for the second distillation. To find this distance, we need to satisfy,
\be
P_{A_{L_2}} = 120 \; d_2 P_{L_2} < P_{A_L} < 10^{-12},
\ee
where the coefficient, 120, stems from 16 logical qubits, two types of logical qubits, three types of error chains and $5d_2/4$ surface code cycles, and 
\be
P_{L_2} \simeq 3 \times 10^{-2} \left(\frac{P}{P_{th}}\right)^\frac{d_2}{2}.
\ee
 We find that for $d_2 = 29$ this requirement is satisfied. Given the rate of $r_2 = 5/2 \times 5d_2^2/4$ physical qubits per logical qubit, we need a total of $n_2 = 4.2 \times 10^ 4$ physical qubits and $t_2 = 10 d_2 = 290 $ surface code cycles for the second round of distillation.

Note that the qubits used in the first round of distillation can be reused during the second round, thus total resources required for two rounds of distillation will be $n_1 = 1.7 \times 10^5$ physical qubits and $t_1 + t_2 = 440$ surface code cycles or $t_{AAA} = 4.4 \times 10^{-5}$ seconds, assuming a surface code cycle time of 100 nanoseconds. Following the argument in~\cite{PhysRevA.86.032324}, this space-time footprint is enough to distill three ancilla states, or an AAA factory. 

If we use only one AAA factory, the time to distill all $10^{12}$ $|A_L\rangle$ states will be $t = 1.5 \times 10^7$ seconds or $\sim 6$ months. To achieve the minimal computation time of $t_c = 10^5$ seconds, we need to create parallel AAA factories. To calculate this number, note that each AAA factory prepares 3 ancilla states in time $t_{AAA} = 4.4 \times 10^{-5}$ seconds. Therefore, in time $t_c = 10^5$ seconds we can create $3 t_c / t_{AAA} = $ states. Thus to achieve the desired number of states $N_T = 10^{12}$ in $t_c = 10^5$ seconds we need to create $N_{AAA} = N_T/(3t_c/t_{AAA})$ or~$\sim 147$ parallel AAA factories. This parallelization will save us time but instead increases our space footprint to $147 \times 1.7 \times 10^5 \simeq 2.5 \times 10^7$ physical qubits. 

If we use the distance of the second layer of distillation for our logical computation, the cost of encoding a logical qubit will be  $5/2 \times 5 \times (16)^2/4 = 2.6 \times 10^6$ physical qubits, which is $\sim 10\%$ of the number of physical qubits needed for the distillation of all the required ancilla states in the desired time. Thus the total footprint for the computation will be $2.8 \times 10^7$ physical qubits.   

If we reduce the error per gate to $P \simeq 10^{-4}$, we can reduce the spacial footprint of the distillation circuit. Following the same logic as before we find that we still need two layers of distillations, where for the first layer we need a code with distance $d_1 = 8$, $n_1 \simeq 4.8 \times 10^4$ physical qubits and  $t_1 = 80$ cycles. For the second round we need $d_2 = 14$, $n_2 \simeq 9.8 \times 10^3$ physical qubits and $t_2 = 140$ cycles. As before the qubits in the first layer can be recycled in the second layer so the total required resources per distilled state will be $n = n_1 = 4.8 \times 10^4$ qubits and $t = t_1 + t_2 = 220$ cycles or $2.2 \times 10^{-5}$ seconds for an AAA factory. If we were to use only a single AAA factory, the time to prepare $N_T = 10^{12}$ state would be $7.3 \times 10^6$ seconds or $\sim 3$ months. To keep pace with the computational time $t_c = 10^{-5}$ seconds, we need to use $10^{12}/(3 \times 10^5 /(2.2 \times 10^{-5})) \sim 74$ parallel AAA factories with the footprint of $\sim 3.5 \times 10^{6}$ physical qubits.

A logical computational qubit in this case will cost $3.125 \times 14^2 = 612$ physical qubits, hence for 1000 logical qubits we need $\sim 6.2 \times 10^5$ physical qubits, which is is~$\sim 17\%$ of the distillation footprint, bringing the total number of physical qubits to $4.2 \times 10^6$. A summary of these results is given in table~\ref{table:FT_sum}.

\begin{table}[h!]
\begin{center}
\begin{tabular}{ |c||c||c| } 
 \hline
 $p/p_{th}$ & $10^{-1}$ & $10^{-2}$ \\ 
 \hline
  \hline
 First distillation distance $d_1$ & 15 & 8 \\ 
 \hline
  Second distillation distance $d_2$ & 29 &  14\\ 
 \hline
Time per AAA (s) & $4.4 \times 10^{-5}$ &  $2.2 \times 10^{-5}$\\ 
 \hline
Total time for $10^{12}$ A states without parallelization (s) & $1.5 \times 10^{7}$&  $7.3 \times 10^{6}$\\ 
 \hline 
 Total time for $10^{12}$ A states with parallelization (s) &$1.0 \times 10^5 $& $1.0 \times 10^5 $\\ 
 \hline 
 Number of parallel AAA factories & 147 &  74\\ 
 \hline
 Number of Physical qubits per AAA factory &  $1.7 \times 10^{5}$ & $4.8 \times 10^{4}$ \\ 
 \hline
Total Physical qubits for distillation with parallelization & $2.5 \times 10^7$ &  $3.5 \times 10^6$ \\
 \hline
 Physical qubits for 1000 logical computational qubits & $2.6 \times 10^6$ &  $7.4 \times 10^5$ \\ 
 \hline
Total physical qubits including computational qubits & $ 2.8 \times 10^7 $ &  $4.2 \times 10^6$ \\ 
 \hline
\end{tabular}
\end{center}
\caption{Summary of the estimated resources required for the fault-tolerant implementation of the algorithm using the surface code. Note that the best possible logical computation time, which is limited by the gate time of $t_m = 10^{-7}$ seconds, is $t_c = 10^{5}$ seconds or $\sim 28$ hours. This time can be achieved by parallelizing magic state factories whose footprint is partially determined by the ratio $p/p_{th}$}. 
\label{table:FT_sum}
\end{table}

The calculations in this section follow closely the arguments of~\cite{fowler2012surface}. Depending on the need to optimize either time or space, the methods discussed in more recent references, e.g.~\cite{Litinski2019}, can be used to modify the estimates. 

\section{Discussion}
\label{sec:Conclusion}
A central challenge that has remained unanswered within the quantum simulation community involves deciding whether simulation of scalar field theories can be practically done on a quantum computer. 
Here we have addressed this by providing a method that uses phase estimation to estimate elements of the $S$-matrix for elastic collisions and designing optimized circuits for Qubitization and Trotter-based simulation of scalar field theory in amplitude or occupation basis.
We find that simulation of scalar field theory in $1+1$D with a occupation cutoff of $9$ or field cutoffs of $20$ with a field volume $\Omega=100$ leads to a number of $T$ gates that is on the order of $10^{12}$ and the number of logical qubits that are on the order of $1000$ for the occupation basis and in on the order of $500$ logical qubits for the qubitization-based amplitude basis algorithm.  We show that the calculation can be performed in just over a day.

These estimates are predicated on $5$ new algorithms considered in this paper. In the occupation basis we describe a Trotter-based simulation algorithm. Of the 4 algorithms described in the amplitude basis, the three qubitization-based ones perform better. Among these, the most efficient one in terms of T-gate-count and number of logical qubits, is the one where the LCU decomposition of operators have been done using binary representation of integers. We can prove that the number of $T$ gates needed for the Trotter-based algorithm, for coupling strength $\lambda$, occupation cutoff $N$, mass $M$, and energy uncertainty $\epsilon_E$, is
\begin{equation}
    N_{T,occ} \in \widetilde{O}\left(\frac{\lambda N^{7}|\Omega|^{3}}{M^{5/2}\epsilon_E^{3/2}}\right).
\end{equation}
We compute this by evaluating commutator bounds on the second-order Trotter formula.  This algorithm performs best in circumstances where the particle mass is large, when the number of particles in occupation basis is dramatically lower than the maximum field or when the coupling strength is relatively weak.
Higher-order formulas can give better scaling, but we focus on the low-order formulas because the tightest bounds available are for the second order formula~\cite{Childs2021,childs2018toward}.  It is worth noting however, that even the tightest bounds on Trotter error are pessimistic~\cite{wecker2014gate,babbush2015chemical,reiher2017elucidating,childs2018toward}. Nonetheless, we note that, if one is interested in high energy scattering, this basis  can be ideally suited for such problems. This results from the fact at high energies the particles participating in scattering events can be thought as being nearly free (i.e. weak coupling), as long as the occupation number is small, while a field-based picture would fail to efficiently describe such a scenario. Further, the occupation or particle based pictures have advantages in the extraction of physical observables such as particle number, while in the field basis such quantities are harder to obtain. For further discussion on these issues, see ~\cite{barata2021single, Li:2021zaw,Barata:2023clv}. 

The most performant field amplitude basis algorithms are based on qubitization and for field cutoff $k$ require a number of $T$-gates that scale as
\begin{equation}
N_{T,amp}=\widetilde{O} \left( \frac{\vert\Omega\vert^{2}}{\epsilon_E} \left[ k^{2}\Lambda + k M^{2} \right] \right)
\end{equation}
where $\Lambda$ is a rescaled interaction strength.  These bounds provide much better scaling with the error tolerance and slightly better scaling with the volume of the simulation.  It is worth noting that this scaling is likely to be close to the empirical performance of the algorithm because the error estimates for qubitization tend to be much tighter than those of the Trotter formula~\cite{childs2018toward}.

This work opens up a number of additional questions.  In particular, the field needs development of improved methods for computation of the elements of the $S$-matrix in higher dimensions or for inelastic collisions.  Further more detailed estimates of the Trotter error may reveal that the cost of these methods are substantially lower than current estimates.  In a similar vein, identification of algorithms that may only require partial error correction can allow these applications to become feasible in early fault-tolerant quantum computers. For example, one possible candidate may be the Trotter-based algorithm in the amplitude basis.
Future conceptual work is also needed for understanding the formal computational complexity of specific regimes of the $S$-matrix.  Similar work could ask about more efficient quantum processes for inverting Eq. \ref{eq:inversion}. This poses an interesting question about the complexity for inverse scattering problems in general. 

Perhaps the most important issue that needs to be addressed involves extending this work to more realistic theories.  Scalar $\phi^4$ theory is often used as a toy theory rather than one that accurately models realistic scattering events in colliders. Previous works have explored a variety of strategies to simulate gauge field theories, including schemes that employ the electric (or representation) basis and impose a cutoff on the maximal value for the electric field \cite{calajò2024digitalquantumsimulation11d,PhysRevD.98.094511,PhysRevD.99.074502,PhysRevD.101.074512,PRXQuantum.5.020315,PhysRevD.109.114510,Illa2024,PhysRevD.103.094501,PhysRevD.92.076003,PhysRevLett.121.223201,PhysRevD.99.114507,bauer2021efficientrepresentationsimulatingu1,grabowska2023overcomingexponentialscalingsize,PhysRevD.102.114514,PhysRevLett.126.172001,kavaki-square-plaq,PhysRevD.106.094504,Zohar_2016,PhysRevLett.110.125304,PhysRevLett.109.125302,PhysRevA.88.023617}, the loop-string hadron formulation, which explicitly enforces gauge invariance on the Hilbert space before truncation \cite{PhysRevD.101.114502,davoudi2024scatteringwavepacketshadrons,PhysRevResearch.2.033039,PhysRevD.104.074505,PhysRevD.106.054510}, the light-front quantization method \cite{PhysRevA.105.032418,PhysRevA.103.062601,e23050597}, the discrete subgroup approximation of the continuous Lie groups relevant to the Standard Model of particle physics \cite{PhysRevD.105.114501,PhysRevD.106.114501,gustafson2023primitivequantumgatessu2,gustafson2024primitivequantumgatessu3,Bender_2018,PhysRevA.99.062341,PhysRevD.100.114501,10.1093/ptep/ptaa171,Haase2021resourceefficient,PhysRevLett.127.250501,PhysRevD.102.114513,PhysRevD.104.094519,PhysRevLett.129.160501}, as well as continuous variable \cite{PhysRevA.92.063825,PhysRevA.109.052412,PhysRevD.97.036004,cv-harmonic-oscillators,Thompson:2023kxz} and qudit based \cite{PhysRevD.103.114505,gustafson2022noiseimprovementsquantumsimulations,PhysRevLett.129.160501,Illa2024,Illa2023QuantumSO,turro2024qutritqubitcircuitsthreeflavor,kurkcuoglu2022quantumsimulationphi4theories,vezvaee2024quantumsimulationfermihubbardmodel,Roy2023QuditbasedQC,meth2024simulating2dlatticegauge,Hanada:2022pps} approaches.

Further studies such as these and ours are needed to consider gate count estimates for more realistic field theories. The  direct extension of this work would be to consider complex scalar theories, or vector field theories. The next step towards realistic theories would then be to develop algorithms for boson-fermion coupled theories.  It is our hope that this will not only reveal that field theory can be practically simulated on a quantum computer, but also help us understand the ultimate goal of figuring out whether quantum computers can simulate all physically realistic processes in polynomial time.

\section{Acknowledgments}
This work was primarily supported by the U.S. Department of Energy,
Office of Science, National Quantum Information Science
Research Centers, Co-design Center for Quantum Advantage
(C2QA) under Contract Number DE-SC0012704. 
AH acknowledges support from a NSERC Graduate Fellowship. MSA, ER, and SH acknowledge support from the U.S. Department of Energy, Office of Science, National Quantum Information Science Research Centers, Superconducting Quantum Materials and Systems Center (SQMS) under contract No. DE-AC02-07CH11359. MSA and SH acknowledge support from USRA NASA Academic Mission Services under contract No. NNA16BD14C with NASA, with this work funded under the NASA-DOE interagency agreement SAA2-403602 governing NASA's work as part of the SQMS center. RK acknowledges support by the U.S. Department of Energy, Office of Basic Energy Sciences, under Contract No. DE-SC0012704.  RV is supported by the U.S. Department of Energy, Office of Science, under contract DE- SC0012704. We thank Erik Gustafson, Vladimir Korepin, Robert Pisarski, Julia Wildeboer, and Ted Yoder for useful discussions.  We thank Marton Lajer both for discussion and for providing the data for the reproduced figures from \cite{Bajnok2016} found in Figs. 1 and 2.

\clearpage

\appendix


\section{Field Occupation Basis}
\label{subsec:algoOcc}
In this Appendix we derive the quantum circuits required to simulate the interacting Hamiltonian $H_{\varphi}$ (Eq. \ref{eqn:Hvarphi}). As explained in Section \ref{subsec:algoOcc}, we divide the terms in the sum into 4 groups, i.e. $H_{\varphi} := H_{1\varphi}+H_{2\varphi}+H_{3\varphi}+H_{4\varphi}$. We map the resulting bosonic expressions into qubit space. We use the following two lemmas repeatedly in order to derive the Pauli expressions for the sum in each of these groups (Eq. \ref{eqn:HphiGroup1}, \ref{eqn:HphiGroup2}, \ref{eqn:HphiGroup3}, \ref{eqn:HphiGroup4}). \par
We provide an overview of the proof structure going into the results presented in Table \ref{tab:compareCost}. To aid in the flow of the argument, we have not provided some technical details in this section. Instead, these can be found in the Supplemental Material \cite{SuppMat}, for example, the proofs of the following two lemmas. 

\begin{lemma}
If $\hat{n}_p$ is the number operator on momentum mode $p$ then for any integer $r\geq 1$ we have,
\begin{equation}
 \left(\hat{n}_p\right)^r=\sum_{n} \frac{n^r}{2} (I_n -Z_n)_p.  \nonumber
\end{equation}
\label{app:lem:npMap}
\end{lemma}

\begin{lemma}
If $m\geq 1$ and $r\geq 0$ are integers then we have,
\begin{equation}
    (a_p^\dagger)^m (n_p)^r + (n_p)^r(a_p)^m = \frac{1}{2}\sum_n \sqrt{\frac{(n+m)!}{n!}} n^r
    (X{_n}X_{n+m}+ Y_{n}Y_{n+m})_p.
\end{equation}

\label{app:lem:npapMap}
\end{lemma}

We first consider the noninteracting term in Eq.\ref{eqn:H_occ}  and derive the circuit required to simulate the unitary exponential, with the following result.
\begin{lemma}[Complexity of noninteracting Hamiltonian Simulation]
We require at most $N|\Omega|$ number of $R_z$ gates to simulate $e^{-iH_0't}$, where $H_0'=:H_0:$.
\label{lem:cktH0}
\end{lemma}
\begin{proof}
The normal ordered $H_0$ i.e. $:H_0:$ is,
\begin{equation}
  :H_0: =    \sum_p \omega_p a^\dagger_p a_p =   \sum_p \omega_p \hat{n}_p =\sum_{n,p}\frac{n\omega_p}{2}(I_n-Z_n)_p,    \label{eqn:H0}
\end{equation}
using Lemm \ref{app:lem:npMap}. The above expression, up to some global phase is equal to a sum of Z operators. And the exponential of each of them can be implemented with a $R_z$ gate, whose angle depends on the coefficient. Thus the lemma follows. 
\end{proof}

\paragraph{Circuits for simulating the interacting Hamiltonian : } We now consider the interacting part of the Hamiltonian i.e. $H_{\varphi}$, which we are rewriting as follows, for convenience. Let $S_{4\bp}=\{\bp=(p_1,p_2,p_3,p_4):p_i\in\Gamma; i=1,2,3,4 \}$, be a set consisting of ordered 4-tuples of the momentum mode, that respects the conservation of momentum constraint ($p_1 + p_2 = \pm (p_3 + p_4)  $). We parameterize this constraint by working in the basis such that such that $\pm p_3=p_1+k$ and $\pm p_4=p_2-k$.

Now, we can divide the terms in the above sum into 4 groups, based on equality of the momentum modes in $\bp$, i.e. $H_{\varphi}:=H_{1\varphi}+H_{2\varphi}+H_{3\varphi}+H_{4\varphi}$. For each such group, we map the resulting bosonic expression into the qubit space using Equation \ref{eqn:ferm2pauli}, Lemma \ref{app:lem:npMap}, Lemma \ref{app:lem:npapMap}, and obtain the Pauli expression. From this, we derive the quantum circuit for the exponentiated sum in each group. We follow the methods described in \cite{2022_MWZ}. Very briefly, we do the following
\begin{enumerate}
    \item Divide the Pauli terms into mutually commuting sets.

    \item For each such set we derive an eigenbasis and diagonalize the operator.
    
    \item From the diagonalized operator we calculate the number of distinct non-zero eigenvalues (ignoring sign), which is equal to the number of (controlled)-rotations we require.

    \item We apply some logical reasoning and optimizations to derive the remaining elements of circuits.
\end{enumerate}
The resulting Trotter error due to such splitting has been calculated in Supplemental Material (Sec. II) \cite{SuppMat}. We would like to emphasize that the quantum circuits and hence resource estimates depend on the grouping into commuting Paulis and we do not claim to give the optimal grouping in this paper.

\subsection*{The commuting groups}

Now we describe the grouping of the Paulis into mutually commuting sets. In the following discussions let $S_{4\vec{n}}=\{\vec{n}=(n_1,n_2,n_3,n_4) : n_i=0,1,\ldots,N; i=1,2,3,4\}$ be an ordered 4-tuple of momentum states.
We reserve $\mathbf{p}$ for vectors of unbolded $p,k, q$ which themselves may be $D$-dimensional vectors.  

\paragraph*{Group I: All distinct momentum modes : $\mathbf{p_1\neq p_2, k\neq 0}$. 
}
We take Eq.\ref{eqn:H_int} with Hermitian terms grouped as 
\begin{equation}
\begin{gathered}
H_\lambda =  \frac{\lambda}{4!|\Omega|^3} \sum_{\lbrace\bp_i\rbrace} \frac{1}{4\sqrt{\omega_{\bp_1} \omega_{\bp_2} \omega_{\bp_3} \omega_{-(\bp_1 + \bp_2 + \bp_3)}}} 
 \left\lbrace \left( a_{\bp_1} + a_{\bp_1}^{\dagger} \right) \left( a_{\bp_2} + a_{\bp_2}^{\dagger} \right)  \left( a_{\bp_3} + a_{\bp_3}^{\dagger} \right) \left( a_{-(\bp_1 + \bp_2 + \bp_3)} + a_{-(\bp_1 + \bp_2 + \bp_3)}^{\dagger} \right)\right\rbrace 
\end{gathered}
\end{equation}
This state divides into a tensor product of terms of the following form. 
\begin{eqnarray}
    :H_{\theta}:=\frac{\lambda}{24 |\Omega|}\sum_{\bp\in S_{4\bp}}\prod_{p_i\in\bp}\frac{1}{\sqrt{2w_{p_i}}}\left(a_{p_i}+a_{p_i}^{\dagger}\right)
    \label{eqn:Hvarphi}
\end{eqnarray}

We denote the sum of the terms having distinct momentum modes by $H_{1\varphi}$. We note that this Hamiltonian is trivially normally order because distinct momentum operators commute.  After the qubit mapping using Eq. \ref{eqn:ferm2pauli}, we get the following.
\begin{eqnarray}
    H_{1\varphi}&=&\frac{\lambda}{24|\Omega|^3}\sum_{\bp\in S_{4\bp}}\prod_{p_i\in\bp}\frac{1}{\sqrt{2w_{p_i}}}\left(\sum_{n}\frac{1}{2}\sqrt{n+1}\left(X_{p_i,n}X_{p_i,n+1}+Y_{p_i,n}Y_{p_i,n+1}\right)\right)   \nonumber \\
    &=&\frac{\lambda}{96\cdot 16|\Omega|^3 }\sum_{\bp\in S_{4\bp}}\sum_{\vec{n}\in S_{4\vec{n}}}\prod_{(p_j,n_j)\in (\bp,\vec{n})}\sqrt{\frac{n_j+1}{w_{p_j}}} \left(X_{p_j,n_j}X_{p_j,n_j+1}+Y_{p_j,n_j}Y_{p_j,n_j+1}\right) 
    \label{eqn:HphiGroup1}
\end{eqnarray}

\paragraph*{Group II : Two distinct momentum modes : $\mathbf{p_1 = p_2, k \neq 0}$. }

Let $H_{2\varphi}$ is the sum of the terms with momentum modes satisfying the given constraint. Then,
\begin{eqnarray}
  : H_{2\varphi} :  
  &=& \frac{\lambda}{24  |\Omega|^2 }\sum_{p,k} \frac{1}{4  \omega_{p} \sqrt{ \omega_{p+k} \omega_{p-k}} }\left((a_{p} + a^\dagger_{p})^2 (a_{p+k} + a^\dagger_{p+k})(a_{p-k} + a^\dagger_{p-k})\right)      \nonumber \\
    &=& \frac{\lambda}{96 } \sum_{p,k} \frac{1}{\omega_{p} \sqrt{ \omega_{p+k} \omega_{p-k}} }  \left( (a^\dagger_p)^2 + (a_p)^2  + 2 n_p \right)  \left(a_{p+k} + a^\dagger_{p+k}\right)\left(a_{p-k} + a^\dagger_{p-k}\right),  \nonumber 
 \end{eqnarray}
and after using after using Eq. \ref{eqn:ferm2pauli}, Lemma \ref{app:lem:npMap}, Lemma \ref{app:lem:npapMap} we get the following.

 \begin{eqnarray}
 :H_{2\varphi} :
    &=&\frac{\lambda}{96 |\Omega|^2}\sum_{p,k}\frac{1}{\omega_p\sqrt{\omega_{p+k}\omega_{p-k}}}\left(\left(\sum_n\frac{\sqrt{(n+2)(n+1)}}{2}(X_nX_{n+2}+Y_nY_{n+2})_p\right)+\left(\sum_nn(I_n-Z_n)_p\right)\right)  \nonumber \\
    &&\left(\sum_n\frac{\sqrt{n+1}}{2}(X_nX_{n+1}+Y_nY_{n+1})_{p+k}\right)\left(\sum_n\frac{\sqrt{n+1}}{2}(X_nX_{n+1}+Y_nY_{n+1})_{p-k}\right)  \nonumber \\
    &=&\frac{\lambda}{96 |\Omega|^2}\sum_{p,k}\frac{1}{\omega_p\sqrt{\omega_{p+k}\omega_{p-k}}}\nonumber\\
    &&\quad\times\left(\sum_{n_1,n_2,n_3}c_{n}^{(1)}(X_{n_1}X_{n_1+2}+Y_{n_1}Y_{n_1+2})_p 
    (X_{n_2}X_{n_2+1}+Y_{n_2}Y_{n_2+1})_{p+k}(X_{n_3}X_{n_3+1}+Y_{n_3}Y_{n_3+1})_{p-k}   \right.    \nonumber \\
    &&\left.\qquad+\sum_{n_1,n_2,n_3}c_n^{(2)}(I_{n_1}-Z_{n_1})_p(X_{n_2}X_{n_2+1}+Y_{n_2}Y_{n_2+1})_{p+k}(X_{n_3}X_{n_3+1}+Y_{n_3}Y_{n_3+1})_{p-k}   \right)
    \label{eqn:HphiGroup2}
\end{eqnarray}

where $c_{n}^{(1)}=\frac{\sqrt{(n_1+2)(n_1+1)(n_2+1)(n_3+1)}}{8}$, $c_{n}^{(2)}=\frac{n_1\sqrt{(n_2+1)(n_3+1)}}{4}$.

\paragraph*{Group III : Two distinct momentum modes : $\mathbf{p_1 \neq p_2, k = 0 }$.} 
We use $H_{3\varphi}$ to denote the sum of the terms with momentum mode satisfying the given constraint. Then, 
\begin{eqnarray}
    : H_{3\varphi} : &=&  \frac{\lambda}{24 |\Omega|^2 }\sum_{p_1,p_2} \frac{1}{4 \omega_{p_1} \omega_{p_2} }\left(a_{p_1} + a^\dagger_{p_1}\right)^2 \left(a_{p_2} + a^\dagger_{p_2}\right)^2   \nonumber \\
  &=&  \frac{\lambda}{96}\sum_{p_1,p_2} \frac{1}{\omega_{p_1} \omega_{p_2} }  \left( (a^\dagger_{p_1})^2 + (a_{p_1})^2  + 2 \hat{n}_{p_1} \right)  \left( (a^\dagger_{p_2})^2 + (a_{p_2})^2  + 2 \hat{n}_{p_2} \right).
\end{eqnarray}
After applying Eq. \ref{eqn:ferm2pauli}, Lemma \ref{app:lem:npMap}, Lemma \ref{app:lem:npapMap} we obtain the following.
\begin{eqnarray}
    && :H_{3\varphi}:   \nonumber \\
    &=&\frac{\lambda}{96 |\Omega|^2}\sum_{p_1,p_2}\frac{1}{\omega_{p_1}\omega_{p_2}} \left(\sum_n\frac{\sqrt{(n+1)(n+2)}}{2}(X_nX_{n+2}+Y_nY_{n+2})_{p_1}+\sum_nn(I_n-Z_n)_{p_1}\right) \nonumber \\
    &&\left(\sum_n\frac{\sqrt{(n+1)(n+2)}}{2}(X_nX_{n+2}+Y_nY_{n+2})_{p_2}+\sum_nn(I_n-Z_n)_{p_2}\right)    \nonumber \\
    &=&\frac{\lambda}{96 |\Omega|^2}\sum_{p_1,p_2}\frac{1}{\omega_{p_1}\omega_{p_2}}\left(\sum_{n_1,n_2} c_n^{(3)}(X_{n_1}X_{n_1+2}+Y_{n_1}Y_{n_1+2})_{p_1}(X_{n_2}X_{n_2+2}+Y_{n_2}Y_{n_2+2})_{p_2}       \right.  \nonumber \\
    &&+\sum_{n_1,n_2}c_n^{(4)}(X_{n_1}X_{n_1+2}+Y_{n_1}Y_{n_1+2})_{p_1}(I_{n_2}-Z_{n_2})_{p_2} 
    +\sum_{n_1,n_2}c_n^{(5)}(X_{n_2}X_{n_2+2}+Y_{n_2}Y_{n_2+2})_{p_2}(I_{n_1}-Z_{n_1})_{p_1} \nonumber \\
    &&\left.+\sum_{n_1,n_2}n_1n_2(I_n-Z_{p_1,n_1}-Z_{p_2,n_2}+Z_{p_1,n_1}Z_{p_2,n_2})   \right)
    \label{eqn:HphiGroup3}
\end{eqnarray} 
where $c_n^{(3)}=\frac{\sqrt{(n_1+2)(n_1+1)(n_2+2)(n_2+1)}}{4}$, $c_n^{(4)}=\frac{n_2\sqrt{(n_1+2)(n_1+1)}}{2}$, $c_n^{(5)}=\frac{n_1\sqrt{(n_2+2)(n_2+1)}}{2}$.

\paragraph*{Group IV: All equal momentum modes : $\mathbf{p_1 = p_2, k = 0}$. } 

The sum of the terms with all four equal momentum mode is denoted by $H_{4\varphi}$. These terms must be normal ordered to be 

\begin{eqnarray}
 : H_{4\varphi} : &=& \frac{\lambda}{24 |\Omega|} \sum_{p} \frac{1}{4 (\omega_{p})^2} \left\{ \left(a^\dagger_p    + a_p\right)^4\right\}  \nonumber \\
 &=&  \frac{\lambda}{24 |\Omega|} \sum_{p} \frac{1}{4 (\omega_{p})^2} \left\{ \left( (a^\dagger_p)^4    + (a_p)^4 \right) + 4\left( (a_p^\dagger)^3 a_p  + a^\dagger_p (a_{p})^3\right) + 6 \left( (a^\dagger_p)^2(a_p)^2\right) \right\}    \nonumber \\
&=& \frac{\lambda}{96 |\Omega|}\sum_{p} \frac{1}{(\omega_p)^2} \left( ( (a^\dagger_p)^4    + (a_p)^4 ) + 
4 \left( (a_p^\dagger)^2 \hat{n}_p  + \hat{n}_p (a_{p})^2 \right) + 6\left(   (\hat{n}_p)^2 -\hat{n}_p \right)  \right);   \nonumber 
\end{eqnarray}
and after applying Eq. \ref{eqn:ferm2pauli}, Lemma \ref{app:lem:npMap}, Lemma \ref{app:lem:npapMap} we get the following.
\begin{eqnarray}
    && :H_{4\varphi}:   \nonumber \\
    &=& \frac{\lambda}{96|\Omega| }\sum_{p,n}\frac{1}{(\omega_p)^2}\left(\frac{\sqrt{(n+4)(n+3)(n+2)(n+1)}}{2} (X_{n}X_{n+4}
    + Y_{n} Y_{n+4})_p \right. \nonumber \\
    &&\left.+2n\sqrt{(n+2)(n+1)} (X_{n}X_{n+2} + Y_{n} Y_{n+2})_p +3(n^2-n)(I_n-Z_n)_p    \right)  
    \label{eqn:HphiGroup4}
\end{eqnarray}

\subsection*{Quantum circuits for exponentiated Hamiltonians}

Before we consider our four Hamiltonian groups, we derive quantum circuits for the exponential of some specific summation of Paulis. We fix some convention and notation. We denote $P_0:=X$, $P_1:=Y$ and for any binary variable $v$ we denote its complement by $\overline{v}:=1\oplus v$.  Consider the following two sums of Paulis that act on $2n$ and $2n+1$ qubits respectively, which we index by $1,2,\ldots, 2n+1$. We denote $Z_{(j)}$ to imply that the operator Z acts on qubit $j$. When we denote a Pauli by $P_j$, $j\in\{0,1\}$ then we do not mention explicitly in the subscript the qubit it acts on, in order to avoid clutter. We assume that the left-most operator is applied on qubit $1$, then next one on qubit $2$ and so on and this should be clear from the context.  
\begin{eqnarray}
    T_1&=& \theta\sum_{a_1,\ldots,a_n\in\{0,1\}}P_{a_1}P_{a_1}P_{a_2}P_{a_2}\ldots P_{a_n}P_{a_n} \nonumber \\
    T_2&=& \theta\left(\sum_{a_1,\ldots,a_n\in\{0,1\}}P_{a_1}P_{a_1}P_{a_2}P_{a_2}\ldots P_{a_n}P_{a_n}\right)\left(I_{(2n+1)}-Z_{(2n+1)}\right)  
    \label{eqn:T12}
\end{eqnarray}
It is quite clear that the above two terms are sum of mutually commuting Paulis belonging to the following two sets, respectively.
\begin{eqnarray}
    \mathcal{G}_1&=&\left\{P_{a_1}P_{a_1}P_{a_2}P_{a_2}\ldots P_{a_n}P_{a_n} : \quad a_j\in\{0,1\}, \quad j=1,\ldots,n  \right\}    \nonumber \\
    \mathcal{G}_2&=&\left\{P_{a_1}P_{a_1}P_{a_2}P_{a_2}\ldots P_{a_n}P_{a_n}Z_{(2n+1)}^b : \quad a_j,b\in\{0,1\}, \quad j=1,\ldots,n  \right\}
    \label{eqn:G12}
\end{eqnarray}
Now we derive the eigenbasis for each of these terms in the following lemma, the proof of which has been given in the Supplemental Material \cite{SuppMat}. 
\begin{lemma}[\textbf{Eigenbasis for $\boldsymbol{\mathcal{G}_1}$ and $\boldsymbol{\mathcal{G}_2}$}]
Let $w,v_2,\ldots v_{2n}\in\{0,1\}$. Then the eigenvectors of the Paulis in $\mathcal{G}_1$ and $\mathcal{G}_2$ are of the following form, respectively.
\begin{eqnarray}
    \ket{\vect{v}_{1,\pm}}&=&\frac{1}{\sqrt{2}}\left(\ket{0v_2\ldots v_{2n}}\pm\ket{1\overline{v_2,\ldots v_{2n}}} \right),\qquad
    \ket{\vect{v}_{2,\pm}}=\frac{1}{\sqrt{2}}\left(\ket{0v_2\ldots v_{2n}w}\pm\ket{1\overline{v_2,\ldots v_{2n}}w} \right)  \nonumber
\end{eqnarray}
Specifically, if $\beta_1=a_1v_2+a_2(v_3+v_4)+\cdots+a_n(v_{2n-1}+v_{2n})$ and $\beta_2=a_1v_2+a_2(v_3+v_4)+\cdots+a_n(v_{2n-1}+v_{2n})+wb$, then we have the following.
\begin{eqnarray}
    P_{a_1}P_{a_1}P_{a_2}P_{a_2}\ldots P_{a_n}P_{a_n}\ket{\vect{v}_{1,\pm}} &=& \pm (-1)^{a_1+a_2+\cdots+a_n+\beta_1}\ket{\vect{v}_{1,\pm}}   \nonumber \\
    P_{a_1}P_{a_1}P_{a_2}P_{a_2}\ldots P_{a_n}P_{a_n}Z_{(2n+1)}^b\ket{\vect{v}_{2,\pm}} &=& \pm (-1)^{a_1+a_2+\cdots+a_n+\beta_2}\ket{\vect{v}_{2,\pm}}   \nonumber
\end{eqnarray}
\label{lem:ebasisG12}
\end{lemma}

It is easy to see that there are $2^{2n-1}\cdot 2=2^{2n}$ mutually orthogonal vectors of the form $\ket{\vect{v}_{1,\pm}}$ and $2^{2n}\cdot 2=2^{2n+1}$ mutually orthogonal vectors of the form $\ket{\vect{v}_{2,\pm}}$, and so we have complete eigenbases for the Paulis in $\mathcal{G}_1$ and $\mathcal{G}_2$. Now, we derive diagonalizing circuits for the set of Paulis in $\mathcal{G}_1$ and $\mathcal{G}_2$.
\begin{theorem}[\textbf{Diagonalizing circuit for $\boldsymbol{\mathcal{G}_1}$ and $\boldsymbol{\mathcal{G}_2}$ }]
Let $W=\left(\prod_{j=2}^{2n}CNOT_{(1,j)}\right)H_{(1)}$ and \\
$\widetilde{\vect{Z}_1}=Z_{(1)}Z_{(2)}^{a_1}Z_{(3)}^{a_2}Z_{(4)}^{a_2}\ldots Z_{(2n-1)}^{a_n}Z_{(2n)}^{a_n}$, $\widetilde{\vect{Z}_2}=Z_{(1)}Z_{(2)}^{a_1}Z_{(3)}^{a_2}Z_{(4)}^{a_2}\ldots Z_{(2n-1)}^{a_n}Z_{(2n)}^{a_n}Z_{(2n+1)}^b$, where $a_1,\ldots a_n,b\in\{0,1\}$. Then,
\begin{eqnarray}
    (-1)^{a_1+\cdots+a_n}W\widetilde{\vect{Z}_1}W^{\dagger}&=&P_{a_1}P_{a_1}P_{a_2}P_{a_2}\ldots P_{a_n}P_{a_n} \in\mathcal{G}_1  \nonumber \\
    (-1)^{a_1+\cdots+a_n}W\widetilde{\vect{Z}_2}W^{\dagger}&=&P_{a_1}P_{a_1}P_{a_2}P_{a_2}\ldots P_{a_n}P_{a_n}Z_{(2n+1)}^b \in\mathcal{G}_2    \nonumber
\end{eqnarray}
\label{thm:diagonalG12}
\end{theorem}

More details, including the proof can be found in the Supplemental Material \cite{SuppMat}.
Using the above theorem, we diagonalize the terms in $T_1$ and $T_2$ and rewrite them as follows.
\begin{eqnarray}
    T_1&=&W\left(\theta Z_{(1)}\sum_{a_2,\ldots,a_n\in\{0,1\}}(-1)^{a_2+\cdots+a_n}Z_{(3)}^{a_2}Z_{(4)}^{a_2}\ldots Z_{(2n)}^{a_n}-\theta Z_{(1)}Z_{(2)}\sum_{a_2,\ldots,a_n\in\{0,1\}}(-1)^{a_2+\cdots+a_n}Z_{(3)}^{a_2}Z_{(4)}^{a_2}\ldots Z_{(2n)}^{a_n}\right)W^{\dagger} \nonumber \\
    T_2&=&W\left(\theta Z_{(1)}\sum_{a_2,\ldots,a_n\in\{0,1\}}(-1)^{a_2+\cdots+a_n}Z_{(3)}^{a_2}Z_{(4)}^{a_2}\ldots Z_{(2n)}^{a_n}(I_{(2n+1)}-Z_{(2n+1)}) \right. \nonumber \\
    &&\left. -\theta Z_{(1)}Z_{(2)}\sum_{a_2,\ldots,a_n\in\{0,1\}}(-1)^{a_2+\cdots+a_n}Z_{(3)}^{a_2}Z_{(4)}^{a_2}\ldots Z_{(2n)}^{a_n} (I_{(2n+1)}-Z_{(2n+1)})\right)W^{\dagger} \nonumber
\end{eqnarray}
Then using Lemma 2.3 in \cite{2022_MWZ}, the eigenvalues of the sum of Z-operators can be expressed as functions of Boolean variables $x_1,\ldots,x_{2n},x_{2n+1}$, as follows.
\begin{eqnarray}
    \phi_1&=&\theta (-1)^{x_1}\sum_{a_2,\ldots,a_n\in\{0,1\}}(-1)^{a_2+\cdots+a_n}(-1)^{x_3^{a_2}+x_4^{a_2}+\cdots+x_{2n}^{a_n}}-\theta (-1)^{x_1+x_2}\sum_{a_2,\ldots,a_n\in\{0,1\}}(-1)^{a_2+\cdots+a_n}(-1)^{x_3^{a_2}+x_4^{a_2}+\cdots+x_{2n}^{a_n}}   \nonumber \\
    &=&\theta (-1)^{x_1}\left(1-(-1)^{x_2}\right)\left(\sum_{a_2,\ldots,a_n\in\{0,1\}}(-1)^{a_2+\cdots+a_n}(-1)^{x_3^{a_2}+x_4^{a_2}+\cdots+x_{2n}^{a_n}}\right)   \nonumber \\
    \phi_2&=&\theta (-1)^{x_1}\left(1-(-1)^{x_2}\right)\left(1-(-1)^{x_{2n+1}}\right)\left(\sum_{a_2,\ldots,a_n\in\{0,1\}}(-1)^{a_2+\cdots+a_n}(-1)^{x_3^{a_2}+x_4^{a_2}+\cdots+x_{2n}^{a_n}}\right)
    \label{eqn:phi12}
\end{eqnarray}

\begin{lemma}
Let $y_1,y_2,\ldots,y_{2m}$ are Boolean variables. Then
\begin{eqnarray}
    \sum_{a_1,\ldots,a_m\in\{0,1\}}(-1)^{a_1+\cdots+a_m}(-1)^{y_1^{a_1}+y_2^{a_1}+\cdots+y_{2m}^{a_m}}=\prod_{j=1}^m\left(1-(-1)^{y_{2j-1}+y_{2j}}\right).   \nonumber
\end{eqnarray}
\label{lem:phiPartial}
\end{lemma}
\begin{proof}
The proof uses induction and can be found in the Supplemental Material \cite{SuppMat}.
\end{proof}

Applying the above lemma in Eq. \ref{eqn:phi12} we obtain,
\begin{eqnarray}
    \phi_1&=&\theta (-1)^{x_1}\left(1-(-1)^{x_2}\right)\prod_{j=2}^n\left(1-(-1)^{x_{2j-1}+x_{2j}}\right)  \nonumber \\
    \phi_2&=&\theta (-1)^{x_1}\left(1-(-1)^{x_2}\right)\left(1-(-1)^{x_{2n+1}}\right)\prod_{j=2}^n\left(1-(-1)^{x_{2j-1}+x_{2j}}\right)  
    \label{eqn:phi12'}
\end{eqnarray}
Now, $\phi_1=(-1)^{x_1}2^n\theta$ when $x_2=x_{2j-1}\oplus x_{2j}=1$, where $j=2,\ldots,n$; else it is $0$. Similarly, $\phi_2=(-1)^{x_1}2^{n+1}\theta$ when $x_2=x_{2n+1}=x_{2j-1}\oplus x_{2j}=1$, where $j=2,\ldots,n$; else it is $0$. Thus, using Lemma 2.4 in \cite{2022_MWZ} we can implement a circuit for $e^{-iT_1t}$ and $e^{-iT_2t}$, using one controlled-$R_z$. The complete circuits (for one time-step) have been shown in Figure \ref{ckt:T1} and \ref{ckt:T2}, respectively. In both these circuits we require $2n-1$ CNOT, controlled on qubit $1$ and target on qubit $2\leq j\leq 2n$, and 2 H gates to implement the diagonalizing circuit $W$. The we use $n-1$ CNOT with control on qubit $2j$ and target on qubit $2j-1$, where $2 \leq j\leq n$, to test the parity constraints. Then we apply the multicontrolled-$R_z$ to implement the rotation when the parity constraints are satisfied. Thus, the total number of gates required for implementing these exponentials per time step, can be summarized in the following lemma.  
\begin{theorem}
Suppose $T_1$ and $T_2$ are sum of Pauli terms, as defined in Eq. \ref{eqn:T12}. Then it is possible to implement $e^{-iT_1t}$ and $e^{-iT_2t}$ using $6n-4$ CNOT and 2 H gates per time step. Additionally, we require one $C^{n}R_z$ gate (per time step) for $e^{-iT_1t}$ and one $C^{n+1}R_z$ gate for $e^{-iT_2t}$.
\label{thm:cktT12}
\end{theorem}
Here $C^kR_z$ refers to a $R_z$ gate controlled on $k$ qubits. We can decompose $C^kR_z$ into compute-uncompute $C^kX$ pairs, $cR_z$ (single-qubit controlled $R_z$) and a single ancilla. We can then further decompose $C^kX$ using $4k-4$ T and $4k-3$ CNOT-gates \cite{2017_HLZetal}.  If we use the logical AND construction in \cite{2018_G} then we require similar number of T and CNOT for both compute-uncompute pair, but at the cost of using measurement and additional classical resources.

\begin{figure}[t!]
    \centering\[
    \Qcircuit @C=1em @R=1em{
    &&&&\lstick{2n} &\qw &\targ &\ctrl{1} &\qw &\ctrl{1} &\targ &\qw \\
    &&&&\lstick{2n-1} &\qw &\targ\qwx &\targ &\ctrl{2} &\targ &\targ\qwx &\qw \\
    &&&&\lstick{2n-2} &\qw &\targ\qwx &\ctrl{1} &\qw &\ctrl{1} &\targ\qwx &\qw \\
    &&&&\lstick{2n-3} &\qw &\targ\qwx &\targ &\ctrl{1} &\targ  &\targ\qwx &\qw \\
    &&&&\lstick{} && \qwx & & & & \qwx\\
    &&&&\lstick{\vdots} && \vdots &\vdots &\vdots &\vdots &\vdots\\
    &&&&\lstick{} && & & & &\\
    &&&&\lstick{2} &\qw &\targ &\qw &\ctrl{1} &\qw &\targ &\qw \\
    &&&&\lstick{1} &\qw &\ctrl{-2} &\gate{H} &\gate{R_z(2^{n+1}\theta)} &\gate{H} &\ctrl{-2} &\qw
    }
    \]
    \caption{Circuit implementing the exponential of the sum $T_1$ (Eq. \ref{eqn:T12}) i.e. $e^{-iT_1t}$.}
    \label{ckt:T1}
\end{figure}

\begin{figure}[h]
    \centering
    \[
    \Qcircuit @C=1em @R=1em{
    &&&&\lstick{2n+1} &\qw &\qw &\qw &\ctrl{2} &\qw &\qw &\qw \\
    &&&&\lstick{2n} &\qw &\targ &\ctrl{1} &\qw &\ctrl{1} &\targ &\qw \\
    &&&&\lstick{2n-1} &\qw &\targ\qwx &\targ &\ctrl{2} &\targ &\targ\qwx &\qw \\
    &&&&\lstick{2n-2} &\qw &\targ\qwx &\ctrl{1} &\qw &\ctrl{1} &\targ\qwx &\qw \\
    &&&&\lstick{2n-3} &\qw &\targ\qwx &\targ &\ctrl{1} &\targ  &\targ\qwx &\qw \\
    &&&&\lstick{} && \qwx & & & & \qwx\\
    &&&&\lstick{\vdots} && \vdots &\vdots &\vdots &\vdots &\vdots\\
    &&&&\lstick{} && & & & &\\
    &&&&\lstick{2} &\qw &\targ &\qw &\ctrl{1} &\qw &\targ &\qw \\
    &&&&\lstick{1} &\qw &\ctrl{-2} &\gate{H} &\gate{R_z(2^{n+2}\theta)} &\gate{H} &\ctrl{-2} &\qw
    }
    \]
    \caption{Circuit implementing the exponential of the sum $T_2$ (Eq. \ref{eqn:T12}) i.e. $e^{-iT_2t}$.}
    \label{ckt:T2}
\end{figure}

We consider another specific sum of Paulis and give the cost of implementing its exponential per time step.
\begin{theorem}
Let $T_3$ is a sum of Paulis, as defined below.
\begin{eqnarray}
    T_3&=&\theta\prod_{j=1}^n(I_{(n)}-Z_{(n)})  \nonumber 
\end{eqnarray}
Then it is possible to implement $e^{-iT_3t}$ using one $C^nR_z$ gate and an extra ancilla, per time step. 
\label{thm:cktT3}
\end{theorem}

\begin{proof}
Using Lemma 2.3 in \cite{2022_MWZ}, the eigenvalues of $T_3$ can be expressed as functions of Boolean variables $x_1,\ldots,x_n$, as follows.
\begin{eqnarray}
    \phi_3&=&\theta\prod_{j=1}^n\left(1-(-1)^{x_j}\right)   \nonumber
\end{eqnarray}
It is easy to see that $\phi_3=2^n\theta$ if and only if $x_j=1$ for $j=1,\ldots, n$; else it is zero. Thus using Lemma 2.4 in \cite{2022_MWZ}, we can implement a circuit for $e^{-iT_3t}$, using one controlled-$R_z$. Specifically, we require a $C^nR_z$ gate and an extra ancilla, initialized to $\ket{0}$. The controls select the proper state of the qubits. The target $R_z$ gate is applied on the ancilla. 
\end{proof}

Now, we consider the 4 Hamiltonian groups - $H_{1\varphi}$ (Eq. \ref{eqn:HphiGroup1}), $H_{2\varphi}$ (Eq. \ref{eqn:HphiGroup2}), $H_{3\varphi}$ (Eq. \ref{eqn:HphiGroup3}) and $H_{4\varphi}$ (Eq. \ref{eqn:HphiGroup4}); and estimate the number of gates and qubits required to implement their exponentials per time step. As discussed, we have two sources of T-gates, one that comes from the approximately implementable (controlled)-rotation gates and the other that comes from the other exactly implementable components, for example, multicontrolled-X gates. When we report the number of T-gates we do not include the ones in the implementation of the rotation gates. The overall T-count can be easily obtained by plugging in the T-count estimates of (controlled)-rotation gates. More discussion on these bounds have been provided in Section \ref{sec:intro}. \par

\paragraph{Group I : $\boldsymbol{H_{1\varphi}}$ : } 
We rewrite Eq. \ref{eqn:HphiGroup1}, for convenience. $S_{4\vec{p}}=\{\vec{p}=(p_1,p_2,p_3,p_4) : p_i\in\Gamma; i=1,\ldots,4 \text{ and }\exists k\in\mathbb{Z}\text{ s.t.} p_3=p_1+k, p_4=p_2-k \}$ and $S_{4\vec{n}}=\{\vec{n}=(n_1,n_2,n_3,n_4) : n_i=1,\ldots,M; i=1,\ldots,4\}$ are ordered 4-tuples of momentum modes and momentum states. 
\begin{eqnarray}
   :H_{1\varphi}:= \frac{\lambda}{24\cdot 64 |\Omega|^3 }\sum_{\vec{p}\in S_{4\vec{p}}}\sum_{\vec{n}\in S_{4\vec{n}}}\prod_{(p_j,n_j)\in (\vec{p},\vec{n})}\sqrt{\frac{n_j+1}{w_{p_j}}} \left(X_{p_j,n_j}X_{p_j,n_j+1}+Y_{p_j,n_j}Y_{p_j,n_j+1}\right)  \nonumber
\end{eqnarray}
We see that for a given $\vec{p},\vec{n}$, the sum of the 16 terms in the innermost summation is of the form $T_1$ (Eq. \ref{eqn:T12}), with $n=4$. Thus we can apply Theorem \ref{thm:cktT12} and conclude that we require 20 CNOT, 2 H and one $C^4R_z$ gate to implement its exponential. The $C^4R_z$ gate can be decomposed further into 1 $cR_z$, 12 T and 13 additional CNOT. In this case $|S_{4\vec{p}}|\leq \binom{V}{2}\cdot V=\frac{V^2(V-1)}{2}$ and $|S_{4\vec{n}}|=M^4$. The $4!$ permutations of the elements in $\vec{n}$ lead to the same coefficient, and hence can be summed together, resulting a form similar to $T_1$, with $n=4$. Hence, to implement $e^{-iH_{1\varphi}t}$ we require at most $\frac{M^4V^2(V-1)}{48}$ $cR_z$, $\frac{12M^4V^2(V-1)}{48}$ T, $\frac{33M^4V^2(V-1)}{48}$ CNOT and $\frac{2M^4V^2(V-1)}{48}$ H gates per time step.

\paragraph{Group II : $\boldsymbol{H_{2\varphi}}$ : } The sum of the terms in this group (Eq. \ref{eqn:HphiGroup2}) is as follows.
\begin{eqnarray}
    &&:H_{2\varphi}:    \nonumber \\
    &=&\frac{\lambda}{96|\Omega|}\sum_{p,k}\frac{1}{\omega_p\sqrt{\omega_{p+k}\omega_{p-k}}}\left(\sum_{n_1,n_2,n_3}c_{n}^{(1)}(X_{n_1}X_{n_1+2}+Y_{n_1}Y_{n_1+2})_p 
    (X_{n_2}X_{n_2+1}+Y_{n_2}Y_{n_2+1})_{p+k}(X_{n_3}X_{n_3+1}+Y_{n_3}Y_{n_3+1})_{p-k}   \right.    \nonumber \\
    &&\left.+\sum_{n_1,n_2,n_3}c_n^{(2)}(I_{n_1}-Z_{n_1})_p(X_{n_2}X_{n_2+1}+Y_{n_2}Y_{n_2+1})_{p+k}(X_{n_3}X_{n_3+1}+Y_{n_3}Y_{n_3+1})_{p-k}   \right) \nonumber
\end{eqnarray}
where $c_{n}^{(1)}=\frac{\sqrt{(n_1+2)(n_1+1)(n_2+1)(n_3+1)}}{8}$, $c_{n}^{(2)}=\frac{n_1\sqrt{(n_2+1)(n_3+1)}}{4}$.

For every $p,k,n_1,n_2,n_3$ there are two sums, one of the form $T_1$, with $n=3$ and the other of the form $T_2$, with $n=2$ (Eq. \ref{eqn:T12}). Using Theorem \ref{thm:cktT12} we require 14+8=22 CNOT, 4 H gates and 2 $C^3R_z$ gates to implement the exponential of  these sums. Each $C^3R_z$ can be implemented with 1 $cR_z$, 8 T and 9 additional CNOT. In this case, $|S_{4\vec{p}}|\leq V^2$, $|S_{4\vec{n}}|\leq M^3$ and there are $3!$ permutations of the momentum states i.e. $n_1,n_2,n_3$ that can be summed together, because they have the same coefficients. Hence, to implement $e^{-iH_{2\varphi}t}$ we require at most $\frac{M^3V^2}{3}$ $cR_z$, $\frac{8M^3V^2}{3}$ T, $\frac{20M^3V^2}{3}$ CNOT and $\frac{2M^3V^2}{3}$ H gates, per time step $t$.

\paragraph{Group III : $\boldsymbol{H_{3\varphi}}$ : } From Eq. \ref{eqn:HphiGroup3} we can rewrite the following.
\begin{eqnarray}
    &&:H_{3\varphi}:    \nonumber \\
    &=&\frac{\lambda}{96 |\Omega| }\sum_{p_1,p_2}\frac{1}{\omega_{p_1}\omega_{p_2}}\left(\sum_{n_1,n_2} c_n^{(3)}(X_{n_1}X_{n_1+2}+Y_{n_1}Y_{n_1+2})_{p_1}(X_{n_2}X_{n_2+2}+Y_{n_2}Y_{n_2+2})_{p_2}       \right.  \nonumber \\
    &&+\sum_{n_1,n_2}c_n^{(4)}(X_{n_1}X_{n_1+2}+Y_{n_1}Y_{n_1+2})_{p_1}(I_{n_2}-Z_{n_2})_{p_2} 
    +\sum_{n_1,n_2}c_n^{(5)}(X_{n_2}X_{n_2+2}+Y_{n_2}Y_{n_2+2})_{p_2}(I_{n_1}-Z_{n_1})_{p_1} \nonumber \\
    &&\left.+\sum_{n_1,n_2}n_1n_2(I_n-Z_{p_1,n_1}-Z_{p_2,n_2}+Z_{p_1,n_1}Z_{p_2,n_2})   \right) \nonumber 
\end{eqnarray}
where $c_n^{(3)}=\frac{\sqrt{(n_1+2)(n_1+1)(n_2+2)(n_2+1)}}{4}$, $c_n^{(4)}=\frac{n_2\sqrt{(n_1+2)(n_1+1)}}{2}$, $c_n^{(5)}=\frac{n_1\sqrt{(n_2+2)(n_2+1)}}{2}$.

Here we see that for a given $p_1,p_2,n_1,n_2$, there is one sum of the form $T_1$ with $n=2$ (Eq. \ref{eqn:T12}), two sums of the form $T_2$ with $n=1$ (Eq. \ref{eqn:T12}) and one sum of the form  $T_3$ with $n=2$ (Theorem \ref{thm:cktT3}). Using Theorem \ref{thm:cktT12} we can implement the exponential of the first sum using 8 CNOT, 2 H and one $C^2 R_z$ gates; each of the second type of exponentiated sums ($T_2$) using 2 CNOT, 2 H and one $C^2R_z$ gates. Using Theorem \ref{thm:cktT3} we can implement the exponential of the last type of sum ($T_3$) using one $C^2R_z$ gate per time step. We can decompose each of the $C^2R_z$ using 4 T-gates, 1 $cR_z$ and 5 CNOT-gates. In this case, $|S_{4\vec{p}}|\leq \frac{V(V-1)}{2}$ and $|S_{4\vec{n}}|\leq M^2$ and the $2!$ permutations of each $(n_1,n_2)$ can be summed together. Thus, overall we require $2M^2V(V-1)$ $cR_z$, $8M^2V(V-1)$ T, $16M^2V(V-1)$ CNOT and $3M^2V(V-1)$ H gates. 

\paragraph{Group IV : $\boldsymbol{H_{4\varphi}}$ : } We rewrite the terms in $H_{4\varphi}$ from Eq. \ref{eqn:HphiGroup4}.
\begin{eqnarray}
    &&:H_{4\varphi}:    \nonumber \\
    &=&\frac{\lambda}{96}\sum_{p,n}\frac{1}{(\omega_p)^2}\left(\frac{\sqrt{(n+4)(n+3)(n+2)(n+1)}}{2} (X_{n}X_{n+4}
    + Y_{n} Y_{n+4})_p \right. \nonumber \\
    &&\left.+2n\sqrt{(n+2)(n+1)} (X_{n}X_{n+2} + Y_{n} Y_{n+2})_p +3(n^2-n)(I_n-Z_n)_p    \right)   \nonumber
\end{eqnarray}
We see that for a given $p,n$ there are two sums of the form $T_1$ with $n=1$ (Eq. \ref{eqn:T12}) and the exponentials of these can be implemented with 2 $cR_z$, 4 H and 4 CNOT (Theorem \ref{thm:cktT12}). The exponential of each term of the form $I_n-Z_n$ can be implemented with one $R_z$ gate. In this case $|S_{4\vec{p}}|\leq V$ and $|S_{4\vec{n}}|\leq M$. Thus, overall we require $3MV$ (c)-$R_z$, $4MV$ H and $4MV$ CNOT-gates.

In summary, we have the following resource estimates for implementing the exponentiated Hamiltonians.
\begin{lemma}
\begin{enumerate}
 \item It is possible to implement $e^{-iH_{1\varphi}t}$ using at most $\frac{N^4|\Omega|^2(|\Omega|-1)}{48}$ $cR_z$, $\frac{N^4|\Omega|^2(|\Omega|-1)}{4}$ T, $\frac{11N^4|\Omega|^2(|\Omega|-1)}{16}$ CNOT and $\frac{N^4|\Omega|^2(|\Omega|-1)}{24}$ H gates per time step $t$.

 \item It is possible to implement $e^{-iH_{2\varphi}t}$ using at most $\frac{N^3|\Omega|^2}{3}$ $cR_z$, $\frac{8N^3|\Omega|^2}{3}$ T, $\frac{20N^3|\Omega|^2}{3}$ CNOT and $\frac{2N^3|\Omega|^2}{3}$ H gates per time step $t$.

 \item It is possible to implement $e^{-iH_{3\varphi}t}$ using at most $2N^2|\Omega|(|\Omega|-1)$ $cR_z$, $8N^2|\Omega|(|\Omega|-1)$ T, $16N^2|\Omega|(|\Omega|-1)$ CNOT and $3N^2|\Omega|(|\Omega|-1)$ H gates per time step $t$.

 \item It is possible to implement $e^{-iH_{4\varphi}t}$ using at most $3N|\Omega|$ (c)-$R_z$, $4N|\Omega|$ CNOT and $4N|\Omega|$ H gates per time step $t$.
 \end{enumerate}
\label{lem:cktH1varphi}
\end{lemma}
 As mentioned earlier, we have separated the T-gates arising from the approximately implementable (controlled)-rotation gates and the exactly implementable Toffoli or multicontrolled-X gates. So we report the number of (controlled)-$R_z$ required and the number of T-gates in Lemma \ref{lem:cktH1varphi} do not include the T-gates from these rotations. The total T-count estimate is obtained by multiplying the number of (controlled)-$R_z$ gates with the T-count of individual rotation gates and adding the product to the T-count arising from the exactly implementable parts. 

An estimate of the total number of gates required to implement $e^{-iHt}$ can be obtained by summing the gate costs in Lemmas \ref{lem:cktH0} and \ref{lem:cktH1varphi}. 
This is summarized in the following theorem, as well as in Table \ref{tab:costOccBasis}.
\begin{table}[t]
\centering
\begin{tabular}{|c|c|c|c|c|c|}
\hline
     & $\mathbf{\# (c)R_z}$ & $\mathbf{\#}$ \textbf{T} & $\mathbf{\#}$ \textbf{CNOT} & $\mathbf{\#}$ \textbf{H} &  \\
    \hline
     $\mathbf{e^{-iH_0t}}$& $N|\Omega|$ & - & - & - & Lemma \ref{lem:cktH0} \\
     \hline
     $\mathbf{e^{-iH_{1\varphi}t}}$& $\frac{N^4|\Omega|^2(|\Omega|-1)}{48}$ & $\frac{N^4|\Omega|^2(|\Omega|-1)}{4}$ & $\frac{11N^4|\Omega|^2(|\Omega|-1)}{16}$ & $\frac{N^4|\Omega|^2(|\Omega|-1)}{24}$ & Lemma \ref{lem:cktH1varphi} \\
     \hline
     $\mathbf{e^{-iH_{2\varphi}t}}$& $\frac{N^3|\Omega|^2}{3}$ & $\frac{8N^3|\Omega|^2}{3}$ & $\frac{20N^3|\Omega|^2}{3}$ & $\frac{2N^3|\Omega|^2}{3}$ & Lemma \ref{lem:cktH1varphi} \\
     \hline
     $\mathbf{e^{-iH_{3\varphi}t}}$& $2N^2|\Omega|(|\Omega|-1)$ & $8N^2|\Omega|(|\Omega|-1)$ & $16N^2|\Omega|(|\Omega|-1)$ & $3N^2|\Omega|(|\Omega|-1)$ & Lemma \ref{lem:cktH1varphi} \\
     \hline
     $\mathbf{e^{-iH_{4\varphi}t}}$& $3N|\Omega|$ & - & $4N|\Omega|$ & $4N|\Omega|$ & Lemma \ref{lem:cktH1varphi} \\
     \hline
\end{tabular}
\caption{Summary of the number of gates required to implement the exponentials of the different Hamiltonian partitions in the occupation basis. }
\label{tab:costOccBasis}
\end{table}

\begin{theorem}
It is possible to implement $e^{-iHt}$ with the following number of gates per time step.
\begin{enumerate}
    \item $\frac{N^4|\Omega|^2(|\Omega|-1)}{48}+\frac{N^3|\Omega|^2}{3}+2N^2|\Omega|(|\Omega|-1)+4N|\Omega|$ (controlled)-$R_z$ gates.

    \item $\frac{N^4|\Omega|^2(|\Omega|-1)}{4}+\frac{8N^3|\Omega|^2}{3}+8N^2|\Omega|(|\Omega|-1)$  additional T-gates.

    \item $\frac{11N^4|\Omega|^2(|\Omega|-1)}{16}+\frac{20N^3|\Omega|^2}{3}+16N^2|\Omega|(|\Omega|-1)+4N|\Omega|$  CNOT-gates.

    \item $\frac{N^4|\Omega|^2(|\Omega|-1)}{24}+\frac{2N^3|\Omega|^2}{3}+3N^2|\Omega|(|\Omega|-1)+4N|\Omega|$ H gates.
\end{enumerate}
\label{thm:costOccBasis}
\end{theorem}

\begin{proof}
The proof follows from  Lemmas \ref{lem:cktH0} and \ref{lem:cktH1varphi}, which are summarized in Table \ref{tab:costOccBasis}. 
\end{proof}

\subsection{Trotter Error Analysis}
\label{subsec:occError}

In this Appendix we discuss and bound the various errors that occur during simulation of the occupation basis Hamiltonian. For convenience, we restate the occupation basis Hamiltonian in the following equation.
\begin{eqnarray}
    H_{occ} &=& H_0 + H_{\lambda}\\
    :H_0: &=& \sum_p\omega_p a_p^{\dagger}a_p \\
     :H_{\lambda} : &=& H_{1\varphi} + H_{2\varphi} +  H_{3\varphi} + H_{4\varphi}
    \label{eqn:HoccRedefine}
\end{eqnarray}

\paragraph{I. Trotter Error :} One primary source of error is the one inherent in the Trotterization procedure, where we express the exponential of sum of operators as product of the exponentiated operators.
This is true if and only if the operators are commuting.
If there are noncommuting parts then the resultant error in the approximation is referred to as the \emph{Trotter error}.
Quite a few bounds on the Trotter error have been derived before \cite{1990_HdR, 2010_WBHS, 2015_WHWetal}, but we use the one in \cite{Childs2021}, which shows the dependence on nested commutators.\par
If a Hamiltonian $H=\sum_{\gamma=1}^{\Gamma}H_{\gamma}$ is a sum of $\Gamma$ fragment Hamiltonians, then $e^{-i\tau H}$ can be approximated by product of exponentials, using the $p^{th}$ order Trotter-Suzuki formula \cite{1991_S}, $$\mathscr{S}_p(\tau)=e^{-i\tau H}+\mathscr{A}(\tau),$$ where $\|\mathscr{A}(\tau)\|\in O\left(\widetilde{\alpha}_{comm} \tau^{p+1}\right)$ if each $H_{\gamma}$ are Hermitian \cite{Childs2021}. 
Here 
$$\widetilde{\alpha}_{comm}=\sum_{\gamma_1,\gamma_2,\ldots,\gamma_{p+1}=1}^{\Gamma}\|[H_{\gamma_{p+1}},\ldots[H_{\gamma_2},H_{\gamma_1}]]\|.$$
In this Appendix, $\| \cdot \|$ refers to the spectral norm \cite{Childs2021}, which is defined as the induced  Euclidean norm on the Hamming-Weight 1 subspace $\mathcal{S}$
\begin{equation}
\|O\|=\max_{ \ket{x} \in \mathcal{S}: \|x\|_2=1} \| O|x\rangle \|_2.
\end{equation}
This definition corresponds with the Schatten infinity-norm which yields the maximum singular value of a matrix.
In most applications, it is quite cumbersome, if not impossible, to derive a rigorous analytic  expression of the nested commutators, in order to tightly bound higher-order Trotter error. 
We use the following bound from \cite{2023_MSW} in this work.
\begin{eqnarray}
    \widetilde{\alpha}_{comm}\leq 2^{p-(p'+1)}\sum_{\gamma_{i_1},\gamma_{i_2},\ldots,\gamma_{i_{p'+1}}} \|[H_{\gamma_{p'+1}},[\ldots[H_{\gamma_3},[H_{\gamma_2},H_{\gamma_1}]]\ldots]]\|  \left(\sum_{\gamma=1}^{\Gamma}\|H_{\gamma}\|\right)^{p-p'}\qquad [1\leq p'\leq p]
    \label{eqn:alphacomm}
\end{eqnarray}
The above formula is specifically useful in scenarios where we can compute tight bounds till level $p'<p$ of nesting. Then we combine the tighter bound till level $p'$ and a looser sum of norm bound for the remaining levels. In this paper we consider the $2^{nd}$ order Trotter error in our implementations. In the following sections we derive rigorous expressions and bounds on the first level commutators and norms. Then we use the above equation in order to bound the higher-order errors. We primarily focus on these low level Trotter errors for two reasons. The symmetric low-level Trotter errors provide tighter bounds \cite{Childs2021}. They also appear to provide more accurate results when compared to explicit calculations \cite{Childs2018}.

In summary we aim to prove the following, by deriving a bound on $\widetilde{\alpha}_{comm}$. The proof has been provided in the Supplemental Material \cite{SuppMat}.
\begin{lemma}
Let $H$ be the occupation basis Hamiltonian derived in Eqs. \ref{eqn:H0}-\ref{eqn:HphiGroup4}. Let $\mathscr{S}_2(\tau)$ be a $2^{\rm nd}$ order Trotter-Suzuki approximation for $e^{-i\tau H}$. Then,
\begin{eqnarray}
\left\|e^{-iH\tau}-\mathscr{S}_p(\tau) \right\| \in O\left(\widetilde{\alpha}_{comm}\tau^{3}\right),    
\end{eqnarray}
where 
\begin{eqnarray}
    \widetilde{\alpha}_{comm}&\in O\left(\frac{\lambda^2 N^{6}}{M^5 }\right)&.
\end{eqnarray}
\label{lem:occTrotter}
\end{lemma}

\subsection{Total T-gate cost estimate}
Here we review the various contributions to the error in phase estimation. Based on previous simulation work, the most efficient distribution of error seems a highly precise phase estimation calculation to offset a single Trotter step \cite{2021_SBWetal}.  In particular, we focus on the scheme of~\cite{Babbush2018}, which allows us to obtain optimal constant factors using Fourier-based phase estimation. We use an approximate quantum Fourier transform (AQFT) \cite{2020_NSM} in the preparation. The number of T-gates required to approximate an $n$-qubit QFT circuit up to error $\epsilon_{QFT}$ is
\begin{eqnarray}
    8n\log_2\left(\frac{n}{\epsilon_{QFT}}\right) + \log_2\left(\frac{n}{\epsilon_{QFT}}\right)\log_2\left(\frac{\log_2\left(\frac{n}{\epsilon_{QFT}}\right)}{\epsilon_{QFT}}\right).
    \label{eqn:TcostAQFT}
\end{eqnarray}

\begin{lemma}
There exists $\epsilon_E'$ such that for all sufficiently small $\epsilon_E \le \epsilon_E'$ the total number of implementations of $\mathscr{S}_2(\tau)$ required to estimate an eigenstate within error $\epsilon_E$ is bounded above by
\begin{equation}
 2^m \leq \frac{\pi^2 \sqrt{\widetilde\alpha_{\rm comm}}}{\epsilon_E^{3/2}}.
\end{equation}

\label{lem:error-phase}
\end{lemma}
\begin{proof}
    The variance on the phase measured from QPE is given as 
\begin{equation}
    \epsilon_V = \langle \cos( (\hat{\theta} - \widetilde{\theta})) \rangle^2 - 1,
\end{equation}
known as the Holevo variance, where $\widetilde{\theta}$ is the true phase and $\hat{\theta}$ is the measured outcome. This allows simple analytic results for a compact phase that may otherwise have artificially high variance when peaked near the boundary. 
 This can be computed using the phase estimation algorithm implemented in \cite{Babbush2018}.  The estimated phase can be written as
 \begin{eqnarray}
     \theta_{\mathrm{est}} = \theta_{\mathrm{true}} + \theta+ \epsilon_{\mathrm{prep}},    
 \end{eqnarray}
 where $\theta$ is a random variable with zero mean and Holevo variance $\epsilon_V$ describing the output of phase estimation and $\epsilon_{\mathrm{prep}}$ represents the systematic errors in the phase that arise because of approximate gate synthesis, approximate QFT (AQFT) and approximation due to Trotterization. In the limit of small variance, with high probability we have the following.
 \begin{equation}
\begin{aligned}
\epsilon_\theta & = \sqrt{\mathbb{E}\left[\left(\theta_{\mathrm{est}}-\theta_{\mathrm{true}}\right)^2\right]}  \approx \sqrt{\epsilon_V(\theta)+\left(\pi \epsilon_{\mathrm{QFT}} + \epsilon_{\mathrm{Trotter}} + \epsilon_{\rm synth}  \right)^2} \\
& \approx \sqrt{\left(\frac{ \pi}{2^{m+1}}\right)^2+\left(\pi \epsilon_{\mathrm{QFT}}+ \epsilon_{\rm synth} + \epsilon_{\mathrm{Trotter}} \right)^2},
\end{aligned}
\label{eqn:error-phase}
\end{equation}
We select the following in order to get the above expression.
\begin{eqnarray}
    \epsilon_{\rm Trotter} \le \frac{\sqrt{2} \epsilon_{\theta} }{4}, \qquad \epsilon_{\rm QFT} \le \frac{\sqrt{2} \epsilon_{\theta} }{8\pi}, \qquad \epsilon_{\rm synth} \le \frac{\sqrt{2} \epsilon_{\theta} }{8}, \qquad 2^{m} \ge \frac{\pi}{\sqrt{2} \epsilon_\theta}
\label{eqn:error-bounds-phase-estimation}
\end{eqnarray}
We require that the error in the Trotter-Suzuki expansion to be at most $\epsilon_{\rm Trotter} = \tilde \alpha_{\rm comm} \tau^3$.  Further, let 
\begin{eqnarray}
\epsilon_\theta = \epsilon_E \tau,  \label{eqn:epsilon_theta}
\end{eqnarray}
where $\epsilon_E$ is the estimate in the energy that comes from rescaling the estimate of the phase by a factor of $\tau$.
Thus it suffices to choose 
\begin{eqnarray}\label{eq:taudef}
    \tau = \sqrt{\frac{\epsilon_E}{2^{3/2}\widetilde\alpha_{\rm comm}} }
\end{eqnarray}
and therefore, the error in the phase is related to the error in the energy estimate as
\begin{equation}
\epsilon_{\theta} = \sqrt{\frac{\epsilon_E^{3}}{2^{3/2}\widetilde\alpha_{\rm comm}} } 
\end{equation}

and therefore for $\pi \sqrt{\widetilde\alpha_{\rm comm}}/\epsilon_E^{3/2} \ge \frac{1}{1-2^{-1/8}}$ at most
\begin{align}
    2^m = \left\lceil\frac{\pi\sqrt{\widetilde\alpha_{\rm comm}}}{2^{1/8} \epsilon_E^{3/2}}\right\rceil \le \frac{\pi\sqrt{\widetilde\alpha_{\rm comm}}}{\epsilon_E^{3/2}}.
\end{align}
Finally, as the analysis in~\cite{Babbush2018} is tight in the limit as $\epsilon_E\rightarrow 0$ let us assume that $\epsilon_E$ is small enough so that 
\begin{eqnarray}
    \left|\sqrt{\mathbb{E}\left[\left(\theta_{\mathrm{est}}-\theta_{\mathrm{true}}\right)^2\right]} - \sqrt{\left(\frac{ \pi}{2^{m+1}}\right)^2+\left(\pi \epsilon_{\mathrm{QFT}}+ \epsilon_{\mathrm{Trotter}} + \epsilon_{\rm synth} \right)^2}\right| \le \epsilon_{\theta}.  
\end{eqnarray}
We then can ensure a sufficient value of $m$ by taking $\epsilon_E \rightarrow \epsilon_E/2$ and using the remaining error budget to accommodate the error due to $m$ being finite.  Using the observation that since $2^{3/2} < \pi$ it suffices to take
\begin{eqnarray}
    \frac{\pi\sqrt{\widetilde\alpha_{\rm comm}}}{(\epsilon_E/2)^{3/2}} \le \frac{\pi^2\sqrt{\widetilde\alpha_{\rm comm}}}{\epsilon_E^{3/2}} =2^m
\end{eqnarray}
And thus this sufficient value is an upper bound on the necessary value.

\end{proof}

We now have sufficient ingredients to prove Theorem \ref{thm:totalTocc}, as follows.

\begin{proof}[Proof of Theorem \ref{thm:totalTocc}]
Suppose we allocate $\epsilon_r$ as an upper bound on the permitted synthesis error per $R_z$ gate. From Theorem \ref{thm:costOccBasis} and Table \ref{tab:costOccBasis} we find that the total number of (controlled)-$R_z$ required for each Trotter step is at most
\begin{eqnarray}
    N_r:=\frac{N^4|\Omega|^2(|\Omega|-1)}{48}+\frac{N^3|\Omega|^2}{3}+2N^2|\Omega|(|\Omega|-1)+4N|\Omega|  \label{eq:Nr}
\end{eqnarray}
and so using the T-count estimate in \cite{2015_KMM} and assuming the T-count of controlled-$R_z$ is at most the T-count of $R_z$ \cite{2022_GMM}, the expected number of T-gates from rotations is at most 
\begin{eqnarray}
    N_{T/R_z}\le N_r\left(3.067\log_2(2/\epsilon_r)-4.327\right).   \nonumber
\end{eqnarray}
Also, from Theorem \ref{thm:costOccBasis} we require the following additional number of T-gates.
\begin{eqnarray}
    N_{+}:= \frac{N^4|\Omega|^2(|\Omega|-1)}{4}+\frac{8N^3|\Omega|^2}{3}+8N^2|\Omega|(|\Omega|-1). 
\end{eqnarray}
So, the total number of T-gates per Trotter step is 
\begin{eqnarray}
    \mathcal{G}_T&\leq&N^4|\Omega|^2(|\Omega|-1)[0.064\log_2(2/\epsilon_r)+0.16]+N^3|\Omega|^2[1.022\log_2(2/\epsilon_r)+1.224]    \\
    &&+N^2|\Omega|(|\Omega|-1)[6.134\log_2(2/\epsilon_r)-0.654]+4N|\Omega|[3.067\log_2(2/\epsilon_r)-4.327]    \nonumber \\
    &\in& O\left(N^4|\Omega|^3\log_2(2/\epsilon_r)\right).
    \label{eqn:TperStep_occ}
\end{eqnarray}

Using Lemmas \ref{lem:occTrotter} and \ref{lem:error-phase}, the total number of T-gates required to achieve an eigenstate within total error $\epsilon_E$ is at most

\begin{eqnarray}
    \mathcal{G}_T\cdot 2^m &\le&\frac{\pi^2\sqrt{\alpha_{\rm comm}}}{\epsilon_E^{3/2}} \left(N^4|\Omega|^2(|\Omega|-1) [0.064\log_2(2/\epsilon_r)+0.16]+N^3|\Omega|^2[1.022\log_2(2/\epsilon_r)+1.224] \right.  \nonumber \\
    &&\left.+N^2|\Omega|(|\Omega|-1)[6.134\log_2(2/\epsilon_r)-0.654]+4N|\Omega|[3.067\log_2(2/\epsilon_r)-4.327] \right) \nonumber\\
    &\in& O\left(\frac{\lambda N^{7}|\Omega|^{3}}{m^{5/2}\epsilon_E^{3/2}}\log(1/\epsilon_r)\right).
    \label{eqn:T_occ2}
\end{eqnarray}
The total synthesis error due to approximation of the rotation gates is $\epsilon_{synth} = \epsilon_r\cdot N_r$, where the value of $\epsilon_{synth}$ is given in Eq. \ref{eqn:error-bounds-phase-estimation}. Plugging in the values of the time step $\tau$ and $\epsilon_{\theta}$ from Eq.~\eqref{eq:taudef} and ~\eqref{eqn:epsilon_theta}, respectively we obtain the following bound on $\epsilon_r$ in order to ensure that the final error in the estimate of the energy is at most $\epsilon_E$.
\begin{equation}
    \epsilon_{r} \leq \frac{\sqrt{2}\epsilon_{E}\tau}{8 N_{r}} = \frac{\sqrt{2}\epsilon_{E}}{8 N_{r}}\sqrt{\frac{\epsilon_E}{2^{3/2} \tilde{\alpha}_{comm}}}
    \label{epsilon_r}
\end{equation}
To obtain the total number of ancillary qubits used, we get
\begin{eqnarray}
m &\in& O \left( \log{\left( \frac{\alpha_{comm}^{1/2}}{\epsilon_{E}^{3/2}} \right)}\right) \nonumber \\
&\in& O \left( \log{\left( \frac{\lambda N}{M \epsilon_E}\right)} \right)
\end{eqnarray}
where we have repeatedly used $\log{(A^m/B^n)} \in O\left( \log{(A/B)}\right)$ for constant $m,n >0$.
Thus the theorem is proved.
\end{proof}

\section{Field Amplitude Basis}
\label{subsec:algoAmp}

We describe, in this Appendix, algorithms to simulate the scalar $\phi^4$ Hamiltonian $H_{amp}$ expressed in the amplitude basis, as given in Eq. \ref{eqn:phi4-hamiltonian}, thus proving the results in Section \ref{subsec:ampBasisMain}. The qubitization-based algorithms have been described in Appendix  \ref{subsecn:equal-weight-lcu} (Algorithm I), \ref{subsec:blockHampZ} (Algorithm IIIa) and \ref{subsec:algoIIIb} (Algorithm IIIb), while the Trotterization based approach has been described in Appendix \ref{subsec:trotterHamp} (Algorithm II). 
Before describing our algorithms we mention the following results which have been used in the LCU-based approaches in order to reduce the gate complexity.

\paragraph{Recursive block encoding : } We use the following theorem to recursively block encode $H_{amp}$ using a divide and conquer approach, as described in \cite{2023_MSW}, where it has been shown that with such an approach it is possible to block encode with a smaller number of gates.
Suppose without loss of generality, we have a Hamiltonian $H_i$ expressed as a LCU i.e. $H_i=\sum_{j=1}^{M_i}h_{ij}U_{ij}$, such that $\lambda_i=\sum_j|h_{ij}|$. In this case, we can have a $\left(\lambda_i,\log M_i,0\right)$-block encoding of $H_i$ using an ancilla preparation subroutine and a unitary selection subroutine, which we denote by $\prep_i$ and $\sel_i$ respectively.
\begin{eqnarray}
    \prep_i\ket{0}^{\log M_i}&=&\sum_{j=1}^{M_i}\sqrt{\frac{h_{ij}}{\lambda_i}}\ket{j}   \label{eqn:prepi} \\
 \sel_i&=&\sum_{j=1}^{M_i}\ket{j}\bra{j}\otimes U_{ij}   \label{eqn:seli} 
\end{eqnarray}
It can be shown that~\cite{2012_CW}
\begin{eqnarray}
    \bra{0}\prep_i^{\dagger}\cdot \sel_i\cdot\prep_i\ket{0}&=&\frac{H_i}{\lambda_i}.    \label{eqn:prepiSeli}
\end{eqnarray}
Suppose we have $M$ Hamiltonians denoted by $H_1,\ldots,H_M$, each of which has an LCU decomposition and for each one of them we define the subroutines as in Eq.~\ref{eqn:prepi} and \ref{eqn:seli}. Now we use these subroutines to define the following,
\begin{eqnarray}
 \prep\ket{0}^{\log M+\sum_i\log M_i}&=&\left(\sum_{i=1}^M\sqrt{\frac{w_i\lambda_i}{\mathcal{\nconst}}}\ket{i}\right)\otimes\bigotimes_{i=1}^M\prep_i    \label{eqn:divPrep} \\
 \sel&=&\sum_{i=1}^M\left(\ket{i}\bra{i}\otimes\bigotimes_{k=1}^{i-1}\id\otimes\sel_i\otimes\bigotimes_{k=i+1}^M\id\right),     \label{eqn:divSel}
\end{eqnarray}
where $w_i>0$ and $\nconst=\sum_{i=1}^Mw_i\lambda_i$. We can use the above two subroutines to block encode a linear combination of Hamiltonians as follows. 
\begin{theorem}[\cite{2023_MSW}]
Let $H=\sum_{i=1}^Mw_iH_i$ be the sum of $M$ Hamiltonians and each of them is expressed as sum of unitaries as : $H_i=\sum_{j=1}^{M_i}h_{ij}U_{ij}$ such that $\lambda_i=\sum_j|h_{ij}|$, $w_i>0$. Each of the summand Hamiltonian is block-encoded using the subroutines defined in Eq.~\ref{eqn:prepi} and \ref{eqn:seli}. Then, we can have an $(\mathcal{A},\lceil \log_2(M) \rceil,0)$-block encoding of $H$, where $\nconst=\sum_{i=1}^Mw_i\lambda_i$, using the ancilla preparation subroutine ($\prep$) defined in Eq.~\ref{eqn:divPrep} and the unitary selection subroutine ($\sel$) defined in Eq.~\ref{eqn:divSel}.
\begin{enumerate}
    \item The PREP subroutine has an implementation cost of $\mathcal{C}_{\prep}=\sum_{i=1}^M\mathcal{C}_{\prep_i}+\mathcal{C}_{w}$, where $\mathcal{C}_{\prep_i}$ is the number of gates to implement $\prep_i$ and $\mathcal{C}_w$ is the cost of preparing the state $\sum_{i=1}^M\sqrt{\frac{w_i\lambda_i}{\nconst}}\ket{i}$.

    \item The $\sel$ subroutine can be implemented with a set of multicontrolled-X gates - \\
    $\{M_i\text{ pairs of }C^{\log_2M_i+1}X\text{ gates }:i=1,\ldots,M\}$, $M$ pairs of $C^{\log M}X$ gates and $\sum_{i=1}^MM_i$ single-controlled unitaries - $\{cU_{ij}: j=1,\ldots,M_i; i=1,\ldots,M\}$. 
\end{enumerate}
 \label{thm:blockEncodeDivConq}
\end{theorem}
Additionally, suppose in the above theorem, all the $H_i$ are the same but they act on disjoint subspaces. In this case, each $\prep_i$ is the same and so it is sufficient to keep only one copy of $\prep_i$ in the $\prep$ subroutine of Eq.~\ref{eqn:divPrep}. We can absorb $w_i$ in the weights of the unitaries obtained in the LCU decomposition of $H_i$. Thus, in this case we have
 \begin{eqnarray}
  \prep\ket{0}^{\log M+\log M_i}=\left(\sqrt{\frac{1}{M}}\sum_{i=1}^M\ket{i}\right)\otimes \prep_i.  \label{eqn:divPrepEq}
 \end{eqnarray}
We require only $\lceil\log M \rceil$ H gates to prepare the superposition in the first register by padding out the number of such subspaces to be a power of $2$.  We also need to make slight modifications in the $\sel$ procedure. This time, we keep an extra ancilla qubit, initialized to 0, in each subspace. Given a particular state of the first register, we select a subspace by flipping the qubit in the corresponding subspace. The unitaries in each subspace are now additionally controlled on this qubit (of its own subspace).

\paragraph{LCU decomposition of integer diagonal matrices :} If a diagonal matrix consists of integers only then we can have an LCU decomposition consisting of $O(\log m_{max})$ signature matrices, where $m_{max}$ is the maximum absolute value of any of its entries. This can be obtained by a binary decomposition of each integer, as stated below.
\begin{lemma}[\cite{2023_MSW}]
    Let $M_I$ is a $N\times N$ diagonal integer matrix, which has $N'$ positive integers whose maximum value is $m_{max}'$ and $N''$ negative integers such that $m_{max}''=\max_i\{|M_I[i,i]|:M_I[i,i]<0\}$. Then, $M=c_0\id+\sum_{i=1}^{N'+N''}c_iD_i$, where $D_i$ are signature matrices and $N'\leq \lceil\log_2(m_{max}'+1)\rceil=\zeta'$ and $N''\leq\lceil\log_2(m_{max}''+1)\rceil=\zeta''$. Also, $\sum_{i=0}^{N'+N''}|c_i|\leq 2^{\zeta'}-1$.
    \label{lem:lcuIntgDiag}
\end{lemma}

\paragraph{Group of multicontrolled-X gate synthesis :} We can further optimize the number of gates by implementing the group of multicontrolled-unitaries in the SELECT subroutines, using the following theorem \cite{2023_MSW}. Here we partition the control qubits into different groups, store intermediate information in some ancillae and then implement the required logic using these intermediate results. 
\begin{theorem}[\cite{2023_MSW}]
Consider the unitary $U = \sum_{j=0}^{M-1} \ketbra{j}{j} \otimes U_j$ for unitary operators $U_j$ that can be implemented controllably. We assume $M$ is a power of 2 for simplicity.
Suppose we have $\log_2M$ qubits and $M$ (compute-uncompute) pairs of $C^{\log_2M}X$ gates for selecting the $M$ basis states. Let $r_1,\ldots,r_n\geq 1$ be positive fractions such that $\sum_{i=1}^n\frac{1}{r_i}=1$ and $\frac{\log_2M}{r_i}$ are integers. Then, $U$ can be implemented with a circuit with $$\sum_{i=1}^nM^{\frac{1}{r_i}}C^{\frac{\log_2M}{r_i}}X + MC^nX$$ 
(compute-uncompute) pairs of gates, $M$ applications of controlled $U_j$ and at most $\sum_{i=1}^nM^{\frac{1}{r_i}}$ ancillae. 
\label{thm:CX}
\end{theorem}
Following the construction in \cite{2017_HLZetal, 2018_G}, the number of T-gates required to implement such multiply controlled gates is
\begin{eqnarray}
   \mathcal{T}_n= \sum_{i=1}^nM^{\frac{1}{r_i}}\left(\frac{4\log_2M}{r_i}-4\right)+M(4n-4).
    \label{eqn:CXT}
\end{eqnarray}
With the help of logical AND gadgets we do not require to use any T-gate for the uncomputation part.

\paragraph{Equal weight LCU:}
Note that the Hamiltonian Eq.~\eqref{eqn:phi4-hamiltonian} consists of 4 different families of terms, the $\Pi^2$, $\Phi\Phi$, $\Phi^2$, and $\Phi^4$ terms, each of which share the same coefficients. In principle, if we can further ensure that the LCU decompositions of each of those terms provides the same weight to every unitary in the LCU, then we can exploit this structure to drastically simplify the $\prep$ and $\sel$ circuits. To achieve such an equal weight LCU, we will make use of several lemmas detailed in the Supplemental Material \cite{SuppMat}.

\paragraph{Error in the scalar field : } We also bound the error in the scalar field in terms of the target energy scale as below.

\begin{theorem}
Let $\vert \phi_{max} \vert$ be the maximum allowed value of the scalar field. Then, it suffices to take 
\begin{equation}
\vert \phi_{max} \vert = \left( \frac{\epsilon E_{max}}{C(M, \Lambda, d) \vert \Omega \vert}\right)^{1/4}
\end{equation}
where $E_{max} > 0$ is the maximum energy scale we wish to allow in our simulation of the Hamiltonian $H_{amp}$ given in Eq.~\eqref{eqn:phi4-hamiltonian}, $\epsilon = \Pr(H_{amp} \geq E_{max})$ is the probability that a measurement of the Hamiltonian exceeds $E_{max}$, $\vert \Omega \vert$ is the lattice size, and $C(M, \Lambda, d) = M^2 + 3d + \frac{\Lambda}{4!} + \frac{3}{2}$.
\label{thm:phi-max-energy}
\end{theorem}
\begin{proof}
Let $\vert \phi_{max} \vert$ be the maximum allowed value of the scalar field in our simulation. Then, in the field amplitude basis defined by $\hat{\Phi} \ket{\phi} = \phi \ket{\phi}$, it readily follows that $\bra{\phi} \hat{\Phi}^{n} \ket{\phi} \leq \vert \phi_{max} \vert^{n}$ at a single site. Similarly, since the momentum operator is related to the field operator via a Fourier transform, $\hat{\Pi} = \mathcal{F}^{\dagger} \hat{\Phi} \mathcal{F}$, then letting $\ket{\tilde{\phi}} = \mathcal{F} \ket{\phi}$, it also readily follows for a single lattice site that
\begin{eqnarray}
\bra{\phi} \hat{\Pi}^{n} \ket{\phi} &=& \bra{\tilde{\phi}} \hat{\Phi}^{n} \ket{\tilde{\phi}} \nonumber \\
&\leq& \vert \phi_{max} \vert^{n}
\end{eqnarray}
Now, the Hamiltonian is given by
\begin{equation}
H_{amp} = \sum_{\vec{x} \in \Omega} \left[ \frac{1}{2} \Pi^{2}(\bx) + (M^2 + d + 1) \Phi^2 (\vec{x}) + \frac{\Lambda}{4!} \Phi^{4}(\vec{x}) - 2 \sum_{i=1}^{d} \Phi(\vec{x}) \Phi(\vec{x} + a \hat{x}_i)\right]
\end{equation}
It therefore follows that
\begin{eqnarray}
\bra{\phi} H_{amp} \ket{\phi} &=& \sum_{\vec{x} \in \Omega} \bra{\phi} \left[ \frac{1}{2} \Pi^{2}(\bx) + (M^2 + d + 1) \Phi^2 (\vec{x}) + \frac{\Lambda}{4!} \Phi^{4}(\vec{x}) - 2 \sum_{i=1}^{d} \Phi(\vec{x}) \Phi(\vec{x} + a \hat{x}_i)\right] \ket{\phi} \nonumber \\
&\leq& \sum_{\vec{x} \in \Omega} \left[ \left( \frac{1}{2} + (M^2 + d + 1) + 2d \right) \vert \phi_{max} \vert^{2} + \frac{\Lambda}{4!} \vert \phi_{max} \vert^{2}  \right] \nonumber \\
&\leq& \sum_{\vec{x} \in \Omega} \left[ \left( \frac{1}{2} + M^2 + 3d + 1 + \frac{\Lambda}{4!} \right) \vert \phi_{max} \vert^{2} \right] \nonumber \\
&=& C(M, \Lambda, d) \vert \Omega \vert \vert \phi_{max}\vert^{4}
\end{eqnarray}
where $C(M, \Lambda, d) = M^2 + 3d + 1 + \frac{\Lambda}{4!} + \frac{1}{2}$. Let $E_{max} > 0$ be the maximum allowed energy scale in our simulation. By Markov's inequality then,
\begin{eqnarray}
\Pr(H_{amp} \geq E_{max}) &\leq& \frac{\bra{\phi} H_{amp} \ket{\phi}}{E_{max}} \nonumber \\
&\leq& \frac{C(M, \Lambda, d) \vert \Omega \vert \vert \phi_{max}\vert^{4}}{E_{max}}
\end{eqnarray}
from which the statement of the theorem readily follows.
\end{proof}

\subsubsection{Algorithm I : Equal weight LCU}
\label{subsecn:equal-weight-lcu}

Here, we describe how to construct the $\prep$ and $\sel$ oracles for the $\Phi^{4}$ Hamiltonian $H_{amp}$ appearing in Eq.~\eqref{eqn:phi4-hamiltonian}, using an equal weight LCU for the operators appearing in the Hamiltonian. By ensuring that we provide the same coefficient to each of the unitaries appearing in the LCU decompositions of $\Phi$, $\Phi^{2}$, $\Phi^{4}$ and $\Pi^{2}$, we essentially ensure that we have a total of only 4 families of terms each sharing the same coefficient within that family of terms. This drastically simplifies the $\prep$ circuit, the only nontrivial part of which is to prepare a 2-qubit state, while the rest of it is composed of transversal Hadamards. We also describe how to efficiently carry out a $\sel$ circuit that does not require selecting each of the unitaries within a family in succession, but can rather apply all of them simultaneously using a comparator.

First, we describe the decompositions of the operators appearing in the Hamiltonian of Eq.~\eqref{eqn:phi4-hamiltonian}. We apply the decomposition lemma from Supplemental Material to obtain these decompositions \cite{SuppMat}. The $\Phi_a\Phi_b$, $\Phi^2$ and $\Phi^4$ terms are all built from the same basic diagonal operator $\frac{\hat{\Phi}}{\Delta\Phi} =  Diag(-k+1, \dots, 0, \dots, k)$.

\paragraph{Decompositions}

First, for the $\hat{\Phi}$ operator, we use $L=2k$, $n_{max} = k$, and $n_j = j-k+1$ to get
\begin{eqnarray}
\frac{\hat{\Phi}}{\Delta\Phi} &=& \frac{1}{2} \sum_{i=0}^{2k-1} U^{(i)}, \quad \text{where} \nonumber \\
U^{(i)} &=& - \sum_{j=0}^{i-1} \vert j \rangle \langle j \vert + \sum_{j=i}^{2k-1} \vert i \rangle \langle i \vert = \sum_{j=0}^{2k-1} \left[ 2\Theta(j-i) - 1\right] \vert j \rangle \langle j \vert
\label{eqn:LCU-phi-cmp}
\end{eqnarray}
On any two sites $a$ and $b$, the operator $-\hat{\Phi}_{a} \hat{\Phi}_{b}$ is then simply given by
\begin{eqnarray}
 -\frac{\hat{\Phi}_{a}\hat{\Phi_{b}}}{\left( \Delta\Phi \right)^2} &=& \frac{1}{4} \sum_{i_a, j_a=0}^{2k-1} \sum_{i_b, j_b=0}^{2k-1} \left[ 2\Theta(j_a - i_a) - 1\right] \left[ 1 - 2\Theta(j_b-i_b)\right] \ket{j_a}\bra{j_a}\otimes \ket{j_b}\bra{j_b}
\end{eqnarray}

For the $\hat{\Phi}^2$ term, we use $L=2k^2$, $n_{max} = k^2$ and $n_j= (k-j-1)^2$ 
to get
\begin{eqnarray}
\left( \frac{\hat{\Phi}}{\Delta\Phi}\right)^2 &=& \frac{1}{2} \sum_{i=0}^{2k^2 - 1} U^{(i)}, \quad \text{where} \nonumber \\
U^{(i)} &=& \sum_{j=0}^{2k-1} \left[ 2\Theta(k^2 + (k-j-1)^2 - i -1) - 1 \right] \vert j \rangle \langle j \vert
\label{eqn:LCU-phisqr}
\end{eqnarray}

Similarly, for the $\hat{\Phi}^4$ term, we use $L = 2k$, $n_{max} = k^4$ and $n_j = (k-j-1)^4$ 
to get
\begin{eqnarray}
\left( \frac{\hat{\Phi}}{\Delta\Phi}\right)^4 &=& \frac{1}{2} \sum_{i=0}^{2k^4 -1} U^{(i)}, \qquad \text{where} \nonumber \\
U^{(i)} &=& \sum_{j=0}^{2k-1} \left[ 2\Theta(k^4 + (k-j-1)^4 -i -1) - 1\right] \ket{j} \bra{j}
\label{eqn:LCU-phipow4}
\end{eqnarray}

\paragraph{$\prep$ circuit}
As the above decompositions illustrate, we have LCU decompositions for each of the terms in $H_{amp}$ such that all the unitaries share the same coefficients within each of 4 different families. In the language of Theorem \ref{thm:blockEncodeDivConq}, this means that we have $M=4$ different families. These equal weight LCU decompositions lead to a much simpler $\prep$ circuit. In particular, we only need to prepare a nontrivial $\log_{2}{M} = 2$ qubit register whose amplitudes encode the shared coefficients of the 4 groups of terms in the Hamiltonian ~\eqref{eqn:phi4-hamiltonian}. The $\prep_i$ (sub-PREPARE) circuits for each of these groups, which would control which of the $U^{(i)}$'s in the LCUs above would be selectively applied, then simply become a layer of transversal Hadamard gates.

In other words, the $\prep$ circuit then simplifies to
\begin{equation}
\prep = \prep_{F} \otimes H^{\otimes \log_{2}{\vert \Omega \vert}} \otimes H^{\otimes \log_{2}{2k^{4}}}
\end{equation}
where
\begin{equation}
\prep_{F} \ket{0}^{\otimes 2} = \alpha_{\pi^{2}} \ket{00} + \alpha_{\phi^2} \ket{01} + \alpha_{\phi^4} \ket{10} + \alpha_{\phi\phi} \ket{11}
\end{equation}
where the amplitudes $\alpha_j$ encode the coefficients of the 4 families of terms in the Hamiltonian~\eqref{eqn:phi4-hamiltonian}
\begin{eqnarray}
\vert \alpha_{\pi^{2}} \vert^2 / \mathcal{N} = \frac{1}{2} &,& \vert \alpha_{\phi^{2}} \vert^2 = \left( M^{2} + d + 1 \right)  / \mathcal{N} \nonumber \\
\vert \alpha_{\phi^{4}} \vert^2 = \frac{\Lambda}{4!}  / \mathcal{N} &,& \vert \alpha_{\phi\phi} \vert^2 = 2  / \mathcal{N}
\end{eqnarray}
where $\mathcal{N} = \sum_{j \in \{\pi^{2}, \phi^2, \phi^4, \phi\phi\}} \vert \alpha_j \vert^{2}$. Note that this is not the same as the coefficient 1-norm of the Hamiltonian, which instead is given by
\begin{equation}
\vert \alpha \vert_1 = \vert \Omega \vert \left[ \frac{k^{4} \Delta_{\phi}^{4} \Lambda}{4!} + k^{2} \Delta_{\phi}^{2} \left( M^{2} + 3d + \frac{3}{2}\right)\right]
\label{eqn:alpha1-Hamp}
\end{equation}
where we take $\Delta_{\phi} = \sqrt{2\pi/2^{n_{Q}}}= \sqrt{\pi/k}$ where $n_{Q} = \log_{2}{2k}$ is the number of qubits required to implement the field register at a single lattice site. This choice ensures that the free part of the Hamiltonian corresponds to the usual eigenspectrum of a harmonic oscillator.

Thus, the $\prep$ circuit will contribute at most $O(1)$ T-gates, by preparing a superposition of the four nontrivial coefficients in the $\phi^{4}$ Hamiltonian, since the LCUs for all the operators appearing in the Hamiltonian are all equal weight. We need only prepare a nontrivial 2-qubit state, with real amplitudes. It was shown by Vatan and Williams \cite{PhysRevA.69.032315} that such an $SO(4)$ operator requires at most 12 RZ gates and 2 CNOTs.

We now use the result of \cite{reiher2017elucidating} that gives us a T-gate cost of $3.067\log_2(2/\epsilon_{synth})-4.327$ to compile an RZ gate to target precision $\epsilon_{synth}$. Other such cost estimates include the result of \cite{2015_BRS}, which estimates that on average, the T-gate cost of a synthesized/compiled unitary $V$ $\epsilon_{synth}$-approximating a single RZ gate $U$ (i.e. $\vert U - V \vert \leq \epsilon_{synth}$) is $1.15 \log_{2}(1/\epsilon_{synth})$, while \cite{selinger2014efficient} estimates a worst case upper bound of $10 + 4 \log_{2}(1/\epsilon_{synth})$ T-gates per RZ gate. Since the errors are additive, the total T-gate cost increases by the number of RZ gates times the compilation cost of a single RZ gate. Thus, the total T-gate cost of the $\prep$ circuit is given by
\begin{eqnarray}
Count(T)_{\prep} &=& 36.804\log_2(2/\epsilon_{synth}) - 51.924
\label{eqn:tgate_cost_prep}
\end{eqnarray}

\paragraph{Simultaneous $\sel$}

\begin{figure}[t!]
        \centering
        \includegraphics[width=0.8\linewidth]{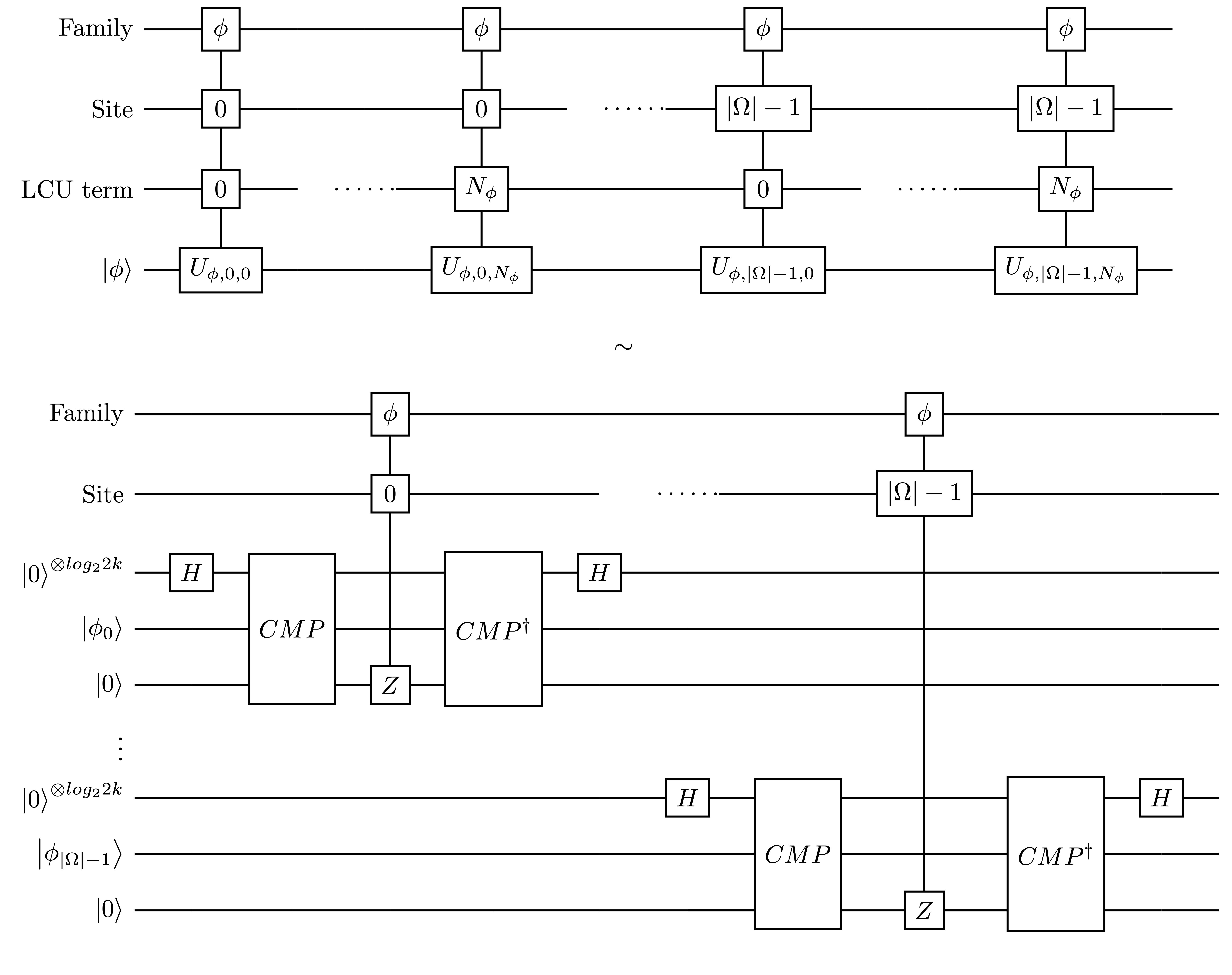}
        \caption{Circuit equivalence demonstrating how the comparator $\texttt{CMP}$ can implement the entire sum of unitaries in Eq.~\eqref{eqn:LCU-phi-cmp}. The ``family" register controls which of the 4 families of terms in the Hamiltonian we want to apply, the ``site" register controls which of the lattice sites we want to apply the operator to, and the ``LCU term" register controls which of the unitaries of Eq.~\eqref{eqn:LCU-phi-cmp} we wish to apply.}
         \label{fig:cmp_advantage_phi}
    \end{figure}

In principle, with the simplification of the $\prep$ circuit described above, we could selectively apply each of the unitaries in a particular LCU by controlling on the state of some register prepared in equal superposition of all bitstrings. However, this decomposition also allows us to simultaneously apply all the $\sel_i$ circuits corresponding to these unitaries. In particular, by making use of the comparator $\texttt{CMP}$ operation, defined as
\begin{equation}
\cmp \ket{i} \ket{j} \ket{0} = \ket{i} \ket{j} \ket{j < i}
\label{eqn:cmp-defn}
\end{equation}

we can apply the following lemma
\begin{lemma}
For integers $i$ and $j$, we have
\begin{equation}
\texttt{CMP}^{\dagger} \left( \mathbb{I} \otimes \mathbb{I} \otimes Z \right) \texttt{CMP} \ket{i} \ket{j} \ket{0} = \left( 2 \Theta(j-i) - 1 \right) \ket{i} \ket{j} \ket{0}
\label{eqn:cmp-z-cmp}
\end{equation}
\label{lemma:cmp-z-cmp}
\end{lemma}
with the proof in the Supplemental Material (Sec. III A) \cite{SuppMat}, from which we readily obtain the following corollary
\begin{corollary}
Let $\ket{\psi} = \sum_{j=0}^{D-1} \alpha_{j} \ket{j}$ denote the state of the scalar field at some lattice point, and $\ket{\overline{+}} = \sum_{i=0}^{D-1} \frac{1}{\sqrt{D}} \ket{i}$ an equal superposition register. Then,
\label{eqn:cmp-phi}
\begin{equation}
\ket{\overline{+}} \otimes \frac{\hat{\Phi}}{\Phi_{max}} \ket{\psi} \otimes \ket{0} = \cmp^{\dagger} \left( \mathbb{I} \otimes \mathbb{I} \otimes Z \right) \cmp \ket{\overline{+}} \ket{\psi} \ket{0}
\end{equation}
where $\Phi_{max} = k\Delta\Phi$.
\label{cor:cmp-phi}
\end{corollary}
In this way, we obviate the need to selectively apply each of the unitaries $U^{(i)}$ appearing in Eq.~\eqref{eqn:LCU-phi-cmp} individually. Instead, we can apply all of them in superposition by a single application of the comparator and its inverse. This simplifies the $\sel_i$ circuits necessary to implement each of the terms in the Hamiltonian. Fig \ref{fig:cmp_advantage_phi} demonstrates this circuit equivalence, and the advantage it confers. Applying each of the equally weighted unitaries individually would require $O(\vert \Omega \vert k)$ many operations controlled on the state of $O(\log{\vert \Omega \vert k})$ many qubits, leading to a $T$ gate count of $O(\vert \Omega \vert k \log{\vert \Omega \vert k})$. Instead, we have $O(\vert \Omega \vert)$ many single-qubit operations controlled on the state of $O(\log{\vert \Omega \vert})$ many qubits, leading to a significantly reduced $T$ gate count of $O(\vert \Omega \vert \log{\vert \Omega \vert})$ for the $\sel$ operation.

\begin{figure}[t!]
    \centering
    \includegraphics[width=0.8\linewidth]{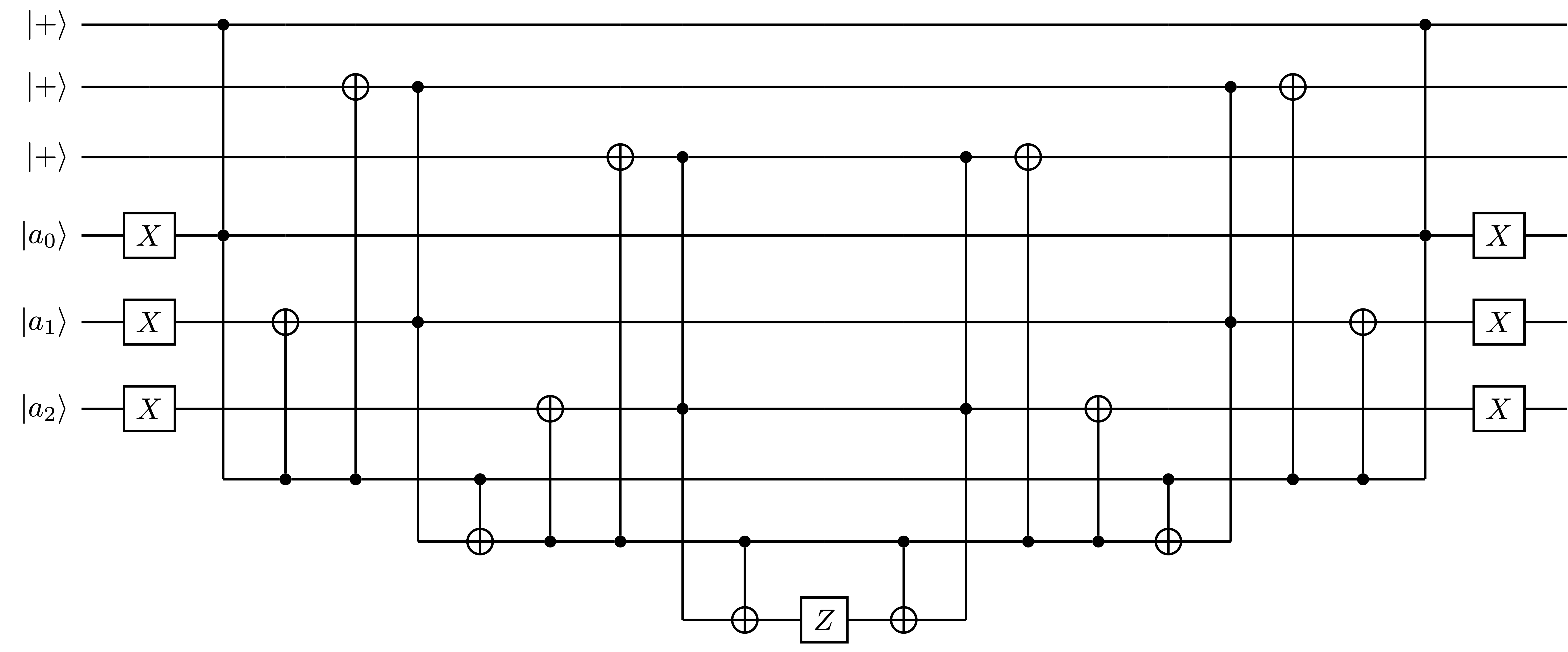}
    \caption{An example circuit for $k=4$ implementing the operation $\cmp'^{\dagger} Z_{anc} \cmp'$ using the logical AND construction of \cite{2018_G}. The field register $\ket{\phi} = \ket{a_2} \ket{a_1} \ket{a_0}$ is compared against an equal superposition of bitstrings using $\cmp'$, the comparison recorded onto an ancilla using $Z_{anc}$, followed by the uncomputation $\cmp'^{\dagger}$ which involves measurements and postmeasurement Clifford operations.}
    \label{fig:cmp-pr-example}
\end{figure}

Note that instead of the comparator, we are also free to apply any other operator $\cmp'$ of the form
\begin{eqnarray}
\texttt{CMP}' \ket{i} \ket{j} \ket{scratch} \ket{0} &=& \ket{\Psi_{i,j,scratch}} \ket{j < i}
\label{eqn:cmp-pr}
\end{eqnarray}
which may significantly reduce the gate cost. An example of such an operation would be to compute the high carry bit of $\overline{\overline{j} + i} = j - i$, which is 1 iff $j < i$, but leaves the state of the $\ket{i}$, $\ket{j}$ and other ancillary qubits $\ket{scratch}$ entangled. Once the appropriate phase has been extracted from the $\ket{j < i}$ qubit, we simply run the inverse of this operation. Noting also that
\begin{eqnarray}
\ket{\bx} \bra{\bx} \otimes \left( U^{\dagger} V U \right) + \ket{\bx^{\perp}} \bra{\bx^{\perp}} \otimes \mathbb{I} &=& \left( \mathbb{I} \otimes U^{\dagger} \right) \left( \ket{\bx} \bra{\bx} \otimes  V + \ket{\bx^{\perp}} \bra{\bx^{\perp}}\otimes \mathbb{I} \right) \left( \mathbb{I} \otimes U \right)
\end{eqnarray}
this means that the controlled version of the entire operation $H^{\otimes m} \texttt{CMP}'^{\dagger} Z_{anc} \texttt{CMP}' H^{\otimes m}$ simply needs to control the $Z_{anc}$ operation. An example circuit involving the $\cmp'$ operation is depicted in Fig. \ref{fig:cmp-pr-example}.

We can now split up the entire $\sel$ circuit as
\begin{equation}
\sel = \sum_{i \in \{ \pi^{2}, \phi^{2}, \phi^{4}, \phi\phi \}} \ket{\alpha_i} \bra{\alpha_i} \otimes \sum_{a=0}^{\vert \Omega \vert - 1} \ket{a} \bra{a} \otimes \sel_{i,a}
\end{equation}
where the left-most register controls the family of terms in the Hamiltonian, the middle register controls which site we apply an operator to, and finally the $\sel_i$ circuits applies the relevant transformations for each of the family of terms as described below, and further detailed in Supplemental Material (Sec. III A) \cite{SuppMat}.

\paragraph{$\sel_{\phi\phi}$ term}
To implement the $-\Phi\Phi$ term, we can simply make use of Corollary ~\eqref{cor:cmp-phi} and the fact that $XZX =-Z$ to obtain
\begin{equation}
\ket{\overline{+}} \otimes \left( - \frac{\hat{\Phi}}{\Phi_{max}} \right) \ket{\psi} \otimes \ket{0} = \cmp^{\dagger} \left( \mathbb{I} \otimes \mathbb{I} \otimes XZX \right) \cmp \ket{\overline{+}} \ket{\psi} \ket{0}
\label{eqn:minus-phi}
\end{equation}
Then, for any two lattice sites $a$ and $b$, the operator $\left( -\hat{\Phi}_a \hat{\Phi}_b \right)/(\Delta\Phi)^2$ is then implemented by the product of operations given in Eqs.~\eqref{eqn:cmp-phi} and ~\eqref{eqn:minus-phi}. In $D$ spatial dimensions, a lattice site has $2D$ neighbors. Since the operation $\left( \cmp^{\dagger}Z_{anc} \cmp\right)$ squares to the identity, this means we only need to apply the operation $\left( \cmp^{\dagger} X_{anc} Z_{anc} X_{anc} \cmp\right)$ on each of the neighbors of a given lattice site, without applying a similar operation on the given lattice site itself. Thus, the entire subselect unitary for this family of terms is then given by
\begin{eqnarray}
\sel_{\phi\phi,a} &=& \sum_{b=0}^{2D-1} \ket{b} \bra{b} \otimes \left( \cmp^{\dagger} Z_{anc} \cmp\right)_{a} \left( \cmp^{\dagger} X_{anc} Z_{anc} X_{anc} \cmp\right)_{b}
\end{eqnarray}
where the $\ket{b}$ register, consisting of $\log_{2}{2D}$ qubits, controls which of the neighbors we apply the operation to. For notational simplicity, in the expression above, the subscripts $a$ and $b$ denote not just the respective lattice sites, but also the associated ancilla required to implement the $\cmp$, as well as the $X_{anc}$ and $Z_{anc}$ operations. As described earlier, the only operation inside the parentheses above that needs to be controlled is the $Z_{anc}$ operation, since all other operations on either side of it are self-conjugate. Any given lattice site will receive one application of $\left( \cmp^{\dagger} Z_{anc} \cmp\right)$ and $2D$ applications of $\left( \cmp^{\dagger} X_{anc} Z_{anc} X_{anc} \cmp\right)$ coming from its neighbors. Between these successive applications, $\cmp$ and its inverse multiply to the identity, as do several applications of the $X$ operator. Therefore, per lattice site, we require one application each of $\cmp$ and its conjugate, two applications of the $X_{anc}$ operator, and $2D+1$ applications of a controlled $Z_{anc}$ operator, each of which is controlled on the state of $\log_{2}{(8D\vert \Omega\vert)}$ ($= 2 + \log_{2}{\vert \Omega \vert} + \log_{2}{(2D)}$) many qubits, where the factor of 2 comes from the $\prep$ circuit, the factor of $\log_{2}{\vert \Omega \vert}$ from the lattice size, and $\log_{2}{(2D)}$ from the number of neighboring lattice sites in $D$ spatial dimensions.

A $C^{n}$-$Z$ gate can be implemented \cite{1995_BBCetal,2018_G} with $n-2$ ancillas initialized in the $\ket{0}$ state, using $n-1$ logical AND computations (and just as many uncomputations) along with 2 Hadamard gates. Each logical AND computation requires 4 T-gates. Therefore, the controlled $Z_{anc}$ operations naively require 
$O(D \vert \Omega \vert \log_{2}{(D \vert \Omega \vert)})$ T-gates
across the entire lattice, and in each of the $D$ spatial dimensions. \par 

A more efficient approach using unary iteration is also described in \cite{Babbush2018}, which brings the cost for implementing such a $\sel$ operation consisting of $L$ operations down to $4L-4$ T-gates, instead of $O(L \log_{2}{L})$. Briefly, this method exploits the fact that successive control patterns differ in at most a constant number of bits, so that the collective circuit can be simplified. In our case, we have $L = \vert \Omega \vert (2D+1)$ many controlled operations to apply across the entire lattice, which incurs a T-gate cost of $4L - 4 = 4\vert \Omega \vert (2D+1) - 4$.

As described before, the $\cmp$ operations may be replaced by $\cmp'$ operations described by Eq.~\eqref{eqn:cmp-pr}. For two $n$-bit numbers $i$ and $j$, the operation $\cmp'$ records the Boolean value of $j < i$ onto an ancillary qubit, and requires $n$ many logical AND computations. Each of these in turn can be implemented using 4 T-gates. The $\cmp'^{\dagger}$ computations are simply the logical AND uncomputations that require measurements on the ancilla and classically conditioned postmeasurement Clifford operations. In our case, we require the comparison of two $\log_{2}{(2k)}$-bit numbers. Thus, the $\cmp'$ operations contribute a total T-gate count of $4\vert \Omega \vert \log_{2}{(2k)}$ across the entire lattice.

The total T-gate count for the $\phi\phi$ part of the Hamiltonian is therefore 
\begin{eqnarray}
\text{Count(T)}_{\phi\phi} &=& 4 \vert \Omega \vert \left( 2D + \log_{2}{k} + 2 \right) - 4
\label{eqn:tgate_cost_phiphi}
\end{eqnarray}

\paragraph{$\sel_{\Phi^2}$ term}
The subselect circuit for $\Phi^2$ term can be constructed quite similarly. The comparator is still used to extract the relevant phase. The only difference is that now we must massage an ancillary register with some initial set of operations that we summarize as $U_{initial}^{\phi^2}$.

Specifically, given $\ket{\psi} = \sum_{j} c_j \ket{j}$, denoting the field register at some lattice point, and another ancillary register $\ket{k}$, we require
\begin{eqnarray}
U_{initial}^{\phi^2} \left( \sum_{j} c_j \ket{j}\right) \ket{k} \ket{scratch} \ket{0}  &=& \ket{\Psi'} \ket{k^2 + (j+1-k)^2}.
\end{eqnarray}
Once the value $k^2 + (j+1-k)^2$ has been written onto a register, we can repeat the procedure described earlier of using a comparator, extracting out the relevant phase, then undoing the comparison. In sum, for each lattice site $a$
\begin{eqnarray}
\sel_{\Phi^2,a} &=& \left( U_{initial}^{\phi^2,\dagger} \texttt{CMP}^{\dagger} Z_{anc} \texttt{CMP} U_{initial}^{\phi^2}\right)_{a}.
\label{eqn:select-phi-sqr}
\end{eqnarray}

As shown in the Appendix \label{app:equal-weight-lcu}, the cost of the $U_{initial}^{\phi^2}$ operation and its inverse incurs a total T-gate cost of $8\vert \Omega \vert \left( 4\left( \log_{2}{2k} \right)^2 + 2\log_{2}{2k} - 6 \right)$, where the leading term comes from multiplying two $\log_{2}{2k}$-bit numbers, while other terms come from other arithmetic operations, such as addition.

The $\texttt{CMP}'$ operation (and its inverse), replacing the $\texttt{CMP}$ operation as before, compares two $\log_2{2k^2}$-qubit numbers and costs $\log_{2}{2k^2}$ many logical ANDs. Thus, the comparisons require $\vert \Omega \vert \log_{2}{2k^2}$ many logical ANDs across the entire lattice, contributing to a T-gate count of $4\vert \Omega \vert \log_{2}{2k^2}$. In addition, we have $\vert \Omega \vert$ many multicontrolled $Z$ gates to apply across the entire lattice, which incurs a cost of $4\vert \Omega \vert -4$ T-gates using the unary iteration method of \cite{Babbush2018}.\\

In all, the total T-gate count for this family of terms is given by
\begin{eqnarray}
Count(T)_{\phi^2} &=& 8\vert \Omega \vert \left( 4\left( \log_{2}{2k} \right)^2 + 2\log_{2}{2k} - 6 \right) + 4 \vert \Omega \vert \left( \log_{2}{2k^2} + 1 \right) - 4 \nonumber \\
&=& 8 \vert \Omega \vert \left( 4 \log_{2}^{2}{k} + 11 \log_{2}{k} + 1 \right) - 4
\label{eqn:tgate_cost_phisqr}
\end{eqnarray}

\paragraph{$\sel_{\pi^2}$ term}
Using the relation $\pi = \mathcal{F}^{\dagger} \phi \mathcal{F} \Rightarrow \pi^2 = \mathcal{F}^{\dagger} \phi^2 \mathcal{F}$, where $\mathcal{F}$ denotes the discrete Fourier transform, we see that the $\sel_{\pi^2}$ circuit is essentially the same as the $\sel_{\phi^2}$ except that at each lattice site, we have additional (uncontrolled) $\log_2{2k}$-qubit Fourier transforms and their inverses at every lattice site. For each lattice site $a$, we have
\begin{eqnarray}
\sel_{\pi^2,a} &=& \left( \mathcal{F}^{\dagger} U_{initial}^{\phi^2,\dagger} \texttt{CMP}^{\dagger} Z_{anc} \texttt{CMP} U_{initial}^{\phi^2} \mathcal{F} \right)_{a}
\label{eqn:select-pi-sqr}
\end{eqnarray}
where $\mathcal{F}$ now represents the QFT circuit. Here, we use the AQFT circuit of \cite{2020_NSM}, which incurs a T-gate cost of approximately
\begin{equation}
Count(T)_{AQFT} \approx 8n \log_{2}\left( \frac{n}{\epsilon_{AQFT}}\right) +  \log_{2}{\left( \frac{n}{\epsilon_{AQFT}} \right)} \log_{2}{\left( \frac{\log_{2}{\left( \frac{n}{\epsilon_{AQFT}}\right)}}{\epsilon_{AQFT}}\right)}
\label{eqn:aqft-cost}
\end{equation}
for a target precision of $\epsilon_{AQFT}$. One application each of the AQFT and its inverse at each lattice site incurs a total cost of $2\vert \Omega\vert$ times the expression in Eq.~\eqref{eqn:aqft-cost} with $n= \log_{2}{2k}$. The total T-gate cost for this family of terms is then
\begin{eqnarray}
Count(T)_{\pi^{2}} &=& Count(T)_{\phi^{2}} + Count(T)_{AQFT} \nonumber \\
&=& 8 \vert \Omega \vert \left( 4 \log_{2}^{2}{k} + 11 \log_{2}{k} + 1 \right) - 4 + 16 \vert \Omega \vert \log_{2}{2k} \log_{2}{\left( \frac{\log_{2}{2k}}{\epsilon_{AQFT}} \right)} + \log_{2}{\left( \frac{\log_{2}{2k}}{\epsilon_{AQFT}}\right)} \log_{2}{\left( \frac{\log_{2}{\left( \frac{\log_{2}{2k}}{\epsilon_{AQFT}}\right)}}{\epsilon_{AQFT}}\right)} \nonumber \\
&& 
\label{eqn:tgate_cost_pisqr}
\end{eqnarray}

\paragraph{$\sel_{\phi^4}$ term}
The subselect circuit for the $\Phi^{4}$ term can be constructed similarly to that of the $\Phi^2$ term. As before, we apply an initial transformation to record an appropriate value to an ancillary register, to which the comparator is applied, and subsequently the appropriate phase extracted to match the entries of the signature matrices of the equal weight LCU Eq.~\eqref{eqn:LCU-phipow4}.

Assuming $\ket{\psi} = \sum_j c_j \ket{j}$ denotes the field register at some lattice point, and we have another ancillary register initialized to $\ket{k}$, along with an entire $\ket{scratch}$ ancillary register to assist with various subroutines, as well as another ancillary register initialized to $\ket{0}$, we require
\begin{eqnarray}
U_{initial}^{\phi^{4}} \left( \sum_j c_j \ket{j} \right) \ket{k} \ket{scratch} \ket{0} &=& \ket{\Psi'} \ket{k^{4} + \left( j + 1 - k\right)^{4}}
\end{eqnarray}
Once the value $k^{4} + \left( j + 1 - k\right)^{4}$ has been recorded in superposition, we can apply the usual comparator technique as described above to extract the relevant phase and produce the desired LCU. In all, we have for each lattice site $a$
\begin{eqnarray}
\sel_{\Phi^4,a} &=& \left( U_{initial}^{\phi^4,\dagger} \texttt{CMP}^{\dagger} Z_{anc} \texttt{CMP} U_{initial}^{\phi^4} \right)_{a}
\label{eqn:select-phi-4}
\end{eqnarray}
As before, the $U_{initial}^{\phi^4}$ circuit requires the implementation of a few basic arithmetic subroutines, including binary multiplication of two $O(\log{k})$-bit numbers, which contributes the dominant cost of $O(\vert \Omega \vert (\log{k})^{2})$ T-gates. As shown in the Supplemental Material (Sec III A Eq. 141) \cite{SuppMat}, this part of the circuit contributes a total T-gate cost of $8 \vert \Omega \vert \left( 20 \log_{2}^{2}{(2k)} - 10\log_{2}{(2k)} + 1\right)$.

The cost of the $\texttt{CMP}'$ (simplifying the $\texttt{CMP}$ operation), which now compares two $\log_{2}{2k^{4}}$-bit numbers, is $4 \vert \Omega \vert \log_{2}{2k^{4}}$. As in the case of the $\phi^{2}$ term, we have $\vert \Omega \vert$ many multicontrolled $Z$ gates to apply across the entire lattice, which contributes a cost of $4 \vert \Omega \vert - 4$ T-gates using the unary iteration method of \cite{Babbush2018}.

In all, the total T-gate count for this family of terms is
\begin{eqnarray}
Count(T)_{\phi^4} &=& 8 \vert \Omega \vert \left( \log_{2}^{2}{k} - 6 \log_{2}{k} - 7 \right) - 4
\label{eqn:tgates_phi4_term}
\end{eqnarray}

The explicit construction of the $U_{initial}^{\phi^4}$ circuit, as well as a detailed counting of the $T$ gates, is provided in Supplemental Material (Sec III A Eq. 138-141)\cite{SuppMat}.

\paragraph{Total cost for Hamiltonian simulation}
The constructions above create a block encoding of the Hamiltonian via $\left( \bra{G} \otimes \mathbb{I} \right) U \left( \ket{G} \otimes \mathbb{I} \right) = H$ where $\ket{G} = U_{PREP} \ket{\overline{0}}$ and $U = U_{SELECT}$. Since $U_{SELECT}^2 = \mathbb{I}$, by Corollary 9 of \cite{Low2019hamiltonian}, we have that the walk operator
\begin{equation}
W = \left( \left( 2 U_{PREP} \ket{0} \bra{0} U_{PREP}^{\dagger} - I \right) \otimes I \right) U_{SELECT}
\label{eqn:walk_operator_phi4}
\end{equation}
also provides a block encoding of the Hamiltonian, i.e. $\left( \bra{G} \otimes \mathbb{I} \right) W \left( \ket{G} \otimes \mathbb{I} \right) = H$ but in an $SU(2)$ invariant subspace containing $\ket{G}$, i.e. in qubitized form. It is this qubitized block encoding that we shall employ to perform phase estimation in order to subsequently estimate scattering matrix elements.

The cost of this block encoding is then given by the sum of twice the cost of the $\prep$ oracle, the cost of the $\sel$ oracle, and the cost of the reflection operator $2 \ket{0} \bra{0} - \mathbb{I}$. The cost of $\prep$ circuit is given by Eq.~\eqref{eqn:tgate_cost_prep}, while the cost of the $\sel$ circuit is given by the sum of Eqs.~\eqref{eqn:tgate_cost_phiphi},~\eqref{eqn:tgate_cost_phisqr},~\eqref{eqn:tgate_cost_pisqr} and ~\eqref{eqn:tgates_phi4_term}. The cost of the reflection operator is at most logarithmic in all parameters, and is exponentially subdominant to the other costs, so we neglect it in our analysis. In all, we have
\begin{eqnarray}
    Cost(T)_{W} &=& 72 \vert \Omega \vert \log_{2}^{2}{k} + 168 \vert \Omega \vert \log_{2}{k} + 8 \vert \Omega \vert D - 32 \vert \Omega \vert + 73.608 \log_{2}{\left( 2/\epsilon_{synth}\right)} \nonumber \\
    && + 16 \vert \Omega \vert \log_{2}{2k} \log_{2}{\left( \frac{\log_{2}{2k}}{\epsilon_{AQFT}} \right)} + \log_{2}{\left( \frac{\log_{2}{2k}}{\epsilon_{AQFT}}\right)} \log_{2}{\left( \frac{\log_{2}{\left( \frac{\log_{2}{2k}}{\epsilon_{AQFT}}\right)}}{\epsilon_{AQFT}}\right)} - 119.848
\end{eqnarray}
In order to employ this block encoding for Hamiltonian simulation using quantum singular value transforms (QSVT), we need
\begin{equation}
O\left( \vert \alpha \vert_1 t + \frac{\log{1/\epsilon}}{\log{\left( e + \frac{\log{(1/\epsilon)}}{\vert \alpha \vert_1 t}\right)}}\right)
\label{eqn:num-queries-qsvt}
\end{equation}
many queries to the walk operator $W$, where the coefficient 1-norm is given by Eq.~\eqref{eqn:alpha1-Hamp}.

\paragraph{Ancilla count} To implement the operations described above, we need several ancillary qubits. First, we need two ancillae for the four families of terms in the Hamiltonian, $\log_{2}{\vert \Omega \vert}$ ancillae for the $\vert \Omega \vert$ sites on the lattice, and $\log_{2}{(2D)}$ ancillae for the neighbors of a given lattice site. These are common across all the $\sel_{i,a}$ operations, even though the last register is only used for the $\sel_{\phi\phi,a}$ operation. 

Second, the comparison operations can only be carried out in succession, and may thus be recycled for use at each lattice site (or lattice site-neighbor pair in the case of the $\sel_{\phi\phi,a}$ operation). The largest comparisons we require are for the $\sel_{\phi^{4},a}$ operation. Each such comparison requires $\log_{2}{(2k^{4})}$ qubits to compare against, another $\log_{2}{(2k^{4})}$ qubits to hold temporary carry values, and 1 more qubit for the $Z_{anc}$ or $(XZX)_{anc}$ operation. The controlled-$Z$ operation further requires at most $\log_{2}{(\vert \Omega \vert D)}$ many ancillary qubits.

Similarly, the ancillae involved in the various $U_{initial}$ operations can be recycled for use across each site and across each family of terms. The dominant cost is for the $\phi^{4}$ term, and provided in Supplemental Material (Sec. III A) below Eq. 139 \cite{SuppMat}. In addition to the ancilla qubits, we also have $\vert \Omega \vert \log_{2}{(2k)}$ many qubits to hold the values of the scalar field across the entire lattice. Adding all these counts together, we find

\begin{equation}
\text{Count(Qubits)} = \vert \Omega \vert \log_{2}{(2k)} + 18 \log^{2}_{2}{k} + 60 \log_{2}{k} + \log_{2}{(\vert \Omega \vert D)} + 29 \in O\left( \vert \Omega \vert \log{k} + \log^{2}{k} \right)
\label{eqn:tot-qubit-count-alg1}
\end{equation}

\subsection{Algorithm II : Trotterization with Z operators}
\label{subsec:trotterHamp}

In this Appendix, we describe a decomposition of the operators $\Phi, \Phi^2,\Phi^4$ as a function of Z operators, thus making it amenable to Trotterization. In the next section we will use these LCU decompositions to estimate the simulation cost using QSVT. The sum of the $\ell_1$ norm of the coefficients in these decompositions is referred to as the $\ell_1$ norm of the decomposition or operator. This is also an upper bound on the spectral norm of the operators and is useful to bound the complexity of LCU-based simulation algorithms.

\paragraph{Decomposition of $\Phi$ : } We decompose $\frac{\Phi}{\Delta\Phi}$ as follows, as done in \cite{shaw2020quantum}.
\begin{eqnarray}
    \frac{\Phi}{\Delta\Phi}&=&\diag(-k,\ldots,-1,0,1,\ldots,k-1)=\frac{1}{2}\diag(-2k+1,\ldots,-1,1,\ldots,2k-1)-\frac{1}{2}\id \nonumber \\
    &=&-\frac{1}{2}\sum_{j=0}^{\log_2k}2^j\Z_j-\frac{1}{2}\id
    \label{app:eqn:phi}
\end{eqnarray}
Number of nonidentity unitaries in the above LCU is $(\log_2k+1)$ and the $\ell_1$ norm is $\frac{1}{2}\left[1+\sum_{j=0}^{\log_2k}2^j\right]=k$.

\paragraph{Decomposition of $\Phi^2$ : } From Eq. ~\eqref{app:eqn:phi}, we have
\begin{eqnarray}
    \left(\frac{\Phi}{\Delta\Phi}\right)^2&=&\left(\frac{\Phi}{\Delta\Phi}+\frac{\id}{2}\right)^2-\left(\frac{\Phi}{\Delta\Phi}+\frac{\id}{2}\right)+\frac{\id}{4}    \nonumber \\
    &=&\frac{1}{12}\left(4^{\log_2k+1}-1\right)\id+\sum_{j=0}^{\log_2k-1}\sum_{k>j}^{\log_2k}2^{j+k-1}\Z_j\Z_k+\frac{1}{2}\sum_{j=0}^{\log_2k}2^j\Z_j+\frac{\id}{4} \nonumber \\
    &=&\frac{1}{6}\left(2^{2\log_2k+1}+1\right)\id+\sum_{j=0}^{\log_2k-1}\sum_{k>j}^{\log_2k}2^{j+k-1}\Z_j\Z_k+\frac{1}{2}\sum_{j=0}^{\log_2k}2^j\Z_j.
    \label{app:eqn:phi2}
\end{eqnarray}
Number of nonidentity unitaries in this expansion is 
\begin{eqnarray}
    1+\frac{\log_2k(\log_2k+1)}{2}+(\log_2k+1)=1+\frac{(\log_2k+1)(\log_2k+2)}{2}.
\end{eqnarray}
If $\log_2k=\zeta$, then the $\ell_1$ norm is
\begin{eqnarray}
&&  \frac{1}{6}\left(2^{2\zeta+1}+1\right)+\frac{1}{2}\sum_{j=0}^{\zeta}2^j\Z_j+\frac{1}{2}\sum_{j=0}^{\zeta-1}2^j\sum_{k>j}^{\zeta} 2^k    \nonumber\\
&=&\frac{1}{6}\left(2k^2+1\right)+\frac{1}{2}\left(2^{\zeta+1}-1\right)+\frac{1}{2}\sum_{j=0}^{\zeta-1}2^{j+j+1}\sum_{k=0}^{\zeta-(j+1)}2^k     \nonumber \\
&=&\frac{k^2}{3}+\frac{1}{6}+k-\frac{1}{2}+\frac{1}{2}\sum_{j=0}^{\zeta-1}2^{2j+1}(2^{\zeta-j}-1)   \nonumber \\
&=&\frac{k^2}{3}+k-\frac{1}{3}+\sum_{j=0}^{\zeta-1}2^{\zeta+j}-\sum_{j=0}^{\zeta-1}4^j  \nonumber \\
&=&\frac{k^2}{3}+k-\frac{1}{3}+2^{\zeta}(2^{\zeta}-1)-\frac{4^{\zeta}-1}{3} \nonumber \\
&=&\frac{k^2}{3}+k-\frac{1}{3}+k^2-k-\frac{k^2}{3}+\frac{1}{3}=k^2.  \label{app:eqn:l1_phi2}
\end{eqnarray}

\paragraph{Decomposition of $\Phi^4$ : }

Since $\left(\frac{\Phi}{\Delta\Phi}\right)^4=\left(\left(\frac{\Phi}{\Delta\Phi}\right)^2\right)^2$, so we can express it as sum of terms with 1, 2, 3 and up to 4 Z gates. So number of unitaries in the decomposition can be at most 
\begin{eqnarray}
&&1+\binom{\zeta}{1}+\binom{\zeta}{2}+\binom{\zeta}{3}+\binom{\zeta}{4}\in \nonumber \\
&=&\frac{1}{24}\left(\zeta^4-2\zeta^3+11\zeta^2+14\zeta+24\right)   \qquad [\zeta=\log_2k]  \nonumber \\
&\in& O(\zeta^4)\quad \text{ or }\quad O\left(\log_2^4k\right).    
\end{eqnarray}
To derive the $\ell_1$ norm of this decomposition we first observe that 
\begin{eqnarray}
    \left(\frac{\Phi}{\Delta\Phi}\right)^4=\left(\left(\frac{\Phi}{\Delta\Phi}\right)^2-\frac{2k^2+1}{6}\id\right)^2+2\frac{2k^2+1}{6}\left(\left(\frac{\Phi}{\Delta\Phi}\right)^2-\frac{2k^2+1}{6}\id\right)+\left(\frac{2k^2+1}{6}\right)^2\id. \nonumber
\end{eqnarray}
From Eq. \ref{app:eqn:l1_phi2} we know that that the $\ell_1$ norm of $\left(\left(\frac{\Phi}{\Delta\Phi}\right)^2-\frac{2k^2+1}{6}\id\right)$ is $k^2-\frac{2k^2+1}{6}=\frac{2k^2}{3}-\frac{1}{6}$. Thus the $\ell_1$ norm of $\left(\frac{\Phi}{\Delta\Phi}\right)^4$ is at most
\begin{eqnarray}
    &&\left(\frac{2k^2}{3}-\frac{1}{6}\right)^2+2\cdot\frac{2k^2+1}{6}\left(\frac{2k^2}{3}-\frac{1}{6}\right)+\left(\frac{2k^2+1}{6}\right)^2=k^4.
    \label{ap:eqn:l1_phi4}
\end{eqnarray} 

Plugging in the above LCU decomposition of the operators into Eq.~\ref{eqn:phi4-hamiltonian}, we obtain the following decomposition of $H_{amp}$. We keep in mind that we have a lattice with $|\Omega|$ vertices, where each vertex has $D$ neighbors. Let us denote the set of edges by $E_D$.
\begin{eqnarray}
    H_{amp}&=&H_{\pi}+H_{\phi}    \label{eqn:HampIII} \\
    H_{\pi}&=&\sum_{\bx\in\Omega} H_{\pi\bx} := \sum_{\bx\in\Omega}\frac{1}{2}\Pi^2(\bx)=\sum_{\bx\in\Omega}\mathcal{F}\left(\gamma_0\id+\sum_j\gamma_jZ_j+\sum_{j,k}\gamma_{jk}Z_jZ_k\right)\mathcal{F}^{\dagger} \label{eqn:Hpi}    \\
    H_{\phi'}&=&\sum_{\bx\in\Omega}\left(\id+\sum_{j}\alpha_jZ_j+\sum_{j,k}\alpha_{jk}Z_jZ_k+\sum_{j,k,l}\alpha_{jkl}Z_jZ_kZ_l+\sum_{j,k,l,m}\alpha_{jklm}Z_jZ_kZ_lZ_m\right) \label{eqn:Hphi'} \\
    H_{\phi''}&=&\sum_{\substack{\bx,\bx'\\(\bx,\bx')\in E_D}}\sum_{j,j'}\beta_{jj'}\left(Z_j\right)_{\bx}\left(Z_{j'}\right)_{\bx'}  \label{eqn:Hphi''} \\
    H_{\phi}&=& H_{\phi'}+H_{\phi''}   \label{eqn:Hphi}
\end{eqnarray}
In the above set of equations the coefficients $\gamma_0, \gamma_j,\gamma_{jk}$, $\alpha_j, \alpha_{jk},\alpha_{jkl}$, $\alpha_{jklm}, \beta_{jj'}$ are determined by the LCU decomposition of the operators as well as Eq. \ref{eqn:phi4-hamiltonian}. We ignore the identity term, since if $H_{amp}=H_{amp'}+\beta_0\id$, where $\beta_0$ is a constant, then $e^{-iH_{amp}'\tau} \propto e^{-iH_{amp}\tau}$ upto some global phase. So the gate complexity to implement the exponentials is the same. 

 We have two noncommuting components, $H_{\phi}$ and $H_{\pi}$. We first analyze the gate complexity to implement $e^{-iH_{\phi}\tau}$ and $e^{-iH_{\pi}\tau}$, where  $\tau$ is one Trotter step. As mentioned, Trotterization becomes quite straightforward with this decomposition. We require 1 $R_z$ gate and few CNOT-gates (to compute the parity) for each exponentiated Z operator. Thus we need at most
\begin{eqnarray}
      |\Omega|\left(\binom{\log_2k+1}{1}+\binom{\log_2k+1}{2}+\binom{\log_2k+1}{3}+\binom{\log_2k+1}{4}\right)+E_D(\log_2k+1)^2
      \label{eqn:rzTrotter}
\end{eqnarray}
number of $R_z$ gates and at most
\begin{eqnarray}
    2|\Omega|\left(\binom{\log_2k+1}{2}+2\binom{\log_2k+1}{3}+3\binom{\log_2k+1}{4}\right)+2E_D(\log_2k+1)^2
    \label{eqn:cnotTrotter}
\end{eqnarray}
number of CNOT-gates. We observe that in order to implement the required exponential of the operators we require a number of CNOT that will realize the necessary paritites i.e. the XOR of all combinations of 2, 3 and 4 qubits. Such a circuit is usually referred to as a 'parity network'. The $R_z$ gates are placed at that point of the circuit where the corresponding parity is realized. The number of CNOT-gates can be optimized using algorithms like \cite{2018_AAM, 2022_GHLetal}. Additionally we require $2|\Omega|$ number of $\log_22k$-qubit QFT. If we use the AQFT circuit of \cite{2020_NSM} then we require 
\begin{eqnarray}
    2|\Omega|\left(8(\log_22k)\log_2\left(\frac{\log_22k}{\epsilon_{QFT}}\right)+\log_2\left(\frac{\log_22k}{\epsilon_{QFT}}\right)\log_2\left(\frac{\log_2\left(\frac{\log_22k}{\epsilon_{QFT}}\right)}{\epsilon_{QFT}} \right)   \right)\label{eqn:approxQFT_ampTrot}   
\end{eqnarray}
number of additional T-gates.

Now we bound the error using the $2^{nd}$ order Trotter formula \cite{1991_S, Childs2021}, in a similar manner as discussed in Section \ref{subsec:occError}. For this we need to estimate the second level nested commutator $\widetilde{\alpha}_{comm}$. In Supplemental Material (Sec. II) \cite{SuppMat} we have proved that
\begin{eqnarray}
    \widetilde{\alpha}_{comm} &\leq& |\Omega|\left( \frac{\Lambda^2}{576}k^{10}\Delta^{10}+\frac{\Lambda}{48}(2M^2+8d^2+2d+3)k^8\Delta^8  + \right. \nonumber \\
    &&\left. \left( \frac{1}{4}(M^2+d+1)(M^2+d+2)+d^2(2M^2+2d+11) \right)k^6\Delta^6  \right) \nonumber \\
    &\in& O\left[|\Omega| \left( \Lambda^2 k^{10}\Delta^{10} + \Lambda M^{2} k^8 \Delta^8 \right) \right],
\label{eqn:alphacomm_ampTrot}
\end{eqnarray}
for fixed spatial dimensionality $d$. We can prove the following bound on the T-gate complexity, using similar reasoning as in Theorem \ref{thm:totalTocc}.

\begin{proof}[Proof of Theorem \ref{thm:ampTrotter}]
Suppose we allocate $\epsilon_r$ as an upper bound on the permitted synthesis error per $R_z$ gate. From Eq. ~\eqref{eqn:rzTrotter} we find that the total number of $R_z$ required for each Trotter step is at most
\begin{eqnarray}
    N_r:=|\Omega|\left(\binom{\log_2k+1}{1}+\binom{\log_2k+1}{2}+\binom{\log_2k+1}{3}+\binom{\log_2k+1}{4}\right)+E_D(\log_2k+1)^2 
    \nonumber
\end{eqnarray}
and so using the T-count estimate in \cite{2015_KMM} gives the expected number of T-gates from rotations to be at most 
\begin{eqnarray}
    N_{T/R_z}\le N_r\left(3.067\log_2(1/\epsilon_r)-4.327\right).   \nonumber
\end{eqnarray}
We also require the following number of additional T-gates due to the QFT performed. Since we are using AQFT, so we allocate an error of $\epsilon_{QFT}$ due to this. Thus from Eq.~\eqref{eqn:approxQFT_ampTrot} we require the following additional number of T-gates.
\begin{eqnarray}
    N_{+}:= 2|\Omega|\left(8(\log_22k)\log_2\left(\frac{\log_22k}{\epsilon_{QFT}}\right)+\log_2\left(\frac{\log_22k}{\epsilon_{QFT}}\right)\log_2\left(\frac{\log_2\left(\frac{\log_22k}{\epsilon_{QFT}}\right)}{\epsilon_{QFT}} \right)   \right). \nonumber 
\end{eqnarray}
So, the total number of T-gates per Trotter step is 
\begin{eqnarray}
    \mathcal{G}_T&=&N_{T/R_z}+N_{+} \in O\left( |\Omega|\log^4(2k)\log(1/\epsilon_r) \right).   \nonumber
\end{eqnarray}
Using Lemmas \ref{lem:occTrotter} and \ref{lem:error-phase} and Eq. ~\eqref{eqn:alphacomm_ampTrot}, the total number of T-gates required to achieve an eigenstate within total error $\epsilon_E$ is at most

\begin{eqnarray}
    \mathcal{G}_T\cdot 2^m &\le&\frac{\pi^2\sqrt{\widetilde{\alpha}_{\rm comm}}}{\epsilon_E^{3/2}} \mathcal{G}_T \in \mathcal{O}\left(\frac{ |\Omega|^{3/2} \sqrt{\Lambda^{2} k^{5} + \Lambda M^{2} k^{4}} \log^4(2k) }{\epsilon_E^{3/2}}\log(1/\epsilon_r)\right) \nonumber 
\end{eqnarray}
where we have used $\Delta = \sqrt{\pi/k}$.
The total synthesis error due to approximation of the rotation gates is $\epsilon_{synth} = \epsilon_r\cdot N_r+\epsilon_{QFT}\geq\epsilon_rN_r$, where the value of $\epsilon_{synth}$ is given in Eq. \ref{eqn:error-bounds-phase-estimation}. Plugging in the values of the time step $\tau$ and $\epsilon_{\theta}$ from Eq. ~\eqref{eq:taudef} and ~\eqref{eqn:epsilon_theta}, respectively we obtain the following bound on $\epsilon_r$ in order to ensure that the final error in the estimate of the energy is at most $\epsilon_E$.
\begin{equation}
    \epsilon_{r} \leq \frac{\sqrt{2}\epsilon_{E}\tau}{8 N_{r}} = \frac{\sqrt{2}\epsilon_{E}}{8 N_{r}}\sqrt{\frac{\epsilon_E}{2^{3/2} \tilde{\alpha}_{comm}}}
    \nonumber
\end{equation}
Note that we get a slightly different prefactor if we also account for the errors arising due to the use of the AQFT, as detailed in Lemma \ref{lemma:trotter-amp-aqft-errors}, but this does not change the asymptotic scaling derived here. Using the above, we obtain
\begin{equation}
\log{\left( \frac{1}{\epsilon_r}\right)} \in O\left( \log{\left( \frac{\vert \Omega \vert k \log{k}}{\epsilon_E} \left( \Lambda^{2}k + \Lambda M^{2} \right)\right)}\right),
\end{equation}
hence proving the theorem.
\end{proof}

 \subsection{Algorithm IIIa : LCU with Z operators}
\label{subsec:blockHampZ}

In this Appendix, we discuss an approach to simulate $H_{amp}$ with qubitization, using Eq. \ref{app:eqn:phi} that expresses $\phi$ as sum of Z operators. Then we repeatedly square it to obtain an LCU decomposition for $\Phi^2$ and $\Phi^4$, as done in previous subsection. We group the terms as follows.
\begin{eqnarray}
    H_{1\bx}&=&\sum_j\alpha_jZ_j+\sum_{j,k}\alpha_{jk}Z_jZ_k+\mathcal{F}\left(\sum_j\gamma_jZ_j+\sum_{j,k}\gamma_{jk}Z_jZ_k\right)\mathcal{F}^{\dagger}  \label{eqn:H1x}   \\
    H_{2\bx}&=&\sum_{j,k,l}\alpha_{jkl}Z_jZ_kZ_l+\sum_{j,k,l,m}\alpha_{jklm}Z_jZ_kZ_lZ_m    \label{eqn:H2x} \\
    H_{12}&=&\sum_{\bx\in\Omega}H_{1\bx}+H_{2\bx}   \label{eqn:H12} \\
    H_{3\bx\bx'}&=&\sum_{j,j'}\beta_{jj'}\left(Z_j\right)_{\bx}\left(Z_{j'}\right)_{\bx'}  \label{eqn:H3x} \\
    H_3&=&\sum_{(\bx,\bx')\in E_D} H_{3\bx\bx'} \label{eqn:H3} \\
    H_{amp}'&=&H_{12}+H_3  \label{eqn:Hamp'}
\end{eqnarray}
Let $L_1=\binom{\log_2k+1}{1}$, $L_2=\binom{\log_2k+1}{2}$, $L_3=\binom{\log_2k+1}{3}$, $L_4=\binom{\log_2k+1}{4}$ and $L_5=(\log_2k+1)^2$. In the following theorem we summarize the number of gates required to block encode $H_{amp}'$ by repeatedly applying Theorem \ref{thm:blockEncodeDivConq}. 

\begin{theorem}
Let $H_{amp}'$ be the Hamiltonian defined in Eq. \ref{eqn:Hamp'} and $\|H_{amp}'\|$ be the $\ell_1$ norm of the coefficients defined in its decomposition (Eq. ~\ref{eqn:H1x}-\ref{eqn:Hamp'}). Then it is possible to have a $\left(\|H_{amp}'\|,\cdot,0\right)$ block encoding of $H_{amp}'$ with $O\left(|\Omega|\log k\right)$
qubits, using the following number of rotation gates
\begin{eqnarray}
    N_r\in O\left(\log^4k\right)   ,\nonumber
\end{eqnarray}
and the following number of additional T-gates (that are not in the decomposition of rotation gates).
\begin{eqnarray}
    N_t\in O\left(|\Omega|\log^4k\right). \nonumber
\end{eqnarray}

\label{thm:blockLcuIIIa}
\end{theorem}

\begin{proof}

We recall the partition of the Hamiltonian $H_{amp}'$ in Eq. \eqref{eqn:H1x}-\eqref{eqn:Hamp'}. $H_{amp}'$ is the nonidentity component of $H_{amp}$. Let $L_1=\binom{\log_2k+1}{1}$, $L_2=\binom{\log_2k+1}{2}$, $L_3=\binom{\log_2k+1}{3}$, $L_4=\binom{\log_2k+1}{4}$ and $L_5=(\log_2k+1)^2$.

\paragraph{Block encoding of $H_{1\vec{x}}$ : } The ancilla preparation subroutine is as follows. We assume a bijective map between $j'=(j,k)$ to some integer in $[L_1+1,L_1+L_2]$.
\begin{eqnarray}
    \prep_{1\vec{x}}\ket{0}^{1+\log_2(L_1+L_2)}&=&\frac{1}{\mathcal{N}_{1\vec{x}}}\left(\sum_{j=1}^{L_1}\sqrt{\alpha_j}\ket{j,0}+\sum_{j'=L_1+1}^{L_1+L_2}\sqrt{\alpha_{j'}}\ket{j',0}+\sum_{j=1}^{L_1}\sqrt{\gamma_j}\ket{j,1}\right.    \nonumber \\
    &&\left.+\sum_{j'=L_1+1}^{L_1+L_2}\sqrt{\gamma_{j'}}\ket{j',0} \right)\qquad [\mathcal{N}_{1\vec{x}} \text{ is normalization constant.}] \label{eqn:prep1x}
\end{eqnarray}
The last qubit is used to select the QFT. We require $1+\log_2(L_1+L_2)$ qubits. For the state preparation we require $1+\log_2(L_1+L_2)$ H, $4(L_1+L_2)-2$ rotation gates and $4(L_1+L_2)+3\log_2(L_1+L_2)-4$ CNOT.

The select subroutine does the following.
\begin{eqnarray}
    \sel_{1\vec{x}}\ket{j,0}\ket{\psi}&\rightarrow&\ket{j,0}Z_j\ket{\psi}   \nonumber \\
    \sel_{1\vec{x}}\ket{j,1}\ket{\psi}&\rightarrow&\ket{j,1}\mathcal{F}Z_j\mathcal{F}^{\dagger}\ket{\psi}   \label{eqn:sel1x}
\end{eqnarray}
If $j>L_1+1$ then we applly two Z gates depending on the mapping. The last qubit is used to selectively apply the pairs of $(\log_2k+1)$-qubit QFT. We require $L_1+L_2$ compute-uncompute pairs of $C^{\log_2(L_1+L_2)}X$ gates, which can be synthesized efficiently using Theorem \ref{thm:CX} \cite{2023_MSW}. If we divide the control qubits into $M_1$ groups such that the $j^{th}$ group has $\frac{1}{r_j}$ fraction of the qubits, then we require
\begin{eqnarray}
    \sum_{i=1}^{M_1}(L_1+L_2)^{\frac{1}{r_i}}C^{\frac{\log (L_1+L_2)}{r_i}}X+(L_1+L_2)\cdot C^{M_1}X \nonumber
\end{eqnarray}
(compute-uncompute) pairs of gates. We assume equal partitioning into two groups i.e. $M_1=2$ and $r_1=\frac{1}{2}$. Then, using the constructions in \cite{2017_HLZetal, 2018_G} , i.e. from Eq. \ref{eqn:CXT} we require
\begin{eqnarray}
    4\sqrt{L_1+L_2}\left(\log_2(L_1+L_2)-2\right)+4(L_1+L_2)    \nonumber
\end{eqnarray}
 T-gates and
\begin{eqnarray}
    \sqrt{L_1+L_2}\left(4\log_2(L_1+L_2)-6\right)+5(L_1+L_2)    \nonumber
\end{eqnarray}
CNOT-gates. Additionally we require $(\log_2k+1)+\frac{\log_2k(\log_2k+1)}{2}=\frac{(\log_2k+1)(\log_2k+2)}{2}$ number of CZ gates and two $(\log_2k+1)$-qubit QFT. To implement the QFTs we require approximately~\cite{2020_NSM}
\begin{eqnarray}
    8(\log_22k)\log_2\left(\frac{\log_22k}{\epsilon}\right)+\log_2\left(\frac{\log_22k}{\epsilon}\right)\log_2\left(\frac{\log_2\left(\frac{\log_22k}{\epsilon}\right)}{\epsilon}\right) \nonumber
\end{eqnarray}
T-gates and almost an equal number of CNOT-gates.

\paragraph{Block encoding of $H_{2\vec{x}}$ : } The ancilla preparation subroutine is as follows.
\begin{eqnarray}
    \prep_{2\vec{x}}\ket{0}^{\log_2(L_3+L_4)}=\frac{1}{\sqrt{\mathcal{N}_{2\vec{x}}}}\sum_{j=1}^{L_3+L_4}\sqrt{\alpha_j''}\ket{j}   \label{eqn:prep2x}
\end{eqnarray}
where $\alpha_j''$ are the weights obtained while expressing $H_{2\vec{x}}$ as sum of Z operators (Eq.~\ref{eqn:H2x}). For the state preparation we require $\log_2(L_3+L_4)$ H, $2(L_3+L_4)+3\log_2(L_3+L_4)-7$ CNOT and $2(L_3+L_4)-2$ rotation gates. 

The select subroutines does the following :
\begin{eqnarray}
    \sel_{2\vec{x}}\ket{j}\ket{\psi}\rightarrow\ket{j}U_j\ket{\psi}, \label{eqn:sel2x}
\end{eqnarray}
where $U_j$ is the $j^{th}$ unitary in the LCU decomposition of $H_{2\vec{x}}$ in Eq.~\ref{eqn:H2x}. To selectively implement the unitaries we require $L_3+L_4$ compute-uncompute pairs of $C^{\log_2(L_3+L_4)}X$ gates. Using Theorem \ref{thm:CX} and assuming that the control qubits have been divided into two equal groups, we require
\begin{eqnarray}
    2(L_3+L_4)^{\frac{1}{2}}C^{\frac{\log(L_3+L_4)}{2}}X+(L_3+L_4)\cdot C^2X
\end{eqnarray}
(compute-uncompute) pairs. These in turn can be decomposed into
\begin{eqnarray}
    4\sqrt{L_3+L_4}\left(\log_2(L_3+L_4)-2\right)+4(L_3+L_4)
\end{eqnarray}
T-gates and
\begin{eqnarray}
    \sqrt{L_3+L_4}\left(4\log_2(L_3+L_4)-6\right)+5(L_3+L_4)
\end{eqnarray}
 CNOT-gates. Additionally, we require 
\begin{eqnarray}
    \binom{\log_2k+1}{3}+\binom{\log_2k+1}{4}=\binom{\log_2k+2}{4}
\end{eqnarray}
CZ gates.

\paragraph{Block encoding of $H_{3\vec{x}\vec{x'}}$ : } The ancilla preparation subroutine is as follows.
\begin{eqnarray}
    \prep_{3\vec{x}\vec{x'}}\ket{0}^{\log_2L_5}=\frac{1}{\mathcal{N}_{3\vec{x}\vec{x'}}}\sum_{j=1}^{L_5}\sqrt{\beta_{j}'}\ket{j}    \label{eqn:prep3x}
\end{eqnarray}
where $\beta_{j}'$ are the weights given in Eq. \ref{eqn:H3x}. For state preparation we require $\log_2L_5$ H gates, $2L_5+3\log_2L_5-7$ CNOT and $2L_5-2$ rotation gates. 

The selection subroutine is as follows.
\begin{eqnarray}
    \sel_{3\vec{x}\vec{x'}}\ket{j}\ket{\psi}\rightarrow\ket{j}U_j\ket{\psi} \label{eqn:sel3x}
\end{eqnarray}
where $U_j$ is the corresponding unitary. We require $L_5$ compute-uncompute pairs of $C^{\log_2L_5}X$ gates. Again we apply Theorem \ref{thm:CX} and assume that the control qubits have been divided into two equal groups. Then we require
\begin{eqnarray}
    2L_5^{\frac{1}{2}} C^{\frac{\log L_5}{2}}X+L_5\cdot C^{2}X
\end{eqnarray}
(compute-uncompute) pairs of gates. These can be further decomposed \cite{2017_HLZetal, 2018_G} into 
\begin{eqnarray}
    4\sqrt{L_5}\left(\log_2L_5-2\right)+4L_5
\end{eqnarray}
T-gates and
\begin{eqnarray}
    \sqrt{L_5}\left(4\log_2L_5-6\right)+5L_5
\end{eqnarray}
CNOT-gates. Additionally we require $\left(\log_2k+1\right)^2$ CZ gates. 

\paragraph{Block encoding of $H_{12}$ : } We use the recursive block encoding Theorem \ref{thm:blockEncodeDivConq}. We can block encode $H_{1\vec{x}}+H_{2\vec{x}}$ using ancilla preparation subroutine that has 1 H and 2 rotation gates. The unitary selection subroutine adds an extra control to each unitary. For $H_{12}$ we prepare an equal superposition of $\log_2|\Omega|$ qubits, using $\log_2|\Omega|$ H gates and use these to select an ancilla of each subspace. The rest of the operations are controlled on this. Thus this adds another control. We require $|\Omega|$ number of $C^{\log_2|\Omega|}X$ compute-uncompute pairs of gates. Using Theorem \ref{thm:CX} and assuming an equal partitioning of the control qubits into two equal groups, we can implement the multicontrolled-X gates using
\begin{eqnarray}
    |\Omega|^{\frac{1}{2}}C^{\frac{\log|\Omega|}{2}}+|\Omega|\cdot C^{M_4}X
\end{eqnarray}
(compute-uncompute) pairs, that can be further decomposed \cite{2017_HLZetal, 2018_G} into
\begin{eqnarray}
    4\sqrt{|\Omega|}\left(\log_2|\Omega|-2\right)+4|\Omega|
\end{eqnarray}
 T-gates and 
\begin{eqnarray}
    \sqrt{|\Omega|}\left(4\log_2|\Omega|-6\right)+5|\Omega|
\end{eqnarray}
 CNOT-gates. 

\paragraph{Block encoding of $H_3$ : } We use Theorem \ref{thm:blockEncodeDivConq}. We prepare $\log_2|E_D|$ qubits in equal superposition, using $\log_2|E_D|$ H gates. We use these to select two subspaces. Specifically, each superimposed state selects an ancilla. From this ancilla we use two CNOTs to select an ancilla in each of the two corresponding subspaces. The rest of the operations are controlled on these ancillae. So, each unitary of $H_{3\vec{x}\vec{x'}}$ has 1 extra control. Using Theorem \ref{thm:CX} and assuming an equal partitioning into two groups, for the selection we require
\begin{eqnarray}
    2|E_D|^{\frac{1}{2}}C^{\frac{\log|E_D|}{2}}X+|E_D|\cdot C^{M_5}X 
\end{eqnarray}
(compute-uncompute) pairs. These, in turn can be decomposed \cite{2017_HLZetal, 2018_G} into
\begin{eqnarray}
    4\sqrt{|E_D|}\left(\log_2|E_D|-2\right)+4|E_D|
\end{eqnarray}
T-gates and
\begin{eqnarray}
    \sqrt{|E_D|}\left(4\log_2|E_D|-6\right)+5|E_D|
\end{eqnarray}
CNOT-gates. 

\paragraph{Block encoding of $H_{amp}'$ : } We prepare 1 qubit in equal superposition using 1 H. The rest of the operations are controlled on this. So it adds an extra control.
Thus overall, unitaries in $H_{1\vec{x}}$, $H_{2\vec{x}}$ have three extra controls and unitaries in $H_{3\vec{x}\vec{x'}}$ have two extra controls. Thus for each unitary in $H_{1\vec{x}}$, $H_{2\vec{x}}$ we require a (compute-uncompute) pair of $C^3X$, that can be implemented with $8$ T and $9$ CNOT-gates. Each unitary in $H_{3\vec{x}\vec{x'}}$ require a (compute-uncompute) pair of $C^2X$, that can be implemented with $4$ T and $5$ CNOT-gates. 

Overall, we require following numbers of T-gates, 
\begin{eqnarray}
   N_t &\leq&|\Omega|\left[4\sqrt{L_1+L_2}\left(\log_2(L_1+L_2)-2\right)+12(L_1+L_2)+4\sqrt{L_3+L_4}\left(\log_2(L_3+L_4)-2\right)+12(L_3+L_4)\right]   \nonumber \\
    &&+|E_D|\left(4\sqrt{L_5}\left(\log_2L_5-2\right)+8L_5\right)+4\sqrt{|\Omega|}\left(\log_2|\Omega|-2\right)+4|\Omega|    
    +4\sqrt{|E_D|}\left(\log_2|E_D|-2\right)+4|E_D| \nonumber \\
    &&+2|\Omega|\left( 8(\log_22k)\log_2\left(\frac{\log_22k}{\epsilon_{QFT}}\right)+\log_2\left(\frac{\log_22k}{\epsilon_{QFT}}\right)\log_2\left(\frac{\log_2\left(\frac{\log_22k}{\epsilon_{QFT}}\right)}{\epsilon_{QFT}}\right)   \right) 
\label{eqn:num-Tplus-aqft-alg3a}
\end{eqnarray}
following number of CNOT-gates,
\begin{eqnarray}
  N_{cx}  &\in&|\Omega|\left[\sqrt{L_1+L_2}\left(4\log_2(L_1+L_2)-6\right)+14(L_1+L_2)+\sqrt{L_3+L_4}\left(4\log_2(L_3+L_4)-6\right)+14(L_3+L_4)\right] \nonumber \\
    &&+|E_D|\left(\sqrt{L_5}\left(4\log_2L_5-6\right)+10L_5\right)+\sqrt{|\Omega|}\left(4\log_2|\Omega|-6\right)+5|\Omega| 
    +\sqrt{|E_D|}\left(4\log_2|E_D|-6\right)+5|E_D| \nonumber \\
    &&+2|\Omega|\left( 8(\log_22k)\log_2\left(\frac{\log_22k}{\epsilon_{QFT}}\right)+\log_2\left(\frac{\log_22k}{\epsilon_{QFT}}\right)\log_2\left(\frac{\log_2\left(\frac{\log_22k}{\epsilon_{QFT}}\right)}{\epsilon_{QFT}}\right) \right)  \nonumber \\
    &&+2(2(L_1+L_2)+L_3+L_4+L_5)+3(\log_2(L_1+L_2)+\log_2(L_3+L_4)+\log_2L_5)-21
\end{eqnarray}
following number of rotation gates,
\begin{eqnarray}    
   N_r\leq 4(L_1+L_2)+2(L_3+L_4)+2L_5-4
\label{eqn:num-rotations-alg3a}
\end{eqnarray}
following number of CZ gates
\begin{eqnarray}
   N_{cz} \leq |\Omega|\left(L_1+L_2+L_3+L_4\right)+|E_D|L_5,
\end{eqnarray}
and the following number of H gates,
\begin{eqnarray}
 N_h\leq   3+\log_2(L_1+L_2)+\log_2(L_3+L_4)+\log_2L_5+\log_2|E_D|+\log_2|\Omega|,
\end{eqnarray}
where $L_1=\log_2k+1$, $L_2=\binom{\log_2k+1}{2}$, $L_3=\binom{\log_2k+1}{3}$, $L_4=\binom{\log_2k+1}{4}$ and $L_5=(\log_2k+1)^2$. Hence we prove Theorem \ref{thm:blockLcuIIIa}.

\end{proof}

 The following lemma gives a bound on the $\ell_1$ norm of $H_{amp'}$. Details of the proof can be found in Supplemental Material (Section III C) \cite{SuppMat}.
 \begin{lemma}
     \begin{eqnarray}
    \|H_{amp}'\|&\leq& |\Omega|\left(\frac{\lambda\Delta^4}{27}k^4+k^2\left(\left(\frac{M^2+7d+1}{3}\right)\Delta^2-0.048611\lambda\Delta^4\right)+k\left(-3d\Delta^2+0.03125\lambda\Delta^4\right) \right. \nonumber \\
    &&\left.+\Delta^2\left(\frac{-M^2+8d-4}{6}\right)-0.0081019\lambda\Delta^4 \right)  
\end{eqnarray}
     \label{lem:l1Hamp'}
 \end{lemma}

 From \cite{2019_GSLW}, we require
 \begin{eqnarray}
     \mathcal{R}\in O\left(\|H_{amp}'\|t+\frac{\log(1/\epsilon)}{\log\log(1/\epsilon)}\right)
 \end{eqnarray}
calls to the block encoding of $\frac{H_{amp}'}{\|H_{amp}'\|}$ in order to implement an $\epsilon$-precise block encoding of $e^{-iH_{amp}'t}$. Thus the total gate complexity is obtained by multiplying $\mathcal{R}$ with the number of gates obtained in Theorem \ref{thm:blockLcuIIIa}.

\subsection{Algorithm IIIb : LCU with binary decomposition of integers }
\label{subsec:algoIIIb}

In this section we discuss a more compact decomposition of the field operators.  The central idea stems from the fact that by using binary representation we can express an integer diagonal matrix as a sum of $O(\log k)$ number of signature matrices, as stated in Conjecture \ref{conj:phi24}. In our case $\Phi^2$ and $\Phi^4$ are diagonal matrices, consisting of $2^{\text{nd}}$ and $4^{\text{th}}$ power of consecutive integers, respectively. If $\zeta'$ is the maximum number of bits required to express the highest integer, then we can express any other integer $n = (b_{\zeta'},\ldots,b_1) =\sum_{j=1}^{\zeta'} b_j 2^{j-1} $. For smaller integers we append leading zeros. The leftmost bit $b_{\zeta'}$ is referred to as the most significant bit, while the rightmost bit is referred to as the least significant bit. The $j^{th}$ signature matrix is obtained by taking the $j^{th}$ bit in the binary expansion of each diagonal integer, replacing 0s by 1 and 1s by -1.

The circuit complexity of the SELECT circuits can be bounded by the sum of gates, qubits, etc required to implement the signature matrices. In order to design efficient circuits for each signature matrix we exploit their structure, which is obtained from the binary decomposition of integers. We first prove the following. 
\begin{lemma}
Suppose $n$ is a positive integer and $(b_m,b_{m-1},\ldots,b_1)$ is its binary expansion. Then,
\begin{eqnarray}
    b_{\ell}&=& 0\qquad\text{if }\qquad n=2^{\ell}k,2^{\ell}k+1,\ldots 2^{\ell}k+2^{\ell-1}-1    \nonumber \\
  &=&1\qquad\text{if }\qquad n=2^{\ell}k+2^{\ell-1},2^{\ell}k+2^{\ell-1}+1,\ldots 2^{\ell}k+2^{\ell}-1    \nonumber \\
  &\qquad\qquad&\qquad\text{where }\qquad k\text{ is a non-negative integer.}  \nonumber
\end{eqnarray}
\label{lem:binIntg}
\end{lemma}

\begin{lemma}
(a) Let $n$ be an integer and $(b_{m'},\ldots,b_1)$ be the binary decomposition of $n^2$. Then $b_1 = 0$ for even $n$ and 1 for odd $n$ and $b_2 = 0$. For $\ell > 2$, $b_{\ell} = 1$ if and only if $n = 2^{\ell-1}j+j'$, where $j$, $j'$ are integers such that $1\leq j'\leq 2^{\ell-1}-1$. 

(b) Let $n$ be an integer and $(b_{m''},\ldots,b_1)$ be the binary decomposition of $n^4$. Then $b_1 = 0$ for even $n$ and 1 for odd $n$ and $b_2 = b_3 = b_4 = 0$. For $\ell\neq 2,3,4$,  $b_{\ell} =1$ if and only if $n = 2^{\ell-2}j+j'$, where $j$, $j'$ are integers such that $1\leq j'\leq 2^{\ell-2}-1$. 
\label{lem:binPattern}
\end{lemma}

Next, let us observe the following. Consider a circuit with $x$ qubits and suppose a ``don't care" condition exists involving $y< x$ of them. This implies that for every possible combination of binary values on $y$ qubits, there is a certain state of the remaining $x-y$ qubits when a -1 phase is implemented. The number of possible states is $2^y$ and to implement this logic in a quantum circuit we require a single Z gate controlled on the state of $x-y$ qubits. Many kinds of ``don't care" conditions exist, for example involving parity. Suppose for every possible binary combinations on $y$ qubits where there are odd number of 1s, there exists a certain state of the $x-y$ qubits when -1 phase is incurred. Here too, there are exponential number of such states and to implement this circuit we require a number of CNOTs to check the parity and a Z-controlled on $x-y+1$ qubits.  In summary, as the number of ``don't care" conditions increases, so  the number of states that satisfy a certain value also increases, usually exponentially. But the increase in the circuit complexity is less. \par

This implies, if we consider Supplemental Material Tables II and IV \cite{SuppMat},
then we can say that as the number of integers in each cell increase with respect to $k= 2^x$ and $\ell$, the probability of the existence of more "don't care" conditions becomes higher and hence $\kappa_{\ell}$ and $\kappa_{\ell}'$ grow more slowly with respect to $\ell$. We have already mentioned that the number of integers in each cell increase exponentially with respect to $\ell$, but the cell at the intersection of $b_{1+2\log k}$ (or $b_{1+4\log k}$) and $k$ (i.e. $2^x$) has value 1. That is why T-count of $U_{1+2\log k}$ and $U_{1+4\log k}'$ is $O(\log k)$. Thus we conjecture the following, based on considerable evidence in Supplemental Material (Sec III D)\cite{SuppMat}. 

\begin{conjecture}

 The T-count needed to exactly implement $U_{\ell}$ and $U_{\ell}'$ is at most $O\left(\min\{\ell, \log k\}\right)$. 
 
    \label{conj:phi24} 
\end{conjecture}
In Supplemental Material (Sec. III D) \cite{SuppMat} we have explicitly constructed some circuits of signature matrices arising in this compact decomposition of $\Phi^2$ and $\Phi^4$. 
Now, we discuss the cost of simulating $H_{amp}$. The procedure is similar to the one described in the previous section for Algorithm IIIa.  In this case we partition $H_{amp}$ as follows.
\begin{eqnarray}
    H_{\phi^2\bx}&=&\frac{1}{2}\Pi^2(\bx)+\frac{1}{2}(M^2+d+1)\Phi^2(\bx) \nonumber \\
    H_{\phi^4\bx}&=&\Phi^4(\bx) \nonumber \\
    H_{\phi\bx\bx'}&=&\Phi(\bx)\Phi(\bx') = H_{3\bx\bx'}  \nonumber \\
    H_3&=&\sum_{\bx,\bx'\in E_D} H_{3\bx\bx'}   \nonumber \\
    H_{12}'&=& \sum_{\bx\in\Omega}H_{\phi^2\bx}+\frac{\Lambda}{4!}H_{\phi^4\bx}    \nonumber \\
    H_{amp}&=&H_{12}'+H_3    \label{eqn:lcuZcz}
\end{eqnarray}

 We summarize the gate complexity for block encoding $H_{amp}$ in the following theorem.
\begin{theorem}
    Let $H_{amp}$ be the Hamiltonian defined in Eq. \ref{eqn:lcuZcz} and $\|H_{amp}\|$ be the $\ell_1$ norm of the coefficients defined in its decomposition Eq. \ref{eqn:lcuZcz}. Then it is possible to have a $\left(\|H_{amp}\|,\cdot,0\right)$ block encoding of $H_{amp}$ with $O\left(|\Omega|\log k\right)$ qubits, using the following number of rotation gates
\begin{eqnarray}
    N_r'\in O\left(\log^2k\right)   ,\nonumber
\end{eqnarray}
 and the following number of additional T-gates (that are not in the decomposition of rotation gates)
\begin{eqnarray}
    N_t'\in O\left(|\Omega|\log^2k\right). \nonumber
\end{eqnarray}
The bound on $N_t'$ is obtained assuming Conjecture \ref{conj:phi24}.
\label{thm:blockLcuIIIb}
\end{theorem}

\begin{proof}
The recursive block encoding is done in a similar fashion as before. We give brief descriptions with gate complexity.

\paragraph{Block encoding of $H_{\phi^2\vec{x}}$ : } Let $\Phi^2=\sum_{j=0}^{2\log_2k}U_j$, where $U_j$ is a signature matrix. These can be obtained using Lemma \ref{lem:lcuIntgDiag}, as described in Section \ref{subsec:algoIIIb}. The ancilla preparation subroutine is,
\begin{eqnarray}
    \prep_{\phi^2\vec{x}}=\frac{1}{\mathcal{N}_{\phi^2\vec{x}}} \left(\sum_{j=0}^{2\log_2k}\sqrt{\alpha_j}\ket{j,0}+\sum_{l=0}^{2\log_2k}\sqrt{\beta_l}\ket{l,1} \right),
\end{eqnarray}
where the weights $\alpha_j$ and $\beta_l$ include the coefficients from LCU decomposition of operators $\Phi^2$ and $\Pi^2=\mathcal{F}\Phi^2\mathcal{F}^{\dagger}$, as well as the coefficients of $\Phi^2$ and $\Pi^2$ in the definition of the Hamiltonian. The last qubit is used to select the QFT. We require $1+\log_2(2\log_2k)$ qubits. For the state preparation we require $1+\log_2(2\log_2k)$ H, $4\log_2(k)-2$ rotations, $4\log_2(k)+3\log_2(2\log_2k)-7$ CNOT-gates.

The SELECT subroutine does the following.
\begin{eqnarray}
    \sel_{\phi^2\vec{x}}\ket{j,0}\ket{\psi} &&\mapsto \ket{j,0}U_j\ket{\psi}    \nonumber \\
    \sel_{\phi^2\vec{x}}\ket{j,1}\ket{\psi} &&\mapsto \ket{j,0}\mathcal{F}U_j\mathcal{F}^{\dagger}\ket{\psi}
\end{eqnarray}
As explained, we require $2\log_2k$ number of $C^{\log(2\log k)}X$ gates to select the unitaries. Using Theorem \ref{thm:CX}, if we divide the control qubits into two equal groups then we require the following number of T-gates to implement the control that selects the signature matrices
\begin{eqnarray}
    4\sqrt{2\log_2k}(\log(2\log_2k)-2)+8\log_2k,
\end{eqnarray}
 and the following number of CNOT-gates
\begin{eqnarray}
    \sqrt{2\log_2k}(4\log(2\log_2k)-6)+10\log_2k.
\end{eqnarray}
Additionally, assuming Conjecture \ref{conj:phi24} we require $O(\log^2 k)$ number of T-gates for the implementation of the signature matrices. We also require two $\log_2k+1$-qubit QFT.

\paragraph{Block encoding of $H_{\phi^4\vec{x}}$ : } Let $\Phi^4=\sum_{j=0}^{4\log_2k}U_j$, where $U_j$ is a signature matrix obtained using Lemma \ref{lem:lcuIntgDiag}, as described in Appendix \ref{subsec:algoIIIb}. The ancilla preparation subroutine is as follows:
\begin{eqnarray}
    \prep_{\phi^4\vec{x}}=\frac{1}{\mathcal{N}_{\phi^4\vec{x}}} \sum_{j=0}^{4\log_2k}\sqrt{\alpha_j}\ket{j}
\end{eqnarray}
where the weights $\alpha_j$ include the coefficients from LCU decomposition of operators $\Phi^4$. We require $1+\log_2(4\log_2k)$ qubits. For the state preparation we require $1+\log_2(4\log_2k)$ H, $8\log_2(k)-2$ rotations, $8\log_2(k)+3\log_2(4\log_2k)-7$ CNOT-gates.

The SELECT subroutine does the following.
\begin{eqnarray}
    \sel_{\phi^4\vec{x}}\ket{j,0}\ket{\psi} &&\mapsto \ket{j}U_j\ket{\psi}    \nonumber \\
\end{eqnarray}
As explained, we require $4\log_2k$ number of $C^{\log(4\log k)}X$ gates to select the unitaries. Using Theorem \ref{thm:CX}, if we divide the control qubits into two equal groups then we require the following number of T-gates
\begin{eqnarray}
    4\sqrt{4\log_2k}(\log(4\log_2k)-2)+16\log_2k,
\end{eqnarray}
 and the following number of CNOT-gates
\begin{eqnarray}
    \sqrt{4\log_2k}(4\log(4\log_2k)-6)+20\log_2k.
\end{eqnarray}
Assuming Conjecture \ref{conj:phi24} we require $O(\log^2 k)$ T-gates to implement the signature matrices.

\paragraph{Block encoding of $H_{12}'$ : } We use the recursive block encoding Theorem \ref{thm:blockEncodeDivConq}. We can block encode $H_{\phi^2\vec{x}}+\frac{\Lambda}{24}H_{\phi^4\vec{x}}$ using ancilla preparation subroutine that has 1 H and 2 rotation gates. For $H_{12}'$ we prepare an equal superposition of $\log_2|\Omega|$ qubits, using $\log_2|\Omega|$ H gates and use these to select an ancilla of each subspace. The rest of the operations are controlled on this. Thus this adds another control. We require $|\Omega|$ number of $C^{\log_2|\Omega|}X$ compute-uncompute pairs of gates. Using Theorem \ref{thm:CX} and assuming an equal partitioning of the control qubits into two equal groups, we can implement the multicontrolled-X gates using
\begin{eqnarray}
    |\Omega|^{\frac{1}{2}}C^{\frac{\log|\Omega|}{2}}X+|\Omega|\cdot C^2X
\end{eqnarray}
(compute-uncompute) pairs, that can be further decomposed \cite{2017_HLZetal, 2018_G} into
\begin{eqnarray}
    4\sqrt{|\Omega|}\left(\log_2|\Omega|-2\right)+4|\Omega|
\end{eqnarray}
T-gates and
\begin{eqnarray}
    \sqrt{|\Omega|}\left(4\log_2|\Omega|-6\right)+5|\Omega|
\end{eqnarray}
CNOT-gates.

\paragraph{Block encoding of $H_{3\vec{x}\vec{x'}}$ and $H_3$ : } The block encoding of $\Phi(\vec{x})\Phi(\vec{x'}) = H_{3\vec{x}\vec{x'}}$ is provided earlier. For the preparation subroutine we require $\log_2|E_D|+\log_2(\log_2k+1)^2$ H, $2(\log_2k+1)^2-2$ rotation gates, $2(\log_2k+1)^2+3\log_2(\log_2k+1)^2-7$ CNOT-gates. For the selection subroutine we require 
\begin{eqnarray}
   && |E_D|\left(8(\log_2k+1)(\log_2(\log_2k+1)-1)+4(\log_2k+1)^2\right) \nonumber \\
    &&+4\sqrt{|E_D|}(\log_2|E_D|-2)+4|E_D|
\end{eqnarray}
T-gates and
\begin{eqnarray}
   && |E_D|\left((\log_2k+1)(8\log_2(\log_2k+1)-6)+5(\log_2k+1)^2\right)    \nonumber \\
   &&+\sqrt{|E_D|}(4\log_2|E_D|-6)+5|E_D|
\end{eqnarray}
CNOT-gates and $(\log_2k+1)^2$ CZ gates.

\paragraph{Block encoding of $H_{amp}$ : } Using Theorem \ref{thm:blockEncodeDivConq} we can block encode $H_{amp}$ using a PREP subroutine with 1 H and 2 rotation gates and a SELECT subroutine that adds an extra control. Thus under the assumptions of Conjecture~\ref{conj:phi24} we require
\begin{eqnarray}
  N_t'&\leq&  |\Omega|\left(4\sqrt{2\log_2k}(\log_2(2\log_2k)-2)+8\log_2k+4\sqrt{4\log_2k}(\log_2(4\log_2k)-2)+16\log_2k  \right) \nonumber \\
  &&+ 4\sqrt{|\Omega|}\left(\log_2|\Omega|-2\right)+4|\Omega| +|E_D| \left(8(\log_2k+1)(\log_2(\log_2k+1)-1)+4(\log_2k+1)^2+4\right)  \nonumber \\
  &&+4\sqrt{|E_D|}(\log_2|E_D|-2)   \nonumber \\
  &&+2|\Omega|( 8(\log_22k)\log_2\left(\frac{\log_22k}{\epsilon_{QFT}}\right)+\log_2\left(\frac{\log_22k}{\epsilon_{QFT}}\right)\log_2\left(\frac{\log_2\left(\frac{\log_22k}{\epsilon_{QFT}}\right)}{\epsilon_{QFT}}\right)  )+O\left(|\Omega|\log_2^2k\right)
  \label{eqn:TlcuIIIb}
\end{eqnarray}
T-gates,
\begin{eqnarray}
N_{cx}'&\leq&|\Omega|\left(\sqrt{2\log_2k}(4\log_2(2\log_2k)-6)+10\log_2k+\sqrt{4\log_2k}(4\log_2(4\log_2k)-6)+20\log_2k+5 \right)   \nonumber \\
&&+\sqrt{|\Omega|}\left(4\log_2|\Omega|-6\right) + |E_D|\left((\log_2k+1)(8\log_2(\log_2k+1)-6)+5(\log_2k+1)^2+5\right)    
   +\sqrt{|E_D|}(4\log_2|E_D|-6)    \nonumber \\
   &&+2|\Omega|\left( 8(\log_22k)\log_2\left(\frac{\log_22k}{\epsilon_{QFT}}\right)+\log_2\left(\frac{\log_22k}{\epsilon_{QFT}}\right)\log_2\left(\frac{\log_2\left(\frac{\log_22k}{\epsilon_{QFT}}\right)}{\epsilon_{QFT}}\right) \right)  
   \label{eqn:CNOTlcuIIIb}
\end{eqnarray}
CNOT-gates, 
\begin{eqnarray}
    N_r\leq 12\log_2k+2(\log_2k+1)^2-3  
\end{eqnarray}
rotation gates. This proves the theorem.
\end{proof}

In this case we have
\begin{eqnarray}
    \|H_{amp}\|&\leq&\frac{|\Omega|}{4}\left(k^2\Delta^2(2+M^2+d)+\frac{\Lambda}{12}k^4\Delta^4\right)+\frac{3}{4}|E_D|\Delta^2k^2.
    \label{eqn:HampL1}
\end{eqnarray}
We again remember that this norm is sum of absolute value of the coefficients from the nonidentity terms only. From \cite{2019_GSLW}, we require
\begin{eqnarray}
    \mathcal{R}'\in O\left(\|H_{amp}\|t+\frac{\log(1/\epsilon)}{\log\log(1/\epsilon)}\right)
\end{eqnarray}
calls to the block encoding of $\frac{H_{amp}}{\|H_{amp}\|}$ in order to implement an $\epsilon$-precise block encoding of $e^{-iH_{amp}t}$. Thus the total gate complexity is obtained by multiplying $\mathcal{R}'$ with the number of gates obtained in Theorem \ref{thm:blockLcuIIIb}.

\subsection{Phase estimation in the amplitude basis}

As described earlier, we can perform phase estimation to compute scattering matrix elements. For simplicity, let us first focus on the $2\rightarrow 2$ particle scattering. In the center of mass frame, the two incoming momenta can be described by $p_1 = p = -p_2$, while the Mandesltam variable is given by $s = E_{CM}^{2} = 2(m^{2} + p^{2})$. The expression for the energy in Eq. ~\eqref{eqn:2_particle_energy} is given as $E = E_{\text{2-particle state}} - E_0$, where $E_0$ is the ground state. Although not equal to the spectral gap of the Hamiltonian, for values of the coupling away from the critical value, i.e. $\lambda \neq \lambda_c$, we can solve for E by extracting the difference between the first excited state and the ground state in the even-particle sector of the model's $\mathbb{Z}_2$ symmetry ($\phi \leftrightarrow -\phi$). Similarly, the difference between the first excited state and the ground state in the odd sector gives us the renormalized mass.

\paragraph{State preparation}

Although ground state preparation is in general a QMA-hard problem, it has been shown that the ground states of free scalar fields, namely a Gaussian in the amplitude basis, is efficient to prepare \cite{kitaev2009wavefunction,Somma2016}. In the presence of interactions, the ground state can in principle be prepared adiabatically. Here, we assume that we are provided with a state that has polynomial overlap with the two lowest interacting eigenstates. Provided such a state, we seek to prepare the even and the odd sectors of the model.
In the field amplitude basis, the even and odd sectors are given by the states of the form
\begin{eqnarray}
\ket{\phi_{even}} &=& \frac{\ket{\phi} + \ket{-\phi}}{\sqrt{2}} \nonumber \\
\ket{\phi_{odd}} &=& \frac{\ket{\phi} - \ket{-\phi}}{\sqrt{2}}
\end{eqnarray}
where given (for a single lattice site) $\ket{\phi} = \sum_{j} \alpha_{j} \ket{j}$, we define $\ket{-\phi} = \sum_{j} \alpha_{-j} \ket{j}$. Given our encoding, where $j \in \{-k+1, \dots k \}$ for some cutoff $k$, we only exchange the coefficients $\alpha_{-j} \leftrightarrow \alpha_{j}$ for $j \in \{-k+1, \dots, k+1 \}$, and assume that $k$ is chosen large enough that $\alpha_k \approx 0$. Thus, the quantum circuit $U$ we use to map $\ket{\phi}$ to $\ket{-\phi}$ leaves the basis states carrying the coefficients for both $\alpha_0$ and $\alpha_k$ unchanged.

\begin{figure}
\centering
\includegraphics[width=0.65\textwidth]{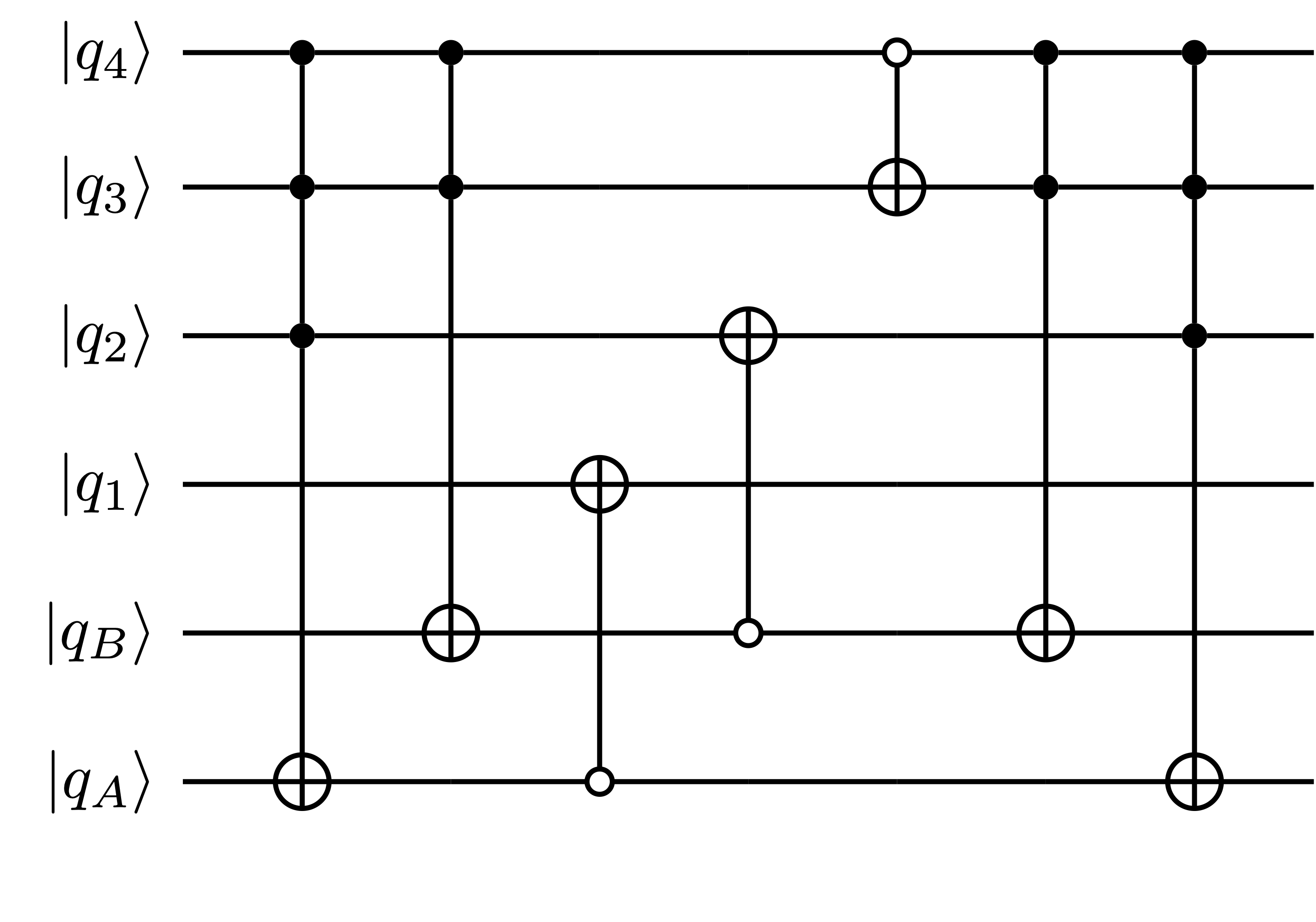}
\caption{Example circuit that maps $\ket{\phi}$ to $\ket{-\phi}$ for $k=4$.}
\label{fig:U_phi_mi_phi}
\end{figure}

The discretized field values $j \in \{-k+1, \dots, k\}$ map to binary numbers $b = j + k - 1$ in our encoding. For simplicity, we assume that the number of basis status used is some power of 2, i.e. $2k = 2^m$ for some positive integer $m$, which gives the number of qubits used in the encoding. The required transformation can then be implemented by flipping a bit only if all the bits to the right of this bit form any bitstring except for the all 1's bitstring. An example circuit that achieves this transformation for $k=4$ is shown in Fig.~\ref{fig:U_phi_mi_phi}.

To achieve the transformation $U: \ket{\phi} \rightarrow \ket{-\phi}$ more generally for some given $k$, setting $m = \log_{2}{(2k)}$, requires $m - 2$ many ancillary qubits. In addition, we need two each of $C^{m-1}X$, $C^{m-2}X$, $\dots$, $C^{2}X$ operations, as well as $m-1$ CNOT and X gates. In turn, each $C^{n}X$ operations, with targets initialized to $\ket{0}$, can be constructed using $n-2$ ancillas and $n-1$ logical AND operations, each of which require four T-gates. The even sector can then be prepared by implementing the projector $I+U$, while the odd sector can similarly be prepared by applying the projector $I-U$.

These projectors can be realized using the Hadamard test. In the LCU picture, this is described simply as initializing an ancillary qubit in the $\ket{+}$ state, and applying a controlled-$U$ operation controlled on the ancilla, and targeted on the field register. Upon measuring the ancilla in the $X$-basis, by applying a Hadamard gate before measurement, an outcome of $0$ would project the field register onto the even sector while an outcome of $1$ would project it onto the odd sector. This operation can be described as
\begin{eqnarray}
\left(H \otimes I \right) \left( \ket{0}\bra{0}\otimes \mathbb{I} + \ket{1}\bra{1} \otimes U \right) \left(H \otimes I \right) \ket{0} \ket{\phi} &=& \frac{1}{\sqrt{2}} \left(\ket{0}\ket{\phi_{even}} + \ket{1}\ket{\phi_{odd}}\right)
\label{eqn:prep-even-odd-sectors}
\end{eqnarray}

We must apply one such operation for each of the lattice sites. With these considerations, we can gate cost the operation that produces an equal superposition of the even and odd sectors.

\begin{lemma}\label{lem:Tcost}
The total number of T-gates required to simulate a controlled-$U$ operation where $U: \ket{\phi} \rightarrow \ket{-\phi}$ is $2 \vert \Omega \vert \left( \log_{2}{k} + 1 \right) \left( \log_{2}{k} + 2 \right) - 8 \in O\left(\vert \Omega \vert \log_{2}^{2}{k} \right)$ where $k = \phi_{max}/\Delta_{\phi}$ and further obeys $k> 3/2$.
\begin{proof}
From the above considerations, taking $m = \log_{2}{2k}$, we have that the total T-gate count for implementing $C$-$U$ for a single lattice site is given by
\begin{eqnarray}
\text{\# T-gates for $C$-$U$} &=& \sum_{n=3}^{m} \text{Cost}(C^{n}X) + (m-1) \text{Cost}(\text{logical AND}) \nonumber \\
&=& \sum_{n=3}^{m} (n-1) \text{Cost}(\text{logical AND}) + 4 (m-1) \nonumber \\
&=& 4 \sum_{n=3}^{m} (n-1) + 4(m-1) \nonumber \\
&=& 2m(m+1) - 8
\end{eqnarray}
The total number of T-gates is then simply the product of the number of lattice sites $\vert \Omega \vert$ with the expression above.
\end{proof}
\end{lemma}
Similarly, we can calculate the total number of ancillary qubits needed to implement such a controlled-$U$ operation
\begin{lemma}
Under the assumptions of Lemma~\ref{lem:Tcost}, the total number of ancillary qubits required to implement a controlled-$U$ operation is 

$$\frac{1}{2} \left( \log_{2}^{2}{2k} + \log_{2}{2k} - 4 \right) \in O \left(\log_{2}^{2}{k} \right)$$ 
if we reuse ancillae, and 
$$\frac{\vert \Omega \vert}{2} \left( \log_{2}^{2}{k} + \log_{2}{k} - 6\right) + 1 \in O \left(\vert \Omega \vert \log_{2}^{2}{k} \right)$$ if we do not reuse ancillae and thereby allow the operations to be executed in parallel.
\begin{proof}
In addition to a single ancilla serving as the control qubit, we also need $m-2$ ancillary qubits to apply a controlled version of $U$ to an $m$-qubit register where $m = \log_{2}{2k}$ for a single lattice site. Moreover, each of the $\{C^{n}X\}_{n=3}^{m}$ operations required to implement $C$-$U$ require $n-2$ ancillae each. This gives a total number of ancillary qubits as
\begin{equation}
1 + (m-2) + \sum_{n=3}^{m} (n-2) = \frac{1}{2} (m^2 - m - 4)
\end{equation}

If the ancillae are reused across all the lattice sites, the total number of ancillary qubits is then simply given by the above expression. If instead the $C$-$U$ is performed in parallel across the entire lattice, then all except 1 of the ancillae, the one used to control the entire operation, have to be associated with each of the lattice sites, giving a total count of
\begin{equation}
\vert \Omega \vert (m-2) + 1 + \frac{\vert \Omega \vert}{2} (m^2 - 3m - 2) = \frac{\vert \Omega \vert}{2} \left( m^2 - m - 6 \right) + 1
\end{equation}
Plugging in $m= \log_{2}{2k}$ into these expressions gives the statement of the lemma.
\end{proof}
\end{lemma}

\paragraph{Eigenvalue extraction for qubitization}
To extract the energy eigenvalues, we perform phase estimation on the walk operator
\begin{equation}
W = \left( 2 \prep \ket{0}\bra{0} \prep^{\dagger} \otimes \mathbb{I} - \mathbb{I}\right)  \cdot \sel
\label{eqn:walk-operator}
\end{equation}
which provides a qubitized block encoding of the Hamiltonian, and furnishes a direct sum of two-dimensional irreducible representations, where each two-dimensional subspace is labelled by an eigenstate of the Hamiltonian.
\begin{equation}
W = \oplus_{k} \begin{pmatrix}
\frac{E_{k}}{\vert \alpha \vert} & \sqrt{1 - \left( \frac{E_k}{\vert \alpha \vert} \right)^{2}} \\
- \sqrt{1 - \left( \frac{E_k}{\vert \alpha \vert} \right)^{2}} & \frac{E_{k}}{\vert \alpha \vert}
\end{pmatrix}  = \oplus_{k} e^{i \arccos{(E_{k}/\vert\alpha\vert)Y}}
\end{equation}
with eigenvalues ${\pm \arccos{(E_k/\vert\alpha\vert)}}$, from which one may readily, with classical post-processing, obtain the desired eigenvalues $E_k$. In practice, we would run $\mathbb{I} \otimes W$ on the input state given in Eq.~\eqref{eqn:prep-even-odd-sectors}, and keep measurements whenever the ancilla measures out to be 0 (1) to infer eigenvalues in the even (odd) sector.

Due to compilation errors, the phase information we extract would be different from the ideal ones. In order to bound this error in the phase, and consequently the energy eigenvalues, we first note a series of observations below.

\begin{lemma}
Let $\prep'$ and $\sel'$ denote the compiled versions of $\prep$ and $\sel$, such that they can be written as
\begin{eqnarray}
\prep' \ket{0}^{\otimes \log_{2}{m}} &=&  \sum_{j=1}^{m} \beta_j \ket{j} \nonumber \\
\sel' &=& \sum_{j=1}^{m} \ket{j} \bra{j} \otimes U_{j}
\end{eqnarray}
where $\beta_j \in \mathbb{C}$. Then, $\prep'^{\dagger} \sel' \prep'$ provides a $\left(\sum_j |\beta_j|^2, \log_{2}{m}, 0 \right)$-block encoding of the operator $H' = \sum_{j=1}^{m} \vert \beta_j \vert^2 U_j$. Furthermore, $H'$ is Hermitian.
\end{lemma}
\begin{proof}
The proof of the statement that 
\begin{equation}
H' = \left( \bra{0} \otimes \mathbb{I} \right) \prep'^{\dagger} \sel' \prep' \left( \ket{0} \otimes \mathbb{I} \right) = \sum_{j=1}^{m} \vert \beta_j \vert^2 U_j
\end{equation}
directly follows from the LCU lemma of~\cite{childs2012hamiltonian,berry2015hamiltonian}. To see that $H'$ is also Hermitian, we note that all the $U_j$'s in the LCU are Hermitian diagonal signature matrices, except those that come from the LCU of the momentum (squared) term. The compilation errors do not affect the Hermiticity of the signature matrices, and only impact the Fourier transforms acting on each of the lattice site to diagonalize the momentum (squared) term. If we denote the true quantum Fourier transform circuit as $F$ and its compiled version as $F'$, then it is straightforward to see that $F' V F'^{\dagger}$ is also Hermitian, given the (Hermitian) signature matrix V. Since the sum of Hermitian operators is Hermitian, $H'$ is also Hermitian.

\end{proof}

The Hermiticity of the block-encoded operator is an important property to ensure that the walk operator we construct using these (compiled) primitives provides a qubitized block encoding of the same operator. Furthermore, we note that the $\sel$ operation, as well as its compiled version $\sel'$, squares to the identity. As a consequence, it follows from Lemma 8 and Corollary 9 of \cite{2019_LC} that the walk operator

\begin{equation}
W' = \left( 2 \prep' \ket{0}\bra{0} \prep'^{\dagger} \otimes \mathbb{I} - \mathbb{I}\right)  \cdot \sel'
\end{equation}
provides a qubitized block encoding of the perturbed Hamiltonian $H'$, and furnishes a direct sum of two-dimensional irreducible representations, where each two-dimensional subspace is labelled by an eigenstate of $H'$.
\begin{equation}
W' = \oplus_{k} \begin{pmatrix}
\frac{E'_{k}}{\vert \alpha \vert} & \sqrt{1 - \left( \frac{E'_k}{\vert \alpha \vert} \right)^{2}} \\
- \sqrt{1 - \left( \frac{E'_k}{\vert \alpha \vert} \right)^{2}} & \frac{E'_{k}}{\vert \alpha \vert}
\end{pmatrix}  = \oplus_{k} e^{i \arccos{(E'_{k}/\vert\alpha\vert)Y}}
\end{equation}
will yield eigenvalues $\pm \arccos{\left( E'_{k} \right)}$, where $E'_{k}$ are the eigenvalues of $H'$. We can now bound the absolute difference between the eigenvalues extracted from the compiled walk operator and the ideal walk operator. The various sources of errors are the number of ancillary qubits used for phase estimation, the QFT part of the phase estimation circuit (which is negligible in gate cost), the synthesis error of the rotation gates and the error in the approximate quantum Fourier transforms (AQFTs) used. We bound these errors in terms of the target error in the estimate of energy eigenstates below, in analogy with Lemma \ref{lem:error-phase}.

\begin{lemma}
To obtain an estimate of an eigenvalue of a Hamiltonian within error $\epsilon_E$, it suffices to perform phase estimation of the compiled walk operator with the following bounds on the contributing errors
\begin{eqnarray}
2^{m} &\geq& \frac{\pi\alpha}{\sqrt{2}\epsilon_E} \nonumber \\
\epsilon_r &\leq& \frac{1}{3\sqrt{2}} \frac{\epsilon_E}{\alpha N_r} \nonumber \\
\epsilon_f &\leq& \frac{1}{3\sqrt{2}} \frac{\epsilon_E}{\alpha N_f}
\label{eqn:error-bounds-qubitization}
\end{eqnarray}
where $m$ is the number of ancillary qubits used for phase estimation, $\epsilon_r$ is the synthesis error per RZ gate, $\epsilon_f$ is the approximation error per individual approximate quantum Fourier transform (AQFT), and $N_r$ and $N_f$ are respectively the total number of rotation gates and AQFTs used in the circuit compiling the walk operator.
\label{lemma:error-bounds-qubitization}
\end{lemma}
\begin{proof}
We can approximate, similarly as in Lemma \ref{lem:error-phase}, the error in the phase estimate as
\begin{equation}
\epsilon_{\theta} \approx \sqrt{\left( \frac{\pi}{2^{m+1}}\right)^{2} + \left( \pi \epsilon_{QFT}^{2} + \epsilon_{synth} + \epsilon_{AQFT}^{2} \right)^{2}}
\end{equation}
\label{eqn:error-phase-with-aqft}
where $m$ is the number of ancillary qubits used for phase estimation, $\epsilon_{QFT}$ is the error in the QFT part of the phase estimation circuit, $\epsilon_{synth}$ is the total synthesis error due to compiling single qubit RZ gates into T-gates, and $\epsilon_{AQFT}$ is the total error due to the approximate quantum Fourier transforms. 
Distributing the errors roughly equally, we bound
\begin{eqnarray}
\left( \frac{\pi}{2^{m+1}}\right)^{2} \leq \frac{\epsilon_{\theta}^{2}}{2}, \qquad \pi \epsilon_{QFT} = \epsilon_{synth} = \epsilon_{AQFT} \leq \frac{1}{3}\frac{\epsilon_{\theta}}{\sqrt{2}}
\end{eqnarray}
Performing phase estimation of the walk operator with error $\epsilon_{\theta}$ induces an error $\epsilon_{E}$ in the estimate of the eigenvalue of the Hamiltonian where the two are related by $\epsilon_{\theta} = \epsilon_{E}/\alpha$ where $\alpha$ is the (coefficient) 1-norm of the Hamiltonian. Using this relation, we can readily solve the bounds for the synthesis error per rotation gate $\epsilon_r = \epsilon_{synth}/N_r$ and the error per approximate quantum Fourier transform (AQFT) $\epsilon_{f} = \epsilon_{AQFT}/N_f$, where $N_r$ and $N_f$ are respectively the total number of rotation gates and AQFTs used in the circuit. These bounds are reported in the statement of the lemma.
\end{proof}

We now use the above results to report the T-gate count of performing phase estimation using the qubitization-based algorithms I, IIIa and IIIb as follows in order to prove Theorem \ref{thm:cost-qpe-qubitization}.

\begin{theorem}
The total cost of performing phase estimation to estimate an eigenvalue of the Hamiltonian to within error $\epsilon_E$ is given by
\begin{eqnarray}
Cost(QPE)^{(I)} &\in& O \left( \frac{\vert\Omega\vert^{2}}{\epsilon_E} \left[ k^{2}\Lambda + k M^{2} \right] \log^{2}{k} \right) \nonumber \\
Cost(QPE)^{(IIIa)} &\in& O \left( \frac{\vert\Omega\vert^{2}}{\epsilon_E} \left[ k^{2}\Lambda + k M^{2} \right] \log^{4}{k} \right) 
\end{eqnarray}
while the total number of logical qubits required, including those employed for phase estimation, are
\begin{eqnarray}
Count(Qubit)^{(I)} &\in& O \left( \vert \Omega \vert \log{k} + \log^{2}{k} + \log{\left( \frac{\vert \Omega \vert \left[ k^{2}\Lambda + k M^{2} \right]}{\epsilon_E}\right)}\right) \nonumber \\
Count(Qubit)^{(IIIa)} &\in& O\left( \vert \Omega \vert \log{k} + \log{\left[ \frac{\vert \Omega \vert \left( k^{2}\Lambda + kM^{2}\right)}{\epsilon_E} \right]} \right)
\end{eqnarray}
where the superscript denotes the algorithm employed.
\end{theorem}

\begin{proof}
The total T-gate cost for performing QPE using any of the qubitization algorithms described above can be approximated as
\begin{equation}
Cost(QPE) \approx 2^{m} \left[ N_r \cdot N^{(s)}(\epsilon_r) + N_f \cdot N^{(f)}(\epsilon_f) + N_{+}\right]
\end{equation}
where $m$ is the total number of ancillary qubits used for QPE, $N_r$ and $N_f$ are  respectively the total number of rotation gates and AQFTs used in the circuit compiling the walk operator, $N_{+}$ is the total number of other T-gates used, and
\begin{equation}
N^{(s)}(\epsilon_r) = 3.067 \log_{2}(2/\epsilon_r) - 4.327
\end{equation}
is the T-gate cost of compiling a single RZ gate from \cite{2015_KMM} to within synthesis error $\epsilon_r$, and
\begin{eqnarray}
    N^{(f)}(\epsilon_f) &=& 8(\log_{2}{(2k)})\log_2\left(\frac{(\log_{2}{(2k)})}{\epsilon_{f}}\right) + \log_2\left(\frac{(\log_{2}{(2k)})}{\epsilon_{f}}\right)\log_2\left(\frac{\log_2\left(\frac{(\log_{2}{(2k)})}{\epsilon_{f}}\right)}{\epsilon_{f}}\right)
\end{eqnarray}
is the cost of compiling a single AQFT to within error $\epsilon_f$ obtained simply by plugging in $n=\log_{2}{(2k)}$ in Eq.~\eqref{eqn:TcostAQFT}.

For Algorithm 1, we have
\begin{eqnarray}
N_r &=& 24 \nonumber \\
N_f &=& 2 \vert \Omega \vert \nonumber \\
N_{+} &=& 72 \vert \Omega \vert \log_{2}^{2}{k} + 168 \vert \Omega \vert \log_{2}{k} + 8 \vert \Omega \vert D - 32 \vert \Omega \vert - 16
\end{eqnarray}
Using the above, as well as the expression for the coefficient 1-norm $\alpha$ of the Hamiltonian provided in Eq.~\eqref{eqn:alpha1-Hamp} for $\Delta_{\phi} = \sqrt{\pi/k}$, and Eq.~\eqref{eqn:error-bounds-qubitization} from Lemma \ref{lemma:error-bounds-qubitization}, we find (for fixed spatial dimensionality)
\begin{eqnarray}
2^m &\in& O\left( \frac{\vert\Omega\vert}{\epsilon_E} \left[ k^{2}\Lambda + k M^{2}\right]\right) \nonumber \\
N_{r}\cdot N^{(s)}(\epsilon_r) &\in& O\left( \log{\left( \frac{\alpha}{\epsilon_E} \right)} \right) \in O\left( \log{\left( \frac{\vert \Omega \vert k (\Lambda + M^{2})}{\epsilon_E}\right)}\right) \nonumber \\
N_{f}\cdot N^{(f)}(\epsilon_f) &\in& O\left( \vert \Omega \vert \log{k} \log{\left[ \frac{\log{k} \cdot \vert \Omega \vert k (\Lambda + M^{2})}{\epsilon_E}\right]}\right) \nonumber \\
N_{+} &\in& O\left( \vert \Omega \vert \log^{2}{k}\right)
\end{eqnarray}
The dominant cost is that of $2^{m}N_{+}$ and so, in all,
\begin{equation}
Cost(QPE)^{(I)} \in O \left( \frac{\vert\Omega\vert^{2}}{\epsilon_E} \left[ k^{2}\Lambda + k M^{2} \right] \log^{2}{k} \right)
\end{equation}
Meanwhile, adding the qubit count from Eq.~\eqref{eqn:tot-qubit-count-alg1} and $m = \log_{2}{\frac{\pi\alpha}{\sqrt{2}\epsilon_E}}$, we get
\begin{equation}
Count(Qubit)^{(I)} \in O \left( \vert \Omega \vert \log{k} + \log^{2}{k} + \log{\left( \frac{\vert \Omega \vert \left[ k^{2}\Lambda + k M^{2} \right]}{\epsilon_E}\right)}\right)
\end{equation}

Similarly, we recall these gate counts for Algorithm 3a, borrowing from Eqs.~\eqref{eqn:num-rotations-alg3a} and ~\eqref{eqn:num-Tplus-aqft-alg3a} here for convenience,
\begin{eqnarray}
N_r &=& 4(L_1+L_2)+2(L_3+L_4)+2L_5-4 \nonumber \\
N_f &=& 2 \vert \Omega \vert \nonumber \\
N_{+} &=& |\Omega|\left[4\sqrt{L_1+L_2}\left(\log_2(L_1+L_2)-2\right)+12(L_1+L_2)+4\sqrt{L_3+L_4}\left(\log_2(L_3+L_4)-2\right)+12(L_3+L_4)\right]   \nonumber \\
    &&+|E_D|\left(4\sqrt{L_5}\left(\log_2L_5-2\right)+8L_5\right)+4\sqrt{|\Omega|}\left(\log_2|\Omega|-2\right)+4|\Omega|    
    +4\sqrt{|E_D|}\left(\log_2|E_D|-2\right)+4|E_D|
\end{eqnarray}
and from the Lemmas in Supplemental Material (Sec III A) \cite{SuppMat}, 
we have that the coefficient 1-norm scales as $\alpha \in O\left( \vert \Omega \vert \left( k^{2}\Lambda + k M^{2}\right)\right)$, similarly as in Algorithm 1. Using the results of Lemma \ref{eqn:error-bounds-qubitization} as before, we find
\begin{eqnarray}
2^m &\in& O\left( \frac{\vert\Omega\vert}{\epsilon_E} \left[ k^{2}\Lambda + k M^{2}\right]\right) \nonumber \\
N_{r}\cdot N^{(s)}(\epsilon_r) &\in& O\left( \log^{4}{k}\cdot \log{\left[ \frac{\vert\Omega\vert \log{k}\left( k^{2}\Lambda + kM^{2}\right)}{\epsilon_E}\right]} \right) \nonumber \\
N_{f}\cdot N^{(f)}(\epsilon_f) &\in& O\left( \vert \Omega \vert \log{k} \log{\left[ \frac{\log{k} \cdot \vert \Omega \vert k (\Lambda + M^{2})}{\epsilon_E}\right]}\right) \nonumber \\
N_{+} &\in& O\left( \vert \Omega \vert \log^{4}{k} \right)
\end{eqnarray}
The dominant cost is that of $2^{m}N_{+}$, and so we have
\begin{equation}
Cost(QPE)^{(IIIa)} \in O \left( \frac{\vert\Omega\vert^{2}}{\epsilon_E} \left[ k^{2}\Lambda + k M^{2} \right] \left[ \log^{4}{k} + \vert \Omega \vert \right] \right)
\end{equation}
We estimate the total number of logical qubits used in Algorithm IIIa as the sum total of $\vert \Omega \vert \log_{2}{2k}$ many qubits used to represent the field itself, $\log_{2}{\vert \Omega \vert}$ used to control the lattice site, $\log_{2}{\log_{2}{k}}$ used to control the unitary in the LCU expansion, and $m = \log_{2}{\frac{\pi \alpha}{\sqrt{2}\epsilon_E}}$ many ancillary qubits used for QPE. Thus, the total qubit count used in Algorithm IIIa is estimated as
\begin{eqnarray}
Count(Qubit)^{(IIIa)} &\approx& \vert \Omega \vert \log_{2}{2k} + \log_{2}{\vert \Omega \vert} + \log_{2}{\log_{2}{k}} + \log_{2}{\frac{\pi \alpha}{\sqrt{2}\epsilon_E}} \nonumber \\
&\in& O\left( \vert \Omega \vert \log{k} + \log{\left[ \frac{\vert \Omega \vert \left( k^{2}\Lambda + kM^{2}\right)}{\epsilon_E} \right]} \right)
\end{eqnarray}
\end{proof}

\begin{proof}[Proof of Proposition \ref{prop:qpe-costs-alg3b}]
As in Theorem \ref{thm:cost-qpe-qubitization}, the total T-gate cost for performing QPE using this qubitization-based algorithm can be approximated as
\begin{equation}
Cost(QPE) \approx 2^{m} \left[ N_r \cdot N^{(s)}(\epsilon_r) + N_f \cdot N^{(f)}(\epsilon_f) + N_{+}\right]
\end{equation}
where $m$ is the total number of ancillary qubits used for QPE, $N_r$ and $N_f$ are  respectively the total number of rotation gates and AQFTs used in the circuit compiling the walk operator, $N_{+}$ is the total number of other T-gates used, and $N^{(s)}(\epsilon_r)$ and $N^{(f)}(\epsilon_f)$ are respectively the T-gate cost of compiling a single RZ gate within synthesis error $\epsilon_r$, and the T-gate cost of compiling a single AQFT within error $\epsilon_f$.

For convenience, we recall the relevant number of gates from Supplemental Material \cite{SuppMat}
\begin{eqnarray}
N_r &=& 12\log_2k+2(\log_2k+1)^2-3  \nonumber \\
N_f &=& 2|\Omega| \nonumber \\
N_{+} &=& |\Omega|\left(4\sqrt{2\log_2k}(\log_2(2\log_2k)-2)+8\log_2k+4\sqrt{4\log_2k}(\log_2(4\log_2k)-2)+16\log_2k  \right) \nonumber \\
  &&+ 4\sqrt{|\Omega|}\left(\log_2|\Omega|-2\right)+4|\Omega| +|E_D| \left(8(\log_2k+1)(\log_2(\log_2k+1)-1)+4(\log_2k+1)^2+4\right)  \nonumber \\
  &&+4\sqrt{|E_D|}(\log_2|E_D|-2)+O\left(|\Omega|\log_2^2k\right)
\end{eqnarray}
Using the expression for the coefficient 1-norm given in Eq.~\eqref{eqn:HampL1}, we find
\begin{eqnarray}
\alpha = \Vert H_{amp} \Vert &\in& O\left( \vert \Omega \vert \left( k^{2}\Lambda + k M^{2} \right) \right) \nonumber \\
\Rightarrow 2^m &\in& O\left( \frac{\vert \Omega \vert \left( k^{2}\Lambda + k M^{2} \right)}{\epsilon_E}\right)
\end{eqnarray}
The dominant cost is now that of $2^m N_{+}$, and so we have
\begin{equation}
Cost(QPE)^{(IIIb)} \in O\left( \frac{\vert \Omega \vert^{2}}{\epsilon_E} \left[ k^{2}\Lambda + k M^{2} \right] \log^{2}{k} \right)
\end{equation}
For the qubit counts, we require a total of $O\left( \vert \Omega \vert \log{k}\right)$ many qubits to hold the field values, and an additional $O\left( \log \log{k}\right)$ qubits for the implementation of the signature matrices in the $O(\log{k})$ LCU decomposition, a cost which is subdominant. In addition, we would require an additional number of $m \in O(\Vert H_{amp}\Vert/\epsilon_E)$ ancillary qubits used for phase estimation, so that in all we have
\begin{equation}
Count(Qubit)^{(IIIb)} \in O\left( \vert \Omega \vert \log{k} + \log{\left[ \frac{\vert \Omega \vert \left( k^{2}\Lambda + kM^{2}\right)}{\epsilon_E} \right]} \right)
\end{equation}
\end{proof}

In order to perform a similar analysis for Algorithm II in the amplitude basis, which is a Trotter-based algorithm that also employs the AQFT, which would incur its own error, we make use of the following lemma.

\begin{lemma}
To obtain an estimate of an eigenvalue of a Hamiltonian within error $\epsilon_E$, in the presence of Trotter, synthesis, and AQFT errors, it suffices to perform phase estimation on a single Trotter step with the following bounds on the contributing errors
\begin{eqnarray}
2^m &\geq& \pi 2^{3/4} \frac{\alpha_{comm}^{1/2}}{\epsilon_{E}^{3/2}} \nonumber \\
\epsilon_r &\leq& 2^{4/15} \frac{\epsilon_{E}^{3/2}}{N_r \alpha_{comm}^{1/2}} \nonumber \\
\epsilon_f &\leq& 2^{4/15} \frac{\epsilon_{E}^{3/2}}{N_f \alpha_{comm}^{1/2}}
\end{eqnarray}
\label{lemma:trotter-amp-aqft-errors}
\end{lemma}

\begin{proof}
The proof proceeds similarly to Lemmas \ref{lem:error-phase} and \ref{eqn:error-bounds-qubitization}, except that we now also account for both Trotter and AQFT errors here. Specifically, we approximate the error in the phase as
\begin{equation}
\epsilon_{\theta} \approx \sqrt{\left( \frac{\pi}{2^{m+1}}\right)^{2} + \left( \pi \epsilon_{QFT}^{2} + \epsilon_{Trotter} + \epsilon_{synth} + \epsilon_{AQFT} \right)^{2}}
\label{eqn:error-trotter-phase-with-aqft}
\end{equation}

The error in the phase error $\epsilon_{\theta}$ is related to the error in the energy estimate $\epsilon_E$ as $\epsilon_{\theta} = \epsilon_E \tau$, where $\tau$ is the total simulation time. Distributing the errors roughly equally, we bound
\begin{eqnarray}
\left( \frac{\pi}{2^{m+1}}\right)^{2} \leq \frac{\epsilon_{\theta}^{2}}{2}, \qquad \pi \epsilon_{QFT} = \epsilon_{Trotter} = \epsilon_{synth} = \epsilon_{AQFT} \leq \frac{1}{4}\frac{\epsilon_{\theta}}{\sqrt{2}}
\end{eqnarray}
Using the Trotter error bound, we obtain
\begin{eqnarray}
\epsilon_{Trotter} &\leq& \frac{1}{4} \frac{\epsilon_E \tau}{\sqrt{2}} \leq \alpha_{comm}\tau^{3} \nonumber \\
\Rightarrow \tau &\geq& \sqrt{\frac{\epsilon_E}{2^{5/2} \alpha_{comm}}}
\end{eqnarray}
Therefore,
\begin{eqnarray}
\left( \frac{\pi}{2^{m+1}}\right)^{2} &\leq& \frac{\epsilon_{\theta}^{2}}{2} \nonumber \\
\Rightarrow 2^m &\geq& \pi 2^{3/4} \frac{\alpha_{comm}^{1/2}}{\epsilon_{E}^{3/2}}
\end{eqnarray}
Similarly, we can derive
\begin{eqnarray}
\epsilon_{synth} = \epsilon_r N_r &\leq& \frac{1}{4}\frac{\epsilon_{\theta}}{\sqrt{2}} \nonumber \\
\Rightarrow \epsilon_r &\leq& 2^{4/15} \frac{\epsilon_{E}^{3/2}}{N_r \alpha_{comm}^{1/2}}
\end{eqnarray}
and similarly,
\begin{eqnarray}
\epsilon_{AQFT} = \epsilon_f N_f &\leq& \frac{1}{4}\frac{\epsilon_{\theta}}{\sqrt{2}} \nonumber \\
\Rightarrow \epsilon_f &\leq& 2^{4/15} \frac{\epsilon_{E}^{3/2}}{N_f \alpha_{comm}^{1/2}}
\end{eqnarray}
\end{proof}

The total number of T-gates for phase estimation using Algorithm II has already been detailed in Theorem \ref{thm:ampTrotter}. The additional number of ancillary qubits required for phase estimation is described in the following lemma.

\begin{lemma}
The total number of qubits required for phase estimation using the amplitude basis Algorithm II is given by
\begin{equation}
Count(Qubit)^{(II)} \approx O\left( \log_{2} \left( \frac{\vert \Omega \vert \Lambda k \left( \Lambda k + N^2 \right)}{\epsilon_E}\right) \right)
\end{equation}
\label{lemma:anc-qubits-alg2}
\end{lemma}
\begin{proof}
As already noted in Theorem \ref{thm:ampTrotter}, the total number of qubits used for block encoding sis $O(\vert \Omega \vert \log{k})$. A further number of ancillary qubits used for phase estimation is given by Lemma \ref{lemma:trotter-amp-aqft-errors} and Eq.~\eqref{eqn:alphacomm_ampTrot} as
\begin{eqnarray}
m &\in& O \left( \log_{2}{\left( \frac{\alpha_{comm}^{1/2}}{\epsilon_{E}^{3/2}}\right)}\right) \nonumber \\
&\in& O\left( \log_{2} \left( \frac{\vert \Omega \vert \Lambda k \left( \Lambda k + N^2 \right)}{\epsilon_E}\right) \right)
\end{eqnarray}
where we have repeatedly used $\log{\left( \frac{A^m}{B^n}\right) } \in O\left( \log{\left( \frac{A}{B}\right)}\right)$ for constant $m,n > 0$.
\end{proof}

\subsubsection{Raw gate costs for phase estimation}
We have reported the asymptotic T-gate costs of all the algorithms discussed above in Table 1. Here, we numerically compute the raw T-gate costs, taking care of pre-factors and other factors missing from the asymptotic expressions. We estimate the total cost of phase estimation as
\begin{equation}
Cost(QPE) \approx 2^{m} \left[ N_r \cdot N^{(s)}(\epsilon_r) + N_f \cdot N^{(f)}(\epsilon_f) + N_{+}\right]
\end{equation}
where $m$ is the total number of ancillary qubits used for QPE, $N_r$ and $N_f$ are  respectively the total number of rotation gates and AQFTs used in the circuit compiling the walk operator, $N_{+}$ is the total number of other T-gates used, and
\begin{equation}
N^{(s)}(\epsilon_r) = 3.067 \log_{2}(2/\epsilon_r) - 4.327
\end{equation}
is the T-gate cost of compiling a single RZ gate from \cite{2015_KMM} to within synthesis error $\epsilon_r$, and
\begin{eqnarray}
    N^{(f)}(\epsilon_f) &=& 8(\log_{2}{(2k)})\log_2\left(\frac{(\log_{2}{(2k)})}{\epsilon_{f}}\right) + \log_2\left(\frac{(\log_{2}{(2k)})}{\epsilon_{f}}\right)\log_2\left(\frac{\log_2\left(\frac{(\log_{2}{(2k)})}{\epsilon_{f}}\right)}{\epsilon_{f}}\right)
\end{eqnarray}
is the cost of compiling a single AQFT to within error $\epsilon_f$ obtained simply by plugging in $n=\log_{2}{(2k)}$ in Eq.~\eqref{eqn:TcostAQFT}.

The bounds for the errors, and the expressions for $N_r$, $N_f$ and $N_{+}$ for the various algorithms have been recalled for convenience in Lemma \ref{lem:error-phase}, Theorem \ref{thm:totalTocc}, Theorem \ref{thm:ampTrotter}, Lemma \ref{lemma:error-bounds-qubitization}, Theorem \ref{thm:cost-qpe-qubitization}, Proposition \ref{prop:qpe-costs-alg3b}, and Lemmas \ref{lemma:trotter-amp-aqft-errors} and \ref{lemma:anc-qubits-alg2} above.


\bibliographystyle{apsrev4-2}
\bibliography{apssamp,lib}

\begin{thebibliography}{172}%
\makeatletter
\providecommand \@ifxundefined [1]{%
 \@ifx{#1\undefined}
}%
\providecommand \@ifnum [1]{%
 \ifnum #1\expandafter \@firstoftwo
 \else \expandafter \@secondoftwo
 \fi
}%
\providecommand \@ifx [1]{%
 \ifx #1\expandafter \@firstoftwo
 \else \expandafter \@secondoftwo
 \fi
}%
\providecommand \natexlab [1]{#1}%
\providecommand \enquote  [1]{``#1''}%
\providecommand \bibnamefont  [1]{#1}%
\providecommand \bibfnamefont [1]{#1}%
\providecommand \citenamefont [1]{#1}%
\providecommand \href@noop [0]{\@secondoftwo}%
\providecommand \href [0]{\begingroup \@sanitize@url \@href}%
\providecommand \@href[1]{\@@startlink{#1}\@@href}%
\providecommand \@@href[1]{\endgroup#1\@@endlink}%
\providecommand \@sanitize@url [0]{\catcode `\\12\catcode `\$12\catcode `\&12\catcode `\#12\catcode `\^12\catcode `\_12\catcode `\%12\relax}%
\providecommand \@@startlink[1]{}%
\providecommand \@@endlink[0]{}%
\providecommand \url  [0]{\begingroup\@sanitize@url \@url }%
\providecommand \@url [1]{\endgroup\@href {#1}{\urlprefix }}%
\providecommand \urlprefix  [0]{URL }%
\providecommand \Eprint [0]{\href }%
\providecommand \doibase [0]{https://doi.org/}%
\providecommand \selectlanguage [0]{\@gobble}%
\providecommand \bibinfo  [0]{\@secondoftwo}%
\providecommand \bibfield  [0]{\@secondoftwo}%
\providecommand \translation [1]{[#1]}%
\providecommand \BibitemOpen [0]{}%
\providecommand \bibitemStop [0]{}%
\providecommand \bibitemNoStop [0]{.\EOS\space}%
\providecommand \EOS [0]{\spacefactor3000\relax}%
\providecommand \BibitemShut  [1]{\csname bibitem#1\endcsname}%
\let\auto@bib@innerbib\@empty
\bibitem [{\citenamefont {Lloyd}(1996)}]{lloyd1996universal}%
  \BibitemOpen
  \bibfield  {author} {\bibinfo {author} {\bibfnamefont {S.}~\bibnamefont {Lloyd}},\ }\href@noop {} {\bibfield  {journal} {\bibinfo  {journal} {Science}\ }\textbf {\bibinfo {volume} {273}},\ \bibinfo {pages} {1073} (\bibinfo {year} {1996})}\BibitemShut {NoStop}%
\bibitem [{\citenamefont {Berry}\ \emph {et~al.}(2007)\citenamefont {Berry}, \citenamefont {Ahokas}, \citenamefont {Cleve},\ and\ \citenamefont {Sanders}}]{berry2007efficient}%
  \BibitemOpen
  \bibfield  {author} {\bibinfo {author} {\bibfnamefont {D.~W.}\ \bibnamefont {Berry}}, \bibinfo {author} {\bibfnamefont {G.}~\bibnamefont {Ahokas}}, \bibinfo {author} {\bibfnamefont {R.}~\bibnamefont {Cleve}},\ and\ \bibinfo {author} {\bibfnamefont {B.~C.}\ \bibnamefont {Sanders}},\ }\href@noop {} {\bibfield  {journal} {\bibinfo  {journal} {Communications in Mathematical Physics}\ }\textbf {\bibinfo {volume} {270}},\ \bibinfo {pages} {359} (\bibinfo {year} {2007})}\BibitemShut {NoStop}%
\bibitem [{\citenamefont {Childs}\ and\ \citenamefont {Wiebe}(2012{\natexlab{a}})}]{childs2012hamiltonian}%
  \BibitemOpen
  \bibfield  {author} {\bibinfo {author} {\bibfnamefont {A.~M.}\ \bibnamefont {Childs}}\ and\ \bibinfo {author} {\bibfnamefont {N.}~\bibnamefont {Wiebe}},\ }\href@noop {} {\bibfield  {journal} {\bibinfo  {journal} {Quantum Information and Computation}\ }\textbf {\bibinfo {volume} {12}},\ \bibinfo {pages} {901} (\bibinfo {year} {2012}{\natexlab{a}})}\BibitemShut {NoStop}%
\bibitem [{\citenamefont {Childs}\ \emph {et~al.}(2017)\citenamefont {Childs}, \citenamefont {Kothari},\ and\ \citenamefont {Somma}}]{childs2017quantum}%
  \BibitemOpen
  \bibfield  {author} {\bibinfo {author} {\bibfnamefont {A.~M.}\ \bibnamefont {Childs}}, \bibinfo {author} {\bibfnamefont {R.}~\bibnamefont {Kothari}},\ and\ \bibinfo {author} {\bibfnamefont {R.~D.}\ \bibnamefont {Somma}},\ }\href@noop {} {\bibfield  {journal} {\bibinfo  {journal} {SIAM Journal on Computing}\ }\textbf {\bibinfo {volume} {46}},\ \bibinfo {pages} {1920} (\bibinfo {year} {2017})}\BibitemShut {NoStop}%
\bibitem [{\citenamefont {Low}\ and\ \citenamefont {Chuang}(2019{\natexlab{a}})}]{Low2019hamiltonian}%
  \BibitemOpen
  \bibfield  {author} {\bibinfo {author} {\bibfnamefont {G.~H.}\ \bibnamefont {Low}}\ and\ \bibinfo {author} {\bibfnamefont {I.~L.}\ \bibnamefont {Chuang}},\ }\href {https://doi.org/10.22331/q-2019-07-12-163} {\bibfield  {journal} {\bibinfo  {journal} {{Quantum}}\ }\textbf {\bibinfo {volume} {3}},\ \bibinfo {pages} {163} (\bibinfo {year} {2019}{\natexlab{a}})}\BibitemShut {NoStop}%
\bibitem [{\citenamefont {Su}\ \emph {et~al.}(2021)\citenamefont {Su}, \citenamefont {Berry}, \citenamefont {Wiebe}, \citenamefont {Rubin},\ and\ \citenamefont {Babbush}}]{2021_SBWetal}%
  \BibitemOpen
  \bibfield  {author} {\bibinfo {author} {\bibfnamefont {Y.}~\bibnamefont {Su}}, \bibinfo {author} {\bibfnamefont {D.~W.}\ \bibnamefont {Berry}}, \bibinfo {author} {\bibfnamefont {N.}~\bibnamefont {Wiebe}}, \bibinfo {author} {\bibfnamefont {N.}~\bibnamefont {Rubin}},\ and\ \bibinfo {author} {\bibfnamefont {R.}~\bibnamefont {Babbush}},\ }\href@noop {} {\bibfield  {journal} {\bibinfo  {journal} {PRX Quantum}\ }\textbf {\bibinfo {volume} {2}},\ \bibinfo {pages} {040332} (\bibinfo {year} {2021})}\BibitemShut {NoStop}%
\bibitem [{\citenamefont {Bauer}\ \emph {et~al.}(2022)\citenamefont {Bauer}, \citenamefont {Davoudi}, \citenamefont {Balantekin}, \citenamefont {Bhattacharya}, \citenamefont {Carena}, \citenamefont {{de Jong}}, \citenamefont {Draper}, \citenamefont {{El-Khadra}}, \citenamefont {Gemelke}, \citenamefont {Hanada}, \citenamefont {Kharzeev}, \citenamefont {Lamm}, \citenamefont {Li}, \citenamefont {Liu}, \citenamefont {Lukin}, \citenamefont {Meurice}, \citenamefont {Monroe}, \citenamefont {Nachman}, \citenamefont {Pagano}, \citenamefont {Preskill}, \citenamefont {Rinaldi}, \citenamefont {Roggero}, \citenamefont {Santiago}, \citenamefont {Savage}, \citenamefont {Siddiqi}, \citenamefont {Siopsis}, \citenamefont {Van~Zanten}, \citenamefont {Wiebe}, \citenamefont {Yamauchi}, \citenamefont {{Yeter-Aydeniz}},\ and\ \citenamefont {Zorzetti}}]{Bauer2022}%
  \BibitemOpen
  \bibfield  {author} {\bibinfo {author} {\bibfnamefont {C.~W.}\ \bibnamefont {Bauer}}, \bibinfo {author} {\bibfnamefont {Z.}~\bibnamefont {Davoudi}}, \bibinfo {author} {\bibfnamefont {A.~B.}\ \bibnamefont {Balantekin}}, \bibinfo {author} {\bibfnamefont {T.}~\bibnamefont {Bhattacharya}}, \bibinfo {author} {\bibfnamefont {M.}~\bibnamefont {Carena}}, \bibinfo {author} {\bibfnamefont {W.~A.}\ \bibnamefont {{de Jong}}}, \bibinfo {author} {\bibfnamefont {P.}~\bibnamefont {Draper}}, \bibinfo {author} {\bibfnamefont {A.}~\bibnamefont {{El-Khadra}}}, \bibinfo {author} {\bibfnamefont {N.}~\bibnamefont {Gemelke}}, \bibinfo {author} {\bibfnamefont {M.}~\bibnamefont {Hanada}}, \bibinfo {author} {\bibfnamefont {D.}~\bibnamefont {Kharzeev}}, \bibinfo {author} {\bibfnamefont {H.}~\bibnamefont {Lamm}}, \bibinfo {author} {\bibfnamefont {Y.-Y.}\ \bibnamefont {Li}}, \bibinfo {author} {\bibfnamefont {J.}~\bibnamefont {Liu}}, \bibinfo {author} {\bibfnamefont {M.}~\bibnamefont {Lukin}}, \bibinfo {author} {\bibfnamefont
  {Y.}~\bibnamefont {Meurice}}, \bibinfo {author} {\bibfnamefont {C.}~\bibnamefont {Monroe}}, \bibinfo {author} {\bibfnamefont {B.}~\bibnamefont {Nachman}}, \bibinfo {author} {\bibfnamefont {G.}~\bibnamefont {Pagano}}, \bibinfo {author} {\bibfnamefont {J.}~\bibnamefont {Preskill}}, \bibinfo {author} {\bibfnamefont {E.}~\bibnamefont {Rinaldi}}, \bibinfo {author} {\bibfnamefont {A.}~\bibnamefont {Roggero}}, \bibinfo {author} {\bibfnamefont {D.~I.}\ \bibnamefont {Santiago}}, \bibinfo {author} {\bibfnamefont {M.~J.}\ \bibnamefont {Savage}}, \bibinfo {author} {\bibfnamefont {I.}~\bibnamefont {Siddiqi}}, \bibinfo {author} {\bibfnamefont {G.}~\bibnamefont {Siopsis}}, \bibinfo {author} {\bibfnamefont {D.}~\bibnamefont {Van~Zanten}}, \bibinfo {author} {\bibfnamefont {N.}~\bibnamefont {Wiebe}}, \bibinfo {author} {\bibfnamefont {Y.}~\bibnamefont {Yamauchi}}, \bibinfo {author} {\bibfnamefont {K.}~\bibnamefont {{Yeter-Aydeniz}}},\ and\ \bibinfo {author} {\bibfnamefont {S.}~\bibnamefont {Zorzetti}},\ }\href@noop {}
  {\bibfield  {journal} {\bibinfo  {journal} {{arXiv}}\ } (\bibinfo {year} {2022})},\ \Eprint {https://arxiv.org/abs/2204.03381} {arxiv:2204.03381 [hep-lat, physics:hep-ph, physics:hep-th, physics:nucl-th, physics:quant-ph]} \BibitemShut {NoStop}%
\bibitem [{\citenamefont {Kogut}\ and\ \citenamefont {Susskind}(1975)}]{Kogut1975}%
  \BibitemOpen
  \bibfield  {author} {\bibinfo {author} {\bibfnamefont {J.}~\bibnamefont {Kogut}}\ and\ \bibinfo {author} {\bibfnamefont {L.}~\bibnamefont {Susskind}},\ }\href {https://doi.org/10.1103/PhysRevD.11.395} {\bibfield  {journal} {\bibinfo  {journal} {Phys. Rev. D}\ }\textbf {\bibinfo {volume} {11}},\ \bibinfo {pages} {395} (\bibinfo {year} {1975})}\BibitemShut {NoStop}%
\bibitem [{\citenamefont {Drell}\ \emph {et~al.}(1979)\citenamefont {Drell}, \citenamefont {Quinn}, \citenamefont {Svetitsky},\ and\ \citenamefont {Weinstein}}]{Drell1979}%
  \BibitemOpen
  \bibfield  {author} {\bibinfo {author} {\bibfnamefont {S.~D.}\ \bibnamefont {Drell}}, \bibinfo {author} {\bibfnamefont {H.~R.}\ \bibnamefont {Quinn}}, \bibinfo {author} {\bibfnamefont {B.}~\bibnamefont {Svetitsky}},\ and\ \bibinfo {author} {\bibfnamefont {M.}~\bibnamefont {Weinstein}},\ }\href {https://doi.org/10.1103/PhysRevD.19.619} {\bibfield  {journal} {\bibinfo  {journal} {Phys. Rev. D}\ }\textbf {\bibinfo {volume} {19}},\ \bibinfo {pages} {619} (\bibinfo {year} {1979})}\BibitemShut {NoStop}%
\bibitem [{\citenamefont {Jordan}\ \emph {et~al.}(2018)\citenamefont {Jordan}, \citenamefont {Krovi}, \citenamefont {Lee},\ and\ \citenamefont {Preskill}}]{jordan2018}%
  \BibitemOpen
  \bibfield  {author} {\bibinfo {author} {\bibfnamefont {S.~P.}\ \bibnamefont {Jordan}}, \bibinfo {author} {\bibfnamefont {H.}~\bibnamefont {Krovi}}, \bibinfo {author} {\bibfnamefont {K.~S.~M.}\ \bibnamefont {Lee}},\ and\ \bibinfo {author} {\bibfnamefont {J.}~\bibnamefont {Preskill}},\ }\href {https://doi.org/10.22331/q-2018-01-08-44} {\bibfield  {journal} {\bibinfo  {journal} {Quantum}\ }\textbf {\bibinfo {volume} {2}},\ \bibinfo {pages} {44} (\bibinfo {year} {2018})}\BibitemShut {NoStop}%
\bibitem [{\citenamefont {Klco}\ and\ \citenamefont {Savage}(2019)}]{klcoDigitizationScalarFields2019}%
  \BibitemOpen
  \bibfield  {author} {\bibinfo {author} {\bibfnamefont {N.}~\bibnamefont {Klco}}\ and\ \bibinfo {author} {\bibfnamefont {M.~J.}\ \bibnamefont {Savage}},\ }\href {https://doi.org/10.1103/PhysRevA.99.052335} {\bibfield  {journal} {\bibinfo  {journal} {Physical Review A}\ }\textbf {\bibinfo {volume} {99}},\ \bibinfo {pages} {052335} (\bibinfo {year} {2019})}\BibitemShut {NoStop}%
\bibitem [{\citenamefont {Alexeev}\ \emph {et~al.}(2021)\citenamefont {Alexeev}, \citenamefont {Bacon}, \citenamefont {Brown}, \citenamefont {Calderbank}, \citenamefont {Carr}, \citenamefont {Chong}, \citenamefont {DeMarco}, \citenamefont {Englund}, \citenamefont {Farhi}, \citenamefont {Fefferman} \emph {et~al.}}]{alexeev2021quantum}%
  \BibitemOpen
  \bibfield  {author} {\bibinfo {author} {\bibfnamefont {Y.}~\bibnamefont {Alexeev}}, \bibinfo {author} {\bibfnamefont {D.}~\bibnamefont {Bacon}}, \bibinfo {author} {\bibfnamefont {K.~R.}\ \bibnamefont {Brown}}, \bibinfo {author} {\bibfnamefont {R.}~\bibnamefont {Calderbank}}, \bibinfo {author} {\bibfnamefont {L.~D.}\ \bibnamefont {Carr}}, \bibinfo {author} {\bibfnamefont {F.~T.}\ \bibnamefont {Chong}}, \bibinfo {author} {\bibfnamefont {B.}~\bibnamefont {DeMarco}}, \bibinfo {author} {\bibfnamefont {D.}~\bibnamefont {Englund}}, \bibinfo {author} {\bibfnamefont {E.}~\bibnamefont {Farhi}}, \bibinfo {author} {\bibfnamefont {B.}~\bibnamefont {Fefferman}}, \emph {et~al.},\ }\href@noop {} {\bibfield  {journal} {\bibinfo  {journal} {PRX quantum}\ }\textbf {\bibinfo {volume} {2}},\ \bibinfo {pages} {017001} (\bibinfo {year} {2021})}\BibitemShut {NoStop}%
\bibitem [{\citenamefont {Bauer}\ \emph {et~al.}(2023)\citenamefont {Bauer}, \citenamefont {Davoudi}, \citenamefont {Balantekin}, \citenamefont {Bhattacharya}, \citenamefont {Carena}, \citenamefont {De~Jong}, \citenamefont {Draper}, \citenamefont {El-Khadra}, \citenamefont {Gemelke}, \citenamefont {Hanada} \emph {et~al.}}]{bauer2023quantum}%
  \BibitemOpen
  \bibfield  {author} {\bibinfo {author} {\bibfnamefont {C.~W.}\ \bibnamefont {Bauer}}, \bibinfo {author} {\bibfnamefont {Z.}~\bibnamefont {Davoudi}}, \bibinfo {author} {\bibfnamefont {A.~B.}\ \bibnamefont {Balantekin}}, \bibinfo {author} {\bibfnamefont {T.}~\bibnamefont {Bhattacharya}}, \bibinfo {author} {\bibfnamefont {M.}~\bibnamefont {Carena}}, \bibinfo {author} {\bibfnamefont {W.~A.}\ \bibnamefont {De~Jong}}, \bibinfo {author} {\bibfnamefont {P.}~\bibnamefont {Draper}}, \bibinfo {author} {\bibfnamefont {A.}~\bibnamefont {El-Khadra}}, \bibinfo {author} {\bibfnamefont {N.}~\bibnamefont {Gemelke}}, \bibinfo {author} {\bibfnamefont {M.}~\bibnamefont {Hanada}}, \emph {et~al.},\ }\href@noop {} {\bibfield  {journal} {\bibinfo  {journal} {PRX quantum}\ }\textbf {\bibinfo {volume} {4}},\ \bibinfo {pages} {027001} (\bibinfo {year} {2023})}\BibitemShut {NoStop}%
\bibitem [{\citenamefont {Nguyen}\ \emph {et~al.}(2022)\citenamefont {Nguyen}, \citenamefont {Tran}, \citenamefont {Zhu}, \citenamefont {Green}, \citenamefont {Alderete}, \citenamefont {Davoudi},\ and\ \citenamefont {Linke}}]{nguyen2022digital}%
  \BibitemOpen
  \bibfield  {author} {\bibinfo {author} {\bibfnamefont {N.~H.}\ \bibnamefont {Nguyen}}, \bibinfo {author} {\bibfnamefont {M.~C.}\ \bibnamefont {Tran}}, \bibinfo {author} {\bibfnamefont {Y.}~\bibnamefont {Zhu}}, \bibinfo {author} {\bibfnamefont {A.~M.}\ \bibnamefont {Green}}, \bibinfo {author} {\bibfnamefont {C.~H.}\ \bibnamefont {Alderete}}, \bibinfo {author} {\bibfnamefont {Z.}~\bibnamefont {Davoudi}},\ and\ \bibinfo {author} {\bibfnamefont {N.~M.}\ \bibnamefont {Linke}},\ }\href@noop {} {\bibfield  {journal} {\bibinfo  {journal} {PRX Quantum}\ }\textbf {\bibinfo {volume} {3}},\ \bibinfo {pages} {020324} (\bibinfo {year} {2022})}\BibitemShut {NoStop}%
\bibitem [{\citenamefont {Shaw}\ \emph {et~al.}(2020)\citenamefont {Shaw}, \citenamefont {Lougovski}, \citenamefont {Stryker},\ and\ \citenamefont {Wiebe}}]{shaw2020quantum}%
  \BibitemOpen
  \bibfield  {author} {\bibinfo {author} {\bibfnamefont {A.~F.}\ \bibnamefont {Shaw}}, \bibinfo {author} {\bibfnamefont {P.}~\bibnamefont {Lougovski}}, \bibinfo {author} {\bibfnamefont {J.~R.}\ \bibnamefont {Stryker}},\ and\ \bibinfo {author} {\bibfnamefont {N.}~\bibnamefont {Wiebe}},\ }\href@noop {} {\bibfield  {journal} {\bibinfo  {journal} {Quantum}\ }\textbf {\bibinfo {volume} {4}},\ \bibinfo {pages} {306} (\bibinfo {year} {2020})}\BibitemShut {NoStop}%
\bibitem [{\citenamefont {Jordan}\ \emph {et~al.}(2012)\citenamefont {Jordan}, \citenamefont {Lee},\ and\ \citenamefont {Preskill}}]{Jordan2012}%
  \BibitemOpen
  \bibfield  {author} {\bibinfo {author} {\bibfnamefont {S.~P.}\ \bibnamefont {Jordan}}, \bibinfo {author} {\bibfnamefont {K.~S.~M.}\ \bibnamefont {Lee}},\ and\ \bibinfo {author} {\bibfnamefont {J.}~\bibnamefont {Preskill}},\ }\href {https://doi.org/10.1126/science.1217069} {\bibfield  {journal} {\bibinfo  {journal} {Science}\ }\textbf {\bibinfo {volume} {336}},\ \bibinfo {pages} {1130} (\bibinfo {year} {2012})},\ \Eprint {https://arxiv.org/abs/https://www.science.org/doi/pdf/10.1126/science.1217069} {https://www.science.org/doi/pdf/10.1126/science.1217069} \BibitemShut {NoStop}%
\bibitem [{\citenamefont {Jordan}\ \emph {et~al.}(2014)\citenamefont {Jordan}, \citenamefont {Lee},\ and\ \citenamefont {Preskill}}]{jordan2014}%
  \BibitemOpen
  \bibfield  {author} {\bibinfo {author} {\bibfnamefont {S.~P.}\ \bibnamefont {Jordan}}, \bibinfo {author} {\bibfnamefont {K.~S.~M.}\ \bibnamefont {Lee}},\ and\ \bibinfo {author} {\bibfnamefont {J.}~\bibnamefont {Preskill}},\ }\href@noop {} {\bibfield  {journal} {\bibinfo  {journal} {arXiv:1404.7115 [hep-th, physics:quant-ph]}\ } (\bibinfo {year} {2014})},\ \Eprint {https://arxiv.org/abs/1404.7115} {arXiv:1404.7115 [hep-th, physics:quant-ph]} \BibitemShut {NoStop}%
\bibitem [{\citenamefont {Kevrekidis}\ and\ \citenamefont {Cuevas-Maraver}(2019)}]{Kevrekidis_2019}%
  \BibitemOpen
  \bibfield  {author} {\bibinfo {author} {\bibfnamefont {P.~G.}\ \bibnamefont {Kevrekidis}}\ and\ \bibinfo {author} {\bibfnamefont {J.}~\bibnamefont {Cuevas-Maraver}},\ }\href@noop {} {\emph {\bibinfo {title} {A dynamical perspective on the $\Phi^4$ model: Past, present and future}}}\ (\bibinfo  {publisher} {Springer},\ \bibinfo {year} {2019})\BibitemShut {NoStop}%
\bibitem [{\citenamefont {Feynman}(2019)}]{Feynman:2019rot}%
  \BibitemOpen
  \bibfield  {author} {\bibinfo {author} {\bibfnamefont {R.~P.}\ \bibnamefont {Feynman}},\ }\href {https://doi.org/10.1201/9780429493331} {\emph {\bibinfo {title} {{Photon-hadron Interactions}}}}\ (\bibinfo  {publisher} {Taylor and Francis},\ \bibinfo {year} {2019})\BibitemShut {NoStop}%
\bibitem [{\citenamefont {Bjorken}\ and\ \citenamefont {Paschos}(1969)}]{Bjorken:1969ja}%
  \BibitemOpen
  \bibfield  {author} {\bibinfo {author} {\bibfnamefont {J.~D.}\ \bibnamefont {Bjorken}}\ and\ \bibinfo {author} {\bibfnamefont {E.~A.}\ \bibnamefont {Paschos}},\ }\href {https://doi.org/10.1103/PhysRev.185.1975} {\bibfield  {journal} {\bibinfo  {journal} {Phys. Rev.}\ }\textbf {\bibinfo {volume} {185}},\ \bibinfo {pages} {1975} (\bibinfo {year} {1969})}\BibitemShut {NoStop}%
\bibitem [{\citenamefont {Barata}\ \emph {et~al.}(2021)\citenamefont {Barata}, \citenamefont {Mueller}, \citenamefont {Tarasov},\ and\ \citenamefont {Venugopalan}}]{barata2021single}%
  \BibitemOpen
  \bibfield  {author} {\bibinfo {author} {\bibfnamefont {J.}~\bibnamefont {Barata}}, \bibinfo {author} {\bibfnamefont {N.}~\bibnamefont {Mueller}}, \bibinfo {author} {\bibfnamefont {A.}~\bibnamefont {Tarasov}},\ and\ \bibinfo {author} {\bibfnamefont {R.}~\bibnamefont {Venugopalan}},\ }\href@noop {} {\bibfield  {journal} {\bibinfo  {journal} {Physical Review A}\ }\textbf {\bibinfo {volume} {103}},\ \bibinfo {pages} {042410} (\bibinfo {year} {2021})}\BibitemShut {NoStop}%
\bibitem [{\citenamefont {Li}\ \emph {et~al.}(2023)\citenamefont {Li}, \citenamefont {Macridin}, \citenamefont {Mrenna},\ and\ \citenamefont {Spentzouris}}]{Li:2022ped}%
  \BibitemOpen
  \bibfield  {author} {\bibinfo {author} {\bibfnamefont {A.~C.~Y.}\ \bibnamefont {Li}}, \bibinfo {author} {\bibfnamefont {A.}~\bibnamefont {Macridin}}, \bibinfo {author} {\bibfnamefont {S.}~\bibnamefont {Mrenna}},\ and\ \bibinfo {author} {\bibfnamefont {P.}~\bibnamefont {Spentzouris}},\ }\href {https://doi.org/10.1103/PhysRevA.107.032603} {\bibfield  {journal} {\bibinfo  {journal} {Phys. Rev. A}\ }\textbf {\bibinfo {volume} {107}},\ \bibinfo {pages} {032603} (\bibinfo {year} {2023})},\ \Eprint {https://arxiv.org/abs/2210.07985} {arXiv:2210.07985 [quant-ph]} \BibitemShut {NoStop}%
\bibitem [{\citenamefont {Kreshchuk}\ \emph {et~al.}(2023)\citenamefont {Kreshchuk}, \citenamefont {Vary},\ and\ \citenamefont {Love}}]{kreshchuk2023simulatingscatteringcompositeparticles}%
  \BibitemOpen
  \bibfield  {author} {\bibinfo {author} {\bibfnamefont {M.}~\bibnamefont {Kreshchuk}}, \bibinfo {author} {\bibfnamefont {J.~P.}\ \bibnamefont {Vary}},\ and\ \bibinfo {author} {\bibfnamefont {P.~J.}\ \bibnamefont {Love}},\ }\href {https://arxiv.org/abs/2310.13742} {\bibinfo {title} {Simulating scattering of composite particles}} (\bibinfo {year} {2023}),\ \Eprint {https://arxiv.org/abs/2310.13742} {arXiv:2310.13742 [quant-ph]} \BibitemShut {NoStop}%
\bibitem [{\citenamefont {Bagherimehrab}\ \emph {et~al.}(2022)\citenamefont {Bagherimehrab}, \citenamefont {Sanders}, \citenamefont {Berry}, \citenamefont {Brennen},\ and\ \citenamefont {Sanders}}]{Bagherimehrab:2021xlp}%
  \BibitemOpen
  \bibfield  {author} {\bibinfo {author} {\bibfnamefont {M.}~\bibnamefont {Bagherimehrab}}, \bibinfo {author} {\bibfnamefont {Y.~R.}\ \bibnamefont {Sanders}}, \bibinfo {author} {\bibfnamefont {D.~W.}\ \bibnamefont {Berry}}, \bibinfo {author} {\bibfnamefont {G.~K.}\ \bibnamefont {Brennen}},\ and\ \bibinfo {author} {\bibfnamefont {B.~C.}\ \bibnamefont {Sanders}},\ }\href {https://doi.org/10.1103/PRXQuantum.3.020364} {\bibfield  {journal} {\bibinfo  {journal} {PRX Quantum}\ }\textbf {\bibinfo {volume} {3}},\ \bibinfo {pages} {020364} (\bibinfo {year} {2022})},\ \Eprint {https://arxiv.org/abs/2110.05708} {arXiv:2110.05708 [quant-ph]} \BibitemShut {NoStop}%
\bibitem [{\citenamefont {Farrelly}\ and\ \citenamefont {Streich}(2020)}]{farrelly2020discretizingquantumfieldtheories}%
  \BibitemOpen
  \bibfield  {author} {\bibinfo {author} {\bibfnamefont {T.}~\bibnamefont {Farrelly}}\ and\ \bibinfo {author} {\bibfnamefont {J.}~\bibnamefont {Streich}},\ }\href {https://arxiv.org/abs/2002.02643} {\bibinfo {title} {Discretizing quantum field theories for quantum simulation}} (\bibinfo {year} {2020}),\ \Eprint {https://arxiv.org/abs/2002.02643} {arXiv:2002.02643 [quant-ph]} \BibitemShut {NoStop}%
\bibitem [{\citenamefont {Klco}\ and\ \citenamefont {Savage}(2020)}]{Klco:2019xro}%
  \BibitemOpen
  \bibfield  {author} {\bibinfo {author} {\bibfnamefont {N.}~\bibnamefont {Klco}}\ and\ \bibinfo {author} {\bibfnamefont {M.~J.}\ \bibnamefont {Savage}},\ }\href {https://doi.org/10.1103/PhysRevA.102.012612} {\bibfield  {journal} {\bibinfo  {journal} {Phys. Rev. A}\ }\textbf {\bibinfo {volume} {102}},\ \bibinfo {pages} {012612} (\bibinfo {year} {2020})},\ \Eprint {https://arxiv.org/abs/1904.10440} {arXiv:1904.10440 [quant-ph]} \BibitemShut {NoStop}%
\bibitem [{\citenamefont {Yeter-Aydeniz}\ \emph {et~al.}(2019)\citenamefont {Yeter-Aydeniz}, \citenamefont {Dumitrescu}, \citenamefont {McCaskey}, \citenamefont {Bennink}, \citenamefont {Pooser},\ and\ \citenamefont {Siopsis}}]{Yeter-Aydeniz:2018mix}%
  \BibitemOpen
  \bibfield  {author} {\bibinfo {author} {\bibfnamefont {K.}~\bibnamefont {Yeter-Aydeniz}}, \bibinfo {author} {\bibfnamefont {E.~F.}\ \bibnamefont {Dumitrescu}}, \bibinfo {author} {\bibfnamefont {A.~J.}\ \bibnamefont {McCaskey}}, \bibinfo {author} {\bibfnamefont {R.~S.}\ \bibnamefont {Bennink}}, \bibinfo {author} {\bibfnamefont {R.~C.}\ \bibnamefont {Pooser}},\ and\ \bibinfo {author} {\bibfnamefont {G.}~\bibnamefont {Siopsis}},\ }\href {https://doi.org/10.1103/PhysRevA.99.032306} {\bibfield  {journal} {\bibinfo  {journal} {Phys. Rev. A}\ }\textbf {\bibinfo {volume} {99}},\ \bibinfo {pages} {032306} (\bibinfo {year} {2019})},\ \Eprint {https://arxiv.org/abs/1811.12332} {arXiv:1811.12332 [quant-ph]} \BibitemShut {NoStop}%
\bibitem [{\citenamefont {Davoudi}\ \emph {et~al.}(2023)\citenamefont {Davoudi}, \citenamefont {Shaw},\ and\ \citenamefont {Stryker}}]{Davoudi_2023}%
  \BibitemOpen
  \bibfield  {author} {\bibinfo {author} {\bibfnamefont {Z.}~\bibnamefont {Davoudi}}, \bibinfo {author} {\bibfnamefont {A.~F.}\ \bibnamefont {Shaw}},\ and\ \bibinfo {author} {\bibfnamefont {J.~R.}\ \bibnamefont {Stryker}},\ }\href {https://doi.org/10.22331/q-2023-12-20-1213} {\bibfield  {journal} {\bibinfo  {journal} {Quantum}\ }\textbf {\bibinfo {volume} {7}},\ \bibinfo {pages} {1213} (\bibinfo {year} {2023})}\BibitemShut {NoStop}%
\bibitem [{\citenamefont {Rhodes}\ \emph {et~al.}(2024)\citenamefont {Rhodes}, \citenamefont {Kreshchuk},\ and\ \citenamefont {Pathak}}]{rhodes2024}%
  \BibitemOpen
  \bibfield  {author} {\bibinfo {author} {\bibfnamefont {M.}~\bibnamefont {Rhodes}}, \bibinfo {author} {\bibfnamefont {M.}~\bibnamefont {Kreshchuk}},\ and\ \bibinfo {author} {\bibfnamefont {S.}~\bibnamefont {Pathak}},\ }\href {https://arxiv.org/abs/2405.10416} {\bibinfo {title} {Exponential improvements in the simulation of lattice gauge theories using near-optimal techniques}} (\bibinfo {year} {2024}),\ \Eprint {https://arxiv.org/abs/2405.10416} {arXiv:2405.10416 [quant-ph]} \BibitemShut {NoStop}%
\bibitem [{\citenamefont {Hariprakash}\ \emph {et~al.}(2024)\citenamefont {Hariprakash}, \citenamefont {Modi}, \citenamefont {Kreshchuk}, \citenamefont {Kane},\ and\ \citenamefont {Bauer}}]{hariprakash2024}%
  \BibitemOpen
  \bibfield  {author} {\bibinfo {author} {\bibfnamefont {S.}~\bibnamefont {Hariprakash}}, \bibinfo {author} {\bibfnamefont {N.~S.}\ \bibnamefont {Modi}}, \bibinfo {author} {\bibfnamefont {M.}~\bibnamefont {Kreshchuk}}, \bibinfo {author} {\bibfnamefont {C.~F.}\ \bibnamefont {Kane}},\ and\ \bibinfo {author} {\bibfnamefont {C.~W.}\ \bibnamefont {Bauer}},\ }\href@noop {} {\bibfield  {journal} {\bibinfo  {journal} {arXiv}\ } (\bibinfo {year} {2024})},\ \Eprint {https://arxiv.org/abs/2312.11637} {2312.11637} \BibitemShut {NoStop}%
\bibitem [{\citenamefont {Bauer}\ \emph {et~al.}(2021)\citenamefont {Bauer}, \citenamefont {Freytsis},\ and\ \citenamefont {Nachman}}]{Bauer:2021gup}%
  \BibitemOpen
  \bibfield  {author} {\bibinfo {author} {\bibfnamefont {C.~W.}\ \bibnamefont {Bauer}}, \bibinfo {author} {\bibfnamefont {M.}~\bibnamefont {Freytsis}},\ and\ \bibinfo {author} {\bibfnamefont {B.}~\bibnamefont {Nachman}},\ }\href {https://doi.org/10.1103/PhysRevLett.127.212001} {\bibfield  {journal} {\bibinfo  {journal} {Phys. Rev. Lett.}\ }\textbf {\bibinfo {volume} {127}},\ \bibinfo {pages} {212001} (\bibinfo {year} {2021})},\ \Eprint {https://arxiv.org/abs/2102.05044} {arXiv:2102.05044 [hep-ph]} \BibitemShut {NoStop}%
\bibitem [{\citenamefont {Lüscher}(1991)}]{Luscher1991}%
  \BibitemOpen
  \bibfield  {author} {\bibinfo {author} {\bibfnamefont {M.}~\bibnamefont {Lüscher}},\ }\href {https://doi.org/https://doi.org/10.1016/0550-3213(91)90366-6} {\bibfield  {journal} {\bibinfo  {journal} {Nuclear Physics B}\ }\textbf {\bibinfo {volume} {354}},\ \bibinfo {pages} {531} (\bibinfo {year} {1991})}\BibitemShut {NoStop}%
\bibitem [{\citenamefont {Hansen}\ and\ \citenamefont {Sharpe}(2019)}]{annurev}%
  \BibitemOpen
  \bibfield  {author} {\bibinfo {author} {\bibfnamefont {M.~T.}\ \bibnamefont {Hansen}}\ and\ \bibinfo {author} {\bibfnamefont {S.~R.}\ \bibnamefont {Sharpe}},\ }\href {https://doi.org/https://doi.org/10.1146/annurev-nucl-101918-023723} {\bibfield  {journal} {\bibinfo  {journal} {Annual Review of Nuclear and Particle Science}\ }\textbf {\bibinfo {volume} {69}},\ \bibinfo {pages} {65} (\bibinfo {year} {2019})}\BibitemShut {NoStop}%
\bibitem [{\citenamefont {Rusetsky}(2019)}]{rusetsky2019particleslattice}%
  \BibitemOpen
  \bibfield  {author} {\bibinfo {author} {\bibfnamefont {A.}~\bibnamefont {Rusetsky}},\ }\href {https://arxiv.org/abs/1911.01253} {\bibinfo {title} {Three particles on the lattice}} (\bibinfo {year} {2019}),\ \Eprint {https://arxiv.org/abs/1911.01253} {arXiv:1911.01253 [hep-lat]} \BibitemShut {NoStop}%
\bibitem [{\citenamefont {Mai}\ \emph {et~al.}(2021)\citenamefont {Mai}, \citenamefont {D\"{o}ring},\ and\ \citenamefont {Rusetsky}}]{Mai2021}%
  \BibitemOpen
  \bibfield  {author} {\bibinfo {author} {\bibfnamefont {M.}~\bibnamefont {Mai}}, \bibinfo {author} {\bibfnamefont {M.}~\bibnamefont {D\"{o}ring}},\ and\ \bibinfo {author} {\bibfnamefont {A.}~\bibnamefont {Rusetsky}},\ }\href {https://doi.org/10.1140/epjs/s11734-021-00146-5} {\bibfield  {journal} {\bibinfo  {journal} {The European Physical Journal Special Topics}\ }\textbf {\bibinfo {volume} {230}},\ \bibinfo {pages} {1623–1643} (\bibinfo {year} {2021})}\BibitemShut {NoStop}%
\bibitem [{\citenamefont {Romero-López}(2023)}]{multihadroninteractionslatticeqcd}%
  \BibitemOpen
  \bibfield  {author} {\bibinfo {author} {\bibfnamefont {F.}~\bibnamefont {Romero-López}},\ }\href {https://arxiv.org/abs/2212.13793} {\bibinfo {title} {Multi-hadron interactions from lattice {QCD}}} (\bibinfo {year} {2023}),\ \Eprint {https://arxiv.org/abs/2212.13793} {arXiv:2212.13793 [hep-lat]} \BibitemShut {NoStop}%
\bibitem [{\citenamefont {Gabai}\ and\ \citenamefont {Yin}(2022)}]{Gabai2022}%
  \BibitemOpen
  \bibfield  {author} {\bibinfo {author} {\bibfnamefont {B.}~\bibnamefont {Gabai}}\ and\ \bibinfo {author} {\bibfnamefont {X.}~\bibnamefont {Yin}},\ }\bibfield  {journal} {\bibinfo  {journal} {Journal of High Energy Physics}\ }\textbf {\bibinfo {volume} {2022}},\ \href {https://doi.org/10.1007/jhep10(2022)168} {10.1007/jhep10(2022)168} (\bibinfo {year} {2022})\BibitemShut {NoStop}%
\bibitem [{\citenamefont {Bajnok}\ and\ \citenamefont {Lajer}(2016)}]{Bajnok2016}%
  \BibitemOpen
  \bibfield  {author} {\bibinfo {author} {\bibfnamefont {Z.}~\bibnamefont {Bajnok}}\ and\ \bibinfo {author} {\bibfnamefont {M.}~\bibnamefont {Lajer}},\ }\bibfield  {journal} {\bibinfo  {journal} {Journal of High Energy Physics}\ }\textbf {\bibinfo {volume} {2016}},\ \href {https://doi.org/10.1007/jhep10(2016)050} {10.1007/jhep10(2016)050} (\bibinfo {year} {2016})\BibitemShut {NoStop}%
\bibitem [{\citenamefont {James}\ \emph {et~al.}(2018)\citenamefont {James}, \citenamefont {Konik}, \citenamefont {Lecheminant}, \citenamefont {Robinson},\ and\ \citenamefont {Tsvelik}}]{James_2018}%
  \BibitemOpen
  \bibfield  {author} {\bibinfo {author} {\bibfnamefont {A.~J.~A.}\ \bibnamefont {James}}, \bibinfo {author} {\bibfnamefont {R.~M.}\ \bibnamefont {Konik}}, \bibinfo {author} {\bibfnamefont {P.}~\bibnamefont {Lecheminant}}, \bibinfo {author} {\bibfnamefont {N.~J.}\ \bibnamefont {Robinson}},\ and\ \bibinfo {author} {\bibfnamefont {A.~M.}\ \bibnamefont {Tsvelik}},\ }\href {https://doi.org/10.1088/1361-6633/aa91ea} {\bibfield  {journal} {\bibinfo  {journal} {Reports on Progress in Physics}\ }\textbf {\bibinfo {volume} {81}},\ \bibinfo {pages} {046002} (\bibinfo {year} {2018})}\BibitemShut {NoStop}%
\bibitem [{\citenamefont {Luu}\ and\ \citenamefont {Savage}(2011)}]{PhysRevD.83.114508}%
  \BibitemOpen
  \bibfield  {author} {\bibinfo {author} {\bibfnamefont {T.}~\bibnamefont {Luu}}\ and\ \bibinfo {author} {\bibfnamefont {M.~J.}\ \bibnamefont {Savage}},\ }\href {https://doi.org/10.1103/PhysRevD.83.114508} {\bibfield  {journal} {\bibinfo  {journal} {Phys. Rev. D}\ }\textbf {\bibinfo {volume} {83}},\ \bibinfo {pages} {114508} (\bibinfo {year} {2011})}\BibitemShut {NoStop}%
\bibitem [{\citenamefont {Dudek}\ \emph {et~al.}(2011)\citenamefont {Dudek}, \citenamefont {Edwards}, \citenamefont {Peardon}, \citenamefont {Richards},\ and\ \citenamefont {Thomas}}]{PhysRevD.83.071504}%
  \BibitemOpen
  \bibfield  {author} {\bibinfo {author} {\bibfnamefont {J.~J.}\ \bibnamefont {Dudek}}, \bibinfo {author} {\bibfnamefont {R.~G.}\ \bibnamefont {Edwards}}, \bibinfo {author} {\bibfnamefont {M.~J.}\ \bibnamefont {Peardon}}, \bibinfo {author} {\bibfnamefont {D.~G.}\ \bibnamefont {Richards}},\ and\ \bibinfo {author} {\bibfnamefont {C.~E.}\ \bibnamefont {Thomas}} (\bibinfo {collaboration} {for the Hadron Spectrum Collaboration}),\ }\href {https://doi.org/10.1103/PhysRevD.83.071504} {\bibfield  {journal} {\bibinfo  {journal} {Phys. Rev. D}\ }\textbf {\bibinfo {volume} {83}},\ \bibinfo {pages} {071504} (\bibinfo {year} {2011})}\BibitemShut {NoStop}%
\bibitem [{\citenamefont {Guo}\ \emph {et~al.}(2018)\citenamefont {Guo}, \citenamefont {Alexandru}, \citenamefont {Molina}, \citenamefont {Mai},\ and\ \citenamefont {D\"oring}}]{PhysRevD.98.014507}%
  \BibitemOpen
  \bibfield  {author} {\bibinfo {author} {\bibfnamefont {D.}~\bibnamefont {Guo}}, \bibinfo {author} {\bibfnamefont {A.}~\bibnamefont {Alexandru}}, \bibinfo {author} {\bibfnamefont {R.}~\bibnamefont {Molina}}, \bibinfo {author} {\bibfnamefont {M.}~\bibnamefont {Mai}},\ and\ \bibinfo {author} {\bibfnamefont {M.}~\bibnamefont {D\"oring}},\ }\href {https://doi.org/10.1103/PhysRevD.98.014507} {\bibfield  {journal} {\bibinfo  {journal} {Phys. Rev. D}\ }\textbf {\bibinfo {volume} {98}},\ \bibinfo {pages} {014507} (\bibinfo {year} {2018})}\BibitemShut {NoStop}%
\bibitem [{\citenamefont {Garofalo}\ \emph {et~al.}(2021)\citenamefont {Garofalo}, \citenamefont {Romero-López}, \citenamefont {Rusetsky},\ and\ \citenamefont {Urbach}}]{Garofalo_2021}%
  \BibitemOpen
  \bibfield  {author} {\bibinfo {author} {\bibfnamefont {M.}~\bibnamefont {Garofalo}}, \bibinfo {author} {\bibfnamefont {F.}~\bibnamefont {Romero-López}}, \bibinfo {author} {\bibfnamefont {A.}~\bibnamefont {Rusetsky}},\ and\ \bibinfo {author} {\bibfnamefont {C.}~\bibnamefont {Urbach}},\ }\bibfield  {journal} {\bibinfo  {journal} {The European Physical Journal C}\ }\textbf {\bibinfo {volume} {81}},\ \href {https://doi.org/10.1140/epjc/s10052-021-09830-1} {10.1140/epjc/s10052-021-09830-1} (\bibinfo {year} {2021})\BibitemShut {NoStop}%
\bibitem [{\citenamefont {Yurov}\ and\ \citenamefont {Zamolodchikov}(1990)}]{yurov1990truncated}%
  \BibitemOpen
  \bibfield  {author} {\bibinfo {author} {\bibfnamefont {V.~P.}\ \bibnamefont {Yurov}}\ and\ \bibinfo {author} {\bibfnamefont {A.~B.}\ \bibnamefont {Zamolodchikov}},\ }\href {https://doi.org/10.1142/S0217751X9000218X} {\bibfield  {journal} {\bibinfo  {journal} {Int. J. Mod. Phys. A}\ }\textbf {\bibinfo {volume} {05}},\ \bibinfo {pages} {3221} (\bibinfo {year} {1990})}\BibitemShut {NoStop}%
\bibitem [{\citenamefont {Yurov}\ and\ \citenamefont {Zamolodchikov}(1991)}]{yurov1991truncated}%
  \BibitemOpen
  \bibfield  {author} {\bibinfo {author} {\bibfnamefont {V.~P.}\ \bibnamefont {Yurov}}\ and\ \bibinfo {author} {\bibfnamefont {A.~B.}\ \bibnamefont {Zamolodchikov}},\ }\href {https://doi.org/10.1142/S0217751X91002161} {\bibfield  {journal} {\bibinfo  {journal} {Int. J. Mod. Phys. A}\ }\textbf {\bibinfo {volume} {06}},\ \bibinfo {pages} {4557} (\bibinfo {year} {1991})}\BibitemShut {NoStop}%
\bibitem [{\citenamefont {Hogervorst}\ \emph {et~al.}(2015)\citenamefont {Hogervorst}, \citenamefont {Rychkov},\ and\ \citenamefont {van Rees}}]{PhysRevD.91.025005}%
  \BibitemOpen
  \bibfield  {author} {\bibinfo {author} {\bibfnamefont {M.}~\bibnamefont {Hogervorst}}, \bibinfo {author} {\bibfnamefont {S.}~\bibnamefont {Rychkov}},\ and\ \bibinfo {author} {\bibfnamefont {B.~C.}\ \bibnamefont {van Rees}},\ }\href {https://doi.org/10.1103/PhysRevD.91.025005} {\bibfield  {journal} {\bibinfo  {journal} {Phys. Rev. D}\ }\textbf {\bibinfo {volume} {91}},\ \bibinfo {pages} {025005} (\bibinfo {year} {2015})}\BibitemShut {NoStop}%
\bibitem [{\citenamefont {Cohen}\ \emph {et~al.}(2022)\citenamefont {Cohen}, \citenamefont {Farnsworth}, \citenamefont {Houtz},\ and\ \citenamefont {Luty}}]{10.21468/SciPostPhys.13.2.011}%
  \BibitemOpen
  \bibfield  {author} {\bibinfo {author} {\bibfnamefont {T.}~\bibnamefont {Cohen}}, \bibinfo {author} {\bibfnamefont {K.}~\bibnamefont {Farnsworth}}, \bibinfo {author} {\bibfnamefont {R.}~\bibnamefont {Houtz}},\ and\ \bibinfo {author} {\bibfnamefont {M.~A.}\ \bibnamefont {Luty}},\ }\href {https://doi.org/10.21468/SciPostPhys.13.2.011} {\bibfield  {journal} {\bibinfo  {journal} {SciPost Phys.}\ }\textbf {\bibinfo {volume} {13}},\ \bibinfo {pages} {011} (\bibinfo {year} {2022})}\BibitemShut {NoStop}%
\bibitem [{\citenamefont {Chen}\ \emph {et~al.}(2022)\citenamefont {Chen}, \citenamefont {Fitzpatrick},\ and\ \citenamefont {Karateev}}]{chen2022}%
  \BibitemOpen
  \bibfield  {author} {\bibinfo {author} {\bibfnamefont {H.}~\bibnamefont {Chen}}, \bibinfo {author} {\bibfnamefont {A.~L.}\ \bibnamefont {Fitzpatrick}},\ and\ \bibinfo {author} {\bibfnamefont {D.}~\bibnamefont {Karateev}},\ }\bibfield  {journal} {\bibinfo  {journal} {Journal of High Energy Physics}\ }\textbf {\bibinfo {volume} {2022}},\ \href {https://doi.org/10.1007/jhep04(2022)109} {10.1007/jhep04(2022)109} (\bibinfo {year} {2022})\BibitemShut {NoStop}%
\bibitem [{\citenamefont {Elias-Mir\'o}\ and\ \citenamefont {Hardy}(2020)}]{PhysRevD.102.065001}%
  \BibitemOpen
  \bibfield  {author} {\bibinfo {author} {\bibfnamefont {J.}~\bibnamefont {Elias-Mir\'o}}\ and\ \bibinfo {author} {\bibfnamefont {E.}~\bibnamefont {Hardy}},\ }\href {https://doi.org/10.1103/PhysRevD.102.065001} {\bibfield  {journal} {\bibinfo  {journal} {Phys. Rev. D}\ }\textbf {\bibinfo {volume} {102}},\ \bibinfo {pages} {065001} (\bibinfo {year} {2020})}\BibitemShut {NoStop}%
\bibitem [{\citenamefont {Liu}\ \emph {et~al.}(2020)\citenamefont {Liu}, \citenamefont {Meltzer}, \citenamefont {Poland},\ and\ \citenamefont {Simmons-Duffin}}]{Liu2020}%
  \BibitemOpen
  \bibfield  {author} {\bibinfo {author} {\bibfnamefont {J.}~\bibnamefont {Liu}}, \bibinfo {author} {\bibfnamefont {D.}~\bibnamefont {Meltzer}}, \bibinfo {author} {\bibfnamefont {D.}~\bibnamefont {Poland}},\ and\ \bibinfo {author} {\bibfnamefont {D.}~\bibnamefont {Simmons-Duffin}},\ }\bibfield  {journal} {\bibinfo  {journal} {Journal of High Energy Physics}\ }\textbf {\bibinfo {volume} {2020}},\ \href {https://doi.org/10.1007/jhep09(2020)115} {10.1007/jhep09(2020)115} (\bibinfo {year} {2020})\BibitemShut {NoStop}%
\bibitem [{\citenamefont {Jarrell}\ and\ \citenamefont {Gubernatis}(1996)}]{Jarrell1996}%
  \BibitemOpen
  \bibfield  {author} {\bibinfo {author} {\bibfnamefont {M.}~\bibnamefont {Jarrell}}\ and\ \bibinfo {author} {\bibfnamefont {J.}~\bibnamefont {Gubernatis}},\ }\href {https://doi.org/https://doi.org/10.1016/0370-1573(95)00074-7} {\bibfield  {journal} {\bibinfo  {journal} {Physics Reports}\ }\textbf {\bibinfo {volume} {269}},\ \bibinfo {pages} {133} (\bibinfo {year} {1996})}\BibitemShut {NoStop}%
\bibitem [{\citenamefont {Sandvik}(1998)}]{Sandvik1998}%
  \BibitemOpen
  \bibfield  {author} {\bibinfo {author} {\bibfnamefont {A.~W.}\ \bibnamefont {Sandvik}},\ }\href {https://doi.org/10.1103/PhysRevB.57.10287} {\bibfield  {journal} {\bibinfo  {journal} {Phys. Rev. B}\ }\textbf {\bibinfo {volume} {57}},\ \bibinfo {pages} {10287} (\bibinfo {year} {1998})}\BibitemShut {NoStop}%
\bibitem [{\citenamefont {Beach}(2004)}]{beach2004}%
  \BibitemOpen
  \bibfield  {author} {\bibinfo {author} {\bibfnamefont {K.~S.~D.}\ \bibnamefont {Beach}},\ }\href {https://arxiv.org/abs/cond-mat/0403055} {\bibinfo {title} {Identifying the maximum entropy method as a special limit of stochastic analytic continuation}} (\bibinfo {year} {2004}),\ \Eprint {https://arxiv.org/abs/cond-mat/0403055} {arXiv:cond-mat/0403055 [cond-mat.str-el]} \BibitemShut {NoStop}%
\bibitem [{\citenamefont {Gustafson}\ \emph {et~al.}(2021)\citenamefont {Gustafson}, \citenamefont {Zhu}, \citenamefont {Dreher}, \citenamefont {Linke},\ and\ \citenamefont {Meurice}}]{PhysRevD.104.054507}%
  \BibitemOpen
  \bibfield  {author} {\bibinfo {author} {\bibfnamefont {E.}~\bibnamefont {Gustafson}}, \bibinfo {author} {\bibfnamefont {Y.}~\bibnamefont {Zhu}}, \bibinfo {author} {\bibfnamefont {P.}~\bibnamefont {Dreher}}, \bibinfo {author} {\bibfnamefont {N.~M.}\ \bibnamefont {Linke}},\ and\ \bibinfo {author} {\bibfnamefont {Y.}~\bibnamefont {Meurice}},\ }\href {https://doi.org/10.1103/PhysRevD.104.054507} {\bibfield  {journal} {\bibinfo  {journal} {Phys. Rev. D}\ }\textbf {\bibinfo {volume} {104}},\ \bibinfo {pages} {054507} (\bibinfo {year} {2021})}\BibitemShut {NoStop}%
\bibitem [{\citenamefont {Carrillo}\ \emph {et~al.}(2024)\citenamefont {Carrillo}, \citenamefont {A.},\ and\ \citenamefont {Sturzu}}]{carrillo2024inclusivereactionsfiniteminkowski}%
  \BibitemOpen
  \bibfield  {author} {\bibinfo {author} {\bibfnamefont {M.~A.}\ \bibnamefont {Carrillo}}, \bibinfo {author} {\bibfnamefont {R.}~\bibnamefont {A.}},\ and\ \bibinfo {author} {\bibfnamefont {A.~M.}\ \bibnamefont {Sturzu}},\ }\href {https://arxiv.org/abs/2406.06877} {\bibinfo {title} {Inclusive reactions from finite minkowski spacetime correlation functions}} (\bibinfo {year} {2024}),\ \Eprint {https://arxiv.org/abs/2406.06877} {arXiv:2406.06877 [hep-lat]} \BibitemShut {NoStop}%
\bibitem [{\citenamefont {Briceño}\ \emph {et~al.}(2021)\citenamefont {Briceño}, \citenamefont {Carrillo}, \citenamefont {Guerrero}, \citenamefont {Hansen},\ and\ \citenamefont {Sturzu}}]{briceño2021accessingscatteringamplitudesusing}%
  \BibitemOpen
  \bibfield  {author} {\bibinfo {author} {\bibfnamefont {R.~A.}\ \bibnamefont {Briceño}}, \bibinfo {author} {\bibfnamefont {M.~A.}\ \bibnamefont {Carrillo}}, \bibinfo {author} {\bibfnamefont {J.~V.}\ \bibnamefont {Guerrero}}, \bibinfo {author} {\bibfnamefont {M.~T.}\ \bibnamefont {Hansen}},\ and\ \bibinfo {author} {\bibfnamefont {A.~M.}\ \bibnamefont {Sturzu}},\ }\href {https://arxiv.org/abs/2112.01968} {\bibinfo {title} {Accessing scattering amplitudes using quantum computers}} (\bibinfo {year} {2021}),\ \Eprint {https://arxiv.org/abs/2112.01968} {arXiv:2112.01968 [hep-lat]} \BibitemShut {NoStop}%
\bibitem [{\citenamefont {Mukhopadhyay}\ \emph {et~al.}(2023)\citenamefont {Mukhopadhyay}, \citenamefont {Wiebe},\ and\ \citenamefont {Zhang}}]{2022_MWZ}%
  \BibitemOpen
  \bibfield  {author} {\bibinfo {author} {\bibfnamefont {P.}~\bibnamefont {Mukhopadhyay}}, \bibinfo {author} {\bibfnamefont {N.}~\bibnamefont {Wiebe}},\ and\ \bibinfo {author} {\bibfnamefont {H.~T.}\ \bibnamefont {Zhang}},\ }\href@noop {} {\bibfield  {journal} {\bibinfo  {journal} {npj Quantum Information}\ }\textbf {\bibinfo {volume} {9}} (\bibinfo {year} {2023})}\BibitemShut {NoStop}%
\bibitem [{\citenamefont {Mukhopadhyay}\ \emph {et~al.}(2024)\citenamefont {Mukhopadhyay}, \citenamefont {Stetina},\ and\ \citenamefont {Wiebe}}]{2023_MSW}%
  \BibitemOpen
  \bibfield  {author} {\bibinfo {author} {\bibfnamefont {P.}~\bibnamefont {Mukhopadhyay}}, \bibinfo {author} {\bibfnamefont {T.~F.}\ \bibnamefont {Stetina}},\ and\ \bibinfo {author} {\bibfnamefont {N.}~\bibnamefont {Wiebe}},\ }\href@noop {} {\bibfield  {journal} {\bibinfo  {journal} {PRX Quantum}\ }\textbf {\bibinfo {volume} {5}},\ \bibinfo {pages} {010345} (\bibinfo {year} {2024})}\BibitemShut {NoStop}%
\bibitem [{\citenamefont {Reiher}\ \emph {et~al.}(2017)\citenamefont {Reiher}, \citenamefont {Wiebe}, \citenamefont {Svore}, \citenamefont {Wecker},\ and\ \citenamefont {Troyer}}]{reiher2017elucidating}%
  \BibitemOpen
  \bibfield  {author} {\bibinfo {author} {\bibfnamefont {M.}~\bibnamefont {Reiher}}, \bibinfo {author} {\bibfnamefont {N.}~\bibnamefont {Wiebe}}, \bibinfo {author} {\bibfnamefont {K.~M.}\ \bibnamefont {Svore}}, \bibinfo {author} {\bibfnamefont {D.}~\bibnamefont {Wecker}},\ and\ \bibinfo {author} {\bibfnamefont {M.}~\bibnamefont {Troyer}},\ }\href@noop {} {\bibfield  {journal} {\bibinfo  {journal} {Proceedings of the national academy of sciences}\ }\textbf {\bibinfo {volume} {114}},\ \bibinfo {pages} {7555} (\bibinfo {year} {2017})}\BibitemShut {NoStop}%
\bibitem [{\citenamefont {Babbush}\ \emph {et~al.}(2018{\natexlab{a}})\citenamefont {Babbush}, \citenamefont {Wiebe}, \citenamefont {McClean}, \citenamefont {McClain}, \citenamefont {Neven},\ and\ \citenamefont {Chan}}]{babbush2018low}%
  \BibitemOpen
  \bibfield  {author} {\bibinfo {author} {\bibfnamefont {R.}~\bibnamefont {Babbush}}, \bibinfo {author} {\bibfnamefont {N.}~\bibnamefont {Wiebe}}, \bibinfo {author} {\bibfnamefont {J.}~\bibnamefont {McClean}}, \bibinfo {author} {\bibfnamefont {J.}~\bibnamefont {McClain}}, \bibinfo {author} {\bibfnamefont {H.}~\bibnamefont {Neven}},\ and\ \bibinfo {author} {\bibfnamefont {G.~K.-L.}\ \bibnamefont {Chan}},\ }\href@noop {} {\bibfield  {journal} {\bibinfo  {journal} {Physical Review X}\ }\textbf {\bibinfo {volume} {8}},\ \bibinfo {pages} {011044} (\bibinfo {year} {2018}{\natexlab{a}})}\BibitemShut {NoStop}%
\bibitem [{\citenamefont {Childs}\ \emph {et~al.}(2018{\natexlab{a}})\citenamefont {Childs}, \citenamefont {Maslov}, \citenamefont {Nam}, \citenamefont {Ross},\ and\ \citenamefont {Su}}]{childs2018toward}%
  \BibitemOpen
  \bibfield  {author} {\bibinfo {author} {\bibfnamefont {A.~M.}\ \bibnamefont {Childs}}, \bibinfo {author} {\bibfnamefont {D.}~\bibnamefont {Maslov}}, \bibinfo {author} {\bibfnamefont {Y.}~\bibnamefont {Nam}}, \bibinfo {author} {\bibfnamefont {N.~J.}\ \bibnamefont {Ross}},\ and\ \bibinfo {author} {\bibfnamefont {Y.}~\bibnamefont {Su}},\ }\href@noop {} {\bibfield  {journal} {\bibinfo  {journal} {Proceedings of the National Academy of Sciences}\ }\textbf {\bibinfo {volume} {115}},\ \bibinfo {pages} {9456} (\bibinfo {year} {2018}{\natexlab{a}})}\BibitemShut {NoStop}%
\bibitem [{\citenamefont {Giles}\ and\ \citenamefont {Selinger}(2013)}]{giles2013exact}%
  \BibitemOpen
  \bibfield  {author} {\bibinfo {author} {\bibfnamefont {B.}~\bibnamefont {Giles}}\ and\ \bibinfo {author} {\bibfnamefont {P.}~\bibnamefont {Selinger}},\ }\href@noop {} {\bibfield  {journal} {\bibinfo  {journal} {Physical Review A—Atomic, Molecular, and Optical Physics}\ }\textbf {\bibinfo {volume} {87}},\ \bibinfo {pages} {032332} (\bibinfo {year} {2013})}\BibitemShut {NoStop}%
\bibitem [{\citenamefont {Kliuchnikov}\ \emph {et~al.}(2015)\citenamefont {Kliuchnikov}, \citenamefont {Maslov},\ and\ \citenamefont {Mosca}}]{2015_KMM}%
  \BibitemOpen
  \bibfield  {author} {\bibinfo {author} {\bibfnamefont {V.}~\bibnamefont {Kliuchnikov}}, \bibinfo {author} {\bibfnamefont {D.}~\bibnamefont {Maslov}},\ and\ \bibinfo {author} {\bibfnamefont {M.}~\bibnamefont {Mosca}},\ }\href@noop {} {\bibfield  {journal} {\bibinfo  {journal} {IEEE Transactions on Computers}\ }\textbf {\bibinfo {volume} {65}},\ \bibinfo {pages} {161} (\bibinfo {year} {2015})}\BibitemShut {NoStop}%
\bibitem [{\citenamefont {Ross}\ and\ \citenamefont {Selinger}(2016)}]{2016_RS}%
  \BibitemOpen
  \bibfield  {author} {\bibinfo {author} {\bibfnamefont {N.~J.}\ \bibnamefont {Ross}}\ and\ \bibinfo {author} {\bibfnamefont {P.}~\bibnamefont {Selinger}},\ }\href@noop {} {\bibfield  {journal} {\bibinfo  {journal} {Quantum Inf. Comput.}\ }\textbf {\bibinfo {volume} {16}},\ \bibinfo {pages} {901} (\bibinfo {year} {2016})}\BibitemShut {NoStop}%
\bibitem [{\citenamefont {Mosca}\ and\ \citenamefont {Mukhopadhyay}(2021)}]{2021_MM}%
  \BibitemOpen
  \bibfield  {author} {\bibinfo {author} {\bibfnamefont {M.}~\bibnamefont {Mosca}}\ and\ \bibinfo {author} {\bibfnamefont {P.}~\bibnamefont {Mukhopadhyay}},\ }\href@noop {} {\bibfield  {journal} {\bibinfo  {journal} {Quantum Science and Technology}\ }\textbf {\bibinfo {volume} {7}},\ \bibinfo {pages} {015003} (\bibinfo {year} {2021})}\BibitemShut {NoStop}%
\bibitem [{\citenamefont {Gheorghiu}\ \emph {et~al.}(2022{\natexlab{a}})\citenamefont {Gheorghiu}, \citenamefont {Mosca},\ and\ \citenamefont {Mukhopadhyay}}]{gheorghiu2022quasi}%
  \BibitemOpen
  \bibfield  {author} {\bibinfo {author} {\bibfnamefont {V.}~\bibnamefont {Gheorghiu}}, \bibinfo {author} {\bibfnamefont {M.}~\bibnamefont {Mosca}},\ and\ \bibinfo {author} {\bibfnamefont {P.}~\bibnamefont {Mukhopadhyay}},\ }\href@noop {} {\bibfield  {journal} {\bibinfo  {journal} {npj Quantum Information}\ }\textbf {\bibinfo {volume} {8}},\ \bibinfo {pages} {110} (\bibinfo {year} {2022}{\natexlab{a}})}\BibitemShut {NoStop}%
\bibitem [{\citenamefont {Gheorghiu}\ \emph {et~al.}(2022{\natexlab{b}})\citenamefont {Gheorghiu}, \citenamefont {Mosca},\ and\ \citenamefont {Mukhopadhyay}}]{2022_GMM}%
  \BibitemOpen
  \bibfield  {author} {\bibinfo {author} {\bibfnamefont {V.}~\bibnamefont {Gheorghiu}}, \bibinfo {author} {\bibfnamefont {M.}~\bibnamefont {Mosca}},\ and\ \bibinfo {author} {\bibfnamefont {P.}~\bibnamefont {Mukhopadhyay}},\ }\href@noop {} {\bibfield  {journal} {\bibinfo  {journal} {npj Quantum Information}\ }\textbf {\bibinfo {volume} {8}},\ \bibinfo {pages} {1} (\bibinfo {year} {2022}{\natexlab{b}})}\BibitemShut {NoStop}%
\bibitem [{\citenamefont {Mukhopadhyay}(2024{\natexlab{a}})}]{mukhopadhyay2024synthesis}%
  \BibitemOpen
  \bibfield  {author} {\bibinfo {author} {\bibfnamefont {P.}~\bibnamefont {Mukhopadhyay}},\ }\href@noop {} {\bibfield  {journal} {\bibinfo  {journal} {Physical Review A}\ }\textbf {\bibinfo {volume} {109}},\ \bibinfo {pages} {052619} (\bibinfo {year} {2024}{\natexlab{a}})}\BibitemShut {NoStop}%
\bibitem [{\citenamefont {Mukhopadhyay}(2024{\natexlab{b}})}]{2024_Mcs}%
  \BibitemOpen
  \bibfield  {author} {\bibinfo {author} {\bibfnamefont {P.}~\bibnamefont {Mukhopadhyay}},\ }\href@noop {} {\bibfield  {journal} {\bibinfo  {journal} {Scientific Reports}\ }\textbf {\bibinfo {volume} {14}},\ \bibinfo {pages} {13916} (\bibinfo {year} {2024}{\natexlab{b}})}\BibitemShut {NoStop}%
\bibitem [{\citenamefont {Mukhopadhyay}(2024{\natexlab{c}})}]{2024_Mtof}%
  \BibitemOpen
  \bibfield  {author} {\bibinfo {author} {\bibfnamefont {P.}~\bibnamefont {Mukhopadhyay}},\ }\href@noop {} {\bibfield  {journal} {\bibinfo  {journal} {arXiv preprint arXiv:2401.08950}\ } (\bibinfo {year} {2024}{\natexlab{c}})}\BibitemShut {NoStop}%
\bibitem [{\citenamefont {Fowler}\ \emph {et~al.}(2012{\natexlab{a}})\citenamefont {Fowler}, \citenamefont {Mariantoni}, \citenamefont {Martinis},\ and\ \citenamefont {Cleland}}]{PhysRevA.86.032324}%
  \BibitemOpen
  \bibfield  {author} {\bibinfo {author} {\bibfnamefont {A.~G.}\ \bibnamefont {Fowler}}, \bibinfo {author} {\bibfnamefont {M.}~\bibnamefont {Mariantoni}}, \bibinfo {author} {\bibfnamefont {J.~M.}\ \bibnamefont {Martinis}},\ and\ \bibinfo {author} {\bibfnamefont {A.~N.}\ \bibnamefont {Cleland}},\ }\href {https://doi.org/10.1103/PhysRevA.86.032324} {\bibfield  {journal} {\bibinfo  {journal} {Phys. Rev. A}\ }\textbf {\bibinfo {volume} {86}},\ \bibinfo {pages} {032324} (\bibinfo {year} {2012}{\natexlab{a}})}\BibitemShut {NoStop}%
\bibitem [{\citenamefont {Fowler}\ and\ \citenamefont {Gidney}(2019)}]{fowler2019}%
  \BibitemOpen
  \bibfield  {author} {\bibinfo {author} {\bibfnamefont {A.~G.}\ \bibnamefont {Fowler}}\ and\ \bibinfo {author} {\bibfnamefont {C.}~\bibnamefont {Gidney}},\ }\href {https://arxiv.org/abs/1808.06709} {\bibinfo {title} {Low overhead quantum computation using lattice surgery}} (\bibinfo {year} {2019}),\ \Eprint {https://arxiv.org/abs/1808.06709} {arXiv:1808.06709 [quant-ph]} \BibitemShut {NoStop}%
\bibitem [{\citenamefont {Litinski}(2019)}]{Litinski2019}%
  \BibitemOpen
  \bibfield  {author} {\bibinfo {author} {\bibfnamefont {D.}~\bibnamefont {Litinski}},\ }\href {https://doi.org/10.22331/q-2019-12-02-205} {\bibfield  {journal} {\bibinfo  {journal} {{Quantum}}\ }\textbf {\bibinfo {volume} {3}},\ \bibinfo {pages} {205} (\bibinfo {year} {2019})}\BibitemShut {NoStop}%
\bibitem [{\citenamefont {Elias-Mir\'o}\ \emph {et~al.}(2017)\citenamefont {Elias-Mir\'o}, \citenamefont {Rychkov},\ and\ \citenamefont {Vitale}}]{PhysRevD.96.065024}%
  \BibitemOpen
  \bibfield  {author} {\bibinfo {author} {\bibfnamefont {J.}~\bibnamefont {Elias-Mir\'o}}, \bibinfo {author} {\bibfnamefont {S.}~\bibnamefont {Rychkov}},\ and\ \bibinfo {author} {\bibfnamefont {L.~G.}\ \bibnamefont {Vitale}},\ }\href {https://doi.org/10.1103/PhysRevD.96.065024} {\bibfield  {journal} {\bibinfo  {journal} {Phys. Rev. D}\ }\textbf {\bibinfo {volume} {96}},\ \bibinfo {pages} {065024} (\bibinfo {year} {2017})}\BibitemShut {NoStop}%
\bibitem [{\citenamefont {collaboration}\ \emph {et~al.}(2009)\citenamefont {collaboration}, \citenamefont {Blossier}, \citenamefont {Morte}, \citenamefont {Hippel}, \citenamefont {Mendes},\ and\ \citenamefont {Sommer}}]{collaboration2009}%
  \BibitemOpen
  \bibfield  {author} {\bibinfo {author} {\bibfnamefont {A.}~\bibnamefont {collaboration}}, \bibinfo {author} {\bibfnamefont {B.}~\bibnamefont {Blossier}}, \bibinfo {author} {\bibfnamefont {M.~D.}\ \bibnamefont {Morte}}, \bibinfo {author} {\bibfnamefont {G.~v.}\ \bibnamefont {Hippel}}, \bibinfo {author} {\bibfnamefont {T.}~\bibnamefont {Mendes}},\ and\ \bibinfo {author} {\bibfnamefont {R.}~\bibnamefont {Sommer}},\ }\href {https://doi.org/10.1088/1126-6708/2009/04/094} {\bibfield  {journal} {\bibinfo  {journal} {Journal of High Energy Physics}\ }\textbf {\bibinfo {volume} {2009}},\ \bibinfo {pages} {094–094} (\bibinfo {year} {2009})}\BibitemShut {NoStop}%
\bibitem [{\citenamefont {Brice\~no}\ \emph {et~al.}(2018)\citenamefont {Brice\~no}, \citenamefont {Dudek},\ and\ \citenamefont {Young}}]{2018reviewLQCDScattering}%
  \BibitemOpen
  \bibfield  {author} {\bibinfo {author} {\bibfnamefont {R.~A.}\ \bibnamefont {Brice\~no}}, \bibinfo {author} {\bibfnamefont {J.~J.}\ \bibnamefont {Dudek}},\ and\ \bibinfo {author} {\bibfnamefont {R.~D.}\ \bibnamefont {Young}},\ }\href {https://doi.org/10.1103/RevModPhys.90.025001} {\bibfield  {journal} {\bibinfo  {journal} {Rev. Mod. Phys.}\ }\textbf {\bibinfo {volume} {90}},\ \bibinfo {pages} {025001} (\bibinfo {year} {2018})}\BibitemShut {NoStop}%
\bibitem [{\citenamefont {Mattuck}\ and\ \citenamefont {Johansson}(1968)}]{mattuck1968quantum}%
  \BibitemOpen
  \bibfield  {author} {\bibinfo {author} {\bibfnamefont {R.~D.}\ \bibnamefont {Mattuck}}\ and\ \bibinfo {author} {\bibfnamefont {B.}~\bibnamefont {Johansson}},\ }\href@noop {} {\bibfield  {journal} {\bibinfo  {journal} {Advances in Physics}\ }\textbf {\bibinfo {volume} {17}},\ \bibinfo {pages} {509} (\bibinfo {year} {1968})}\BibitemShut {NoStop}%
\bibitem [{\citenamefont {Espinosa}\ \emph {et~al.}(1992)\citenamefont {Espinosa}, \citenamefont {Quiros},\ and\ \citenamefont {Zwirner}}]{espinosa1992phase}%
  \BibitemOpen
  \bibfield  {author} {\bibinfo {author} {\bibfnamefont {J.}~\bibnamefont {Espinosa}}, \bibinfo {author} {\bibfnamefont {M.}~\bibnamefont {Quiros}},\ and\ \bibinfo {author} {\bibfnamefont {F.}~\bibnamefont {Zwirner}},\ }\href@noop {} {\bibfield  {journal} {\bibinfo  {journal} {Physics Letters B}\ }\textbf {\bibinfo {volume} {291}},\ \bibinfo {pages} {115} (\bibinfo {year} {1992})}\BibitemShut {NoStop}%
\bibitem [{\citenamefont {Hagan}\ and\ \citenamefont {Wiebe}(2023)}]{hagan2023composite}%
  \BibitemOpen
  \bibfield  {author} {\bibinfo {author} {\bibfnamefont {M.}~\bibnamefont {Hagan}}\ and\ \bibinfo {author} {\bibfnamefont {N.}~\bibnamefont {Wiebe}},\ }\href@noop {} {\bibfield  {journal} {\bibinfo  {journal} {Quantum}\ }\textbf {\bibinfo {volume} {7}},\ \bibinfo {pages} {1181} (\bibinfo {year} {2023})}\BibitemShut {NoStop}%
\bibitem [{\citenamefont {Rajput}\ \emph {et~al.}(2022)\citenamefont {Rajput}, \citenamefont {Roggero},\ and\ \citenamefont {Wiebe}}]{rajput2022hybridized}%
  \BibitemOpen
  \bibfield  {author} {\bibinfo {author} {\bibfnamefont {A.}~\bibnamefont {Rajput}}, \bibinfo {author} {\bibfnamefont {A.}~\bibnamefont {Roggero}},\ and\ \bibinfo {author} {\bibfnamefont {N.}~\bibnamefont {Wiebe}},\ }\href@noop {} {\bibfield  {journal} {\bibinfo  {journal} {Quantum}\ }\textbf {\bibinfo {volume} {6}},\ \bibinfo {pages} {780} (\bibinfo {year} {2022})}\BibitemShut {NoStop}%
\bibitem [{\citenamefont {Low}\ and\ \citenamefont {Chuang}(2017)}]{2017_LC}%
  \BibitemOpen
  \bibfield  {author} {\bibinfo {author} {\bibfnamefont {G.~H.}\ \bibnamefont {Low}}\ and\ \bibinfo {author} {\bibfnamefont {I.~L.}\ \bibnamefont {Chuang}},\ }\href@noop {} {\bibfield  {journal} {\bibinfo  {journal} {Physical Review Letters}\ }\textbf {\bibinfo {volume} {118}},\ \bibinfo {pages} {010501} (\bibinfo {year} {2017})}\BibitemShut {NoStop}%
\bibitem [{\citenamefont {Low}\ and\ \citenamefont {Chuang}(2019{\natexlab{b}})}]{2019_LC}%
  \BibitemOpen
  \bibfield  {author} {\bibinfo {author} {\bibfnamefont {G.~H.}\ \bibnamefont {Low}}\ and\ \bibinfo {author} {\bibfnamefont {I.~L.}\ \bibnamefont {Chuang}},\ }\href@noop {} {\bibfield  {journal} {\bibinfo  {journal} {Quantum}\ }\textbf {\bibinfo {volume} {3}},\ \bibinfo {pages} {163} (\bibinfo {year} {2019}{\natexlab{b}})}\BibitemShut {NoStop}%
\bibitem [{\citenamefont {Gily{\'e}n}\ \emph {et~al.}(2019)\citenamefont {Gily{\'e}n}, \citenamefont {Su}, \citenamefont {Low},\ and\ \citenamefont {Wiebe}}]{2019_GSLW}%
  \BibitemOpen
  \bibfield  {author} {\bibinfo {author} {\bibfnamefont {A.}~\bibnamefont {Gily{\'e}n}}, \bibinfo {author} {\bibfnamefont {Y.}~\bibnamefont {Su}}, \bibinfo {author} {\bibfnamefont {G.~H.}\ \bibnamefont {Low}},\ and\ \bibinfo {author} {\bibfnamefont {N.}~\bibnamefont {Wiebe}},\ }in\ \href@noop {} {\emph {\bibinfo {booktitle} {Proceedings of the 51st Annual ACM SIGACT Symposium on Theory of Computing}}}\ (\bibinfo {year} {2019})\ pp.\ \bibinfo {pages} {193--204}\BibitemShut {NoStop}%
\bibitem [{\citenamefont {Suzuki}(1991)}]{1991_S}%
  \BibitemOpen
  \bibfield  {author} {\bibinfo {author} {\bibfnamefont {M.}~\bibnamefont {Suzuki}},\ }\href@noop {} {\bibfield  {journal} {\bibinfo  {journal} {Journal of Mathematical Physics}\ }\textbf {\bibinfo {volume} {32}},\ \bibinfo {pages} {400} (\bibinfo {year} {1991})}\BibitemShut {NoStop}%
\bibitem [{\citenamefont {Somma}(2016)}]{Somma2016}%
  \BibitemOpen
  \bibfield  {author} {\bibinfo {author} {\bibfnamefont {R.~D.}\ \bibnamefont {Somma}},\ }\href@noop {} {\bibfield  {journal} {\bibinfo  {journal} {{arXiv}}\ } (\bibinfo {year} {2016})},\ \Eprint {https://arxiv.org/abs/1503.06319} {arxiv:1503.06319 [quant-ph]} \BibitemShut {NoStop}%
\bibitem [{\citenamefont {Macridin}\ \emph {et~al.}(2018{\natexlab{a}})\citenamefont {Macridin}, \citenamefont {Spentzouris}, \citenamefont {Amundson},\ and\ \citenamefont {Harnik}}]{Macridin_A2018}%
  \BibitemOpen
  \bibfield  {author} {\bibinfo {author} {\bibfnamefont {A.}~\bibnamefont {Macridin}}, \bibinfo {author} {\bibfnamefont {P.}~\bibnamefont {Spentzouris}}, \bibinfo {author} {\bibfnamefont {J.}~\bibnamefont {Amundson}},\ and\ \bibinfo {author} {\bibfnamefont {R.}~\bibnamefont {Harnik}},\ }\href {https://doi.org/10.1103/PhysRevA.98.042312} {\bibfield  {journal} {\bibinfo  {journal} {Physical Review A}\ }\textbf {\bibinfo {volume} {98}},\ \bibinfo {pages} {042312} (\bibinfo {year} {2018}{\natexlab{a}})},\ \Eprint {https://arxiv.org/abs/1805.09928} {arxiv:1805.09928 [cond-mat, physics:hep-ph, physics:quant-ph]} \BibitemShut {NoStop}%
\bibitem [{\citenamefont {Macridin}\ \emph {et~al.}(2018{\natexlab{b}})\citenamefont {Macridin}, \citenamefont {Spentzouris}, \citenamefont {Amundson},\ and\ \citenamefont {Harnik}}]{Macridin_B2018}%
  \BibitemOpen
  \bibfield  {author} {\bibinfo {author} {\bibfnamefont {A.}~\bibnamefont {Macridin}}, \bibinfo {author} {\bibfnamefont {P.}~\bibnamefont {Spentzouris}}, \bibinfo {author} {\bibfnamefont {J.}~\bibnamefont {Amundson}},\ and\ \bibinfo {author} {\bibfnamefont {R.}~\bibnamefont {Harnik}},\ }\href {https://doi.org/10.1103/PhysRevLett.121.110504} {\bibfield  {journal} {\bibinfo  {journal} {Physical Review Letters}\ }\textbf {\bibinfo {volume} {121}},\ \bibinfo {pages} {110504} (\bibinfo {year} {2018}{\natexlab{b}})}\BibitemShut {NoStop}%
\bibitem [{\citenamefont {Rinaldi}\ \emph {et~al.}(2022)\citenamefont {Rinaldi}, \citenamefont {Han}, \citenamefont {Hassan}, \citenamefont {Feng}, \citenamefont {Nori}, \citenamefont {McGuigan},\ and\ \citenamefont {Hanada}}]{Rinaldi2022}%
  \BibitemOpen
  \bibfield  {author} {\bibinfo {author} {\bibfnamefont {E.}~\bibnamefont {Rinaldi}}, \bibinfo {author} {\bibfnamefont {X.}~\bibnamefont {Han}}, \bibinfo {author} {\bibfnamefont {M.}~\bibnamefont {Hassan}}, \bibinfo {author} {\bibfnamefont {Y.}~\bibnamefont {Feng}}, \bibinfo {author} {\bibfnamefont {F.}~\bibnamefont {Nori}}, \bibinfo {author} {\bibfnamefont {M.}~\bibnamefont {McGuigan}},\ and\ \bibinfo {author} {\bibfnamefont {M.}~\bibnamefont {Hanada}},\ }\href {https://doi.org/10.1103/PRXQuantum.3.010324} {\bibfield  {journal} {\bibinfo  {journal} {PRX Quantum}\ }\textbf {\bibinfo {volume} {3}},\ \bibinfo {pages} {010324} (\bibinfo {year} {2022})},\ \Eprint {https://arxiv.org/abs/2108.02942} {arxiv:2108.02942 [hep-lat, physics:hep-th, physics:quant-ph]} \BibitemShut {NoStop}%
\bibitem [{\citenamefont {Hanada}\ \emph {et~al.}(2023{\natexlab{a}})\citenamefont {Hanada}, \citenamefont {Liu}, \citenamefont {Rinaldi},\ and\ \citenamefont {Tezuka}}]{Hanada2023}%
  \BibitemOpen
  \bibfield  {author} {\bibinfo {author} {\bibfnamefont {M.}~\bibnamefont {Hanada}}, \bibinfo {author} {\bibfnamefont {J.}~\bibnamefont {Liu}}, \bibinfo {author} {\bibfnamefont {E.}~\bibnamefont {Rinaldi}},\ and\ \bibinfo {author} {\bibfnamefont {M.}~\bibnamefont {Tezuka}},\ }\href@noop {} {\bibfield  {journal} {\bibinfo  {journal} {{arXiv}}\ } (\bibinfo {year} {2023}{\natexlab{a}})},\ \Eprint {https://arxiv.org/abs/2212.08546} {arxiv:2212.08546 [hep-lat, physics:hep-th, physics:quant-ph]} \BibitemShut {NoStop}%
\bibitem [{\citenamefont {Akahoshi}\ \emph {et~al.}(2024)\citenamefont {Akahoshi}, \citenamefont {Maruyama}, \citenamefont {Oshima}, \citenamefont {Sato},\ and\ \citenamefont {Fujii}}]{2024_AMOetal}%
  \BibitemOpen
  \bibfield  {author} {\bibinfo {author} {\bibfnamefont {Y.}~\bibnamefont {Akahoshi}}, \bibinfo {author} {\bibfnamefont {K.}~\bibnamefont {Maruyama}}, \bibinfo {author} {\bibfnamefont {H.}~\bibnamefont {Oshima}}, \bibinfo {author} {\bibfnamefont {S.}~\bibnamefont {Sato}},\ and\ \bibinfo {author} {\bibfnamefont {K.}~\bibnamefont {Fujii}},\ }\href@noop {} {\bibfield  {journal} {\bibinfo  {journal} {PRX Quantum}\ }\textbf {\bibinfo {volume} {5}},\ \bibinfo {pages} {010337} (\bibinfo {year} {2024})}\BibitemShut {NoStop}%
\bibitem [{\citenamefont {Campbell}\ \emph {et~al.}(2017)\citenamefont {Campbell}, \citenamefont {Terhal},\ and\ \citenamefont {Vuillot}}]{Campbell2017}%
  \BibitemOpen
  \bibfield  {author} {\bibinfo {author} {\bibfnamefont {E.~T.}\ \bibnamefont {Campbell}}, \bibinfo {author} {\bibfnamefont {B.~M.}\ \bibnamefont {Terhal}},\ and\ \bibinfo {author} {\bibfnamefont {C.}~\bibnamefont {Vuillot}},\ }\href {https://doi.org/10.1038/nature23460} {\bibfield  {journal} {\bibinfo  {journal} {Nature}\ }\textbf {\bibinfo {volume} {549}},\ \bibinfo {pages} {172–179} (\bibinfo {year} {2017})}\BibitemShut {NoStop}%
\bibitem [{\citenamefont {Fowler}\ \emph {et~al.}(2012{\natexlab{b}})\citenamefont {Fowler}, \citenamefont {Mariantoni}, \citenamefont {Martinis},\ and\ \citenamefont {Cleland}}]{fowler2012surface}%
  \BibitemOpen
  \bibfield  {author} {\bibinfo {author} {\bibfnamefont {A.~G.}\ \bibnamefont {Fowler}}, \bibinfo {author} {\bibfnamefont {M.}~\bibnamefont {Mariantoni}}, \bibinfo {author} {\bibfnamefont {J.~M.}\ \bibnamefont {Martinis}},\ and\ \bibinfo {author} {\bibfnamefont {A.~N.}\ \bibnamefont {Cleland}},\ }\href@noop {} {\bibfield  {journal} {\bibinfo  {journal} {Physical Review A—Atomic, Molecular, and Optical Physics}\ }\textbf {\bibinfo {volume} {86}},\ \bibinfo {pages} {032324} (\bibinfo {year} {2012}{\natexlab{b}})}\BibitemShut {NoStop}%
\bibitem [{\citenamefont {Childs}\ \emph {et~al.}(2021)\citenamefont {Childs}, \citenamefont {Su}, \citenamefont {Tran}, \citenamefont {Wiebe},\ and\ \citenamefont {Zhu}}]{Childs2021}%
  \BibitemOpen
  \bibfield  {author} {\bibinfo {author} {\bibfnamefont {A.~M.}\ \bibnamefont {Childs}}, \bibinfo {author} {\bibfnamefont {Y.}~\bibnamefont {Su}}, \bibinfo {author} {\bibfnamefont {M.~C.}\ \bibnamefont {Tran}}, \bibinfo {author} {\bibfnamefont {N.}~\bibnamefont {Wiebe}},\ and\ \bibinfo {author} {\bibfnamefont {S.}~\bibnamefont {Zhu}},\ }\href {https://doi.org/10.1103/PhysRevX.11.011020} {\bibfield  {journal} {\bibinfo  {journal} {Phys. Rev. X}\ }\textbf {\bibinfo {volume} {11}},\ \bibinfo {pages} {011020} (\bibinfo {year} {2021})}\BibitemShut {NoStop}%
\bibitem [{\citenamefont {Wecker}\ \emph {et~al.}(2014)\citenamefont {Wecker}, \citenamefont {Bauer}, \citenamefont {Clark}, \citenamefont {Hastings},\ and\ \citenamefont {Troyer}}]{wecker2014gate}%
  \BibitemOpen
  \bibfield  {author} {\bibinfo {author} {\bibfnamefont {D.}~\bibnamefont {Wecker}}, \bibinfo {author} {\bibfnamefont {B.}~\bibnamefont {Bauer}}, \bibinfo {author} {\bibfnamefont {B.~K.}\ \bibnamefont {Clark}}, \bibinfo {author} {\bibfnamefont {M.~B.}\ \bibnamefont {Hastings}},\ and\ \bibinfo {author} {\bibfnamefont {M.}~\bibnamefont {Troyer}},\ }\href {https://doi.org/10.1103/PhysRevA.90.022305} {\bibfield  {journal} {\bibinfo  {journal} {Phys. Rev. A}\ }\textbf {\bibinfo {volume} {90}},\ \bibinfo {pages} {022305} (\bibinfo {year} {2014})}\BibitemShut {NoStop}%
\bibitem [{\citenamefont {Babbush}\ \emph {et~al.}(2015)\citenamefont {Babbush}, \citenamefont {McClean}, \citenamefont {Wecker}, \citenamefont {Aspuru-Guzik},\ and\ \citenamefont {Wiebe}}]{babbush2015chemical}%
  \BibitemOpen
  \bibfield  {author} {\bibinfo {author} {\bibfnamefont {R.}~\bibnamefont {Babbush}}, \bibinfo {author} {\bibfnamefont {J.}~\bibnamefont {McClean}}, \bibinfo {author} {\bibfnamefont {D.}~\bibnamefont {Wecker}}, \bibinfo {author} {\bibfnamefont {A.}~\bibnamefont {Aspuru-Guzik}},\ and\ \bibinfo {author} {\bibfnamefont {N.}~\bibnamefont {Wiebe}},\ }\href@noop {} {\bibfield  {journal} {\bibinfo  {journal} {Physical Review A}\ }\textbf {\bibinfo {volume} {91}},\ \bibinfo {pages} {022311} (\bibinfo {year} {2015})}\BibitemShut {NoStop}%
\bibitem [{\citenamefont {Li}\ \emph {et~al.}(2021)\citenamefont {Li}, \citenamefont {Lappi},\ and\ \citenamefont {Zhao}}]{Li:2021zaw}%
  \BibitemOpen
  \bibfield  {author} {\bibinfo {author} {\bibfnamefont {M.}~\bibnamefont {Li}}, \bibinfo {author} {\bibfnamefont {T.}~\bibnamefont {Lappi}},\ and\ \bibinfo {author} {\bibfnamefont {X.}~\bibnamefont {Zhao}},\ }\href {https://doi.org/10.1103/PhysRevD.104.056014} {\bibfield  {journal} {\bibinfo  {journal} {Phys. Rev. D}\ }\textbf {\bibinfo {volume} {104}},\ \bibinfo {pages} {056014} (\bibinfo {year} {2021})},\ \Eprint {https://arxiv.org/abs/2107.02225} {arXiv:2107.02225 [hep-ph]} \BibitemShut {NoStop}%
\bibitem [{\citenamefont {Barata}\ \emph {et~al.}(2023)\citenamefont {Barata}, \citenamefont {Du}, \citenamefont {Li}, \citenamefont {Qian},\ and\ \citenamefont {Salgado}}]{Barata:2023clv}%
  \BibitemOpen
  \bibfield  {author} {\bibinfo {author} {\bibfnamefont {J.~a.}\ \bibnamefont {Barata}}, \bibinfo {author} {\bibfnamefont {X.}~\bibnamefont {Du}}, \bibinfo {author} {\bibfnamefont {M.}~\bibnamefont {Li}}, \bibinfo {author} {\bibfnamefont {W.}~\bibnamefont {Qian}},\ and\ \bibinfo {author} {\bibfnamefont {C.~A.}\ \bibnamefont {Salgado}},\ }\href {https://doi.org/10.1103/PhysRevD.108.056023} {\bibfield  {journal} {\bibinfo  {journal} {Phys. Rev. D}\ }\textbf {\bibinfo {volume} {108}},\ \bibinfo {pages} {056023} (\bibinfo {year} {2023})},\ \Eprint {https://arxiv.org/abs/2307.01792} {arXiv:2307.01792 [hep-ph]} \BibitemShut {NoStop}%
\bibitem [{\citenamefont {Calajò}\ \emph {et~al.}(2024)\citenamefont {Calajò}, \citenamefont {Magnifico}, \citenamefont {Edmunds}, \citenamefont {Ringbauer}, \citenamefont {Montangero},\ and\ \citenamefont {Silvi}}]{calajò2024digitalquantumsimulation11d}%
  \BibitemOpen
  \bibfield  {author} {\bibinfo {author} {\bibfnamefont {G.}~\bibnamefont {Calajò}}, \bibinfo {author} {\bibfnamefont {G.}~\bibnamefont {Magnifico}}, \bibinfo {author} {\bibfnamefont {C.}~\bibnamefont {Edmunds}}, \bibinfo {author} {\bibfnamefont {M.}~\bibnamefont {Ringbauer}}, \bibinfo {author} {\bibfnamefont {S.}~\bibnamefont {Montangero}},\ and\ \bibinfo {author} {\bibfnamefont {P.}~\bibnamefont {Silvi}},\ }\href {https://arxiv.org/abs/2402.07987} {\bibinfo {title} {{Digital quantum simulation of a (1+1)D SU(2) lattice gauge theory with ion qudits}}} (\bibinfo {year} {2024}),\ \Eprint {https://arxiv.org/abs/2402.07987} {arXiv:2402.07987 [quant-ph]} \BibitemShut {NoStop}%
\bibitem [{\citenamefont {Unmuth-Yockey}\ \emph {et~al.}(2018)\citenamefont {Unmuth-Yockey}, \citenamefont {Zhang}, \citenamefont {Bazavov}, \citenamefont {Meurice},\ and\ \citenamefont {Tsai}}]{PhysRevD.98.094511}%
  \BibitemOpen
  \bibfield  {author} {\bibinfo {author} {\bibfnamefont {J.}~\bibnamefont {Unmuth-Yockey}}, \bibinfo {author} {\bibfnamefont {J.}~\bibnamefont {Zhang}}, \bibinfo {author} {\bibfnamefont {A.}~\bibnamefont {Bazavov}}, \bibinfo {author} {\bibfnamefont {Y.}~\bibnamefont {Meurice}},\ and\ \bibinfo {author} {\bibfnamefont {S.-W.}\ \bibnamefont {Tsai}},\ }\href {https://doi.org/10.1103/PhysRevD.98.094511} {\bibfield  {journal} {\bibinfo  {journal} {Phys. Rev. D}\ }\textbf {\bibinfo {volume} {98}},\ \bibinfo {pages} {094511} (\bibinfo {year} {2018})}\BibitemShut {NoStop}%
\bibitem [{\citenamefont {Unmuth-Yockey}(2019)}]{PhysRevD.99.074502}%
  \BibitemOpen
  \bibfield  {author} {\bibinfo {author} {\bibfnamefont {J.~F.}\ \bibnamefont {Unmuth-Yockey}},\ }\href {https://doi.org/10.1103/PhysRevD.99.074502} {\bibfield  {journal} {\bibinfo  {journal} {Phys. Rev. D}\ }\textbf {\bibinfo {volume} {99}},\ \bibinfo {pages} {074502} (\bibinfo {year} {2019})}\BibitemShut {NoStop}%
\bibitem [{\citenamefont {Klco}\ \emph {et~al.}(2020)\citenamefont {Klco}, \citenamefont {Savage},\ and\ \citenamefont {Stryker}}]{PhysRevD.101.074512}%
  \BibitemOpen
  \bibfield  {author} {\bibinfo {author} {\bibfnamefont {N.}~\bibnamefont {Klco}}, \bibinfo {author} {\bibfnamefont {M.~J.}\ \bibnamefont {Savage}},\ and\ \bibinfo {author} {\bibfnamefont {J.~R.}\ \bibnamefont {Stryker}},\ }\href {https://doi.org/10.1103/PhysRevD.101.074512} {\bibfield  {journal} {\bibinfo  {journal} {Phys. Rev. D}\ }\textbf {\bibinfo {volume} {101}},\ \bibinfo {pages} {074512} (\bibinfo {year} {2020})}\BibitemShut {NoStop}%
\bibitem [{\citenamefont {Farrell}\ \emph {et~al.}(2024{\natexlab{a}})\citenamefont {Farrell}, \citenamefont {Illa}, \citenamefont {Ciavarella},\ and\ \citenamefont {Savage}}]{PRXQuantum.5.020315}%
  \BibitemOpen
  \bibfield  {author} {\bibinfo {author} {\bibfnamefont {R.~C.}\ \bibnamefont {Farrell}}, \bibinfo {author} {\bibfnamefont {M.}~\bibnamefont {Illa}}, \bibinfo {author} {\bibfnamefont {A.~N.}\ \bibnamefont {Ciavarella}},\ and\ \bibinfo {author} {\bibfnamefont {M.~J.}\ \bibnamefont {Savage}},\ }\href {https://doi.org/10.1103/PRXQuantum.5.020315} {\bibfield  {journal} {\bibinfo  {journal} {PRX Quantum}\ }\textbf {\bibinfo {volume} {5}},\ \bibinfo {pages} {020315} (\bibinfo {year} {2024}{\natexlab{a}})}\BibitemShut {NoStop}%
\bibitem [{\citenamefont {Farrell}\ \emph {et~al.}(2024{\natexlab{b}})\citenamefont {Farrell}, \citenamefont {Illa}, \citenamefont {Ciavarella},\ and\ \citenamefont {Savage}}]{PhysRevD.109.114510}%
  \BibitemOpen
  \bibfield  {author} {\bibinfo {author} {\bibfnamefont {R.~C.}\ \bibnamefont {Farrell}}, \bibinfo {author} {\bibfnamefont {M.}~\bibnamefont {Illa}}, \bibinfo {author} {\bibfnamefont {A.~N.}\ \bibnamefont {Ciavarella}},\ and\ \bibinfo {author} {\bibfnamefont {M.~J.}\ \bibnamefont {Savage}},\ }\href {https://doi.org/10.1103/PhysRevD.109.114510} {\bibfield  {journal} {\bibinfo  {journal} {Phys. Rev. D}\ }\textbf {\bibinfo {volume} {109}},\ \bibinfo {pages} {114510} (\bibinfo {year} {2024}{\natexlab{b}})}\BibitemShut {NoStop}%
\bibitem [{\citenamefont {Illa}\ \emph {et~al.}(2024{\natexlab{a}})\citenamefont {Illa}, \citenamefont {Robin},\ and\ \citenamefont {Savage}}]{illa2024qu8itsquantumsimulationslattice}%
  \BibitemOpen
  \bibfield  {author} {\bibinfo {author} {\bibfnamefont {M.}~\bibnamefont {Illa}}, \bibinfo {author} {\bibfnamefont {C.~E.~P.}\ \bibnamefont {Robin}},\ and\ \bibinfo {author} {\bibfnamefont {M.~J.}\ \bibnamefont {Savage}},\ }\href {https://arxiv.org/abs/2403.14537} {\bibinfo {title} {{Qu8its for Quantum Simulations of Lattice Quantum Chromodynamics}}} (\bibinfo {year} {2024}{\natexlab{a}}),\ \Eprint {https://arxiv.org/abs/2403.14537} {arXiv:2403.14537 [quant-ph]} \BibitemShut {NoStop}%
\bibitem [{\citenamefont {Ciavarella}\ \emph {et~al.}(2021)\citenamefont {Ciavarella}, \citenamefont {Klco},\ and\ \citenamefont {Savage}}]{PhysRevD.103.094501}%
  \BibitemOpen
  \bibfield  {author} {\bibinfo {author} {\bibfnamefont {A.}~\bibnamefont {Ciavarella}}, \bibinfo {author} {\bibfnamefont {N.}~\bibnamefont {Klco}},\ and\ \bibinfo {author} {\bibfnamefont {M.~J.}\ \bibnamefont {Savage}},\ }\href {https://doi.org/10.1103/PhysRevD.103.094501} {\bibfield  {journal} {\bibinfo  {journal} {Phys. Rev. D}\ }\textbf {\bibinfo {volume} {103}},\ \bibinfo {pages} {094501} (\bibinfo {year} {2021})}\BibitemShut {NoStop}%
\bibitem [{\citenamefont {Bazavov}\ \emph {et~al.}(2015)\citenamefont {Bazavov}, \citenamefont {Meurice}, \citenamefont {Tsai}, \citenamefont {Unmuth-Yockey},\ and\ \citenamefont {Zhang}}]{PhysRevD.92.076003}%
  \BibitemOpen
  \bibfield  {author} {\bibinfo {author} {\bibfnamefont {A.}~\bibnamefont {Bazavov}}, \bibinfo {author} {\bibfnamefont {Y.}~\bibnamefont {Meurice}}, \bibinfo {author} {\bibfnamefont {S.-W.}\ \bibnamefont {Tsai}}, \bibinfo {author} {\bibfnamefont {J.}~\bibnamefont {Unmuth-Yockey}},\ and\ \bibinfo {author} {\bibfnamefont {J.}~\bibnamefont {Zhang}},\ }\href {https://doi.org/10.1103/PhysRevD.92.076003} {\bibfield  {journal} {\bibinfo  {journal} {Phys. Rev. D}\ }\textbf {\bibinfo {volume} {92}},\ \bibinfo {pages} {076003} (\bibinfo {year} {2015})}\BibitemShut {NoStop}%
\bibitem [{\citenamefont {Zhang}\ \emph {et~al.}(2018)\citenamefont {Zhang}, \citenamefont {Unmuth-Yockey}, \citenamefont {Zeiher}, \citenamefont {Bazavov}, \citenamefont {Tsai},\ and\ \citenamefont {Meurice}}]{PhysRevLett.121.223201}%
  \BibitemOpen
  \bibfield  {author} {\bibinfo {author} {\bibfnamefont {J.}~\bibnamefont {Zhang}}, \bibinfo {author} {\bibfnamefont {J.}~\bibnamefont {Unmuth-Yockey}}, \bibinfo {author} {\bibfnamefont {J.}~\bibnamefont {Zeiher}}, \bibinfo {author} {\bibfnamefont {A.}~\bibnamefont {Bazavov}}, \bibinfo {author} {\bibfnamefont {S.-W.}\ \bibnamefont {Tsai}},\ and\ \bibinfo {author} {\bibfnamefont {Y.}~\bibnamefont {Meurice}},\ }\href {https://doi.org/10.1103/PhysRevLett.121.223201} {\bibfield  {journal} {\bibinfo  {journal} {Phys. Rev. Lett.}\ }\textbf {\bibinfo {volume} {121}},\ \bibinfo {pages} {223201} (\bibinfo {year} {2018})}\BibitemShut {NoStop}%
\bibitem [{\citenamefont {Bazavov}\ \emph {et~al.}(2019)\citenamefont {Bazavov}, \citenamefont {Catterall}, \citenamefont {Jha},\ and\ \citenamefont {Unmuth-Yockey}}]{PhysRevD.99.114507}%
  \BibitemOpen
  \bibfield  {author} {\bibinfo {author} {\bibfnamefont {A.}~\bibnamefont {Bazavov}}, \bibinfo {author} {\bibfnamefont {S.}~\bibnamefont {Catterall}}, \bibinfo {author} {\bibfnamefont {R.~G.}\ \bibnamefont {Jha}},\ and\ \bibinfo {author} {\bibfnamefont {J.}~\bibnamefont {Unmuth-Yockey}},\ }\href {https://doi.org/10.1103/PhysRevD.99.114507} {\bibfield  {journal} {\bibinfo  {journal} {Phys. Rev. D}\ }\textbf {\bibinfo {volume} {99}},\ \bibinfo {pages} {114507} (\bibinfo {year} {2019})}\BibitemShut {NoStop}%
\bibitem [{\citenamefont {Bauer}\ and\ \citenamefont {Grabowska}(2021)}]{bauer2021efficientrepresentationsimulatingu1}%
  \BibitemOpen
  \bibfield  {author} {\bibinfo {author} {\bibfnamefont {C.~W.}\ \bibnamefont {Bauer}}\ and\ \bibinfo {author} {\bibfnamefont {D.~M.}\ \bibnamefont {Grabowska}},\ }\href {https://arxiv.org/abs/2111.08015} {\bibinfo {title} {{Efficient Representation for Simulating U(1) Gauge Theories on Digital Quantum Computers at All Values of the Coupling}}} (\bibinfo {year} {2021}),\ \Eprint {https://arxiv.org/abs/2111.08015} {arXiv:2111.08015 [hep-ph]} \BibitemShut {NoStop}%
\bibitem [{\citenamefont {Grabowska}\ \emph {et~al.}(2023)\citenamefont {Grabowska}, \citenamefont {Kane}, \citenamefont {Nachman},\ and\ \citenamefont {Bauer}}]{grabowska2023overcomingexponentialscalingsize}%
  \BibitemOpen
  \bibfield  {author} {\bibinfo {author} {\bibfnamefont {D.~M.}\ \bibnamefont {Grabowska}}, \bibinfo {author} {\bibfnamefont {C.}~\bibnamefont {Kane}}, \bibinfo {author} {\bibfnamefont {B.}~\bibnamefont {Nachman}},\ and\ \bibinfo {author} {\bibfnamefont {C.~W.}\ \bibnamefont {Bauer}},\ }\href {https://arxiv.org/abs/2208.03333} {\bibinfo {title} {{Overcoming exponential scaling with system size in Trotter-Suzuki implementations of constrained Hamiltonians: 2+1 U(1) lattice gauge theories}}} (\bibinfo {year} {2023}),\ \Eprint {https://arxiv.org/abs/2208.03333} {arXiv:2208.03333 [quant-ph]} \BibitemShut {NoStop}%
\bibitem [{\citenamefont {Buser}\ \emph {et~al.}(2020)\citenamefont {Buser}, \citenamefont {Bhattacharya}, \citenamefont {Cincio},\ and\ \citenamefont {Gupta}}]{PhysRevD.102.114514}%
  \BibitemOpen
  \bibfield  {author} {\bibinfo {author} {\bibfnamefont {A.~J.}\ \bibnamefont {Buser}}, \bibinfo {author} {\bibfnamefont {T.}~\bibnamefont {Bhattacharya}}, \bibinfo {author} {\bibfnamefont {L.}~\bibnamefont {Cincio}},\ and\ \bibinfo {author} {\bibfnamefont {R.}~\bibnamefont {Gupta}},\ }\href {https://doi.org/10.1103/PhysRevD.102.114514} {\bibfield  {journal} {\bibinfo  {journal} {Phys. Rev. D}\ }\textbf {\bibinfo {volume} {102}},\ \bibinfo {pages} {114514} (\bibinfo {year} {2020})}\BibitemShut {NoStop}%
\bibitem [{\citenamefont {Bhattacharya}\ \emph {et~al.}(2021)\citenamefont {Bhattacharya}, \citenamefont {Buser}, \citenamefont {Chandrasekharan}, \citenamefont {Gupta},\ and\ \citenamefont {Singh}}]{PhysRevLett.126.172001}%
  \BibitemOpen
  \bibfield  {author} {\bibinfo {author} {\bibfnamefont {T.}~\bibnamefont {Bhattacharya}}, \bibinfo {author} {\bibfnamefont {A.~J.}\ \bibnamefont {Buser}}, \bibinfo {author} {\bibfnamefont {S.}~\bibnamefont {Chandrasekharan}}, \bibinfo {author} {\bibfnamefont {R.}~\bibnamefont {Gupta}},\ and\ \bibinfo {author} {\bibfnamefont {H.}~\bibnamefont {Singh}},\ }\href {https://doi.org/10.1103/PhysRevLett.126.172001} {\bibfield  {journal} {\bibinfo  {journal} {Phys. Rev. Lett.}\ }\textbf {\bibinfo {volume} {126}},\ \bibinfo {pages} {172001} (\bibinfo {year} {2021})}\BibitemShut {NoStop}%
\bibitem [{\citenamefont {Kavaki}\ and\ \citenamefont {Lewis}(2024)}]{kavaki-square-plaq}%
  \BibitemOpen
  \bibfield  {author} {\bibinfo {author} {\bibfnamefont {A.~H.~Z.}\ \bibnamefont {Kavaki}}\ and\ \bibinfo {author} {\bibfnamefont {R.}~\bibnamefont {Lewis}},\ }\href@noop {} {\bibfield  {journal} {\bibinfo  {journal} {Communications Physics}\ }\textbf {\bibinfo {volume} {7}},\ \bibinfo {pages} {208} (\bibinfo {year} {2024})}\BibitemShut {NoStop}%
\bibitem [{\citenamefont {Murairi}\ \emph {et~al.}(2022)\citenamefont {Murairi}, \citenamefont {Cervia}, \citenamefont {Kumar}, \citenamefont {Bedaque},\ and\ \citenamefont {Alexandru}}]{PhysRevD.106.094504}%
  \BibitemOpen
  \bibfield  {author} {\bibinfo {author} {\bibfnamefont {E.~M.}\ \bibnamefont {Murairi}}, \bibinfo {author} {\bibfnamefont {M.~J.}\ \bibnamefont {Cervia}}, \bibinfo {author} {\bibfnamefont {H.}~\bibnamefont {Kumar}}, \bibinfo {author} {\bibfnamefont {P.~F.}\ \bibnamefont {Bedaque}},\ and\ \bibinfo {author} {\bibfnamefont {A.}~\bibnamefont {Alexandru}},\ }\href {https://doi.org/10.1103/PhysRevD.106.094504} {\bibfield  {journal} {\bibinfo  {journal} {Phys. Rev. D}\ }\textbf {\bibinfo {volume} {106}},\ \bibinfo {pages} {094504} (\bibinfo {year} {2022})}\BibitemShut {NoStop}%
\bibitem [{\citenamefont {Zohar}\ \emph {et~al.}(2015)\citenamefont {Zohar}, \citenamefont {Cirac},\ and\ \citenamefont {Reznik}}]{Zohar_2016}%
  \BibitemOpen
  \bibfield  {author} {\bibinfo {author} {\bibfnamefont {E.}~\bibnamefont {Zohar}}, \bibinfo {author} {\bibfnamefont {J.~I.}\ \bibnamefont {Cirac}},\ and\ \bibinfo {author} {\bibfnamefont {B.}~\bibnamefont {Reznik}},\ }\href {https://doi.org/10.1088/0034-4885/79/1/014401} {\bibfield  {journal} {\bibinfo  {journal} {Reports on Progress in Physics}\ }\textbf {\bibinfo {volume} {79}},\ \bibinfo {pages} {014401} (\bibinfo {year} {2015})}\BibitemShut {NoStop}%
\bibitem [{\citenamefont {Zohar}\ \emph {et~al.}(2013{\natexlab{a}})\citenamefont {Zohar}, \citenamefont {Cirac},\ and\ \citenamefont {Reznik}}]{PhysRevLett.110.125304}%
  \BibitemOpen
  \bibfield  {author} {\bibinfo {author} {\bibfnamefont {E.}~\bibnamefont {Zohar}}, \bibinfo {author} {\bibfnamefont {J.~I.}\ \bibnamefont {Cirac}},\ and\ \bibinfo {author} {\bibfnamefont {B.}~\bibnamefont {Reznik}},\ }\href {https://doi.org/10.1103/PhysRevLett.110.125304} {\bibfield  {journal} {\bibinfo  {journal} {Phys. Rev. Lett.}\ }\textbf {\bibinfo {volume} {110}},\ \bibinfo {pages} {125304} (\bibinfo {year} {2013}{\natexlab{a}})}\BibitemShut {NoStop}%
\bibitem [{\citenamefont {Zohar}\ \emph {et~al.}(2012)\citenamefont {Zohar}, \citenamefont {Cirac},\ and\ \citenamefont {Reznik}}]{PhysRevLett.109.125302}%
  \BibitemOpen
  \bibfield  {author} {\bibinfo {author} {\bibfnamefont {E.}~\bibnamefont {Zohar}}, \bibinfo {author} {\bibfnamefont {J.~I.}\ \bibnamefont {Cirac}},\ and\ \bibinfo {author} {\bibfnamefont {B.}~\bibnamefont {Reznik}},\ }\href {https://doi.org/10.1103/PhysRevLett.109.125302} {\bibfield  {journal} {\bibinfo  {journal} {Phys. Rev. Lett.}\ }\textbf {\bibinfo {volume} {109}},\ \bibinfo {pages} {125302} (\bibinfo {year} {2012})}\BibitemShut {NoStop}%
\bibitem [{\citenamefont {Zohar}\ \emph {et~al.}(2013{\natexlab{b}})\citenamefont {Zohar}, \citenamefont {Cirac},\ and\ \citenamefont {Reznik}}]{PhysRevA.88.023617}%
  \BibitemOpen
  \bibfield  {author} {\bibinfo {author} {\bibfnamefont {E.}~\bibnamefont {Zohar}}, \bibinfo {author} {\bibfnamefont {J.~I.}\ \bibnamefont {Cirac}},\ and\ \bibinfo {author} {\bibfnamefont {B.}~\bibnamefont {Reznik}},\ }\href {https://doi.org/10.1103/PhysRevA.88.023617} {\bibfield  {journal} {\bibinfo  {journal} {Phys. Rev. A}\ }\textbf {\bibinfo {volume} {88}},\ \bibinfo {pages} {023617} (\bibinfo {year} {2013}{\natexlab{b}})}\BibitemShut {NoStop}%
\bibitem [{\citenamefont {Raychowdhury}\ and\ \citenamefont {Stryker}(2020{\natexlab{a}})}]{PhysRevD.101.114502}%
  \BibitemOpen
  \bibfield  {author} {\bibinfo {author} {\bibfnamefont {I.}~\bibnamefont {Raychowdhury}}\ and\ \bibinfo {author} {\bibfnamefont {J.~R.}\ \bibnamefont {Stryker}},\ }\href {https://doi.org/10.1103/PhysRevD.101.114502} {\bibfield  {journal} {\bibinfo  {journal} {Phys. Rev. D}\ }\textbf {\bibinfo {volume} {101}},\ \bibinfo {pages} {114502} (\bibinfo {year} {2020}{\natexlab{a}})}\BibitemShut {NoStop}%
\bibitem [{\citenamefont {Davoudi}\ \emph {et~al.}(2024)\citenamefont {Davoudi}, \citenamefont {Hsieh},\ and\ \citenamefont {Kadam}}]{davoudi2024scatteringwavepacketshadrons}%
  \BibitemOpen
  \bibfield  {author} {\bibinfo {author} {\bibfnamefont {Z.}~\bibnamefont {Davoudi}}, \bibinfo {author} {\bibfnamefont {C.-C.}\ \bibnamefont {Hsieh}},\ and\ \bibinfo {author} {\bibfnamefont {S.~V.}\ \bibnamefont {Kadam}},\ }\href {https://arxiv.org/abs/2402.00840} {\bibinfo {title} {{Scattering wave packets of hadrons in gauge theories: Preparation on a quantum computer}}} (\bibinfo {year} {2024}),\ \Eprint {https://arxiv.org/abs/2402.00840} {arXiv:2402.00840 [quant-ph]} \BibitemShut {NoStop}%
\bibitem [{\citenamefont {Raychowdhury}\ and\ \citenamefont {Stryker}(2020{\natexlab{b}})}]{PhysRevResearch.2.033039}%
  \BibitemOpen
  \bibfield  {author} {\bibinfo {author} {\bibfnamefont {I.}~\bibnamefont {Raychowdhury}}\ and\ \bibinfo {author} {\bibfnamefont {J.~R.}\ \bibnamefont {Stryker}},\ }\href {https://doi.org/10.1103/PhysRevResearch.2.033039} {\bibfield  {journal} {\bibinfo  {journal} {Phys. Rev. Res.}\ }\textbf {\bibinfo {volume} {2}},\ \bibinfo {pages} {033039} (\bibinfo {year} {2020}{\natexlab{b}})}\BibitemShut {NoStop}%
\bibitem [{\citenamefont {Davoudi}\ \emph {et~al.}(2021)\citenamefont {Davoudi}, \citenamefont {Raychowdhury},\ and\ \citenamefont {Shaw}}]{PhysRevD.104.074505}%
  \BibitemOpen
  \bibfield  {author} {\bibinfo {author} {\bibfnamefont {Z.}~\bibnamefont {Davoudi}}, \bibinfo {author} {\bibfnamefont {I.}~\bibnamefont {Raychowdhury}},\ and\ \bibinfo {author} {\bibfnamefont {A.}~\bibnamefont {Shaw}},\ }\href {https://doi.org/10.1103/PhysRevD.104.074505} {\bibfield  {journal} {\bibinfo  {journal} {Phys. Rev. D}\ }\textbf {\bibinfo {volume} {104}},\ \bibinfo {pages} {074505} (\bibinfo {year} {2021})}\BibitemShut {NoStop}%
\bibitem [{\citenamefont {Mathew}\ and\ \citenamefont {Raychowdhury}(2022)}]{PhysRevD.106.054510}%
  \BibitemOpen
  \bibfield  {author} {\bibinfo {author} {\bibfnamefont {E.}~\bibnamefont {Mathew}}\ and\ \bibinfo {author} {\bibfnamefont {I.}~\bibnamefont {Raychowdhury}},\ }\href {https://doi.org/10.1103/PhysRevD.106.054510} {\bibfield  {journal} {\bibinfo  {journal} {Phys. Rev. D}\ }\textbf {\bibinfo {volume} {106}},\ \bibinfo {pages} {054510} (\bibinfo {year} {2022})}\BibitemShut {NoStop}%
\bibitem [{\citenamefont {Kreshchuk}\ \emph {et~al.}(2022)\citenamefont {Kreshchuk}, \citenamefont {Kirby}, \citenamefont {Goldstein}, \citenamefont {Beauchemin},\ and\ \citenamefont {Love}}]{PhysRevA.105.032418}%
  \BibitemOpen
  \bibfield  {author} {\bibinfo {author} {\bibfnamefont {M.}~\bibnamefont {Kreshchuk}}, \bibinfo {author} {\bibfnamefont {W.~M.}\ \bibnamefont {Kirby}}, \bibinfo {author} {\bibfnamefont {G.}~\bibnamefont {Goldstein}}, \bibinfo {author} {\bibfnamefont {H.}~\bibnamefont {Beauchemin}},\ and\ \bibinfo {author} {\bibfnamefont {P.~J.}\ \bibnamefont {Love}},\ }\href {https://doi.org/10.1103/PhysRevA.105.032418} {\bibfield  {journal} {\bibinfo  {journal} {Phys. Rev. A}\ }\textbf {\bibinfo {volume} {105}},\ \bibinfo {pages} {032418} (\bibinfo {year} {2022})}\BibitemShut {NoStop}%
\bibitem [{\citenamefont {Kreshchuk}\ \emph {et~al.}(2021{\natexlab{a}})\citenamefont {Kreshchuk}, \citenamefont {Jia}, \citenamefont {Kirby}, \citenamefont {Goldstein}, \citenamefont {Vary},\ and\ \citenamefont {Love}}]{PhysRevA.103.062601}%
  \BibitemOpen
  \bibfield  {author} {\bibinfo {author} {\bibfnamefont {M.}~\bibnamefont {Kreshchuk}}, \bibinfo {author} {\bibfnamefont {S.}~\bibnamefont {Jia}}, \bibinfo {author} {\bibfnamefont {W.~M.}\ \bibnamefont {Kirby}}, \bibinfo {author} {\bibfnamefont {G.}~\bibnamefont {Goldstein}}, \bibinfo {author} {\bibfnamefont {J.~P.}\ \bibnamefont {Vary}},\ and\ \bibinfo {author} {\bibfnamefont {P.~J.}\ \bibnamefont {Love}},\ }\href {https://doi.org/10.1103/PhysRevA.103.062601} {\bibfield  {journal} {\bibinfo  {journal} {Phys. Rev. A}\ }\textbf {\bibinfo {volume} {103}},\ \bibinfo {pages} {062601} (\bibinfo {year} {2021}{\natexlab{a}})}\BibitemShut {NoStop}%
\bibitem [{\citenamefont {Kreshchuk}\ \emph {et~al.}(2021{\natexlab{b}})\citenamefont {Kreshchuk}, \citenamefont {Jia}, \citenamefont {Kirby}, \citenamefont {Goldstein}, \citenamefont {Vary},\ and\ \citenamefont {Love}}]{e23050597}%
  \BibitemOpen
  \bibfield  {author} {\bibinfo {author} {\bibfnamefont {M.}~\bibnamefont {Kreshchuk}}, \bibinfo {author} {\bibfnamefont {S.}~\bibnamefont {Jia}}, \bibinfo {author} {\bibfnamefont {W.~M.}\ \bibnamefont {Kirby}}, \bibinfo {author} {\bibfnamefont {G.}~\bibnamefont {Goldstein}}, \bibinfo {author} {\bibfnamefont {J.~P.}\ \bibnamefont {Vary}},\ and\ \bibinfo {author} {\bibfnamefont {P.~J.}\ \bibnamefont {Love}},\ }\bibfield  {journal} {\bibinfo  {journal} {Entropy}\ }\textbf {\bibinfo {volume} {23}},\ \href {https://doi.org/10.3390/e23050597} {10.3390/e23050597} (\bibinfo {year} {2021}{\natexlab{b}})\BibitemShut {NoStop}%
\bibitem [{\citenamefont {Alam}\ \emph {et~al.}(2022)\citenamefont {Alam}, \citenamefont {Hadfield}, \citenamefont {Lamm},\ and\ \citenamefont {Li}}]{PhysRevD.105.114501}%
  \BibitemOpen
  \bibfield  {author} {\bibinfo {author} {\bibfnamefont {M.~S.}\ \bibnamefont {Alam}}, \bibinfo {author} {\bibfnamefont {S.}~\bibnamefont {Hadfield}}, \bibinfo {author} {\bibfnamefont {H.}~\bibnamefont {Lamm}},\ and\ \bibinfo {author} {\bibfnamefont {A.~C.~Y.}\ \bibnamefont {Li}} (\bibinfo {collaboration} {SQMS Collaboration}),\ }\href {https://doi.org/10.1103/PhysRevD.105.114501} {\bibfield  {journal} {\bibinfo  {journal} {Phys. Rev. D}\ }\textbf {\bibinfo {volume} {105}},\ \bibinfo {pages} {114501} (\bibinfo {year} {2022})}\BibitemShut {NoStop}%
\bibitem [{\citenamefont {Gustafson}\ \emph {et~al.}(2022)\citenamefont {Gustafson}, \citenamefont {Lamm}, \citenamefont {Lovelace},\ and\ \citenamefont {Musk}}]{PhysRevD.106.114501}%
  \BibitemOpen
  \bibfield  {author} {\bibinfo {author} {\bibfnamefont {E.~J.}\ \bibnamefont {Gustafson}}, \bibinfo {author} {\bibfnamefont {H.}~\bibnamefont {Lamm}}, \bibinfo {author} {\bibfnamefont {F.}~\bibnamefont {Lovelace}},\ and\ \bibinfo {author} {\bibfnamefont {D.}~\bibnamefont {Musk}},\ }\href {https://doi.org/10.1103/PhysRevD.106.114501} {\bibfield  {journal} {\bibinfo  {journal} {Phys. Rev. D}\ }\textbf {\bibinfo {volume} {106}},\ \bibinfo {pages} {114501} (\bibinfo {year} {2022})}\BibitemShut {NoStop}%
\bibitem [{\citenamefont {Gustafson}\ \emph {et~al.}(2023)\citenamefont {Gustafson}, \citenamefont {Lamm},\ and\ \citenamefont {Lovelace}}]{gustafson2023primitivequantumgatessu2}%
  \BibitemOpen
  \bibfield  {author} {\bibinfo {author} {\bibfnamefont {E.~J.}\ \bibnamefont {Gustafson}}, \bibinfo {author} {\bibfnamefont {H.}~\bibnamefont {Lamm}},\ and\ \bibinfo {author} {\bibfnamefont {F.}~\bibnamefont {Lovelace}},\ }\href {https://arxiv.org/abs/2312.10285} {\bibinfo {title} {{Primitive Quantum Gates for an $SU(2)$ Discrete Subgroup: Binary Octahedral}}} (\bibinfo {year} {2023}),\ \Eprint {https://arxiv.org/abs/2312.10285} {arXiv:2312.10285 [hep-lat]} \BibitemShut {NoStop}%
\bibitem [{\citenamefont {Gustafson}\ \emph {et~al.}(2024)\citenamefont {Gustafson}, \citenamefont {Ji}, \citenamefont {Lamm}, \citenamefont {Murairi},\ and\ \citenamefont {Zhu}}]{gustafson2024primitivequantumgatessu3}%
  \BibitemOpen
  \bibfield  {author} {\bibinfo {author} {\bibfnamefont {E.~J.}\ \bibnamefont {Gustafson}}, \bibinfo {author} {\bibfnamefont {Y.}~\bibnamefont {Ji}}, \bibinfo {author} {\bibfnamefont {H.}~\bibnamefont {Lamm}}, \bibinfo {author} {\bibfnamefont {E.~M.}\ \bibnamefont {Murairi}},\ and\ \bibinfo {author} {\bibfnamefont {S.}~\bibnamefont {Zhu}},\ }\href {https://arxiv.org/abs/2405.05973} {\bibinfo {title} {{Primitive Quantum Gates for an SU(3) Discrete Subgroup: $\Sigma(36\times3)$}}} (\bibinfo {year} {2024}),\ \Eprint {https://arxiv.org/abs/2405.05973} {arXiv:2405.05973 [hep-lat]} \BibitemShut {NoStop}%
\bibitem [{\citenamefont {Bender}\ \emph {et~al.}(2018)\citenamefont {Bender}, \citenamefont {Zohar}, \citenamefont {Farace},\ and\ \citenamefont {Cirac}}]{Bender_2018}%
  \BibitemOpen
  \bibfield  {author} {\bibinfo {author} {\bibfnamefont {J.}~\bibnamefont {Bender}}, \bibinfo {author} {\bibfnamefont {E.}~\bibnamefont {Zohar}}, \bibinfo {author} {\bibfnamefont {A.}~\bibnamefont {Farace}},\ and\ \bibinfo {author} {\bibfnamefont {J.~I.}\ \bibnamefont {Cirac}},\ }\href {https://doi.org/10.1088/1367-2630/aadb71} {\bibfield  {journal} {\bibinfo  {journal} {New Journal of Physics}\ }\textbf {\bibinfo {volume} {20}},\ \bibinfo {pages} {093001} (\bibinfo {year} {2018})}\BibitemShut {NoStop}%
\bibitem [{\citenamefont {Hackett}\ \emph {et~al.}(2019)\citenamefont {Hackett}, \citenamefont {Howe}, \citenamefont {Hughes}, \citenamefont {Jay}, \citenamefont {Neil},\ and\ \citenamefont {Simone}}]{PhysRevA.99.062341}%
  \BibitemOpen
  \bibfield  {author} {\bibinfo {author} {\bibfnamefont {D.~C.}\ \bibnamefont {Hackett}}, \bibinfo {author} {\bibfnamefont {K.}~\bibnamefont {Howe}}, \bibinfo {author} {\bibfnamefont {C.}~\bibnamefont {Hughes}}, \bibinfo {author} {\bibfnamefont {W.}~\bibnamefont {Jay}}, \bibinfo {author} {\bibfnamefont {E.~T.}\ \bibnamefont {Neil}},\ and\ \bibinfo {author} {\bibfnamefont {J.~N.}\ \bibnamefont {Simone}},\ }\href {https://doi.org/10.1103/PhysRevA.99.062341} {\bibfield  {journal} {\bibinfo  {journal} {Phys. Rev. A}\ }\textbf {\bibinfo {volume} {99}},\ \bibinfo {pages} {062341} (\bibinfo {year} {2019})}\BibitemShut {NoStop}%
\bibitem [{\citenamefont {Alexandru}\ \emph {et~al.}(2019)\citenamefont {Alexandru}, \citenamefont {Bedaque}, \citenamefont {Harmalkar}, \citenamefont {Lamm}, \citenamefont {Lawrence},\ and\ \citenamefont {Warrington}}]{PhysRevD.100.114501}%
  \BibitemOpen
  \bibfield  {author} {\bibinfo {author} {\bibfnamefont {A.}~\bibnamefont {Alexandru}}, \bibinfo {author} {\bibfnamefont {P.~F.}\ \bibnamefont {Bedaque}}, \bibinfo {author} {\bibfnamefont {S.}~\bibnamefont {Harmalkar}}, \bibinfo {author} {\bibfnamefont {H.}~\bibnamefont {Lamm}}, \bibinfo {author} {\bibfnamefont {S.}~\bibnamefont {Lawrence}},\ and\ \bibinfo {author} {\bibfnamefont {N.~C.}\ \bibnamefont {Warrington}} (\bibinfo {collaboration} {NuQS Collaboration}),\ }\href {https://doi.org/10.1103/PhysRevD.100.114501} {\bibfield  {journal} {\bibinfo  {journal} {Phys. Rev. D}\ }\textbf {\bibinfo {volume} {100}},\ \bibinfo {pages} {114501} (\bibinfo {year} {2019})}\BibitemShut {NoStop}%
\bibitem [{\citenamefont {Yamamoto}(2020)}]{10.1093/ptep/ptaa171}%
  \BibitemOpen
  \bibfield  {author} {\bibinfo {author} {\bibfnamefont {A.}~\bibnamefont {Yamamoto}},\ }\href {https://doi.org/10.1093/ptep/ptaa171} {\bibfield  {journal} {\bibinfo  {journal} {Progress of Theoretical and Experimental Physics}\ }\textbf {\bibinfo {volume} {2021}},\ \bibinfo {pages} {013B06} (\bibinfo {year} {2020})},\ \Eprint {https://arxiv.org/abs/https://academic.oup.com/ptep/article-pdf/2021/1/013B06/36161140/ptaa171.pdf} {https://academic.oup.com/ptep/article-pdf/2021/1/013B06/36161140/ptaa171.pdf} \BibitemShut {NoStop}%
\bibitem [{\citenamefont {Haase}\ \emph {et~al.}(2021)\citenamefont {Haase}, \citenamefont {Dellantonio}, \citenamefont {Celi}, \citenamefont {Paulson}, \citenamefont {Kan}, \citenamefont {Jansen},\ and\ \citenamefont {Muschik}}]{Haase2021resourceefficient}%
  \BibitemOpen
  \bibfield  {author} {\bibinfo {author} {\bibfnamefont {J.~F.}\ \bibnamefont {Haase}}, \bibinfo {author} {\bibfnamefont {L.}~\bibnamefont {Dellantonio}}, \bibinfo {author} {\bibfnamefont {A.}~\bibnamefont {Celi}}, \bibinfo {author} {\bibfnamefont {D.}~\bibnamefont {Paulson}}, \bibinfo {author} {\bibfnamefont {A.}~\bibnamefont {Kan}}, \bibinfo {author} {\bibfnamefont {K.}~\bibnamefont {Jansen}},\ and\ \bibinfo {author} {\bibfnamefont {C.~A.}\ \bibnamefont {Muschik}},\ }\href {https://doi.org/10.22331/q-2021-02-04-393} {\bibfield  {journal} {\bibinfo  {journal} {{Quantum}}\ }\textbf {\bibinfo {volume} {5}},\ \bibinfo {pages} {393} (\bibinfo {year} {2021})}\BibitemShut {NoStop}%
\bibitem [{\citenamefont {Armon}\ \emph {et~al.}(2021)\citenamefont {Armon}, \citenamefont {Ashkenazi}, \citenamefont {Garcia-Moreno}, \citenamefont {Gonzalez-Tudela},\ and\ \citenamefont {Zohar}}]{PhysRevLett.127.250501}%
  \BibitemOpen
  \bibfield  {author} {\bibinfo {author} {\bibfnamefont {T.}~\bibnamefont {Armon}}, \bibinfo {author} {\bibfnamefont {S.}~\bibnamefont {Ashkenazi}}, \bibinfo {author} {\bibfnamefont {G.}~\bibnamefont {Garcia-Moreno}}, \bibinfo {author} {\bibfnamefont {A.}~\bibnamefont {Gonzalez-Tudela}},\ and\ \bibinfo {author} {\bibfnamefont {E.}~\bibnamefont {Zohar}},\ }\href {https://doi.org/10.1103/PhysRevLett.127.250501} {\bibfield  {journal} {\bibinfo  {journal} {Phys. Rev. Lett.}\ }\textbf {\bibinfo {volume} {127}},\ \bibinfo {pages} {250501} (\bibinfo {year} {2021})}\BibitemShut {NoStop}%
\bibitem [{\citenamefont {Ji}\ \emph {et~al.}(2020)\citenamefont {Ji}, \citenamefont {Lamm},\ and\ \citenamefont {Zhu}}]{PhysRevD.102.114513}%
  \BibitemOpen
  \bibfield  {author} {\bibinfo {author} {\bibfnamefont {Y.}~\bibnamefont {Ji}}, \bibinfo {author} {\bibfnamefont {H.}~\bibnamefont {Lamm}},\ and\ \bibinfo {author} {\bibfnamefont {S.}~\bibnamefont {Zhu}} (\bibinfo {collaboration} {NuQS Collaboration}),\ }\href {https://doi.org/10.1103/PhysRevD.102.114513} {\bibfield  {journal} {\bibinfo  {journal} {Phys. Rev. D}\ }\textbf {\bibinfo {volume} {102}},\ \bibinfo {pages} {114513} (\bibinfo {year} {2020})}\BibitemShut {NoStop}%
\bibitem [{\citenamefont {Carena}\ \emph {et~al.}(2021)\citenamefont {Carena}, \citenamefont {Lamm}, \citenamefont {Li},\ and\ \citenamefont {Liu}}]{PhysRevD.104.094519}%
  \BibitemOpen
  \bibfield  {author} {\bibinfo {author} {\bibfnamefont {M.}~\bibnamefont {Carena}}, \bibinfo {author} {\bibfnamefont {H.}~\bibnamefont {Lamm}}, \bibinfo {author} {\bibfnamefont {Y.-Y.}\ \bibnamefont {Li}},\ and\ \bibinfo {author} {\bibfnamefont {W.}~\bibnamefont {Liu}},\ }\href {https://doi.org/10.1103/PhysRevD.104.094519} {\bibfield  {journal} {\bibinfo  {journal} {Phys. Rev. D}\ }\textbf {\bibinfo {volume} {104}},\ \bibinfo {pages} {094519} (\bibinfo {year} {2021})}\BibitemShut {NoStop}%
\bibitem [{\citenamefont {Gonz\'alez-Cuadra}\ \emph {et~al.}(2022)\citenamefont {Gonz\'alez-Cuadra}, \citenamefont {Zache}, \citenamefont {Carrasco}, \citenamefont {Kraus},\ and\ \citenamefont {Zoller}}]{PhysRevLett.129.160501}%
  \BibitemOpen
  \bibfield  {author} {\bibinfo {author} {\bibfnamefont {D.}~\bibnamefont {Gonz\'alez-Cuadra}}, \bibinfo {author} {\bibfnamefont {T.~V.}\ \bibnamefont {Zache}}, \bibinfo {author} {\bibfnamefont {J.}~\bibnamefont {Carrasco}}, \bibinfo {author} {\bibfnamefont {B.}~\bibnamefont {Kraus}},\ and\ \bibinfo {author} {\bibfnamefont {P.}~\bibnamefont {Zoller}},\ }\href {https://doi.org/10.1103/PhysRevLett.129.160501} {\bibfield  {journal} {\bibinfo  {journal} {Phys. Rev. Lett.}\ }\textbf {\bibinfo {volume} {129}},\ \bibinfo {pages} {160501} (\bibinfo {year} {2022})}\BibitemShut {NoStop}%
\bibitem [{\citenamefont {Marshall}\ \emph {et~al.}(2015)\citenamefont {Marshall}, \citenamefont {Pooser}, \citenamefont {Siopsis},\ and\ \citenamefont {Weedbrook}}]{PhysRevA.92.063825}%
  \BibitemOpen
  \bibfield  {author} {\bibinfo {author} {\bibfnamefont {K.}~\bibnamefont {Marshall}}, \bibinfo {author} {\bibfnamefont {R.}~\bibnamefont {Pooser}}, \bibinfo {author} {\bibfnamefont {G.}~\bibnamefont {Siopsis}},\ and\ \bibinfo {author} {\bibfnamefont {C.}~\bibnamefont {Weedbrook}},\ }\href {https://doi.org/10.1103/PhysRevA.92.063825} {\bibfield  {journal} {\bibinfo  {journal} {Phys. Rev. A}\ }\textbf {\bibinfo {volume} {92}},\ \bibinfo {pages} {063825} (\bibinfo {year} {2015})}\BibitemShut {NoStop}%
\bibitem [{\citenamefont {Jha}\ \emph {et~al.}(2024)\citenamefont {Jha}, \citenamefont {Ringer}, \citenamefont {Siopsis},\ and\ \citenamefont {Thompson}}]{PhysRevA.109.052412}%
  \BibitemOpen
  \bibfield  {author} {\bibinfo {author} {\bibfnamefont {R.~G.}\ \bibnamefont {Jha}}, \bibinfo {author} {\bibfnamefont {F.}~\bibnamefont {Ringer}}, \bibinfo {author} {\bibfnamefont {G.}~\bibnamefont {Siopsis}},\ and\ \bibinfo {author} {\bibfnamefont {S.}~\bibnamefont {Thompson}},\ }\href {https://doi.org/10.1103/PhysRevA.109.052412} {\bibfield  {journal} {\bibinfo  {journal} {Phys. Rev. A}\ }\textbf {\bibinfo {volume} {109}},\ \bibinfo {pages} {052412} (\bibinfo {year} {2024})}\BibitemShut {NoStop}%
\bibitem [{\citenamefont {Yeter-Aydeniz}\ and\ \citenamefont {Siopsis}(2018)}]{PhysRevD.97.036004}%
  \BibitemOpen
  \bibfield  {author} {\bibinfo {author} {\bibfnamefont {K.}~\bibnamefont {Yeter-Aydeniz}}\ and\ \bibinfo {author} {\bibfnamefont {G.}~\bibnamefont {Siopsis}},\ }\href {https://doi.org/10.1103/PhysRevD.97.036004} {\bibfield  {journal} {\bibinfo  {journal} {Phys. Rev. D}\ }\textbf {\bibinfo {volume} {97}},\ \bibinfo {pages} {036004} (\bibinfo {year} {2018})}\BibitemShut {NoStop}%
\bibitem [{\citenamefont {Deng}\ \emph {et~al.}(2016)\citenamefont {Deng}, \citenamefont {Hao}, \citenamefont {Guo}, \citenamefont {Xie},\ and\ \citenamefont {Su}}]{cv-harmonic-oscillators}%
  \BibitemOpen
  \bibfield  {author} {\bibinfo {author} {\bibfnamefont {X.}~\bibnamefont {Deng}}, \bibinfo {author} {\bibfnamefont {S.}~\bibnamefont {Hao}}, \bibinfo {author} {\bibfnamefont {H.}~\bibnamefont {Guo}}, \bibinfo {author} {\bibfnamefont {C.}~\bibnamefont {Xie}},\ and\ \bibinfo {author} {\bibfnamefont {X.}~\bibnamefont {Su}},\ }\href@noop {} {\bibfield  {journal} {\bibinfo  {journal} {Scientific Reports}\ }\textbf {\bibinfo {volume} {6}},\ \bibinfo {pages} {22914} (\bibinfo {year} {2016})}\BibitemShut {NoStop}%
\bibitem [{\citenamefont {Thompson}\ and\ \citenamefont {Siopsis}(2023)}]{Thompson:2023kxz}%
  \BibitemOpen
  \bibfield  {author} {\bibinfo {author} {\bibfnamefont {S.}~\bibnamefont {Thompson}}\ and\ \bibinfo {author} {\bibfnamefont {G.}~\bibnamefont {Siopsis}},\ }\href {https://doi.org/10.1007/s11128-023-04149-0} {\bibfield  {journal} {\bibinfo  {journal} {Quant. Inf. Proc.}\ }\textbf {\bibinfo {volume} {22}},\ \bibinfo {pages} {396} (\bibinfo {year} {2023})},\ \Eprint {https://arxiv.org/abs/2303.02425} {arXiv:2303.02425 [quant-ph]} \BibitemShut {NoStop}%
\bibitem [{\citenamefont {Gustafson}(2021)}]{PhysRevD.103.114505}%
  \BibitemOpen
  \bibfield  {author} {\bibinfo {author} {\bibfnamefont {E.~J.}\ \bibnamefont {Gustafson}},\ }\href {https://doi.org/10.1103/PhysRevD.103.114505} {\bibfield  {journal} {\bibinfo  {journal} {Phys. Rev. D}\ }\textbf {\bibinfo {volume} {103}},\ \bibinfo {pages} {114505} (\bibinfo {year} {2021})}\BibitemShut {NoStop}%
\bibitem [{\citenamefont {Gustafson}(2022)}]{gustafson2022noiseimprovementsquantumsimulations}%
  \BibitemOpen
  \bibfield  {author} {\bibinfo {author} {\bibfnamefont {E.}~\bibnamefont {Gustafson}},\ }\href {https://arxiv.org/abs/2201.04546} {\bibinfo {title} {Noise improvements in quantum simulations of sqed using qutrits}} (\bibinfo {year} {2022}),\ \Eprint {https://arxiv.org/abs/2201.04546} {arXiv:2201.04546 [quant-ph]} \BibitemShut {NoStop}%
\bibitem [{\citenamefont {Illa}\ \emph {et~al.}(2024{\natexlab{b}})\citenamefont {Illa}, \citenamefont {Robin},\ and\ \citenamefont {Savage}}]{Illa2024Qu8itsFQ}%
  \BibitemOpen
  \bibfield  {author} {\bibinfo {author} {\bibfnamefont {M.}~\bibnamefont {Illa}}, \bibinfo {author} {\bibfnamefont {C.~E.~P.}\ \bibnamefont {Robin}},\ and\ \bibinfo {author} {\bibfnamefont {M.~J.}\ \bibnamefont {Savage}},\ }\href {https://api.semanticscholar.org/CorpusID:268553953} {\bibfield  {journal} {\bibinfo  {journal} {Physical Review D}\ } (\bibinfo {year} {2024}{\natexlab{b}})}\BibitemShut {NoStop}%
\bibitem [{\citenamefont {Illa}\ \emph {et~al.}(2023)\citenamefont {Illa}, \citenamefont {Robin},\ and\ \citenamefont {Savage}}]{Illa2023QuantumSO}%
  \BibitemOpen
  \bibfield  {author} {\bibinfo {author} {\bibfnamefont {M.}~\bibnamefont {Illa}}, \bibinfo {author} {\bibfnamefont {C.}~\bibnamefont {Robin}},\ and\ \bibinfo {author} {\bibfnamefont {M.~J.}\ \bibnamefont {Savage}},\ }\href {https://api.semanticscholar.org/CorpusID:258832818} {\bibfield  {journal} {\bibinfo  {journal} {Physical Review C}\ } (\bibinfo {year} {2023})}\BibitemShut {NoStop}%
\bibitem [{\citenamefont {Turro}\ \emph {et~al.}(2024)\citenamefont {Turro}, \citenamefont {Chernyshev}, \citenamefont {Bhaskar},\ and\ \citenamefont {Illa}}]{turro2024qutritqubitcircuitsthreeflavor}%
  \BibitemOpen
  \bibfield  {author} {\bibinfo {author} {\bibfnamefont {F.}~\bibnamefont {Turro}}, \bibinfo {author} {\bibfnamefont {I.~A.}\ \bibnamefont {Chernyshev}}, \bibinfo {author} {\bibfnamefont {R.}~\bibnamefont {Bhaskar}},\ and\ \bibinfo {author} {\bibfnamefont {M.}~\bibnamefont {Illa}},\ }\href {https://arxiv.org/abs/2407.13914} {\bibinfo {title} {Qutrit and qubit circuits for three-flavor collective neutrino oscillations}} (\bibinfo {year} {2024}),\ \Eprint {https://arxiv.org/abs/2407.13914} {arXiv:2407.13914 [quant-ph]} \BibitemShut {NoStop}%
\bibitem [{\citenamefont {Kurkcuoglu}\ \emph {et~al.}(2022)\citenamefont {Kurkcuoglu}, \citenamefont {Alam}, \citenamefont {Job}, \citenamefont {Li}, \citenamefont {Macridin}, \citenamefont {Perdue},\ and\ \citenamefont {Providence}}]{kurkcuoglu2022quantumsimulationphi4theories}%
  \BibitemOpen
  \bibfield  {author} {\bibinfo {author} {\bibfnamefont {D.~M.}\ \bibnamefont {Kurkcuoglu}}, \bibinfo {author} {\bibfnamefont {M.~S.}\ \bibnamefont {Alam}}, \bibinfo {author} {\bibfnamefont {J.~A.}\ \bibnamefont {Job}}, \bibinfo {author} {\bibfnamefont {A.~C.~Y.}\ \bibnamefont {Li}}, \bibinfo {author} {\bibfnamefont {A.}~\bibnamefont {Macridin}}, \bibinfo {author} {\bibfnamefont {G.~N.}\ \bibnamefont {Perdue}},\ and\ \bibinfo {author} {\bibfnamefont {S.}~\bibnamefont {Providence}},\ }\href {https://arxiv.org/abs/2108.13357} {\bibinfo {title} {Quantum simulation of $\phi^4$ theories in qudit systems}} (\bibinfo {year} {2022}),\ \Eprint {https://arxiv.org/abs/2108.13357} {arXiv:2108.13357 [quant-ph]} \BibitemShut {NoStop}%
\bibitem [{\citenamefont {Vezvaee}\ \emph {et~al.}(2024)\citenamefont {Vezvaee}, \citenamefont {Earnest-Noble},\ and\ \citenamefont {Najafi}}]{vezvaee2024quantumsimulationfermihubbardmodel}%
  \BibitemOpen
  \bibfield  {author} {\bibinfo {author} {\bibfnamefont {A.}~\bibnamefont {Vezvaee}}, \bibinfo {author} {\bibfnamefont {N.}~\bibnamefont {Earnest-Noble}},\ and\ \bibinfo {author} {\bibfnamefont {K.}~\bibnamefont {Najafi}},\ }\href {https://arxiv.org/abs/2402.01243} {\bibinfo {title} {Quantum simulation of fermi-hubbard model based on transmon qudit interaction}} (\bibinfo {year} {2024}),\ \Eprint {https://arxiv.org/abs/2402.01243} {arXiv:2402.01243 [quant-ph]} \BibitemShut {NoStop}%
\bibitem [{\citenamefont {Roy}(2023)}]{Roy2023QuditbasedQC}%
  \BibitemOpen
  \bibfield  {author} {\bibinfo {author} {\bibfnamefont {T.}~\bibnamefont {Roy}},\ }\href {https://api.semanticscholar.org/CorpusID:263296135} {\bibfield  {journal} {\bibinfo  {journal} {Qudit-based quantum computing with SRF cavities at Fermilab}\ } (\bibinfo {year} {2023})}\BibitemShut {NoStop}%
\bibitem [{\citenamefont {Meth}\ \emph {et~al.}(2024)\citenamefont {Meth}, \citenamefont {Haase}, \citenamefont {Zhang}, \citenamefont {Edmunds}, \citenamefont {Postler}, \citenamefont {Steiner}, \citenamefont {Jena}, \citenamefont {Dellantonio}, \citenamefont {Blatt}, \citenamefont {Zoller}, \citenamefont {Monz}, \citenamefont {Schindler}, \citenamefont {Muschik},\ and\ \citenamefont {Ringbauer}}]{meth2024simulating2dlatticegauge}%
  \BibitemOpen
  \bibfield  {author} {\bibinfo {author} {\bibfnamefont {M.}~\bibnamefont {Meth}}, \bibinfo {author} {\bibfnamefont {J.~F.}\ \bibnamefont {Haase}}, \bibinfo {author} {\bibfnamefont {J.}~\bibnamefont {Zhang}}, \bibinfo {author} {\bibfnamefont {C.}~\bibnamefont {Edmunds}}, \bibinfo {author} {\bibfnamefont {L.}~\bibnamefont {Postler}}, \bibinfo {author} {\bibfnamefont {A.}~\bibnamefont {Steiner}}, \bibinfo {author} {\bibfnamefont {A.~J.}\ \bibnamefont {Jena}}, \bibinfo {author} {\bibfnamefont {L.}~\bibnamefont {Dellantonio}}, \bibinfo {author} {\bibfnamefont {R.}~\bibnamefont {Blatt}}, \bibinfo {author} {\bibfnamefont {P.}~\bibnamefont {Zoller}}, \bibinfo {author} {\bibfnamefont {T.}~\bibnamefont {Monz}}, \bibinfo {author} {\bibfnamefont {P.}~\bibnamefont {Schindler}}, \bibinfo {author} {\bibfnamefont {C.}~\bibnamefont {Muschik}},\ and\ \bibinfo {author} {\bibfnamefont {M.}~\bibnamefont {Ringbauer}},\ }\href {https://arxiv.org/abs/2310.12110} {\bibinfo {title} {Simulating 2d lattice gauge theories on a qudit
  quantum computer}} (\bibinfo {year} {2024}),\ \Eprint {https://arxiv.org/abs/2310.12110} {arXiv:2310.12110 [quant-ph]} \BibitemShut {NoStop}%
\bibitem [{\citenamefont {Hanada}\ \emph {et~al.}(2023{\natexlab{b}})\citenamefont {Hanada}, \citenamefont {Liu}, \citenamefont {Rinaldi},\ and\ \citenamefont {Tezuka}}]{Hanada:2022pps}%
  \BibitemOpen
  \bibfield  {author} {\bibinfo {author} {\bibfnamefont {M.}~\bibnamefont {Hanada}}, \bibinfo {author} {\bibfnamefont {J.}~\bibnamefont {Liu}}, \bibinfo {author} {\bibfnamefont {E.}~\bibnamefont {Rinaldi}},\ and\ \bibinfo {author} {\bibfnamefont {M.}~\bibnamefont {Tezuka}},\ }\href {https://doi.org/10.1088/2632-2153/ad035c} {\bibfield  {journal} {\bibinfo  {journal} {Mach. Learn. Sci. Tech.}\ }\textbf {\bibinfo {volume} {4}},\ \bibinfo {pages} {045021} (\bibinfo {year} {2023}{\natexlab{b}})},\ \Eprint {https://arxiv.org/abs/2212.08546} {arXiv:2212.08546 [quant-ph]} \BibitemShut {NoStop}%
\bibitem [{Sup()}]{SuppMat}%
  \BibitemOpen
  \href@noop {} {}\bibinfo {note} {See Supplemental Material at https://journals.aps.org/authors/supplemental-material-instructions for additional derivations of theorems in the main text and explicit circuit constructions.}\BibitemShut {Stop}%
\bibitem [{\citenamefont {He}\ \emph {et~al.}(2017)\citenamefont {He}, \citenamefont {Luo}, \citenamefont {Zhang}, \citenamefont {Wang},\ and\ \citenamefont {Wang}}]{2017_HLZetal}%
  \BibitemOpen
  \bibfield  {author} {\bibinfo {author} {\bibfnamefont {Y.}~\bibnamefont {He}}, \bibinfo {author} {\bibfnamefont {M.-X.}\ \bibnamefont {Luo}}, \bibinfo {author} {\bibfnamefont {E.}~\bibnamefont {Zhang}}, \bibinfo {author} {\bibfnamefont {H.-K.}\ \bibnamefont {Wang}},\ and\ \bibinfo {author} {\bibfnamefont {X.-F.}\ \bibnamefont {Wang}},\ }\href@noop {} {\bibfield  {journal} {\bibinfo  {journal} {International Journal of Theoretical Physics}\ }\textbf {\bibinfo {volume} {56}},\ \bibinfo {pages} {2350} (\bibinfo {year} {2017})}\BibitemShut {NoStop}%
\bibitem [{\citenamefont {Gidney}(2018)}]{2018_G}%
  \BibitemOpen
  \bibfield  {author} {\bibinfo {author} {\bibfnamefont {C.}~\bibnamefont {Gidney}},\ }\href@noop {} {\bibfield  {journal} {\bibinfo  {journal} {Quantum}\ }\textbf {\bibinfo {volume} {2}},\ \bibinfo {pages} {74} (\bibinfo {year} {2018})}\BibitemShut {NoStop}%
\bibitem [{\citenamefont {Huyghebaert}\ and\ \citenamefont {De~Raedt}(1990)}]{1990_HdR}%
  \BibitemOpen
  \bibfield  {author} {\bibinfo {author} {\bibfnamefont {J.}~\bibnamefont {Huyghebaert}}\ and\ \bibinfo {author} {\bibfnamefont {H.}~\bibnamefont {De~Raedt}},\ }\href@noop {} {\bibfield  {journal} {\bibinfo  {journal} {Journal of Physics A: Mathematical and General}\ }\textbf {\bibinfo {volume} {23}},\ \bibinfo {pages} {5777} (\bibinfo {year} {1990})}\BibitemShut {NoStop}%
\bibitem [{\citenamefont {Wiebe}\ \emph {et~al.}(2010)\citenamefont {Wiebe}, \citenamefont {Berry}, \citenamefont {H{\o}yer},\ and\ \citenamefont {Sanders}}]{2010_WBHS}%
  \BibitemOpen
  \bibfield  {author} {\bibinfo {author} {\bibfnamefont {N.}~\bibnamefont {Wiebe}}, \bibinfo {author} {\bibfnamefont {D.}~\bibnamefont {Berry}}, \bibinfo {author} {\bibfnamefont {P.}~\bibnamefont {H{\o}yer}},\ and\ \bibinfo {author} {\bibfnamefont {B.~C.}\ \bibnamefont {Sanders}},\ }\href@noop {} {\bibfield  {journal} {\bibinfo  {journal} {Journal of Physics A: Mathematical and Theoretical}\ }\textbf {\bibinfo {volume} {43}},\ \bibinfo {pages} {065203} (\bibinfo {year} {2010})}\BibitemShut {NoStop}%
\bibitem [{\citenamefont {Wecker}\ \emph {et~al.}(2015)\citenamefont {Wecker}, \citenamefont {Hastings}, \citenamefont {Wiebe}, \citenamefont {Clark}, \citenamefont {Nayak},\ and\ \citenamefont {Troyer}}]{2015_WHWetal}%
  \BibitemOpen
  \bibfield  {author} {\bibinfo {author} {\bibfnamefont {D.}~\bibnamefont {Wecker}}, \bibinfo {author} {\bibfnamefont {M.~B.}\ \bibnamefont {Hastings}}, \bibinfo {author} {\bibfnamefont {N.}~\bibnamefont {Wiebe}}, \bibinfo {author} {\bibfnamefont {B.~K.}\ \bibnamefont {Clark}}, \bibinfo {author} {\bibfnamefont {C.}~\bibnamefont {Nayak}},\ and\ \bibinfo {author} {\bibfnamefont {M.}~\bibnamefont {Troyer}},\ }\href@noop {} {\bibfield  {journal} {\bibinfo  {journal} {Physical Review A}\ }\textbf {\bibinfo {volume} {92}},\ \bibinfo {pages} {062318} (\bibinfo {year} {2015})}\BibitemShut {NoStop}%
\bibitem [{\citenamefont {Childs}\ \emph {et~al.}(2018{\natexlab{b}})\citenamefont {Childs}, \citenamefont {Maslov}, \citenamefont {Nam}, \citenamefont {Ross},\ and\ \citenamefont {Su}}]{Childs2018}%
  \BibitemOpen
  \bibfield  {author} {\bibinfo {author} {\bibfnamefont {A.~M.}\ \bibnamefont {Childs}}, \bibinfo {author} {\bibfnamefont {D.}~\bibnamefont {Maslov}}, \bibinfo {author} {\bibfnamefont {Y.}~\bibnamefont {Nam}}, \bibinfo {author} {\bibfnamefont {N.~J.}\ \bibnamefont {Ross}},\ and\ \bibinfo {author} {\bibfnamefont {Y.}~\bibnamefont {Su}},\ }\href {https://doi.org/10.1073/pnas.1801723115} {\bibfield  {journal} {\bibinfo  {journal} {Proceedings of the National Academy of Sciences}\ }\textbf {\bibinfo {volume} {115}},\ \bibinfo {pages} {9456} (\bibinfo {year} {2018}{\natexlab{b}})},\ \Eprint {https://arxiv.org/abs/https://www.pnas.org/doi/pdf/10.1073/pnas.1801723115} {https://www.pnas.org/doi/pdf/10.1073/pnas.1801723115} \BibitemShut {NoStop}%
\bibitem [{\citenamefont {Babbush}\ \emph {et~al.}(2018{\natexlab{b}})\citenamefont {Babbush}, \citenamefont {Gidney}, \citenamefont {Berry}, \citenamefont {Wiebe}, \citenamefont {McClean}, \citenamefont {Paler}, \citenamefont {Fowler},\ and\ \citenamefont {Neven}}]{Babbush2018}%
  \BibitemOpen
  \bibfield  {author} {\bibinfo {author} {\bibfnamefont {R.}~\bibnamefont {Babbush}}, \bibinfo {author} {\bibfnamefont {C.}~\bibnamefont {Gidney}}, \bibinfo {author} {\bibfnamefont {D.~W.}\ \bibnamefont {Berry}}, \bibinfo {author} {\bibfnamefont {N.}~\bibnamefont {Wiebe}}, \bibinfo {author} {\bibfnamefont {J.}~\bibnamefont {McClean}}, \bibinfo {author} {\bibfnamefont {A.}~\bibnamefont {Paler}}, \bibinfo {author} {\bibfnamefont {A.}~\bibnamefont {Fowler}},\ and\ \bibinfo {author} {\bibfnamefont {H.}~\bibnamefont {Neven}},\ }\href {https://doi.org/10.1103/PhysRevX.8.041015} {\bibfield  {journal} {\bibinfo  {journal} {Phys. Rev. X}\ }\textbf {\bibinfo {volume} {8}},\ \bibinfo {pages} {041015} (\bibinfo {year} {2018}{\natexlab{b}})}\BibitemShut {NoStop}%
\bibitem [{\citenamefont {Nam}\ \emph {et~al.}(2020)\citenamefont {Nam}, \citenamefont {Su},\ and\ \citenamefont {Maslov}}]{2020_NSM}%
  \BibitemOpen
  \bibfield  {author} {\bibinfo {author} {\bibfnamefont {Y.}~\bibnamefont {Nam}}, \bibinfo {author} {\bibfnamefont {Y.}~\bibnamefont {Su}},\ and\ \bibinfo {author} {\bibfnamefont {D.}~\bibnamefont {Maslov}},\ }\href@noop {} {\bibfield  {journal} {\bibinfo  {journal} {npj Quantum Information}\ }\textbf {\bibinfo {volume} {6}},\ \bibinfo {pages} {1} (\bibinfo {year} {2020})}\BibitemShut {NoStop}%
\bibitem [{\citenamefont {Childs}\ and\ \citenamefont {Wiebe}(2012{\natexlab{b}})}]{2012_CW}%
  \BibitemOpen
  \bibfield  {author} {\bibinfo {author} {\bibfnamefont {A.~M.}\ \bibnamefont {Childs}}\ and\ \bibinfo {author} {\bibfnamefont {N.}~\bibnamefont {Wiebe}},\ }\href@noop {} {\bibfield  {journal} {\bibinfo  {journal} {arXiv preprint arXiv:1202.5822}\ } (\bibinfo {year} {2012}{\natexlab{b}})}\BibitemShut {NoStop}%
\bibitem [{\citenamefont {Vatan}\ and\ \citenamefont {Williams}(2004)}]{PhysRevA.69.032315}%
  \BibitemOpen
  \bibfield  {author} {\bibinfo {author} {\bibfnamefont {F.}~\bibnamefont {Vatan}}\ and\ \bibinfo {author} {\bibfnamefont {C.}~\bibnamefont {Williams}},\ }\href {https://doi.org/10.1103/PhysRevA.69.032315} {\bibfield  {journal} {\bibinfo  {journal} {Phys. Rev. A}\ }\textbf {\bibinfo {volume} {69}},\ \bibinfo {pages} {032315} (\bibinfo {year} {2004})}\BibitemShut {NoStop}%
\bibitem [{\citenamefont {Bocharov}\ \emph {et~al.}(2015)\citenamefont {Bocharov}, \citenamefont {Roetteler},\ and\ \citenamefont {Svore}}]{2015_BRS}%
  \BibitemOpen
  \bibfield  {author} {\bibinfo {author} {\bibfnamefont {A.}~\bibnamefont {Bocharov}}, \bibinfo {author} {\bibfnamefont {M.}~\bibnamefont {Roetteler}},\ and\ \bibinfo {author} {\bibfnamefont {K.~M.}\ \bibnamefont {Svore}},\ }\href {https://doi.org/10.1103/PhysRevLett.114.080502} {\bibfield  {journal} {\bibinfo  {journal} {Phys. Rev. Lett.}\ }\textbf {\bibinfo {volume} {114}},\ \bibinfo {pages} {080502} (\bibinfo {year} {2015})}\BibitemShut {NoStop}%
\bibitem [{\citenamefont {Selinger}(2014)}]{selinger2014efficient}%
  \BibitemOpen
  \bibfield  {author} {\bibinfo {author} {\bibfnamefont {P.}~\bibnamefont {Selinger}},\ }\href@noop {} {\bibinfo {title} {Efficient {Clifford+T} approximation of single-qubit operators}} (\bibinfo {year} {2014}),\ \Eprint {https://arxiv.org/abs/1212.6253} {arXiv:1212.6253 [quant-ph]} \BibitemShut {NoStop}%
\bibitem [{\citenamefont {Barenco}\ \emph {et~al.}(1995)\citenamefont {Barenco}, \citenamefont {Bennett}, \citenamefont {Cleve}, \citenamefont {DiVincenzo}, \citenamefont {Margolus}, \citenamefont {Shor}, \citenamefont {Sleator}, \citenamefont {Smolin},\ and\ \citenamefont {Weinfurter}}]{1995_BBCetal}%
  \BibitemOpen
  \bibfield  {author} {\bibinfo {author} {\bibfnamefont {A.}~\bibnamefont {Barenco}}, \bibinfo {author} {\bibfnamefont {C.~H.}\ \bibnamefont {Bennett}}, \bibinfo {author} {\bibfnamefont {R.}~\bibnamefont {Cleve}}, \bibinfo {author} {\bibfnamefont {D.~P.}\ \bibnamefont {DiVincenzo}}, \bibinfo {author} {\bibfnamefont {N.}~\bibnamefont {Margolus}}, \bibinfo {author} {\bibfnamefont {P.}~\bibnamefont {Shor}}, \bibinfo {author} {\bibfnamefont {T.}~\bibnamefont {Sleator}}, \bibinfo {author} {\bibfnamefont {J.~A.}\ \bibnamefont {Smolin}},\ and\ \bibinfo {author} {\bibfnamefont {H.}~\bibnamefont {Weinfurter}},\ }\href@noop {} {\bibfield  {journal} {\bibinfo  {journal} {Physical review A}\ }\textbf {\bibinfo {volume} {52}},\ \bibinfo {pages} {3457} (\bibinfo {year} {1995})}\BibitemShut {NoStop}%
\bibitem [{\citenamefont {Amy}\ \emph {et~al.}(2018)\citenamefont {Amy}, \citenamefont {Azimzadeh},\ and\ \citenamefont {Mosca}}]{2018_AAM}%
  \BibitemOpen
  \bibfield  {author} {\bibinfo {author} {\bibfnamefont {M.}~\bibnamefont {Amy}}, \bibinfo {author} {\bibfnamefont {P.}~\bibnamefont {Azimzadeh}},\ and\ \bibinfo {author} {\bibfnamefont {M.}~\bibnamefont {Mosca}},\ }\href@noop {} {\bibfield  {journal} {\bibinfo  {journal} {Quantum Science and Technology}\ }\textbf {\bibinfo {volume} {4}},\ \bibinfo {pages} {015002} (\bibinfo {year} {2018})}\BibitemShut {NoStop}%
\bibitem [{\citenamefont {Gheorghiu}\ \emph {et~al.}(2022{\natexlab{c}})\citenamefont {Gheorghiu}, \citenamefont {Huang}, \citenamefont {Li}, \citenamefont {Mosca},\ and\ \citenamefont {Mukhopadhyay}}]{2022_GHLetal}%
  \BibitemOpen
  \bibfield  {author} {\bibinfo {author} {\bibfnamefont {V.}~\bibnamefont {Gheorghiu}}, \bibinfo {author} {\bibfnamefont {J.}~\bibnamefont {Huang}}, \bibinfo {author} {\bibfnamefont {S.~M.}\ \bibnamefont {Li}}, \bibinfo {author} {\bibfnamefont {M.}~\bibnamefont {Mosca}},\ and\ \bibinfo {author} {\bibfnamefont {P.}~\bibnamefont {Mukhopadhyay}},\ }\href@noop {} {\bibfield  {journal} {\bibinfo  {journal} {IEEE Transactions on Computer-Aided Design of Integrated Circuits and Systems}\ } (\bibinfo {year} {2022}{\natexlab{c}})}\BibitemShut {NoStop}%
\bibitem [{\citenamefont {Kitaev}\ and\ \citenamefont {Webb}(2009)}]{kitaev2009wavefunction}%
  \BibitemOpen
  \bibfield  {author} {\bibinfo {author} {\bibfnamefont {A.}~\bibnamefont {Kitaev}}\ and\ \bibinfo {author} {\bibfnamefont {W.~A.}\ \bibnamefont {Webb}},\ }\href@noop {} {\bibinfo {title} {Wavefunction preparation and resampling using a quantum computer}} (\bibinfo {year} {2009}),\ \Eprint {https://arxiv.org/abs/0801.0342} {arXiv:0801.0342 [quant-ph]} \BibitemShut {NoStop}%
\bibitem [{\citenamefont {Berry}\ \emph {et~al.}(2015)\citenamefont {Berry}, \citenamefont {Childs},\ and\ \citenamefont {Kothari}}]{berry2015hamiltonian}%
  \BibitemOpen
  \bibfield  {author} {\bibinfo {author} {\bibfnamefont {D.~W.}\ \bibnamefont {Berry}}, \bibinfo {author} {\bibfnamefont {A.~M.}\ \bibnamefont {Childs}},\ and\ \bibinfo {author} {\bibfnamefont {R.}~\bibnamefont {Kothari}},\ }in\ \href@noop {} {\emph {\bibinfo {booktitle} {2015 IEEE 56th annual symposium on foundations of computer science}}}\ (\bibinfo {organization} {IEEE},\ \bibinfo {year} {2015})\ pp.\ \bibinfo {pages} {792--809}\BibitemShut {NoStop}%
\end{thebibliography}%
\newpage
\newpage
\foreach \x in {1,...,57}
{
\clearpage
\includepdf[pages={\x}]{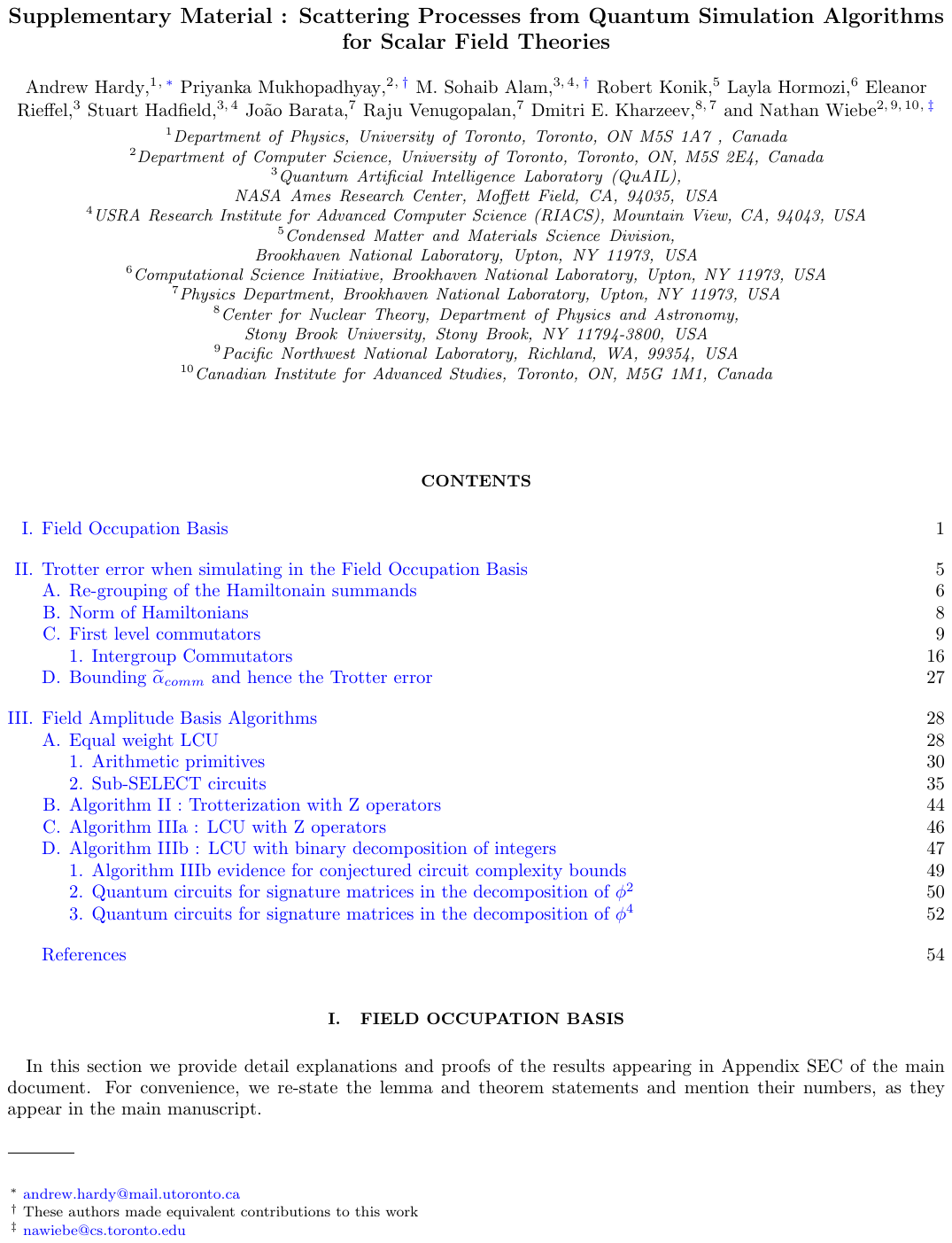} 
}
\end{document}